\documentclass[fleqn,iop,numberedappendix]{emulateapj}
\usepackage{morefloats}
\usepackage{amsmath}
\usepackage{multirow}

\slugcomment{Accepted for publication in The Astronomical Journal}
\begin{document}
\title{THE SHAPES OF THE HI VELOCITY PROFILES OF THE THINGS GALAXIES}
\author{R. Ianjamasimanana\altaffilmark{1}, W.J.G. de
  Blok\altaffilmark{2,1}, Fabian Walter\altaffilmark{3}, George
  H. Heald\altaffilmark{2}} \altaffiltext{1}{Astrophysics, Cosmology
  and Gravity Centre, Department of Astronomy, University of Cape
  Town, Private Bag X3, Rondebosch 7701, South Africa;
  ianja@ast.uct.ac.za} \altaffiltext{2}{Netherlands Institute for
  Radio Astronomy (ASTRON), Postbus 2, 7990 AA Dwingeloo, the
  Netherlands} \altaffiltext{3}{Max-Planck Institut
  f$\rm{\ddot{u}}$r Astronomie, K\"onigstuhl 17, 69117,
  Heidelberg, Germany} 
\begin{abstract}
  We analyze the shapes of the HI velocity profiles of The HI Nearby
  Galaxy Survey (THINGS) to study the phase structure of the neutral
  interstellar medium (ISM) and its relation to global galaxy
  properties. We use a method analogous to the stacking method
  sometimes used in high redshift HI observations to construct high
  signal-to-noise (S/N) profiles. We call these high S/N profiles
  \textit{super profiles}. We analyze and discuss possible systematics
  that may change the observed shapes of the super
  profiles. After quantifying these effects and selecting a sub-sample of unaffected galaxies, 
  we find that the super profiles are best described by a narrow and a broad Gaussian
  component, which are evidence of the presence of the Cold Neutral
  Medium (CNM) and the Warm Neutral Medium (WNM). The velocity dispersion 
  of the narrow component range from $\sim$3.4 to $\sim$8.6 $\rm{km~s^{-1}}$  
  with an average of 6.5$\pm$1.5 $\rm{km~s^{-1}}$, 
  whereas that of the broad component range from $\sim$10.1 to $\sim$24.3 $\rm{km~s^{-1}}$ 
  with an average of 16.8$\pm$4.3 $\rm{km~s^{-1}}$. 
  We find that the super profile parameters correlate with star formation indicators
  such as metallicity, FUV-NUV colors and $\rm{H\alpha}$
  luminosities. The flux ratio between the narrow and broad
  components tends to be highest for high metallicity,
  high star formation rate (SFR) galaxies. We show that the narrow component 
  identified in the super profiles is associated with the presence of star formation, 
  and possibly with molecular hydrogen.
\end{abstract}


\section{Introduction}

Gas in the interstellar medium (ISM) in disk galaxies can exist in
  a molecular, atomic or ionised phase. Most of it will be in the form
  of neutral atomic hydrogen (HI) whose hyperfine transition can be
  detected at the 21-cm wavelength. Since the early work of
  \citet{clark65}, the neutral atomic ISM has been known to exist in
  two phases, known as the Cold Neutral Medium (CNM) and the Warm
  Neutral Medium (WNM). \citet{clark65} first suggested that HI seen
  in emission and absorption had different line widths because it
  arose from two distinct components of the ISM. The narrow absorption
  spectra are caused by a cold component of the ISM (the CNM) with a
  temperature of $\sim 100~\rm{K}$ whereas the broad emission spectra
  arose from a warm component with a temperature as high as
  $10^{4}~\rm{K}$ (the WNM). The existence of these two components
was later confirmed by \citet{radhakrishnanetal72} using
interferometric observations of HI emission and absorption spectra
towards 35 extragalactic sources situated at intermediate and high
Galactic latitude. The CNM and the WNM are known to coexist in
  pressure equilibrium over a narrow range of pressure
  \citep{wolfireetal03}. The CNM dominates at high densities and
  pressures whereas the WNM prevails at low densities and
  pressures. These two components have been identified in the Milky Way,
  in nearby dwarfs, and spiral galaxies by comparing HI seen in
  emission and absorption as well as from observations of HI emission
  alone.

\citet{younglo96,younglo97} and \citet{youngetal03} analyzed HI
emission velocity profiles of seven nearby dwarf galaxies. By fitting
these profiles with Gaussian components, they found evidence for a
broad component with a dispersion ranging from about 8 to 13
$\rm{km~s^{-1}}$ as well as a much narrower component with a
dispersion ranging from 3 to 5 $\rm{km~s^{-1}}$. These velocity
dispersions are larger than the predicted thermal line widths for the
CNM and the WNM, showing that other energy sources are also playing
role in broadening the line profiles. Moreover, they found that the
narrow component shows a clumpy distribution and tends to be located
near star formation regions, whereas the broad component shows a more
ubiquitous distribution. Because of the similarities of the observed
properties of these two components with those of the CNM and WNM in M
31 and in our Galaxy, they associated the narrow component with the
CNM and the broad component with the WNM phases of the ISM.  A similar
study was made by \citet{deblokwalter06} for the Local Group dwarf
galaxy NGC 6822.  They also found narrow and broad HI components with
mean velocity dispersions of 4 $\rm{km~s^{-1}}$ and 8
$\rm{km~s^{-1}}$, respectively. Here also, the narrow component is
usually located near star forming regions, whereas the broad component
tends to be found along every line of sight.  \citet{braun97} analyzed
HI emission of 11 nearby spiral galaxies and found what he called a
high brightness filamentary network (HBN) of HI emission which he
associated with the CNM and a diffuse low brightness emission which he
identified with the WNM. The HBN dominates inside the optical radius
$r_{25}$ where it accounts for 60-90 \% of the HI emission, whereas
the diffuse component dominates outside this radius. By averaging
spectra with peak brightness temperature higher than 4$\sigma$ within
annular ellipses in his sample, he found that the combined spectra
were characterized by a narrow core superposed on broad Lorentzian
wings. The width of the narrow line cores allowed \citet{braun97} to
give an upper limit of 300 K for the HBN. \citet{braun97} also found that 
the kinetic temperature of the HBN increases with increasing radius. 
\citet{braun97} interpreted this as a result of the decrease in the mid-plane thermal 
pressure towards the outer disks.

Understanding the properties of the gas in the ISM is crucial
  as it is the fuel for star formation. Most of this star formation
occurs in Giant Molecular Clouds (GMCs). These are self gravitating
clouds with masses $\sim10^{5}~M_{\odot}$ and sizes $\sim40~\rm{pc}$
\citep[e.g.][]{tielens05}. Hydrogen molecules ($\rm{H_{2}}$) are the
main constituents of GMCs, but unfortunately $\rm{H_{2}}$ is not
easily observable. Carbon monoxide (CO) is the second most abundant
molecule in GMCs and can be detected at millimeter
wavelengths. Typically, CO line emission is used to infer the amount
and distribution of $\rm{H_{2}}$ in galaxies using some
CO-to-$\rm{H_{2}}$ conversion factor
\citep[e.g.,][]{heracles}. However, the strength of the CO emission
depends on the metallicity of the ISM \citep{bolattoetal11,leroyetal11}
and it is therefore difficult to detect in low metallicity dwarf
galaxies. An alternative approach is thus necessary to investigate the
distribution of $\rm{H_{2}}$ there. One example of this is to
  use the dust emission at infrared wavelengths as a tracer of the
  molecular gas \citep[e.g.,][]{leroyetal11}.  Another approach, which
  will be the topic of this paper, is to try and identify a precursor
  of the molecular gas. If the H$_2$ component forms from the cooling
  of the CNM, then identifying the CNM in galaxies can yield
  independent information on the properties of a component of the gas
  that is closely linked to star formation.
  
In this paper, we attempt to detect the CNM and WNM components of the
ISM by analyzing the shapes of stacked HI emission profiles of
galaxies observed as part of The HI Nearby Galaxies Survey
\citep[THINGS,][]{ walteretal08}. THINGS is a large survey of 34
nearby spiral and dwarf galaxies performed with the NRAO\footnote{The
  National Radio Astronomy Observatory is a facility of the National
  Science Foundation operated under cooperative agreement by
  Associated Universities, Inc.} Very Large Array (VLA). The purpose
of THINGS was to obtain high spectral and spatial resolution maps of
the neutral hydrogen in these galaxies. The THINGS galaxies have
distances between 2 and 15 $\rm{Mpc}$, star formation rates of $\sim
10^{-3}$ to $10^{-6}~M_\odot~\rm{yr^{-1}}$, absolute $B$-luminosities
of $-11.5$ to $-21.7$ mag, total HI masses of (0.01---14) $\times
10^9~M_\odot$, and metallicities of 7.5 to 9.2 in units of [12 + 
log(O/H)]. Using the VLA B, C and D arrays, the THINGS targets were
observed at a spatial resolution of 
$\sim 6^{\prime\prime}$ (robust weighting) or
$\sim 11^{\prime\prime}$ (natural weighting), respectively. Two of 
the THINGS galaxies were observed at a spectral resolution of 1.3 
$\rm{km~s^{-1}}$ and the rest at a resolution of either 2.6 or 5.2 
$\rm{km~s^{-1}}$ (see Table 2 of \citet{walteretal08}). We use the
natural-weighted (and non-residual scaled) cubes in this paper, though
our results are independent of the weighting scheme used. We refer the
reader to \citet{walteretal08} for an explanation on how the data
cubes were produced. We use the task PBCORR in the Groningen Image
Processing System (GIPSY) to correct for the effect of the primary
beam of the telescope in the image cubes. The publicly
  available THINGS moment maps already are corrected for primary beam
  attenuation.
 
This paper is organized as follows. In Section \ref{sec:method}, we
describe the principles of our method for obtaining high
signal-to-noise (S/N) stacked profiles. Section \ref{sec:fitting}
describes the fitting and decomposition of the profiles. In Section
\ref{sec:sys_effects}, we discuss various effects that may influence
the shapes of the profiles. In Section
\ref{sec:cleansample}, we describe how we define a clean sample
based on the studies of these effects. In Section \ref{sub:asymm_effect}, 
we check the reliability of the profile shapes of the clean sample. 
In Section \ref{sec:globaltrend}, we
check possible correlations between the shapes of the profiles and
properties of galaxies or morphology. In Section
\ref{sec:moleculargas}, we investigate whether the narrow component is
associated with molecular gas. We summarize our results in Section
\ref{sec:summary}.

\section{Generating high signal-to-noise profiles}\label{sec:method} 
 
Individual velocity profiles in typical HI observations generally have
a low ($\leq5$) to moderate ($\leq10$) S/N. In the THINGS
observations, fitting individual, separate profiles in a data cube
needs to be done carefully to avoid skewed results due to the effects
of noise. Here, we use a method analogous to the \textit{stacking}
method sometimes used in high-redshift HI observations
\citep[e.g.,][]{fabelloetal11} to construct high S/N
profiles. Note that we stack individual spectra within 
galaxies rather than combining global galaxy spectra. Since
individual profiles located at different positions in a galaxy have
different radial velocities, the profiles cannot be summed as they
are, but need to be shifted in velocity to the same reference
velocity.  We use the GIPSY task SHUFFLE to do this. This task uses a
velocity field to define the amount by which a profile in the
corresponding data cube needs to be shifted. We use Gauss-Hermite
$h_3$ polynomial velocity fields as input for SHUFFLE. Gauss
Hermite polynomial velocity fields give robust estimates for the
radial velocity of the peak of the profile, even in the presence of
profile asymmetries. For an extensive description see  \citet{debloketal08}.

To minimize the effects of noise, we only include individual profiles
whose peak fluxes exceed the 3$\sigma$ noise level in the data
cubes. For each galaxy, we sum the shuffled profiles to produce what
we call \textit{super profiles}. These are high quality profiles,
which due to their increased S/N allow us to reliably explore the
presence of the cold and warm neutral components of the
ISM. However, as information on the individual profiles is
  lost during the stacking, care needs to be taken that systematic
  effects and possible asymmetries in the input profiles are
  quantified, to prevent these from mimicking the presence of multiple
  components in the super profiles.  This is discussed in detail in
  Section \ref{sub:asymm_effect}.
 
\begin{figure*}
\begin{center}
    \begin{tabular}{l l l}
  \rotatebox{0}{\resizebox{58mm}{!}{\includegraphics[width = 0.6in,height = 0.6in]{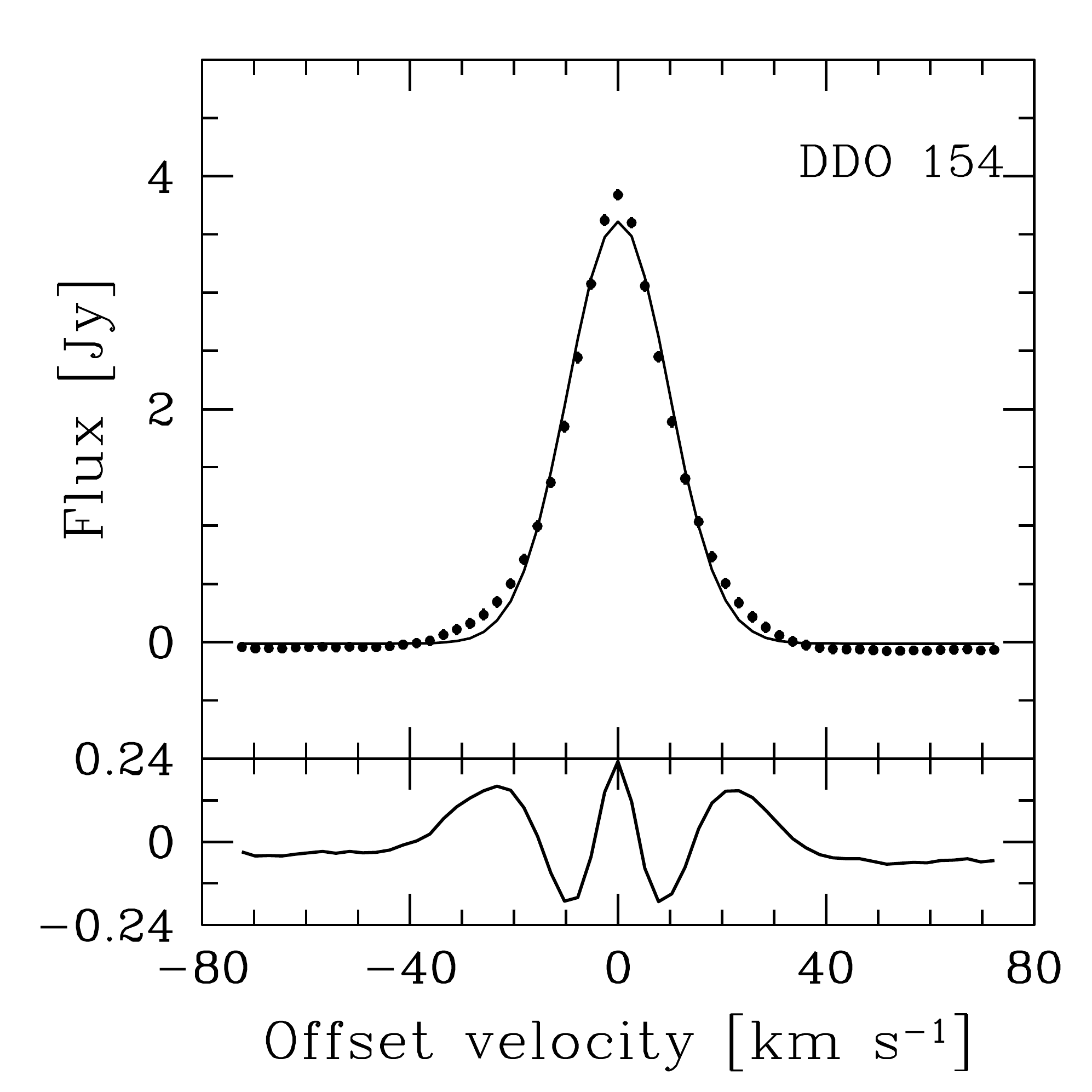}}}&
\rotatebox{0}{\resizebox{58mm}{!}{\includegraphics[width = 0.6in,height = 0.6in]{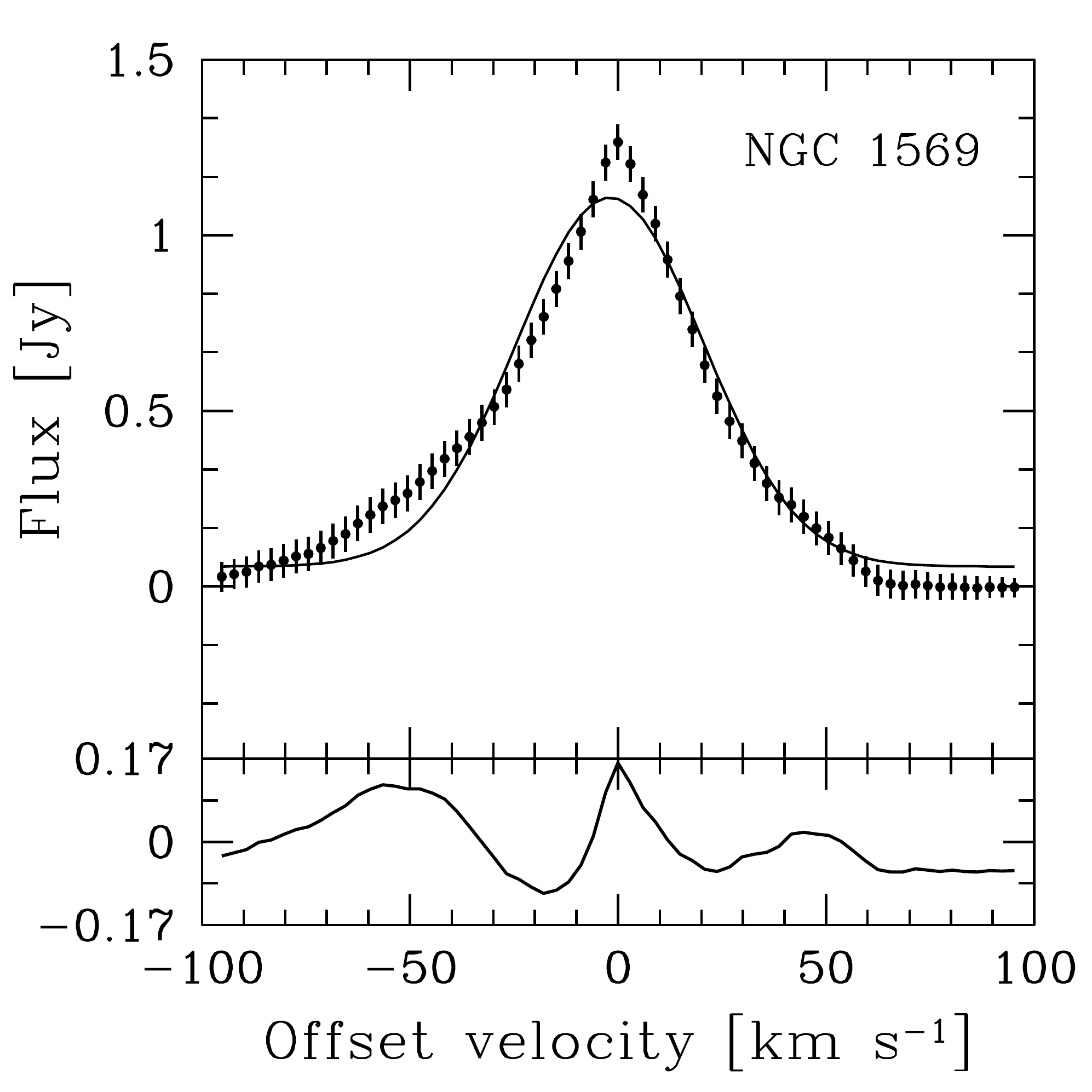}}}&
  \rotatebox{0}{\resizebox{58mm}{!}{\includegraphics[width = 0.6in,height = 0.6in]{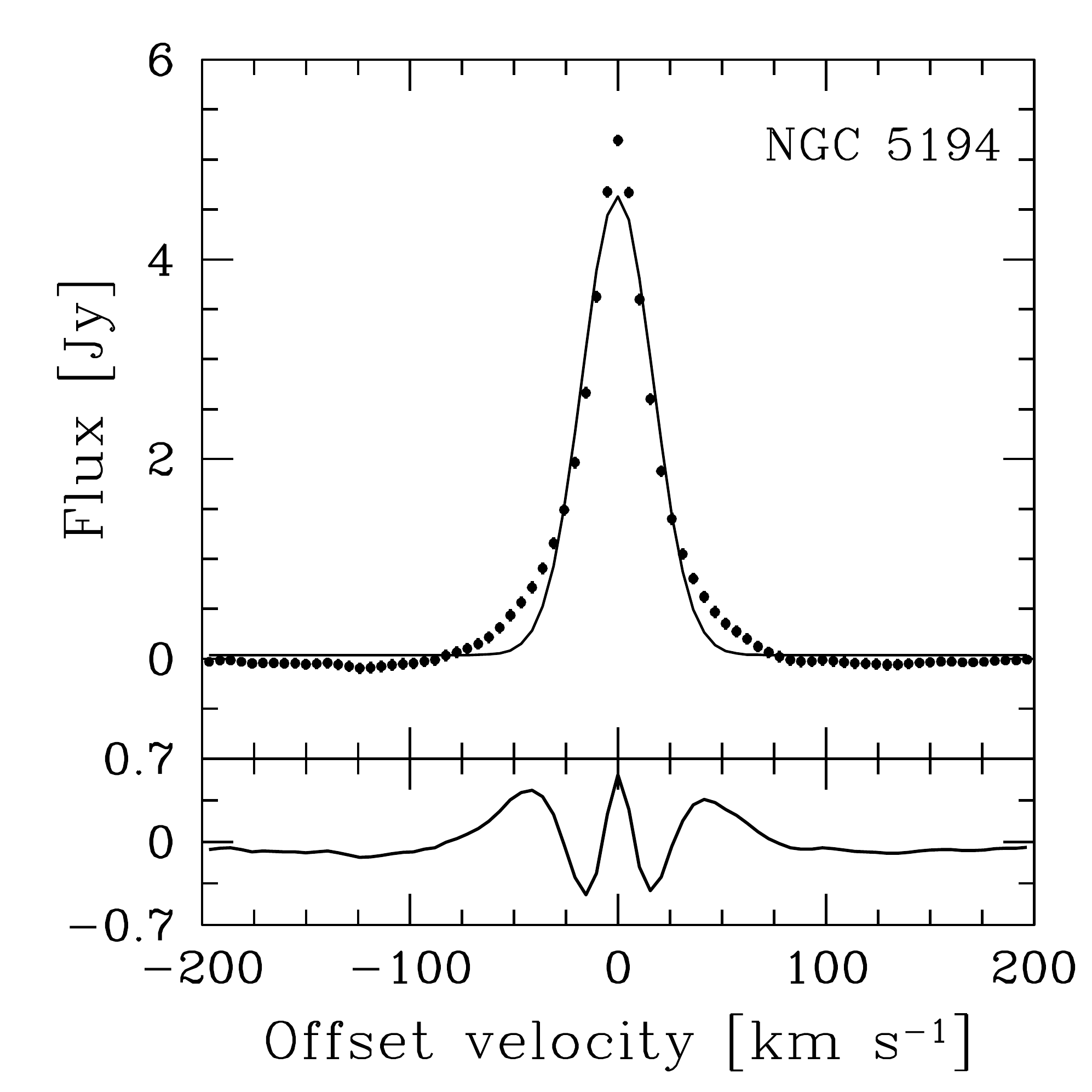}}}\\
\rotatebox{0}{\resizebox{58mm}{!}{\includegraphics[width = 0.6in,height = 0.6in]{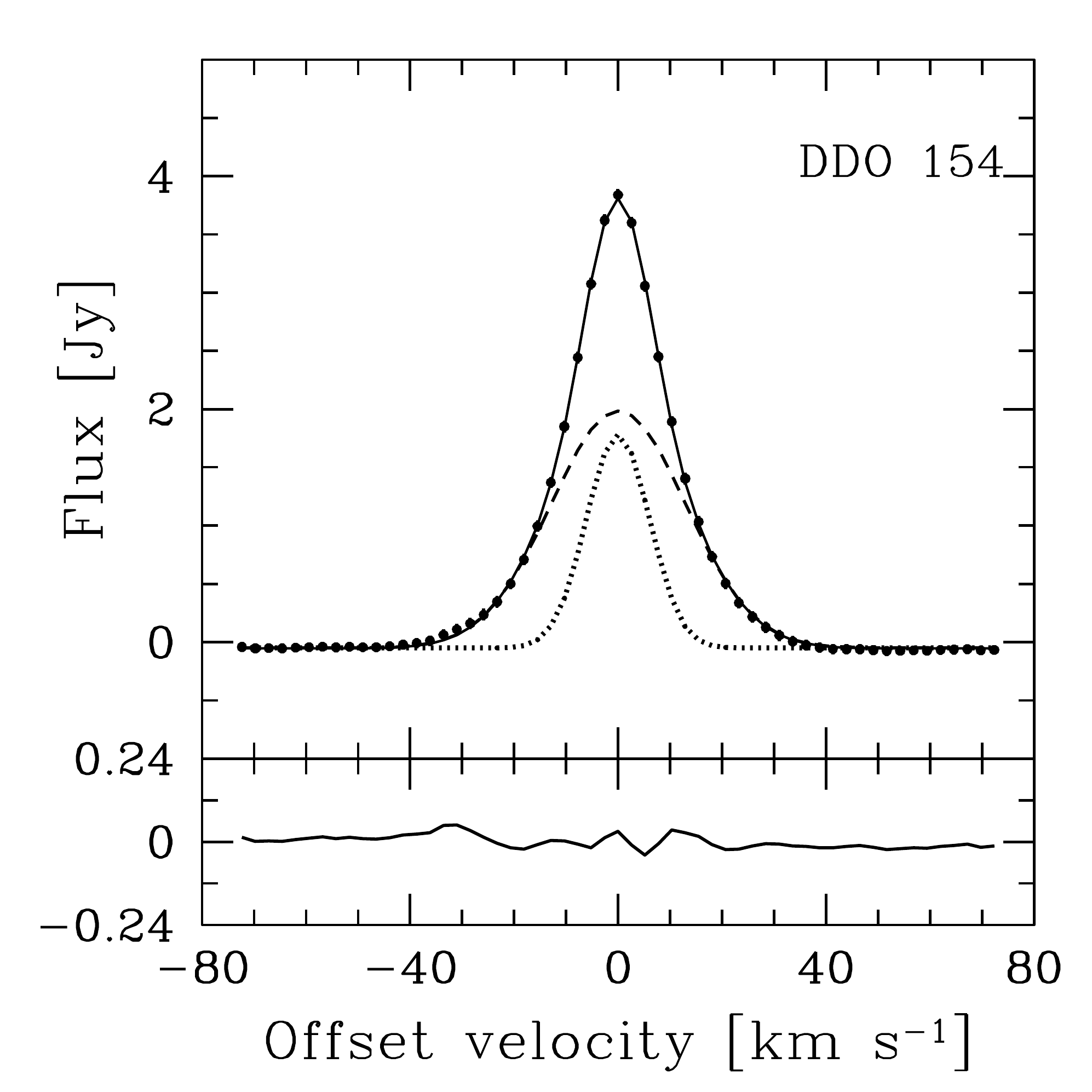}}}&
  \rotatebox{0}{\resizebox{58mm}{!}{\includegraphics[width = 0.6in,height = 0.6in]{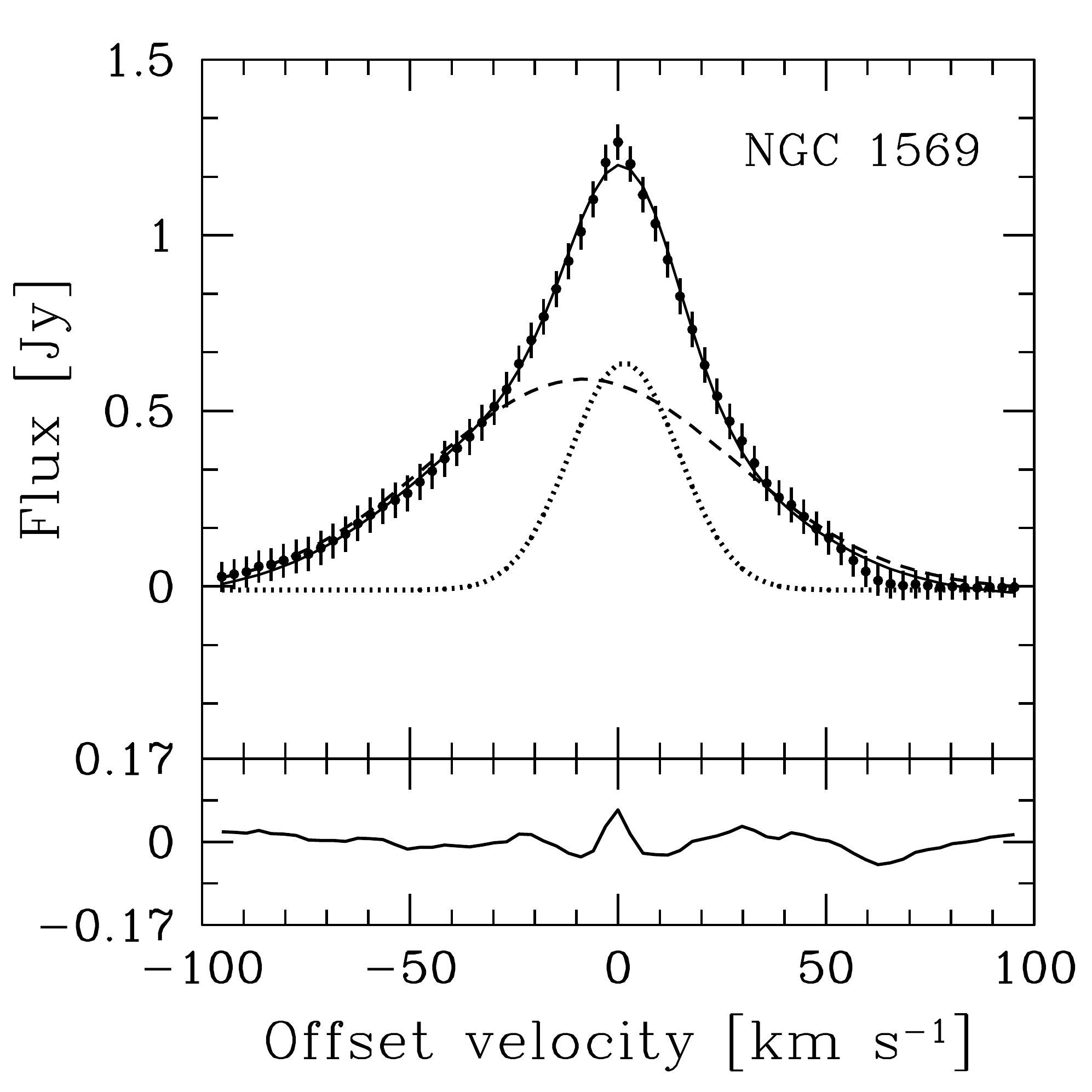}}}&
\rotatebox{0}{\resizebox{58mm}{!}{\includegraphics[width = 0.6in,height = 0.6in]{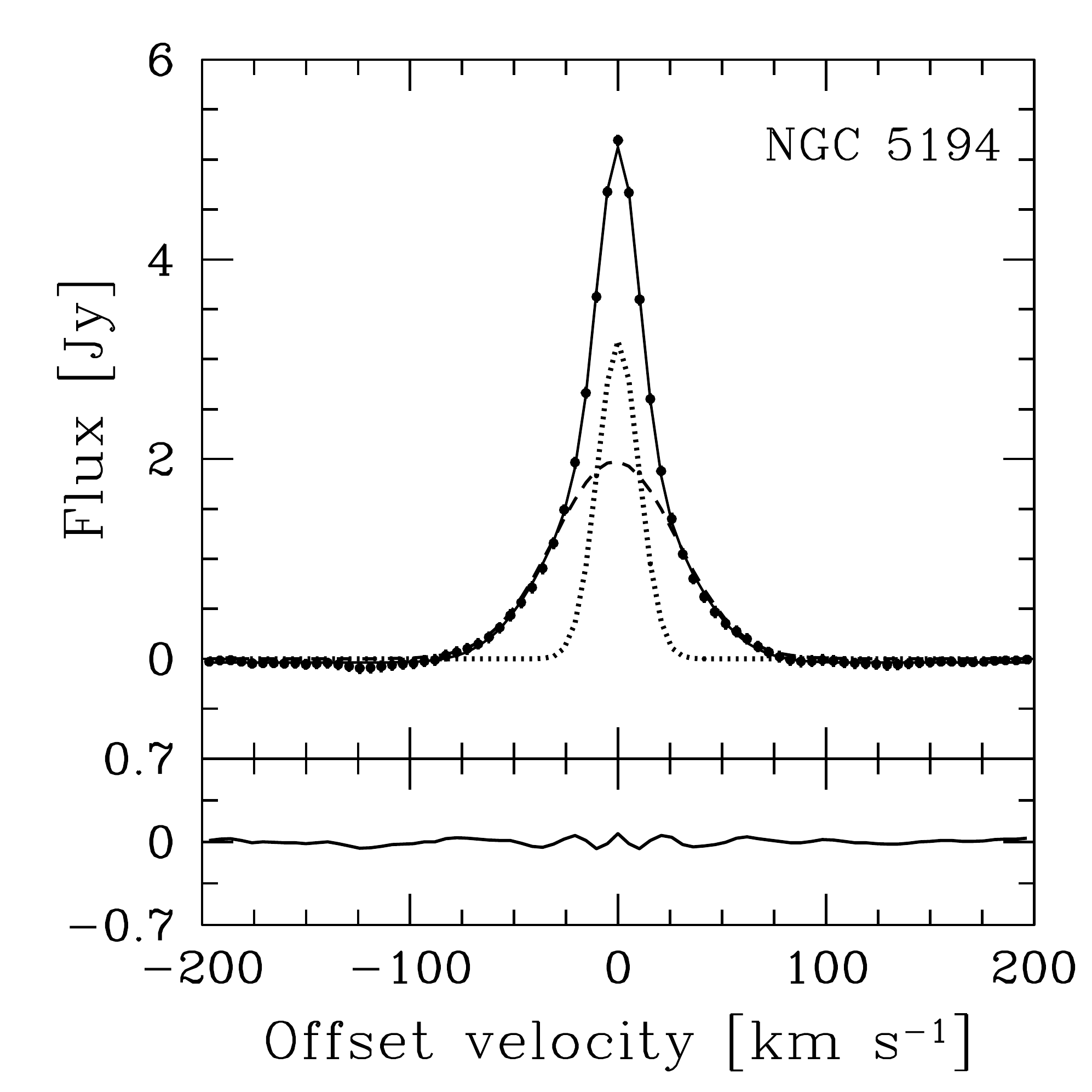}}}
\end{tabular}
\end{center}
\caption{Three examples of super profiles for the THINGS galaxies (the
  rest are shown in Appendix A). For each galaxy, we show the results
  from the single (top panel) and double (bottom panel) Gaussian
  fitting. The bottom panel of each plot show the residuals
  from the fits. Filled circles indicate the data. The solid 
  lines represent the results from the single and double Gaussian
  fitting. The dotted and the dashed lines represent the narrow and
  broad components required in the double Gaussian fitting. We plot
  error bars as 3$\sigma$ error bars, though in most cases they are
  smaller than the symbols plotted.}
\label{fig:superpro_ex}

\end{figure*}

We show in Figure \ref{fig:superpro_ex} examples of the super profiles
of three galaxies (one quiescent galaxy: DDO154, one starburst galaxy:
NGC 1569, and an interacting galaxy: NGC 5194; the rest are shown in
Appendix A). These super profiles are summed over the entire HI disks
of the galaxies using the 3$\sigma$ cut-off mentioned
previously. Typically, each super profile is the sum of a few hundred
to a few thousand independent individual profiles (see Table
\ref{tab:table_superpro}). We also plot $3\sigma$ uncertainties (as
defined in Section \ref{sec:fitting}), though in many cases these are
smaller than the symbols used to plot the profiles. The super profiles
are characterized by broader wings and narrower peaks than purely
Gaussian profiles. Most of the super profiles are symmetrical. These
properties are all similar to what was found for the individual
profiles of seven nearby dwarf galaxies studied by
\citet{younglo96,younglo97} and \citet{youngetal03}, but observed at a
much higher S/N.
 
In the following, we will analyze the shapes of the super profiles and
see if they can be used to infer the presence of the CNM and WNM
components of the ISM.

\begin{deluxetable*}{l r r c c c c c r}
\centering
\tabletypesize{\scriptsize}
\tablewidth{0pt}
\tablecaption{Fitted parameters of the THINGS super profiles \label{tab:table_superpro}}
\tablehead{
\multicolumn{1}{c}{Galaxy}&\multicolumn{1}{c}{$\sigma_{1g}$}
&\multicolumn{1}{c}{$\sigma_{n}$}&$\sigma_{b}$ &   deg. & $A_{n}/A_{b}$ 
& $\chi^{2}_{2G}/\chi^{2}_{1G}$& N &\\
&     &     &     &   asym. &   & & &\\   
&$(\rm{km~ s^{-1}})$&$(\rm{km~ s^{-1})}$&$\rm{(km~ s^{-1})}$&$(\rm{km~ s^{-1}})$ & & & &\\
\multicolumn{1}{c}{1}&\multicolumn{1}{c}{2}&\multicolumn{1}{c}{3}& 4 &5 &  6    &   7   & 8 &
}
\startdata
DDO 53	&9.9$\rm{\pm}0.2$&6.2$\rm{\pm}0.1$&15.0$\rm{\pm}0.2$&0.2$\rm{\pm}$0.1&0.46$\rm{\pm}0.02$&0.03&272\\
DDO 154	&9.7$\rm{\pm}0.1$&6.0$\rm{\pm}0.1$&13.0$\rm{\pm}0.2$&0.1$\rm{\pm}$0.1&0.42$\rm{\pm}0.03$&0.04&1116 \\
Ho I&8.9$\rm{\pm}0.2$&5.4$\rm{\pm}0.1$&13.3$\rm{\pm}0.2$&0.3$\rm{\pm}$0.1&0.42$\rm{\pm}0.02$&0.04&380\\
Ho II&9.1$\rm{\pm}0.2$&5.2$\rm{\pm}0.1$&12.6$\rm{\pm}0.2$&0.3$\rm{\pm}$0.1&0.38$\rm{\pm}0.02$& 0.04&2400\\
IC 2574	&10.1$\rm{\pm}0.2$&5.9$\rm{\pm}0.2$&14.6$\rm{\pm}0.3$&0.5$\rm{\pm}$0.2&0.44$\rm{\pm}0.03$&0.08&4224\\
M81 dwB	&11.3$\rm{\pm}0.2$&6.6$\rm{\pm}0.1$&15.2$\rm{\pm}0.2$&0.2$\rm{\pm}$0.1&0.28$\rm{\pm}0.01$&0.04& 40 \\
M81 dwA	&8.2$\rm{\pm}0.2$&3.4$\rm{\pm}0.2$&10.4$\rm{\pm}0.3$&0.1$\rm{\pm}$0.2&0.14$\rm{\pm}0.01$&0.12& 88\\
NGC 3031&11.2$\rm{\pm}0.2$&6.7$\rm{\pm}0.1$&19.1$\rm{\pm}0.2$&2.9$\rm{\pm}$0.2&0.59$\rm{\pm}0.02$&0.05&8668\\
NGC 628	&9.0$\rm{\pm}0.2$&4.4$\rm{\pm}0.1$&11.6$\rm{\pm}0.1$&0.0$\rm{\pm}$0.0&0.25$\rm{\pm}0.02$&0.02&7964\\
NGC 925	&12.7$\rm{\pm}0.2$&8.6$\rm{\pm}0.2$&22.9$\rm{\pm}0.6$&0.2$\rm{\pm}$0.2&0.75$\rm{\pm}0.05$&0.03&4208\\
NGC 1569&22.5$\rm{\pm}0.7$&13.0$\rm{\pm}0.4$&36.6$\rm{\pm}1.2$&6.9$\rm{\pm}$0.7&0.38$\rm{\pm}0.03$&0.09&440 \\
NGC 2366&12.1$\rm{\pm}0.3$&7.9$\rm{\pm}0.1$&20.0$\rm{\pm}0.3$&1.0$\rm{\pm}$0.2&0.59$\rm{\pm}0.02$&0.03&1776\\
NGC 2403&11.1$\rm{\pm}0.3$&6.6$\rm{\pm}0.1$&18.4$\rm{\pm}0.2$&0.2$\rm{\pm}$0.1&0.54$\rm{\pm}0.04$&0.06&19796\\
NGC 2841&16.3$\rm{\pm}0.4$&10.4$\rm{\pm}0.1$&43.2$\rm{\pm}0.9$&0.8$\rm{\pm}$0.7&0.72$\rm{\pm}0.03$&0.06&2600\\
NGC 2903&13.3$\rm{\pm}0.3$&8.8$\rm{\pm}0.1$&28.9$\rm{\pm}0.7$&1.7$\rm{\pm}$0.5&0.77$\rm{\pm}0.04$& 0.06&2600\\
NGC 2976&11.9$\rm{\pm}0.3$&8.4$\rm{\pm}0.1$&20.9$\rm{\pm}0.4$&1.7$\rm{\pm}$0.3&0.83$\rm{\pm}0.04$&0.03&640\\
NGC 3077&12.6$\rm{\pm}0.3$&5.6$\rm{\pm}0.2$&17.2$\rm{\pm}0.2$&0.0$\rm{\pm}$0.2&0.27$\rm{\pm}0.01$&0.04&2956\\
NGC 3184&11.3$\rm{\pm}0.3$&6.1$\rm{\pm}0.1$&18.8$\rm{\pm}0.2$&0.6$\rm{\pm}$0.2&0.43$\rm{\pm}0.01$&0.02&4860\\
NGC 3198&13.1$\rm{\pm}0.3$&8.8$\rm{\pm}0.1$&22.6$\rm{\pm}0.4$&0.1$\rm{\pm}$0.2&0.76$\rm{\pm}0.03$&0.03&2356\\
NGC 3351&10.2$\rm{\pm}0.3$&7.0$\rm{\pm}0.1$&22.4$\rm{\pm}0.6$&0.3$\rm{\pm}$0.4&0.86$\rm{\pm}0.05$&0.06&2296 \\
NGC 3521&17.4$\rm{\pm}0.4$&12.4$\rm{\pm}0.2$&50.6$\rm{\pm}1.3$&0.4$\rm{\pm}$0.9&0.89$\rm{\pm}0.04$&0.08&1448\\
NGC 3621&11.4$\rm{\pm}0.3$&8.7$\rm{\pm}0.1$&29.6$\rm{\pm}1.1$&4.1$\rm{\pm}$0.7&1.20$\rm{\pm}0.08$&0.08&4740\\
NGC 3627&20.8$\rm{\pm}0.5$&14.3$\rm{\pm}0.2$&43.2$\rm{\pm}0.8$&0.5$\rm{\pm}$0.5&0.82$\rm{\pm}0.03$&0.04&344\\
NGC 4214&8.8$\rm{\pm}0.1$&4.5$\rm{\pm}0.0$&11.9$\rm{\pm}0.1$&0.0$\rm{\pm}$0.1&0.31$\rm{\pm}0.01$&0.02&2596\\
NGC 4449&13.8$\rm{\pm}0.2$&6.9$\rm{\pm}0.4$&16.8$\rm{\pm}0.3$&0.4$\rm{\pm}$0.2&0.21$\rm{\pm}0.02$&0.04&2016\\
NGC 4736&11.0$\rm{\pm}0.3$&7.4$\rm{\pm}0.1$&22.3$\rm{\pm}0.6$&0.2$\rm{\pm}$0.4&0.79$\rm{\pm}0.05$&0.03&2312\\
NGC 4826&13.1$\rm{\pm}$0.5&7.9$\rm{\pm}0.4$&32.0$\rm{\pm}0.8$&1.6$\rm{\pm}$0.6&0.61$\rm{\pm}0.03$&0.04&264\\
NGC 5055&14.0$\rm{\pm}0.4$&8.3$\rm{\pm}0.2$&25.2$\rm{\pm}0.6$&0.1$\rm{\pm}$0.4&0.59$\rm{\pm}0.03$&0.06&6212\\
NGC 5194&17.0$\rm{\pm}0.4$&9.9$\rm{\pm}0.1$&29.6$\rm{\pm}0.4$&1.2$\rm{\pm}$0.2&0.54$\rm{\pm}0.02$&0.02&2616\\
NGC 5236&11.2$\rm{\pm}0.2$&5.6$\rm{\pm}0.1$&16.8$\rm{\pm}0.2$&0.7$\rm{\pm}$0.1&0.38$\rm{\pm}0.03$&0.04&1912 \\
NGC 5457&13.1$\rm{\pm}0.2$&6.4$\rm{\pm}0.1$&15.9$\rm{\pm}0.0$&0.1$\rm{\pm}$0.1&0.21$\rm{\pm}0.02$&0.02&19000\\
NGC 6946&10.4$\rm{\pm}0.2$&6.1$\rm{\pm}0.1$&18.3$\rm{\pm}0.3$&0.4$\rm{\pm}$0.2&0.55$\rm{\pm}0.03$&0.07&7252\\
NGC 7331&19.5$\rm{\pm}0.4$&11.5$\rm{\pm}0.1$&35.3$\rm{\pm}0.3$&0.2$\rm{\pm}$0.2&0.58$\rm{\pm}0.01$&0.01&2360\\
NGC 7793&10.6$\rm{\pm}0.2$&6.6$\rm{\pm}0.1$&17.9$\rm{\pm}0.2$&0.0$\rm{\pm}$0.1&0.60$\rm{\pm}0.02$&0.03&1996\\
\enddata
\tablecomments{Column 1: Name of galaxy; Column 2: Velocity dispersions derived from the 
	one component Gaussian fit; Column 3: Velocity dispersions of the narrow 
	component; Column 4: Velocity dispersions of the broad component; Column 5: 
	Degree of asymmetry of the super profiles defined as the offset between the 
	peak velocities  of the narrow and broad components; Column 6: Ratio of the 
	area of the narrow component to the area under the broad component. Column 7: 
	Ratio of the $\rm{\chi^{2}}$ values from the single and double Gaussian fitting; 
Column 8: Number of independent resolution elements used to make super profiles.}
\end{deluxetable*}
\section{Profile fitting and decomposition}\label{sec:fitting}
 
Following \citet{deblokwalter06}, we fit the super profiles with
single Gaussian and double Gaussian components using the GIPSY task
XGAUPROF.  All parameters (dispersion, central velocity,
amplitude, constant background term) are left free in the fits. 
The uncertainties in the data points of each super profiles are
defined as
 \begin{equation}\label{eq:equation1}
 \sigma = \sigma_{ch.map} \times \sqrt{N_{prof}/N_{prof,beam}}
 \end{equation}
where $\sigma_{ch.map}$ is the rms noise level in one channel map,
$N_{prof}$ is the number of stacked profiles at a certain velocity
$V$, and $N_{prof,beam}$ is the number of profiles in one resolution
element (i.e., the number of pixels or profiles per beam). The inverse
square of these uncertainties is used as the weight during the
fitting. Note that in Eq.~\ref{eq:equation1}, we assume that the noise
is uniform throughout each data cube.  Though strictly speaking
incorrect (due to the primary beam correction), this is a reasonable
assumption as most of the galaxies only occupy the inner quarter or so
of the area of the primary beam. The primary beam correction factors
are therefore on average small.  For 4 galaxies, we correct
for the presence of a negative baseline level (caused by missing
zero-spacings) prior to the Gaussian fitting procedure. We fit the baseline 
with a polynomial and substract it from the super profiles. To illustrate this method, 
we show in the Appendix the super profiles before and after the correction as well as the 
polynomial fit to the baseline. In general the correction is of the order of 10\% for the 
velocity dispersions and 20\% for the fitted area of the super profiles. More 
details about this can be found in the Appendix.

\begin{figure}%
\includegraphics[width=3in, height=3in]{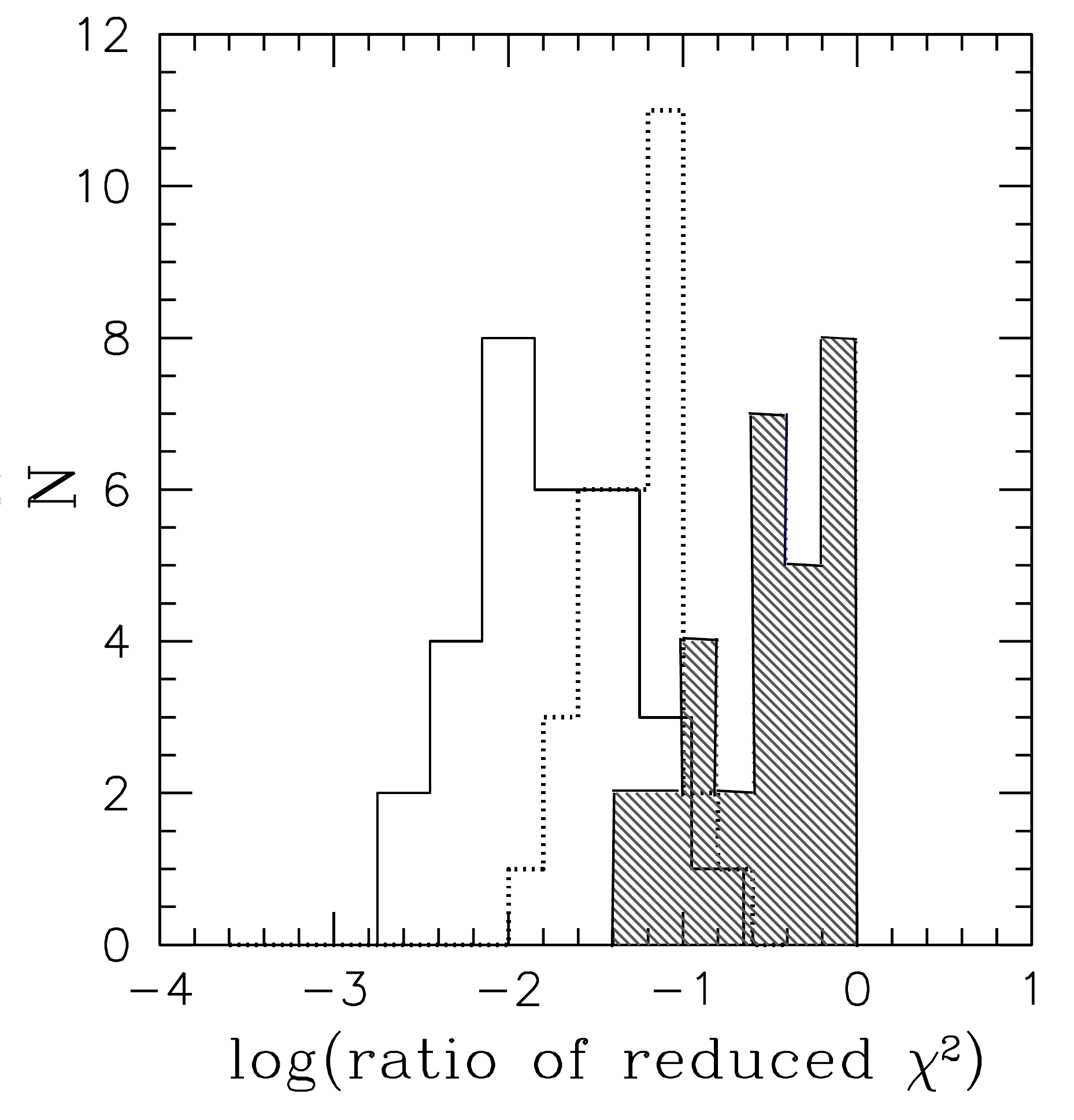}

\caption{\textit{Solid line histogram:} ratio of the reduced $\chi^{2}$
  from the triple and single Gaussian fits
  ($\chi^{2}_{3G}/\chi^{2}_{1G}$). \textit{Dotted line histogram:} ratio of
  the reduced $\chi^{2}$ from the double and single Gaussian fits
  ($\chi^{2}_{2G}/\chi^{2}_{1G}$). \textit{Hatched histogram:} ratio
  of the reduced $\chi^{2}$ from the triple and double Gaussian
  fits ($\chi^{2}_{3G}/\chi^{2}_{2G}$).}
\label{fig:chisquare}%
\end{figure}

\begin{figure}
\includegraphics[width=3in, height=3in]{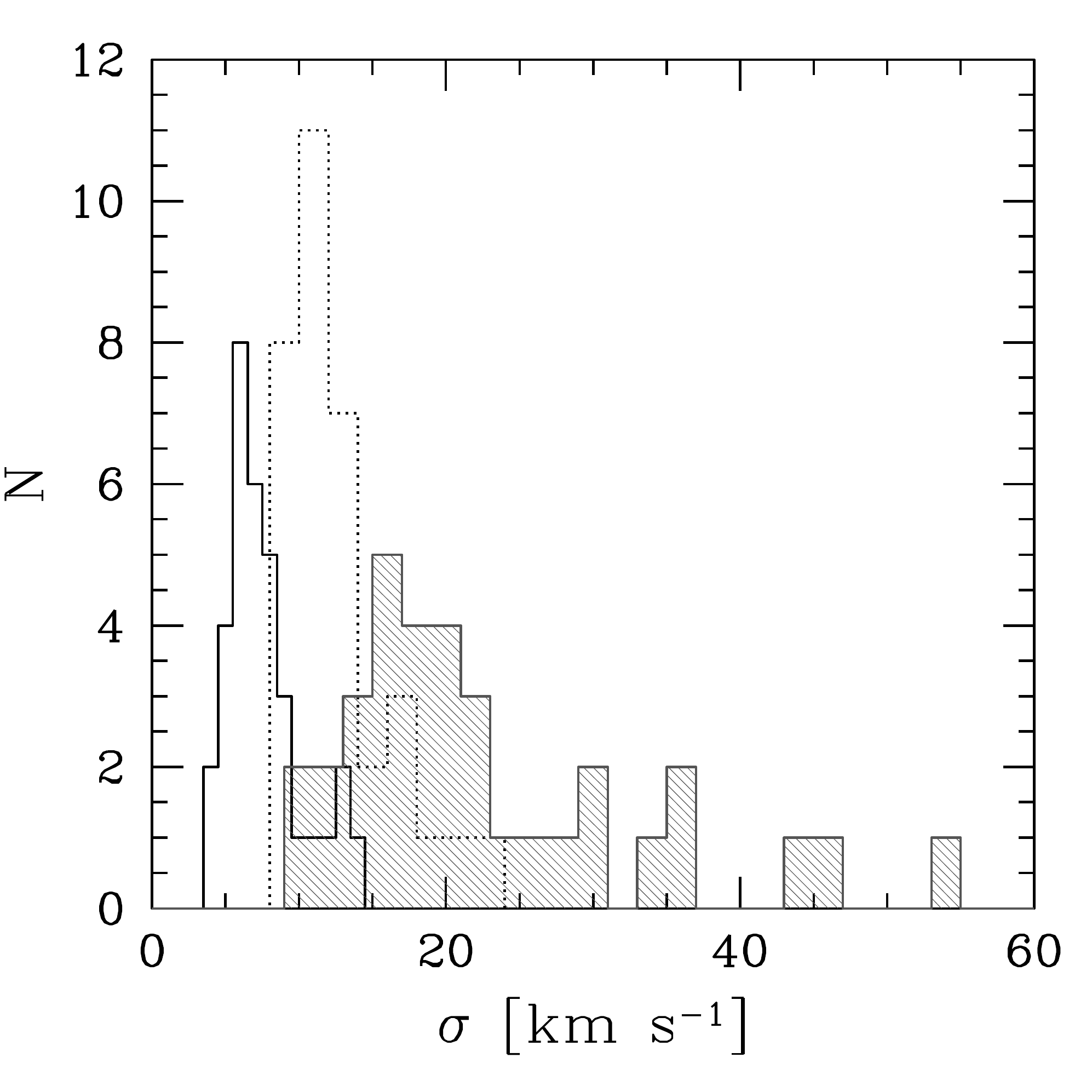}
\caption{Histograms of velocity dispersions derived from both one and
  two-component Gaussian fits. The dotted line histogram represents the
  velocity dispersions derived from the single Gaussian fit. The solid
  line and hatched histograms represent the velocity dispersions of the
  double-Gaussian narrow and broad components, respectively.}
\label{fig:histog_veldisp}%
\end{figure}

In addition to the one and two component fits, we have also tried
three-component Gaussian fits to check if this further improves the
quality of the fits. Comparison of the quality of the different fits
in Fig.~\ref{fig:chisquare} and Table~\ref{tab:table_superpro}, as
well as the amplitudes of the residuals in Fig.~\ref{fig:superpro_ex},
convincingly shows that the super profiles are non-Gaussian and best
described by two Gaussian components.  Following \citet{braun97}, we
also fit the super profiles with a Lorentzian function. These results
are presented in the Appendix. Figure \ref{fig:histog_veldisp} shows
histograms of the derived velocity dispersions from both the single
and the double Gaussian fitting. The distribution of the narrow
component in Fig.~\ref{fig:histog_veldisp} has a mean of $7.6\pm2.5~
\rm{km~s^{-1}}$, whereas that of the broad component has a mean value
of $22.4 \pm 10.0~\rm{km~s^{-1}}$. Also, the broad component has a 
long tail of up to 60 
$\rm{km~s^{-1}}$. 
The single Gaussian velocity
dispersion has a mean of $12.5\pm3.5~\rm{km~s^{-1}}$. If we only
consider galaxies with inclinations $i < 60^\circ$, the single Gaussian
velocity dispersion has a mean of $10.9\pm 2.1~\rm{km~s^{-1}}$. This
is in agreement with the mean second-moment value of
$11\pm3~\rm{km~s^{-1}}$ which \citet{leroyetal08} derived for the same
galaxies.  Table \ref{tab:table_superpro} summarizes the various fit
parameters.

\section{Velocity dispersion or bulk motions}\label{sec:sys_effects}
 
Even though we have established that the super profiles are
  best described by two Gaussian components, interpretation of this is
  complicated by the possible presence of streaming or radial motions
  and asymmetrical input profiles, as well as the limited sensitivity and
  resolution of the observations. All of this could potentially affect
  the shapes of the super profiles.  The brief analysis presented in
  the previous section did not take these effects into account. Here
  we investigate these effects, and attempt to identify a sample of
  galaxies where the super profiles trace the intrinsic velocity
  dispersions of the ISM. In Appendix B we also present a number of
  tests dealing with the accuracy of the shuffling method itself.

  \subsection{Modeling the effects of resolution and inclination}\label{sub:resolution}
 
The gradient in rotation velocity across the finite beam of a
telescope can broaden velocity profiles. Although the THINGS galaxies
have been observed at relatively high resolution, this could still
affect the shapes of the profiles, especially for highly inclined
galaxies.

To quantify these effects, we construct model data cubes
of  IC 2574 and NGC 2403 using the GIPSY task GALMOD. IC 2574 
is a dwarf galaxy with a solid body rotation curve, 
a distance of $\sim$4 Mpc and an inclination of 53$^\circ$.
NGC 2403 
is a well resolved, rotation dominated galaxy typical 
of spirals found in the THINGS sample. 

For the NGC 2403 models we use the observed position angle and inclination ($63^\circ$) 
as given in \citet{debloketal08}. We adopt a
vertical Gaussian scale height of 100 pc, a \emph{single} component
velocity dispersion of 8 $\rm{km~s^{-1}}$, a constant HI surface
density and a constant rotation curve of 130 $\rm{km~s^{-1}}$ (the 
maximum observed rotation velocity of NGC 2403). We adopt a channel 
spacing of 5.2 $\rm{km~s^{-1}}$ as in the THINGS observations.
We also create an almost identical model, but with an
inclination of $80^\circ$ (note that the highest inclination in our
sample is $76^\circ$).  We smooth these model data cubes to various
resolutions and create the corresponding super profiles. We fit the
artificial super profiles both with a single and a double Gaussian to
see how the dispersion changes with resolution and to check whether
resolution and inclination can create broad and narrow components from
a single input component. We have also tested a case where 
the input rotation curve rises linearly up to 130 $\rm{km~s^{-1}}$ at 1 kpc and flattens. 
The results from the constant and the linearly rising plus flat rotation curve models are identical. 

For the IC 2574 models, we use the average position angle derived by \citet{debloketal08}, but 
assume an inclination of
$63^\circ$ and $80^\circ$ to facilitate a direct comparison with the NGC 2403 models. Here we also use a 
\emph{single} component velocity dispersion of 8 $\rm{km~s^{-1}}$, a constant HI surface density 
but an input rotation curve that rises linearly to 80 $\rm{km~s^{-1}}$ at the edge of the HI 
disk. We use a channel spacing of 2.6 $\rm{km~s^{-1}}$ as used by THINGS for IC 2574. We adopt a
vertical Gaussian scale height of 0.1 kpc. We  fit the model super profiles of the IC 2574 
with both single and double Gaussians  but the results of the double Gaussian decomposition 
are not meaningful as the super profiles are 
  almost perfectly Gaussian. So, for the IC 2574 models, we only present 
  the single Gaussian fitting results.  We have also tested the case where 
  the scale height is 0.5 kpc but the results 
are similar to the 0.1 kpc models.

\begin{figure*}
    \begin{tabular}{l l l}
  \includegraphics[scale=0.28]{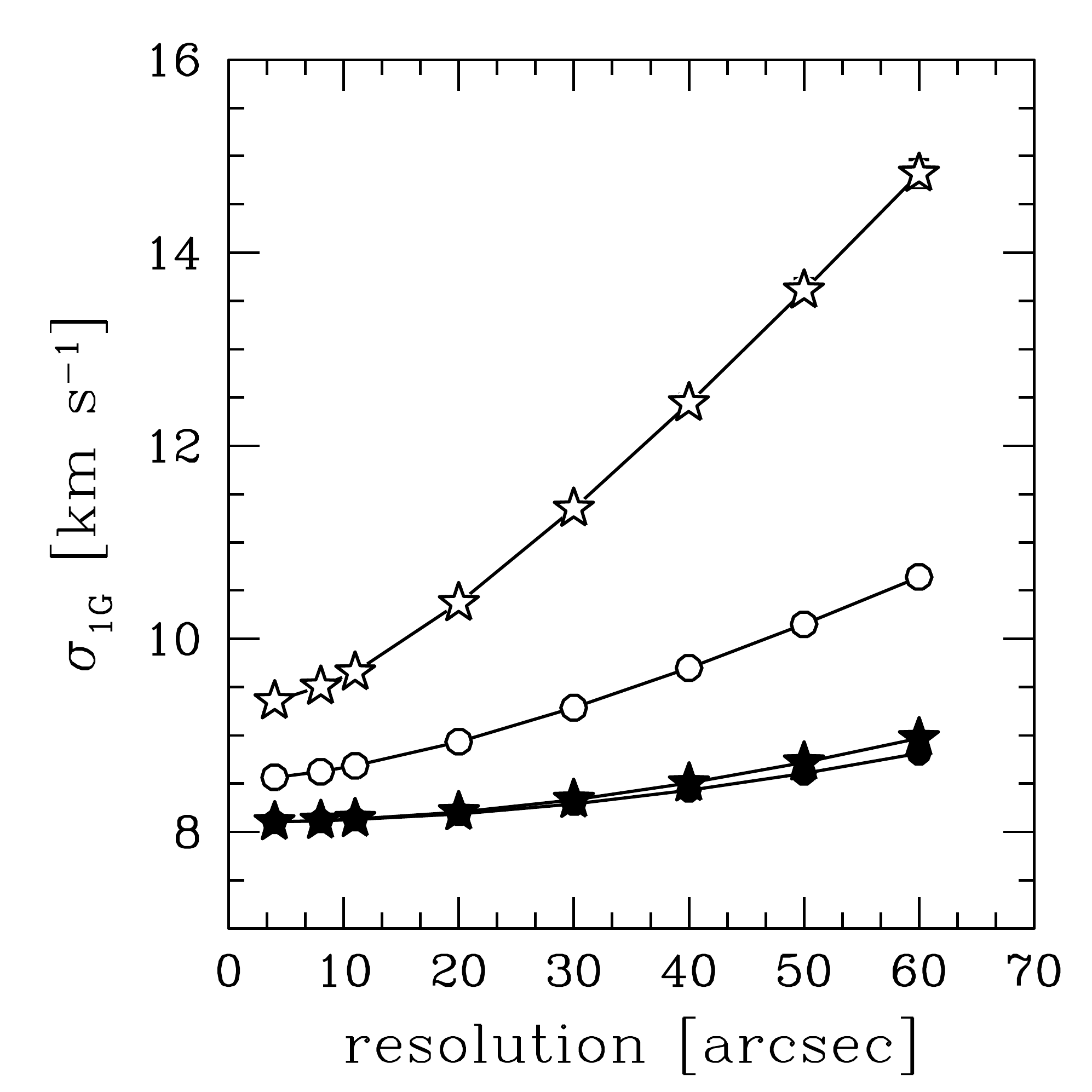}&
\includegraphics[scale=0.28]{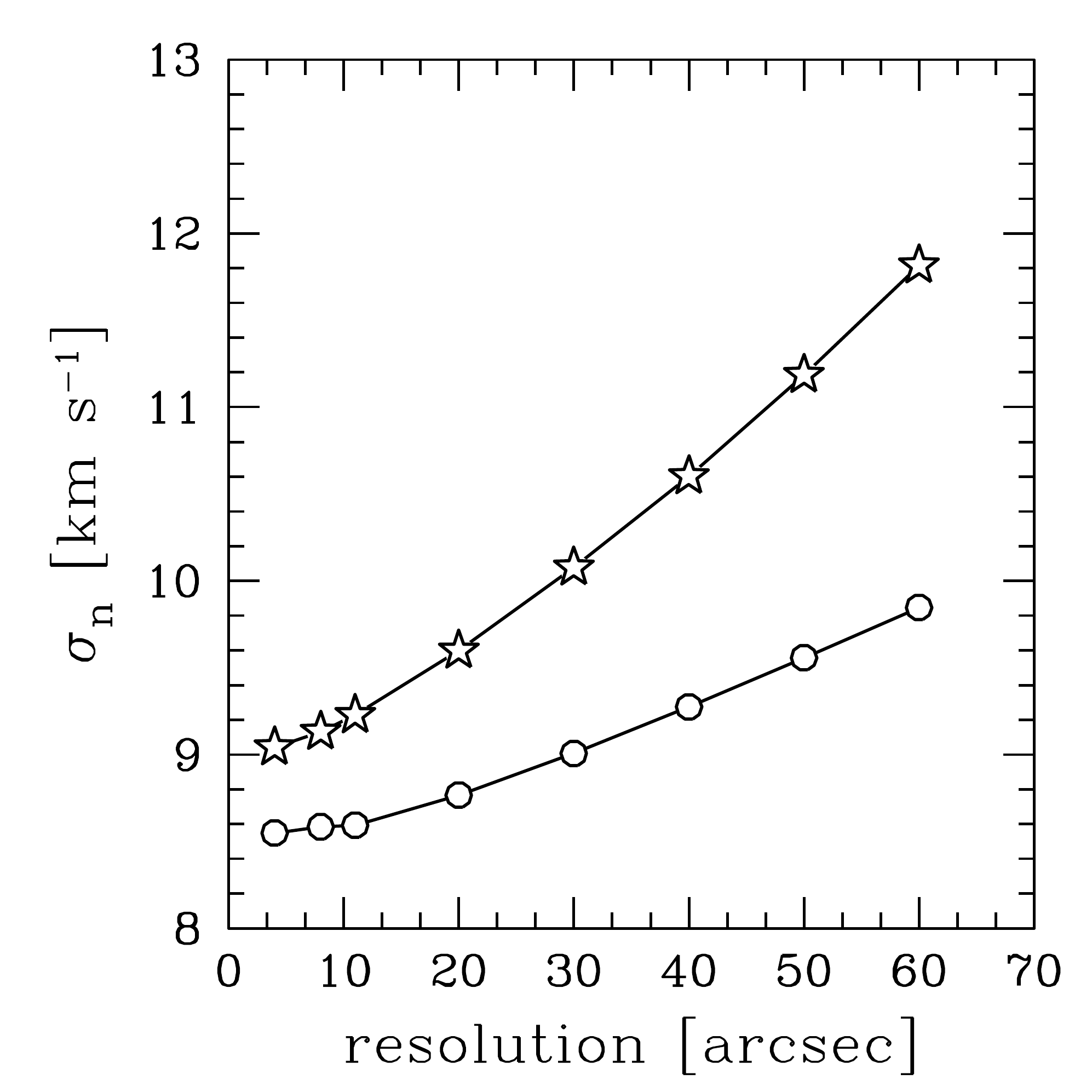}&
 \includegraphics[scale=0.28]{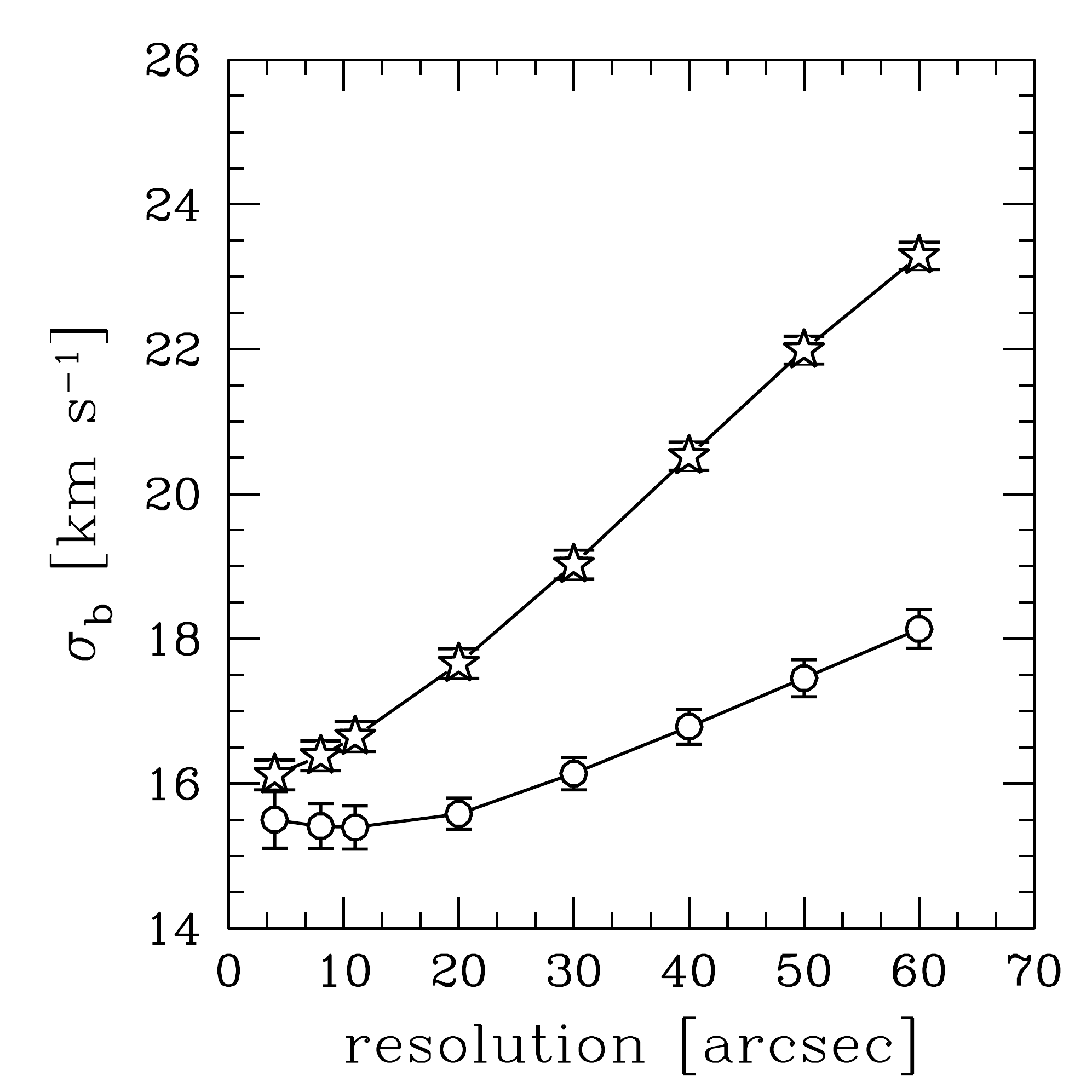}
\end{tabular}
\caption{Super profile parameters derived from the model data cubes of
  IC 2574 and NGC 2403 as a function of the resolution. Solid and open symbols 
  represent the models of IC 2574 and NGC 2403, respectively.
  The circle symbols represent the $63^\circ$ models, whereas the star symbols represent the 
  $80^\circ$ inclination
  models. We only present the single Gaussian fitting results of the IC 2574 models as these profiles 
   do not show a double Gaussian function signature. \textit{Left panel:} single Gaussian
  velocity dispersion. \textit{Middle panel:} Narrow component
  velocity dispersion. \textit{Right panel}: Broad component
  velocity dispersion.}
\label{fig:galmod_param}
\end{figure*}
 
Figure \ref{fig:galmod_param} and Table \ref{tab:effect_of_res}
summarize the model super profile parameters as a function of the
resolution. It is clear that profiles become broader with decreasing
resolution, and more so for a higher inclination.  For example, at
$11^{\prime\prime}$ (the average resolution of THINGS), the difference
between the velocity dispersion measured from the $80^\circ$ and the
$63^\circ$ inclination model data cubes of NGC 2403 is $\sim 9 \%$. At a
resolution of $60^{\prime\prime}$, the difference is $\sim 29\%$. 
From Figure \ref{fig:galmod_param}, it is apparent that we do
  not recover the input 8 $\rm{km~s^{-1}}$ dispersion even at very
  high resolution. This is a result of the finite channel spacing of
  the NGC 2403 data cubes used (5.2 $\rm{km~s^{-1}}$) in combination with the
  high inclination. However, for the IC 2574 models, the velocity dispersions 
  approach the input 8 $\rm{km~s^{-1}}$ value, and this is due to the smaller 
  channel spacing used (2.6 $\rm{km~s^{-1}}$). 

\begin{figure*}
  \centering
    \begin{tabular}{l l}
  \includegraphics[width = 3in,height = 3in]{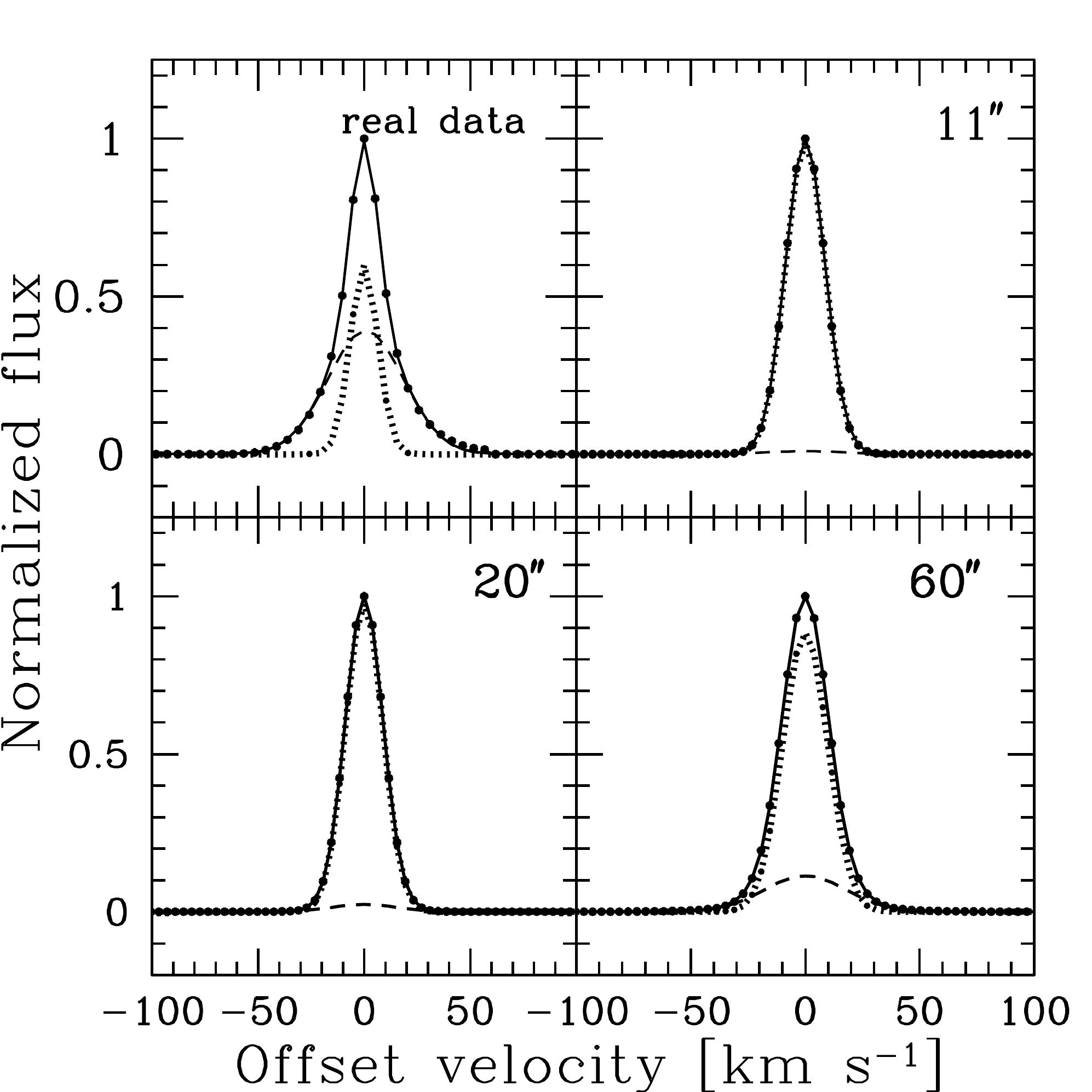}&
  \includegraphics[width = 3in,height = 3in]{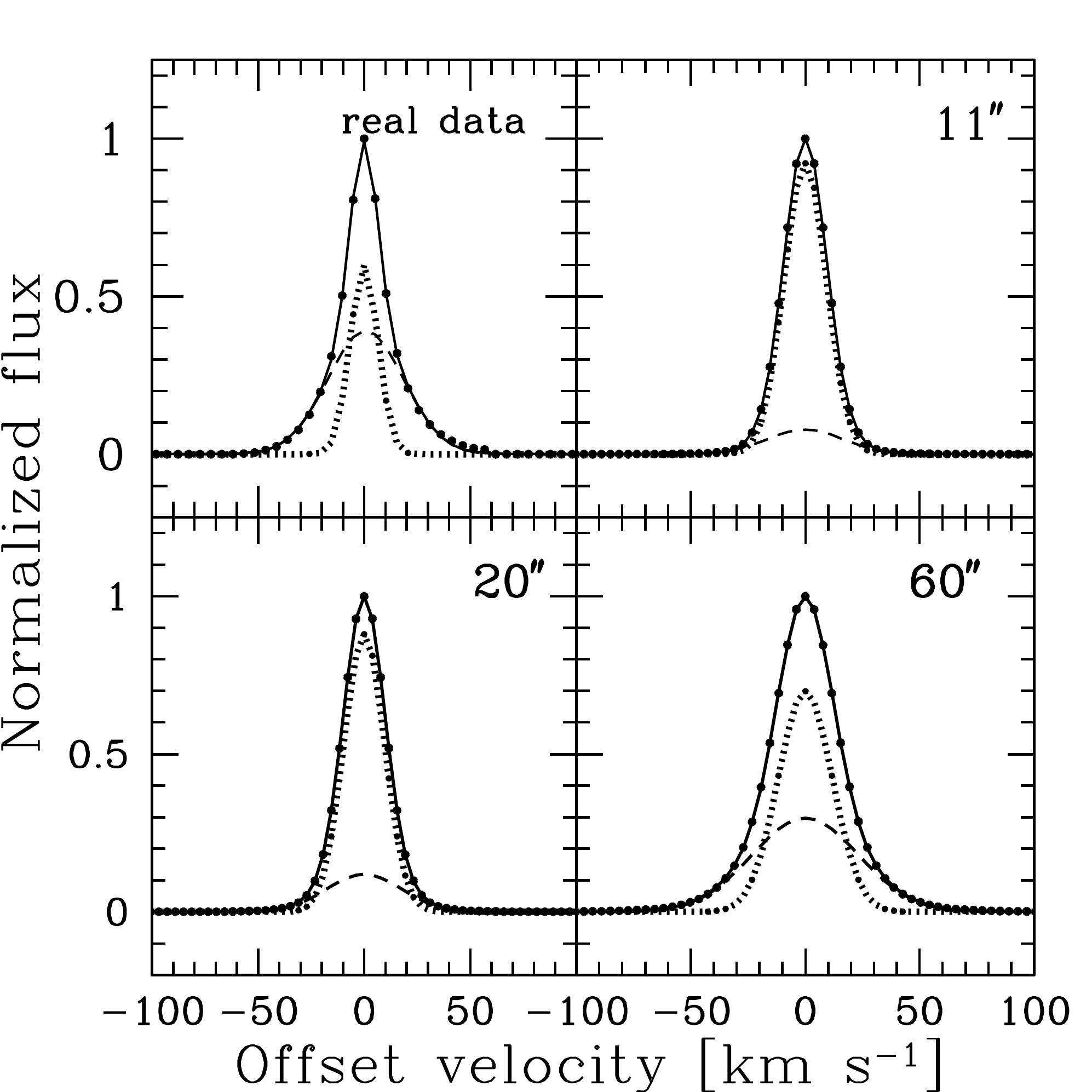}
    \end{tabular}
\caption{Observed and simulated super profiles of NGC 2403 derived
  after smoothing the model data cubes to the resolution indicated on
  the plots. The left and right panels represent the results from the
  model data cubes of NGC 2403 with input inclination of $63^\circ$
  and $80^\circ$, respectively. The dotted and the dashed lines
  represent the narrow and broad components required in the double
  Gaussian fitting. The solid lines represent the results of the
  double Gaussian fitting.}
\label{fig:obs_simulated_sup}
\end{figure*}

\begin{figure}
\centering
 \begin{tabular}{l}
\includegraphics[width = 3in,height = 3in]{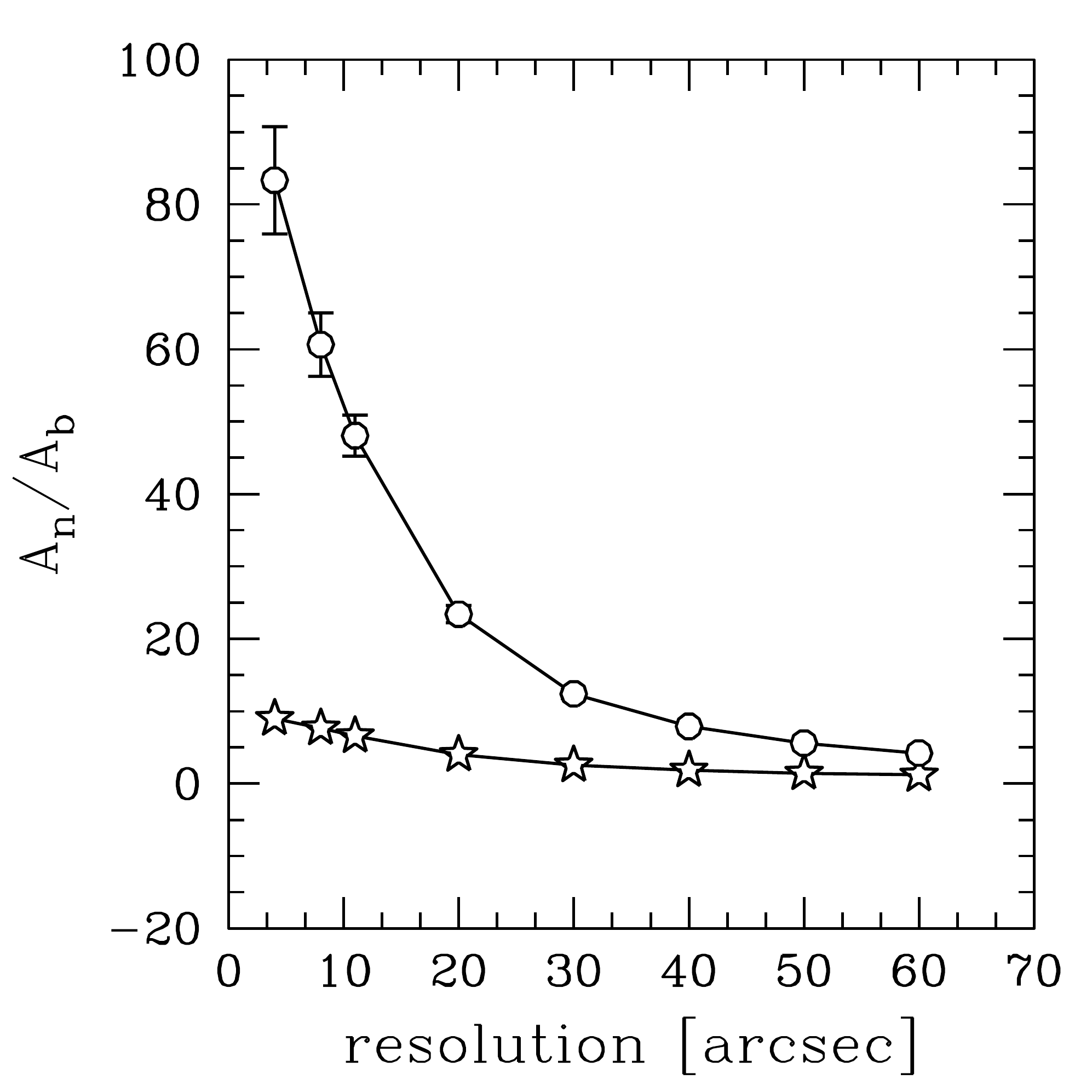}
  \end{tabular}
\caption{ Flux ratio of the narrow and broad components, $A_{n}/A_{b}$, derived from 
the model data cubes of NGC 2403 as a function of the resolution. The open circle and 
  star symbols represent the $63^\circ$ and the $80^\circ$ inclination
  models, respectively.}
\label{fig:flux_model1}
\end{figure}

Figure \ref{fig:obs_simulated_sup} shows a selection of the modelled
super profiles of NGC 2403. It is clear that resolution effects can introduce a
spurious broad component. We use the ratio between the areas of
the narrow and broad components $A_n/A_b$ (i.e., the flux or mass
ratio) to gauge whether any of the models resemble the data. Figure
\ref{fig:flux_model1} shows the $A_n/A_b$ ratio as a function of
resolution. This ratio decreases with resolution, showing the
increasing importance of the spurious broad component. The extremely
high ratios at high resolution should not be over-interpreted. These
are simply due to the fitting routine being forced to fit an almost
perfectly Gaussian profile with two components.  At the lowest
resolution in the models, the broad component created by the low
resolution accounts for $\sim 13\%$ of the total flux. In the THINGS
data, the broad component accounts for up to $\sim 70\%$ of the total
flux. This discrepancy also indicates that resolution effects are not
solely responsible for the observed broad components.
Comparison of the models and data shows that at $11^{\prime\prime}$
(the typical THINGS resolution), the resulting super profile is well
fitted by a single Gaussian function. It also shows that the shape of
the super profile only starts to significantly deviate from a single
Gaussian at a resolution about six times worse than the THINGS
resolution. In summary, models with single Gaussian input profiles are 
well recovered by the THINGS resolution.

\begin{deluxetable*}{c c c c c c c c}[htb]
\centering
\tabletypesize{\scriptsize}
\tablecaption{Effect of the resolution on the shapes of the super profiles 
	\label{tab:effect_of_res}}
\tablewidth{0pt}
	\tablehead{
	\multicolumn{6}{c}{NGC 2403 models}&\multicolumn{2}{c}{IC 2574 models}
}
	\startdata
	\multicolumn{2}{c}{Resolution}& \multicolumn{3}{c}{Velocity}
	&Flux ratio&Resolution& Velocity\\
	&  & &   dispersion  &    &   & & dispersion\\
	\multicolumn{2}{c}{}&$\sigma_{1G}$&$\sigma_{n}$ &$\sigma_{b}$&$A_{n}/A_{b}$& &$\sigma_{1G}$\\
	$( ^{\prime\prime} )~~~~~~$&(pc)$~~~$ & $(\rm{km~s^{-1}})$&$(\rm{km~s^{-1}})$&
	$(\rm{km~s^{-1}})$ & &(pc)&$(\rm{km~s^{-1}})$\\
	\cutinhead{Inclination: 63$^\circ$}
	4 $~~~~~~~~~$&62 $~~~~$&8.6 &8.5& 15.5&83.3&78  &8.1\\
	8 $~~~~~~~~~$&124$~~~~$&8.6 &8.6&15.4 &60.6&155 &8.1\\
	11$~~~~~~~~~$&171$~~~~$&8.6 &8.6&15.4 &56.6&213 &8.1\\
	20$~~~~~~~~~$&310$~~~~$&8.9 &8.8&15.6 &23.4&388 &8.2\\
	30$~~~~~~~~~$&465$~~~~$&9.3 &9.0&16.1 &12.4&582 &8.3\\
	40$~~~~~~~~~$&621$~~~~$&9.8 &9.3&16.8 &7.9 &776 &8.4\\
	50$~~~~~~~~~$&776$~~~~$&10.1&9.6&17.5 &5.6 &970 &8.6\\
	60$~~~~~~~~~$&931$~~~~$&10.6&9.8&18.1 &4.2 &1164&8.8\\
	\cutinhead{Inclination: 80$^\circ$}
	4$~~~~~~~~~$ &62 $~~~~$&9.4 &9.0 & 16.1& 9.0&78  &8.1\\
	8$~~~~~~~~~$ &124$~~~~$&9.5 &9.1 & 16.4& 7.7&155 &8.1\\
	11$~~~~~~~~~$&171$~~~~$&9.7 &9.2 & 16.6& 6.6&213 &8.1\\
	20$~~~~~~~~~$&310$~~~~$&10.4&9.6 & 17.7& 4.0&388 &8.2\\
	30$~~~~~~~~~$&465$~~~~$&11.4&10.0& 19.0& 2.6&582 &8.3\\
	40$~~~~~~~~~$&621$~~~~$&12.5&10.6& 20.5& 1.8&776 &8.5\\
	50$~~~~~~~~~$&776$~~~~$&13.6&11.2& 21.9& 1.4&970 &8.7\\
	60$~~~~~~~~~$&931$~~~~$&14.9&11.8& 23.3& 1.2&1164&9.0\\
\enddata
\tablecomments{$\sigma_{1G}$: Velocity dispersion
   derived from one-component Gaussian fit. $\sigma_{n}$: Velocity
   dispersion of the narrow component. $\sigma_{b}$: Velocity
   dispersion of the broad component. $A_{n}/A_{b}$: Ratio of the
   fluxes or masses associated with the narrow and broad components. Here, only 
   the single Gaussian results of the IC 2574 models are shown (see Section 
\ref{sub:resolution}).}
\end{deluxetable*}

\begin{figure}
\centering
    \begin{tabular}{l}
\includegraphics[width = 3in,height = 3in]{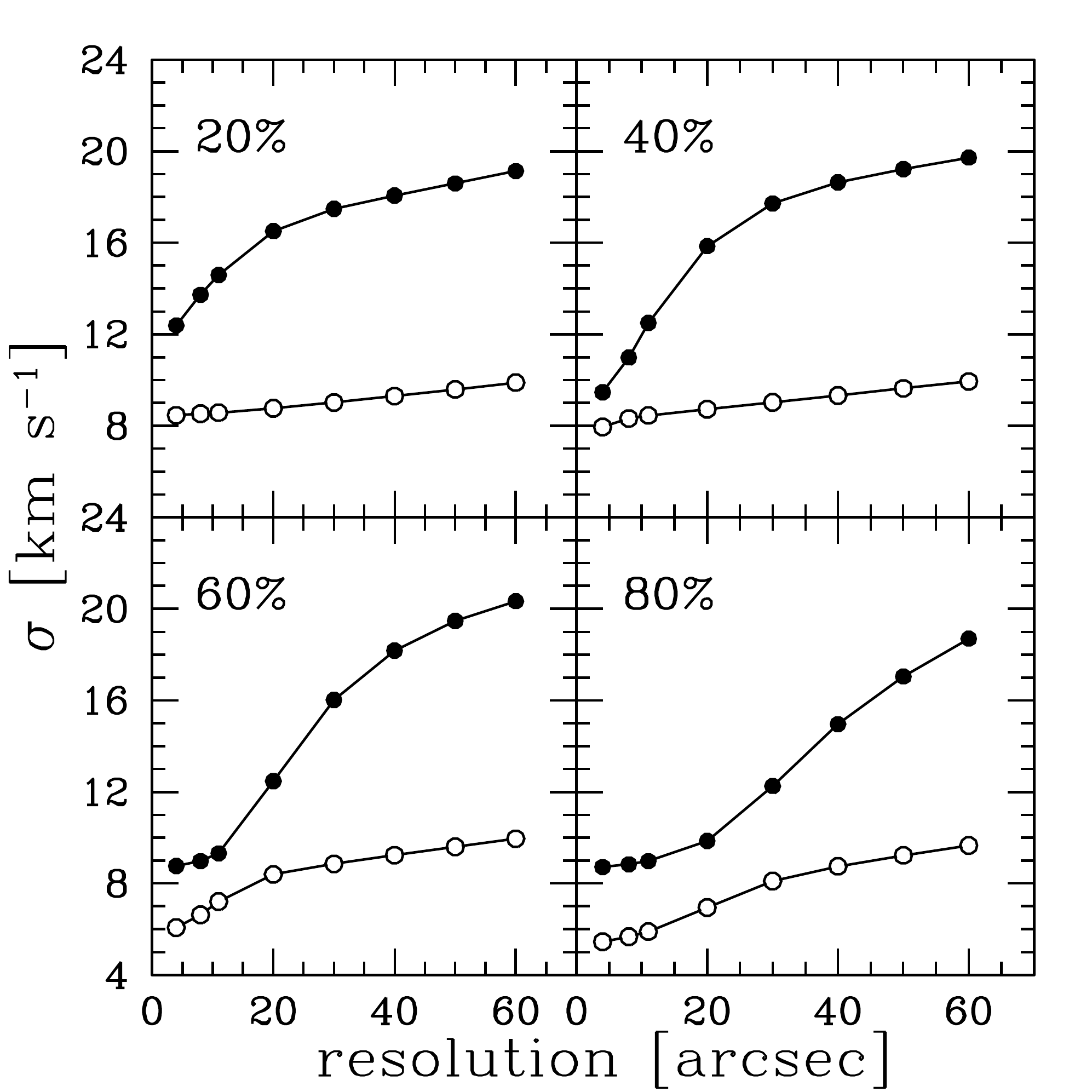}
    \end{tabular}
\caption{Velocity dispersions derived from the modelled narrow and
  broad disk models (see Sect.~\ref{sub:resolution}) as a
  function of the resolution. Open circles represent the narrow
  component. Filled circles  represent the broad
  component. The percentage indicates the narrow profile disk filling factor.}
\label{fig:modelone}

\end{figure}

The previous test investigated whether double components can be
spuriously created using single component input models. We now
investigate whether input double components can be recovered from
limited resolution models.  We use the $63^\circ$ inclination model of
NGC 2403 described before. Recall that this model assumes a filled HI
disk with a Gaussian velocity dispersion of $8~\rm{km~s^{-1}}$. We
will here refer to this as the broad component model.  We also construct a
second set of cubes, assuming a $4~\rm{km~s^{-1}}$ dispersion and
disks that are only partially filled (we assume 20\%, 40\%, 60\% and
80\% area filling factors). We refer to these as the narrow component
models.  We add each of the narrow component models to the broad
component model, smooth the resulting cubes to different resolutions
and then derive the corresponding super profiles. The narrow component
models are scaled to give input $A_n/A_b$ ratios equal to unity
in the regions where both narrow and broad components are present.
Figure \ref{fig:modelone} shows the measured narrow and the broad
components velocity dispersions derived from the smoothed models, as a
function of resolution.  At high resolution, the super profile shapes
are sensitive to area filling factor of the narrow component.  As we
increase the number of narrow profiles in the disk, the super
profiles' narrow and broad component velocity dispersions approach the
input dispersions of the narrow and broad component disks,
respectively.  At low resolutions, the super profile parameters become
insensitive to the narrow component filling factor. The models thus
show that at the THINGS resolution we are able to recover the presence
of narrow and broad components from the models.

\subsection{Effect of high intensity profiles}\label{sub:highintensity}

Another possible explanation for the non-Gaussianity could be that the
shapes of the super profiles are dominated by a few lines of
  sight with high intensities and narrow dispersions.  This implies
some correlation between peak brightness of the input profiles and
velocity dispersion.  To test this, we create super profiles using the
50\% brightest and 50\% faintest (in terms of peak brightness)
profiles in each galaxy.


\begin{figure*}
\centering
    \begin{tabular}{l l l}
  \rotatebox{0}{\resizebox{58mm}{!}{\includegraphics[width = 0.6in,height = 0.6in]{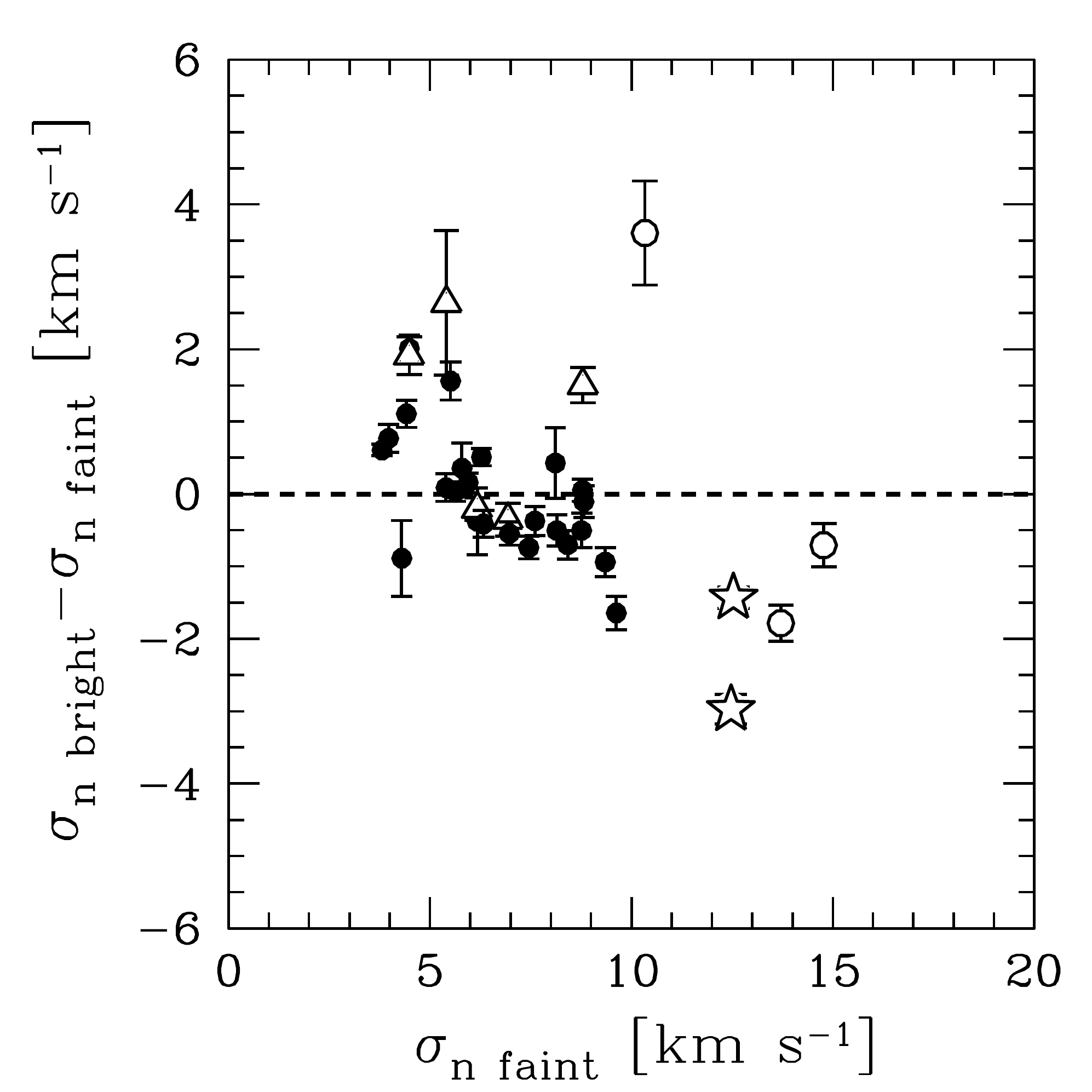}}}&
 \rotatebox{0}{\resizebox{58mm}{!}{\includegraphics[width = 0.6in,height = 0.6in]{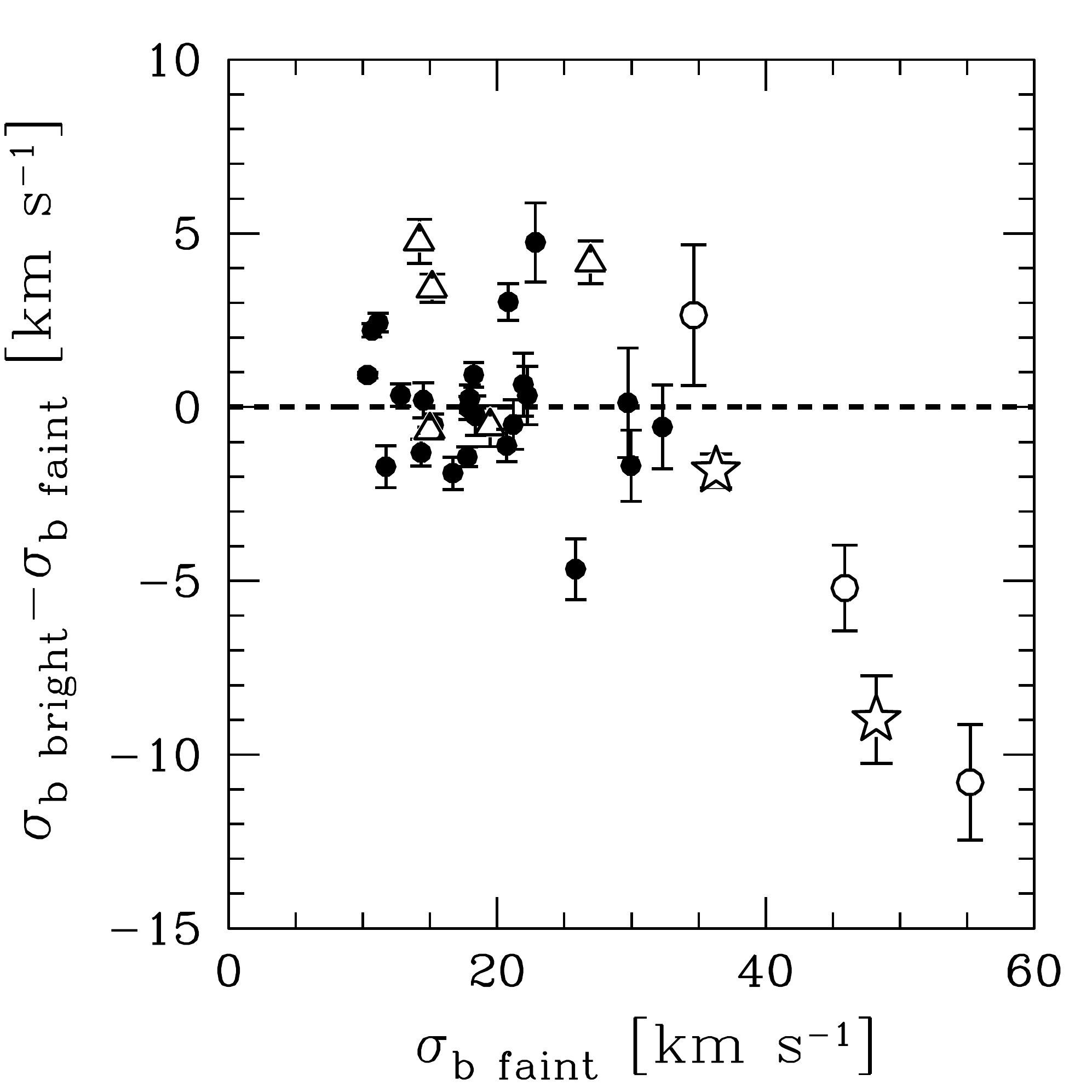}}}&
 \rotatebox{0}{\resizebox{58mm}{!}{\includegraphics[width = 0.6in,height = 0.6in]{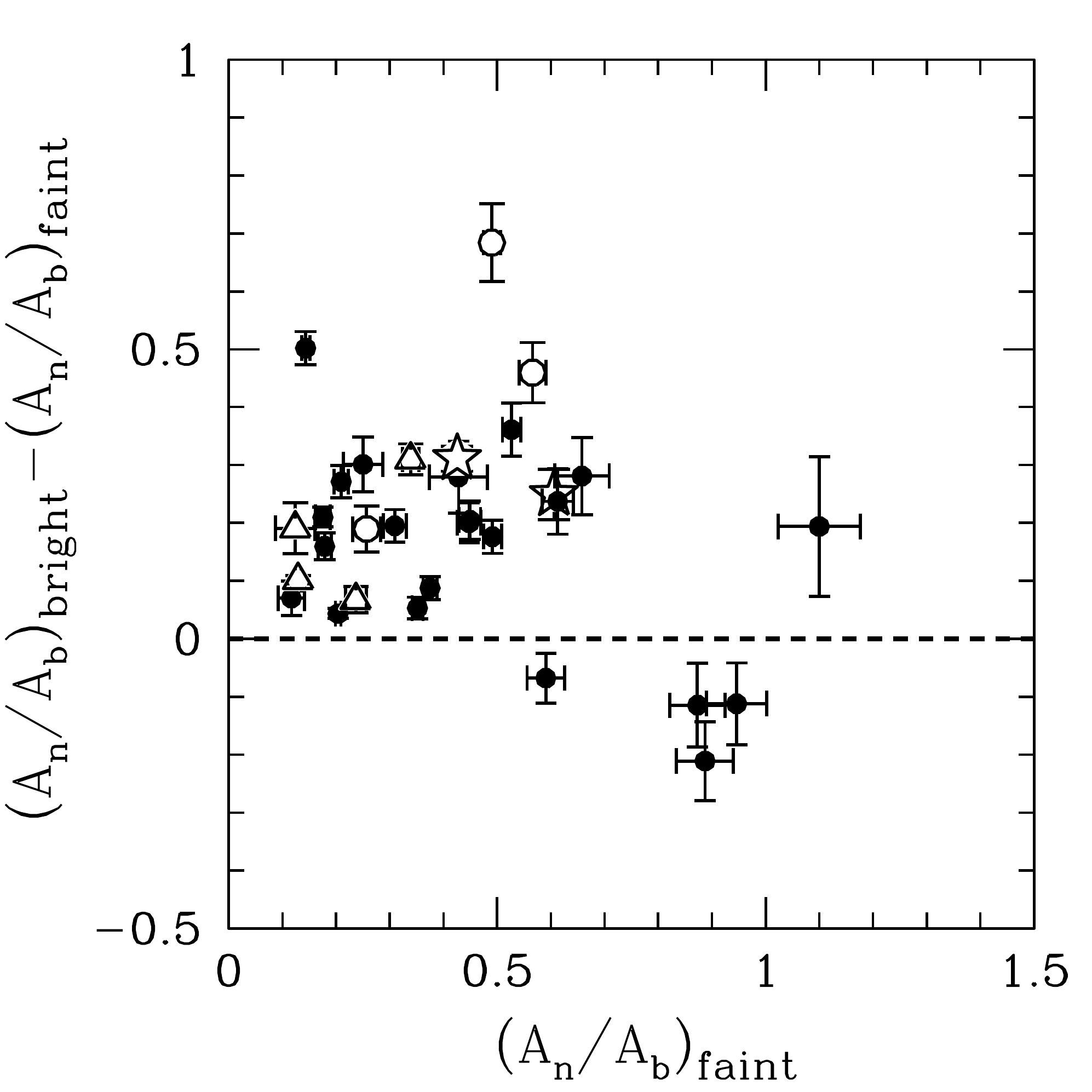}}}\\
\rotatebox{0}{\resizebox{58mm}{!}{\includegraphics[width = 0.6in,height = 0.6in]{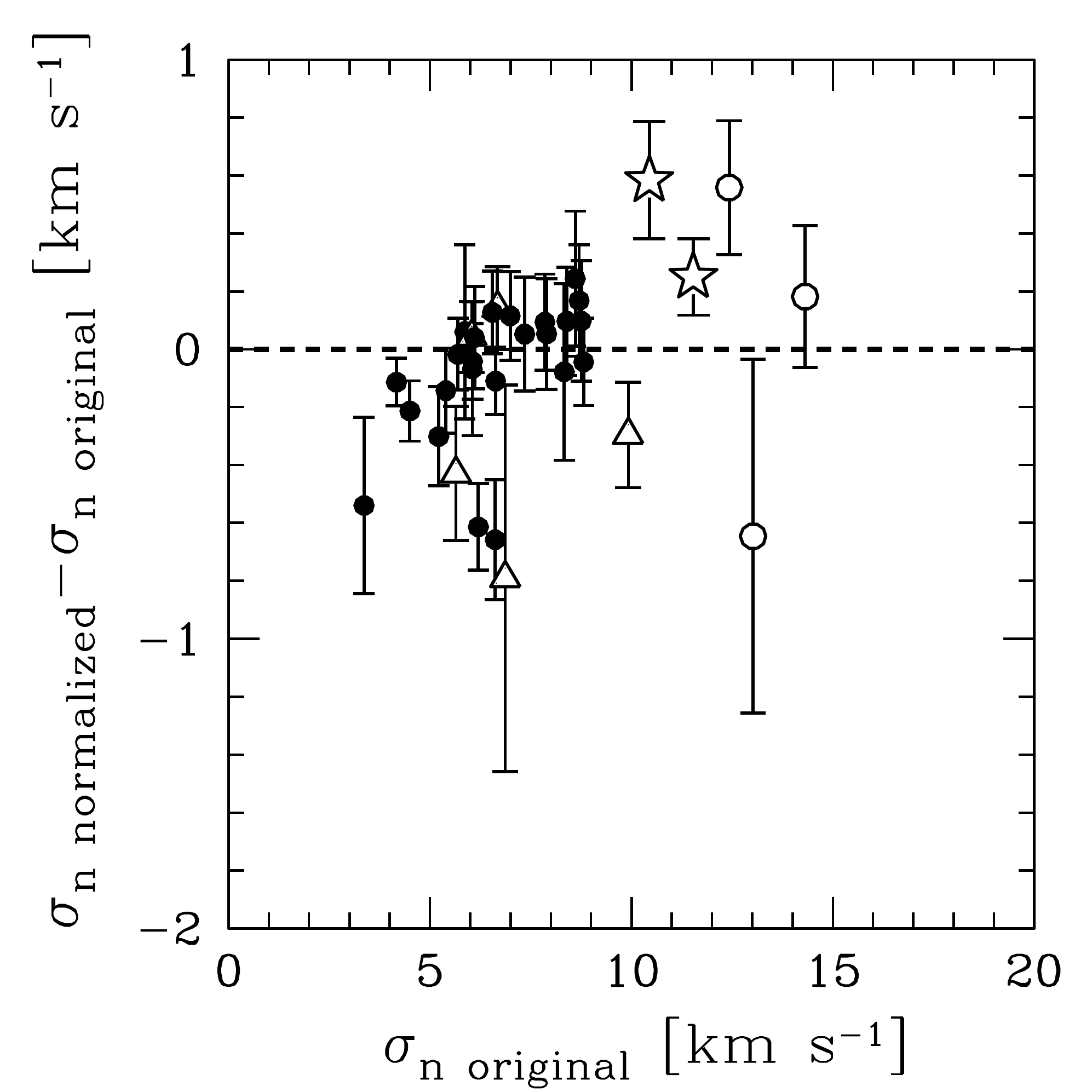}}}&
\rotatebox{0}{\resizebox{58mm}{!}{\includegraphics[width = 0.6in,height = 0.6in]{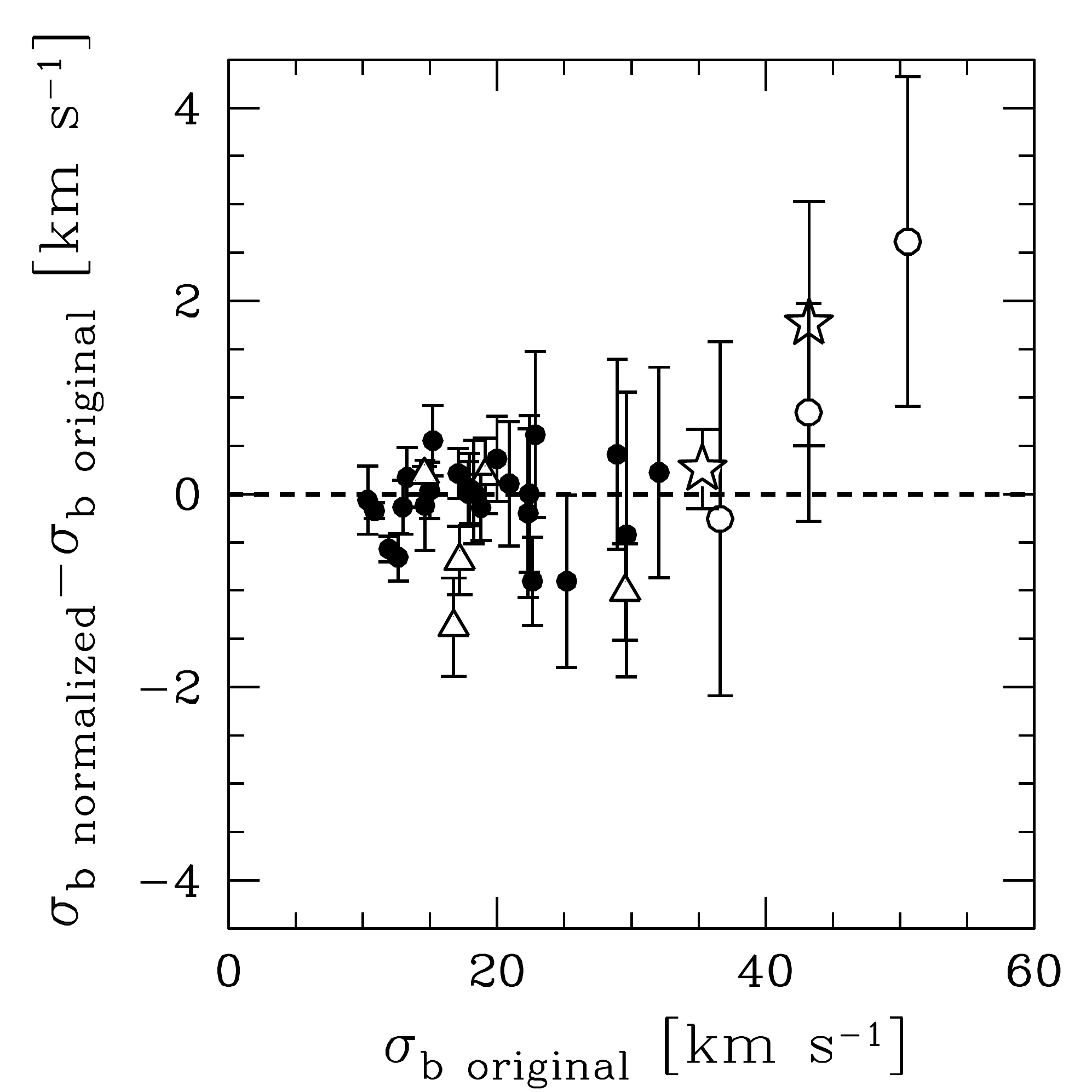}}}&
\rotatebox{0}{\resizebox{58mm}{!}{\includegraphics[width = 0.6in,height = 0.6in]{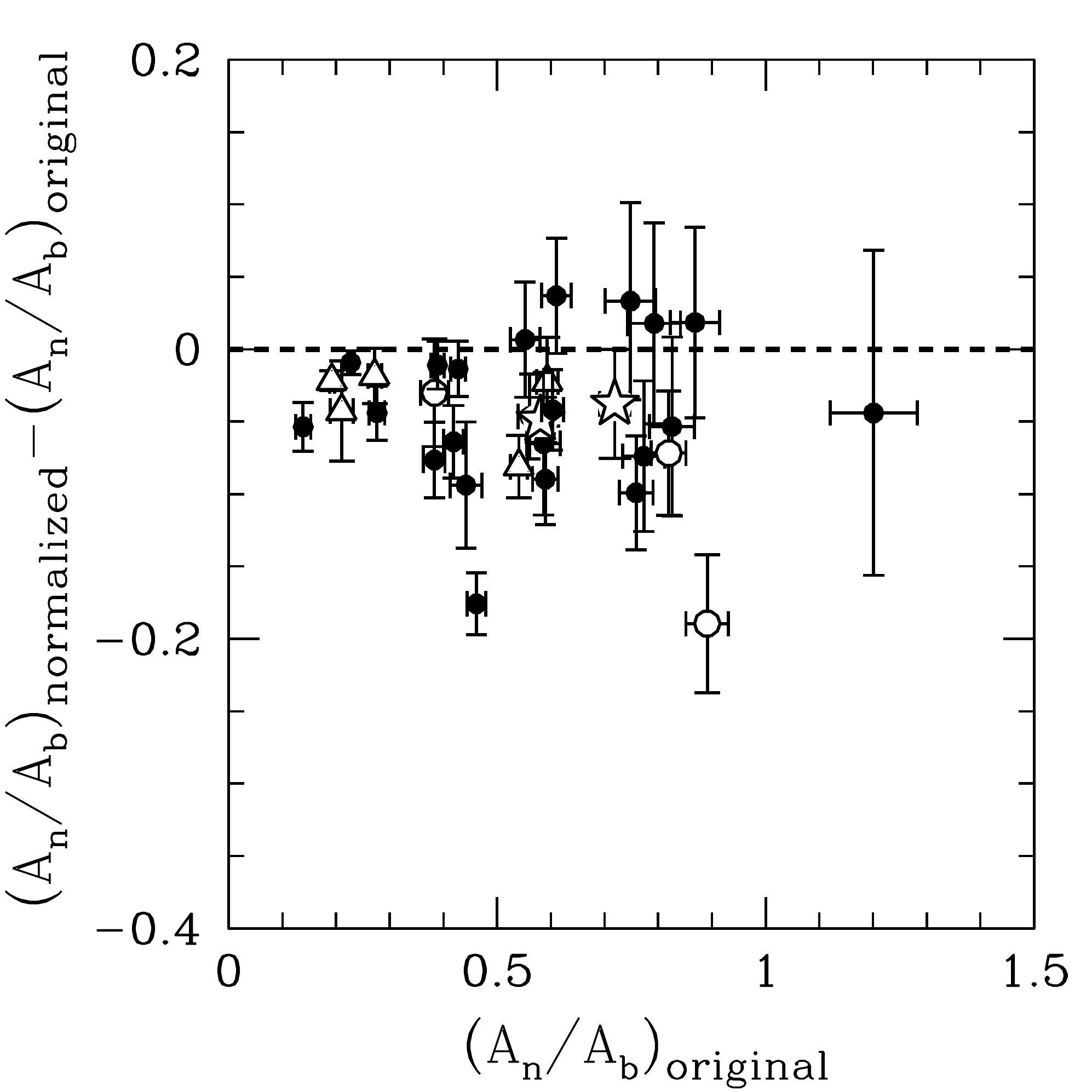}}}
    \end{tabular}
\caption{\textit{Top panel:} comparison of super profile parameters
  derived from the faint and bright parts of galaxies. \textit{Bottom
    panel:} comparison of super profile parameters from the normalized
  and the original super profiles. The triangle
  shaped symbols represent interacting or disturbed galaxies (NGC
  3031, NGC 4449, NGC 5194, NGC 5457, NGC 3077). The open circle
  symbols represent galaxies with star bursts or mild AGN activity
  (NGC 1569, NGC 3521, NGC 3627). The star symbols indicate
  non-interacting galaxies that have an anomalously high velocity
  dispersion (NGC 7331, NGC 2841). The rest of the galaxies are represented by the filled circle symbols.}
\label{fig:faint_brigh_norm}
 \end{figure*}

We compare the parameters of the faint and the bright super profiles
of each galaxy in Figure \ref{fig:faint_brigh_norm}. Most velocity
dispersion points scatter around the line of equality. However, four
high dispersion galaxies systematically deviate.  These are NGC 2841,
NGC 3521, NGC 3627 and NGC 925 (in order from highest to lowest
difference). This may indicate some kind of bulk motion of gas caused
by, e.g., interaction or tidal effects in these galaxies. In terms of
area, the faint super profiles show $A_{n}/A_{b}$ ratios that tend to
be smaller than those of the bright super profiles. The mean
difference between the bright and faint $A_{n}/A_{b}$ ratios is
$\sim$36\%. As most of the faint profiles are found in the outer disks
of the galaxies, this seems to indicate that the $A_{n}/A_{b}$ ratio
in the outer HI disks is smaller than in the inner disks. This result
will be discussed further in a future paper.  On the whole though, it
is clear that the dispersions of the super profiles of the bright and
faint parts are not significantly different.

As an additional test, we also create super profiles by normalizing
individual profiles by their peak intensity before summing them. This
way, any possible effects of a small number of bright input profiles
are removed. Figure \ref{fig:faint_brigh_norm} shows that, in terms of
velocity dispersion, the normalized super profiles are very similar to
the original ones. The correlation coefficient is $R^{2}\simeq0.99$
for both the narrow and the broad components. In terms of area, the
$A_n/A_b$ ratios derived from the normalized super profiles tend to be
smaller than those from the original super profiles. The difference
between the mean values is, however, only $\sim$1.3\%. In summary, the
shapes of the super profiles are not dominated by high intensity, narrow 
dispersion input profiles.

\subsection{Effects of radial motions}\label{sub:radialmotion}

Having ruled out resolution effects and high-intensity input profiles
as the cause for the non-Gaussianity of the observed super profiles,
we now investigate the effects of large-scale radial motions.  A net
radial motion in one direction will create an asymmetric tail in the
velocity profiles.  More specifically, for a given radial motion, the
location of the tail with respect to the profile depends on the
spatial position of the profile: a radial motion causing a tail on the
high-velocity side of a profile that is located on one side of the
major axis of the galaxy, will create a tail on the low-velocity side
of the profile when it is located on the opposite side of the major
axis. See \citet{warneretal73} and \citet{vanderkruitallen78} for a
detailed discussion of these kind of patterns typically observed in
velocity fields and profiles.

\begin{figure*}
\begin{tabular}{l}
\centerline{\rotatebox{0}{\resizebox{170mm}{!}{\includegraphics{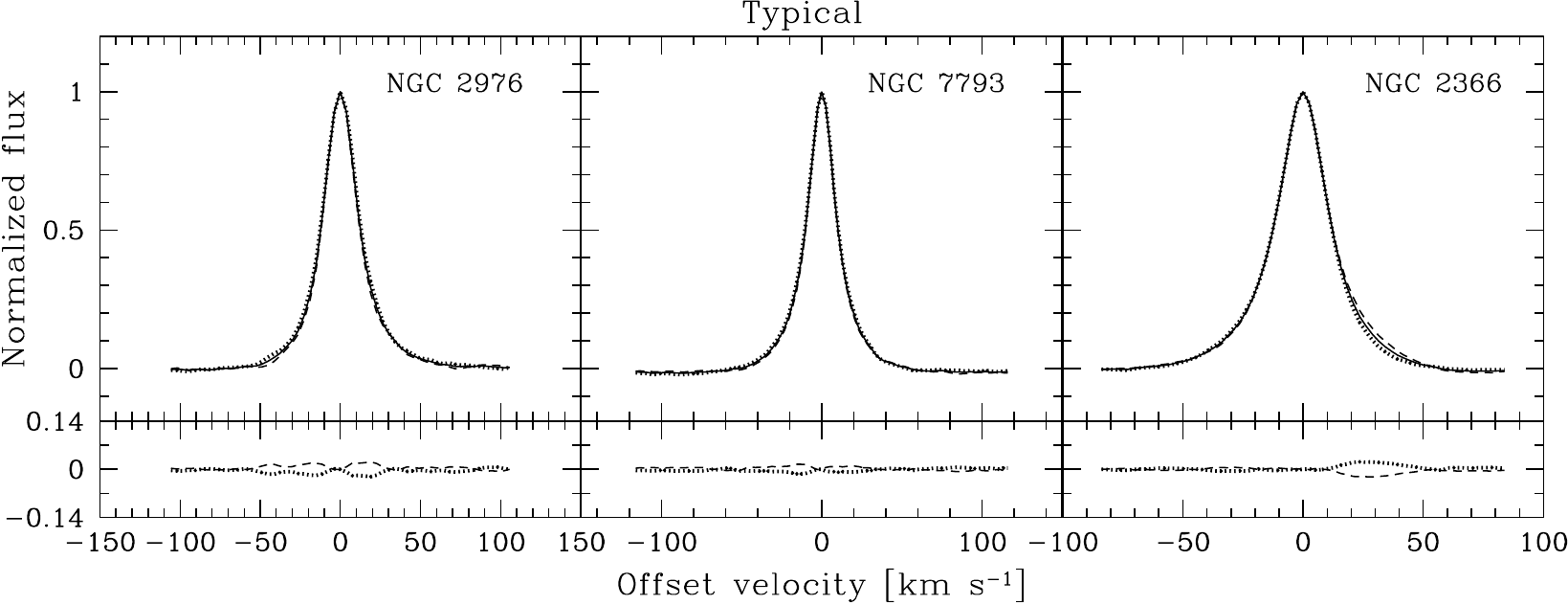}}}}\\\\
\centerline{\rotatebox{0}{\resizebox{170mm}{!}{\includegraphics{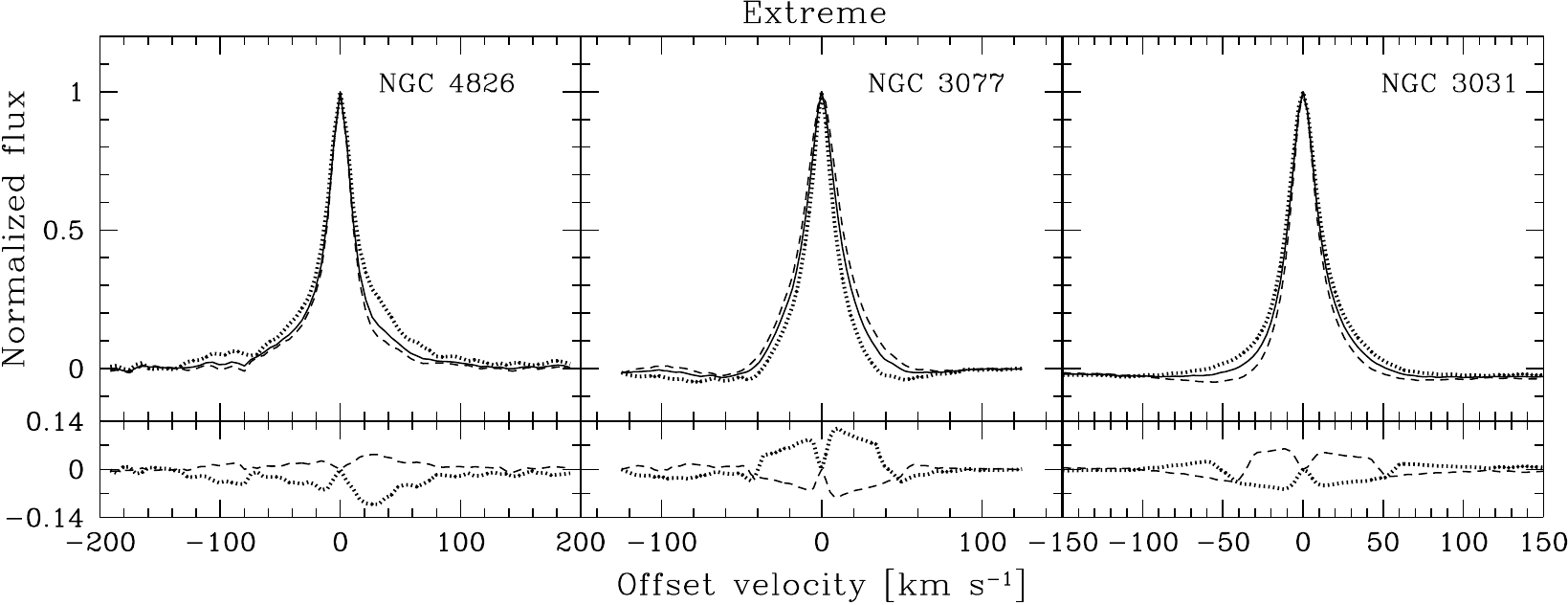}}}}
\end{tabular}
\caption{Super profiles extracted from the opposite sides of the
  major axes of each galaxy. The super profiles from the two halves of
  each galaxy are represented as dashed and dotted lines. The overall
  super profiles are represented by the solid lines. At the bottom of
  each panel is the difference between the super profiles from the two
  sides and the total super profiles. Note that the super profiles are
  normalized to their peak values.}
\label{fig:above_below}
\end{figure*}

\begin{figure}
\includegraphics[width=3in,height=3in]{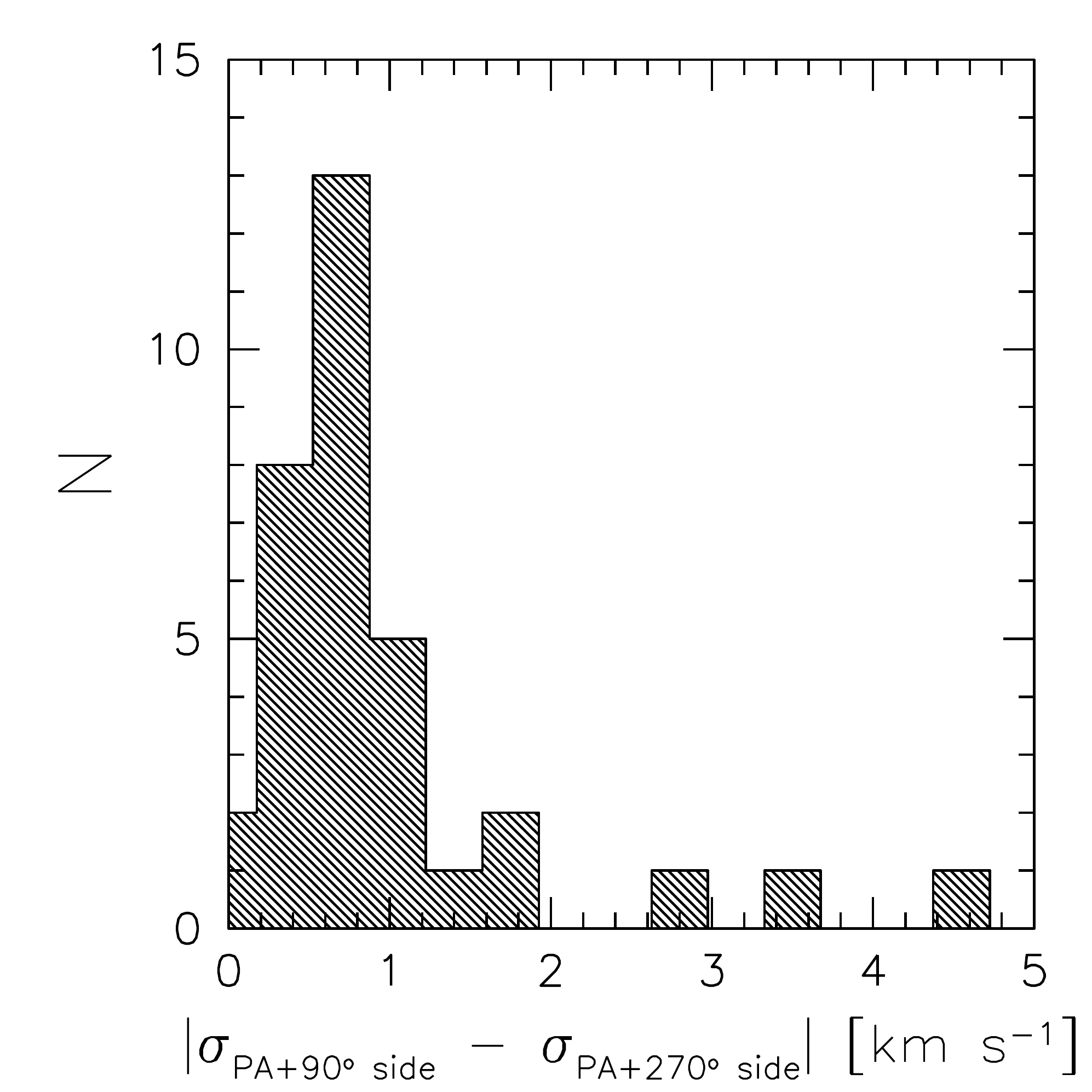}
\caption{Histogram of the difference between the velocity dispersion
  derived from the opposite sides of the major axis, where PA+$90^{\circ}$ side refers to 
  the position 
  corresponding to a position angle (PA) of $90^{\circ}$ with respect to the PA of the 
  receding major axis and PA+$270^{\circ}$ side is the position corresponding 
  to a PA offset of $270^{\circ}$ from the receding major axis.}
\label{fig:com_large_small_MAJOR}
\end{figure}

\begin{figure}
    \begin{tabular}{l}
\includegraphics[width = 3in,height = 3in]{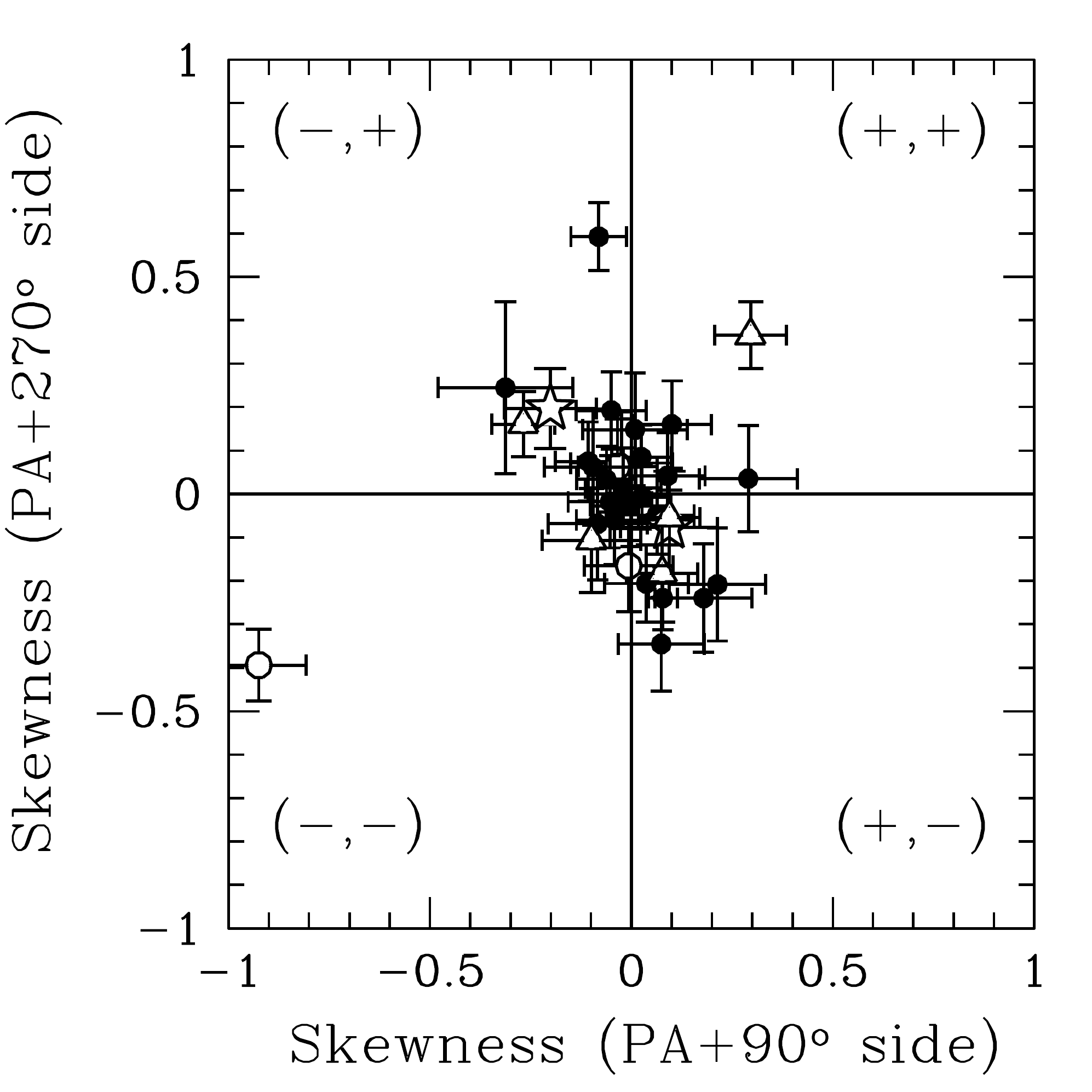}
    \end{tabular}
\caption{Skewness (defined as $4\sqrt{3}h_{3}$, where $h_{3}$ is the third 
coefficient of the Gauss-Hermite function.) 
of the super profiles derived from the opposite sides of the major axis. \textit{PA+$90^{\circ}$ side}: 
position corresponding to a position angle (PA) of $90^{\circ}$ with respect to 
the PA of the receding major axis. \textit{PA+$270^{\circ}$ side:} position corresponding to a PA 
offset of $270^{\circ}$ from the receding major axis.The three outlying galaxies are, from left 
to right, NGC 1569, DDO 53 and NGC 3031. The plus and minus signs 
represent the signs of the skewness parameters. Symbols follow Fig.
  \ref{fig:faint_brigh_norm}.}
\label{fig:rad_skew}
\end{figure}

Figure \ref{fig:above_below} shows examples of super profiles
extracted from the opposite sides of the major axes of each galaxy overplotted on
top of the overall super profiles of these galaxies. We use a consistent definition of sides 
corresponding to their average position angle (PA). One side of the galaxy is 
chosen as the one with an average position angle of 90$^{\circ}$ with respect to the 
receding semi-major axis (denoted as the PA+90$^{\circ}$ side); the other 
side has an average PA of 270$^{\circ}$ 
with respect to the receding semi-major axis (the PA+270$^{\circ}$ side). 
We show in Figure
\ref{fig:com_large_small_MAJOR} a histogram of the difference between
the velocity dispersion derived from the opposite sides with respect to the 
major axes of each galaxy. For about 90\% of the studied galaxies, the
difference is less than 1.5 $\rm{km~s^{-1}}$. There are four galaxies
with higher velocity dispersion differences. These are NGC 3031, NGC
3077 (both interacting), NGC 4826 (counter-rotating HI disk) and NGC
7331. The latter has the highest inclination (76$^{\circ}$) in the
sample and the difference in velocity dispersion for this galaxy may
be a projection effect. The velocity dispersion comparison 
presented above only works if the distortion due to radial motions is asymmetric. Therefore, 
we also measure the skewness of
each of the super profiles on the opposite sides of the major axis. 
We fit the super profiles with a 
Gauss-Hermite $h_{3}$ function. The skewness coefficient is 
defined as $4\sqrt{3}h_{3}$ and the results are
shown in Fig.~\ref{fig:rad_skew}. The apparent trend visible in the  
figure is mostly caused by highly inclined, interacting or disturbed, 
as well as starburst galaxies. It is possible that these galaxies suffer from radial motions, 
and thus they will not be part of our subsequent analysis, 
as described in detail in Sect.~\ref{sec:cleansample}. 
The remaining outliers have very symmetrical profiles with only some faint extended wings 
showing up at levels well below 20\% of the peak intensity. While these affect the skewness, 
they contain negligible flux, and have no impact on the dispersions or flux ratios. Most likely these 
apparent wings are caused by residual baseline effects.

In the following, we develop simple models to understand how radial motions 
affect the shapes of the super profiles. We do this by creating models which have different 
ratios of rotation and radial motions. We again use the $63^\circ$
inclination NGC 2403 model discussed in Sect.~\ref{sub:resolution}.
We create four new model data cubes by adding outward radial motions
of 5, 10, 15, and 20 $\rm{km~s^{-1}}$ respectively to the original 
model (thus shifting the input profiles in velocity). We again assume
a single-component Gaussian velocity dispersion of $8~\rm{km~s^{-1}}$, 
a constant rotation curve and a constant surface density 
HI disk for all models. 

For each of the 5, 10, 15 and 20 $\rm{km~s^{-1}}$ model data cubes, we
randomly choose 20\%, 40\%, 60\%, 80\% of the profiles and add these to the 
original NGC 2403 model (which contains no radial
motions). We thus create 16 data cubes with different
fractions of the disk area containing asymmetric profiles with different amounts
of radial motion. We smooth these model data cubes to the THINGS
$11^{\prime\prime}$ resolution. The velocity fields derived from the combined, 
smoothed cubes are then used to shuffle profiles and derive the corresponding 
super profiles, which we then fit with single Gaussian functions.

\begin{figure*}
\begin{tabular}{l l}
\includegraphics[width=3in,height=3in]{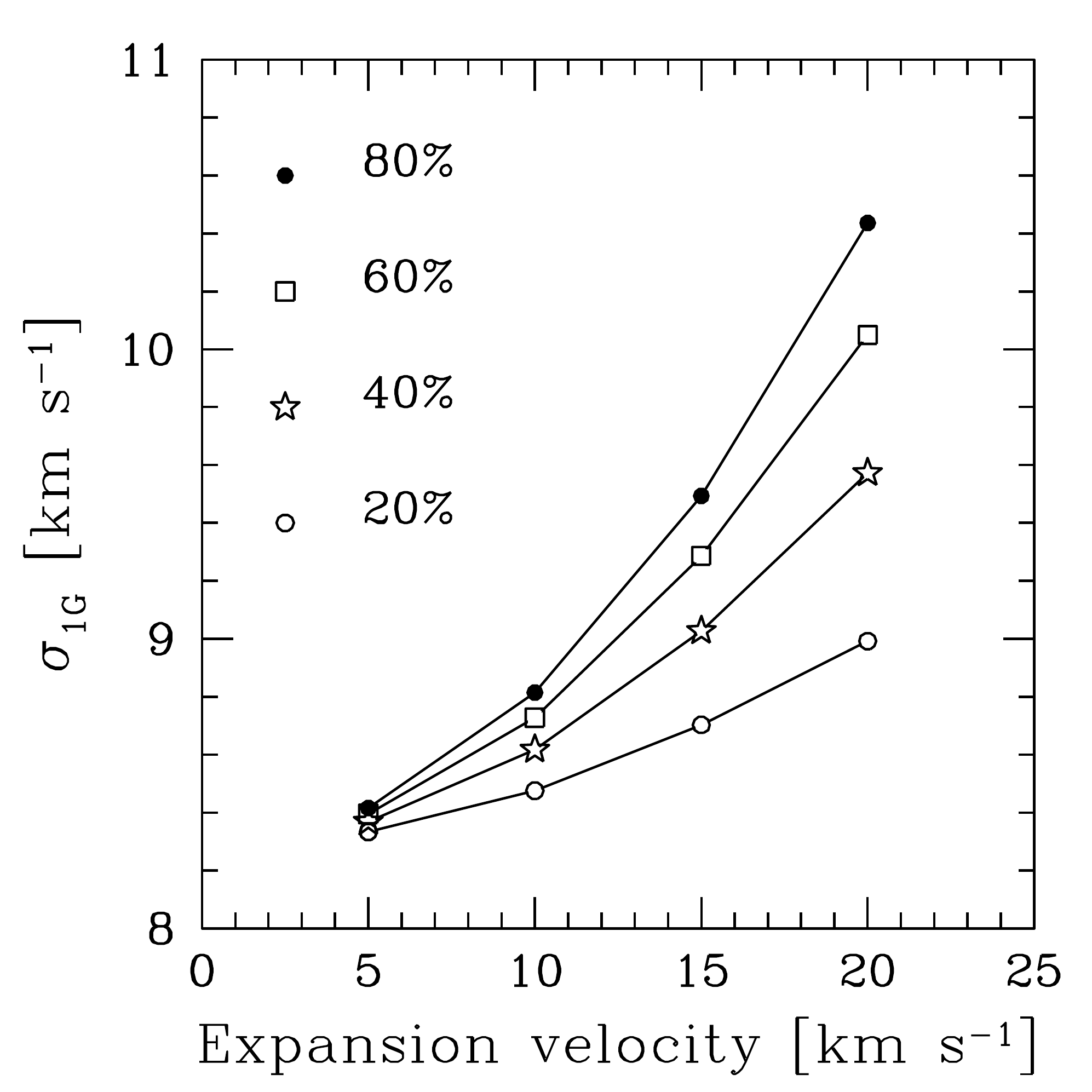} &
\includegraphics[width=3in,height=3in]{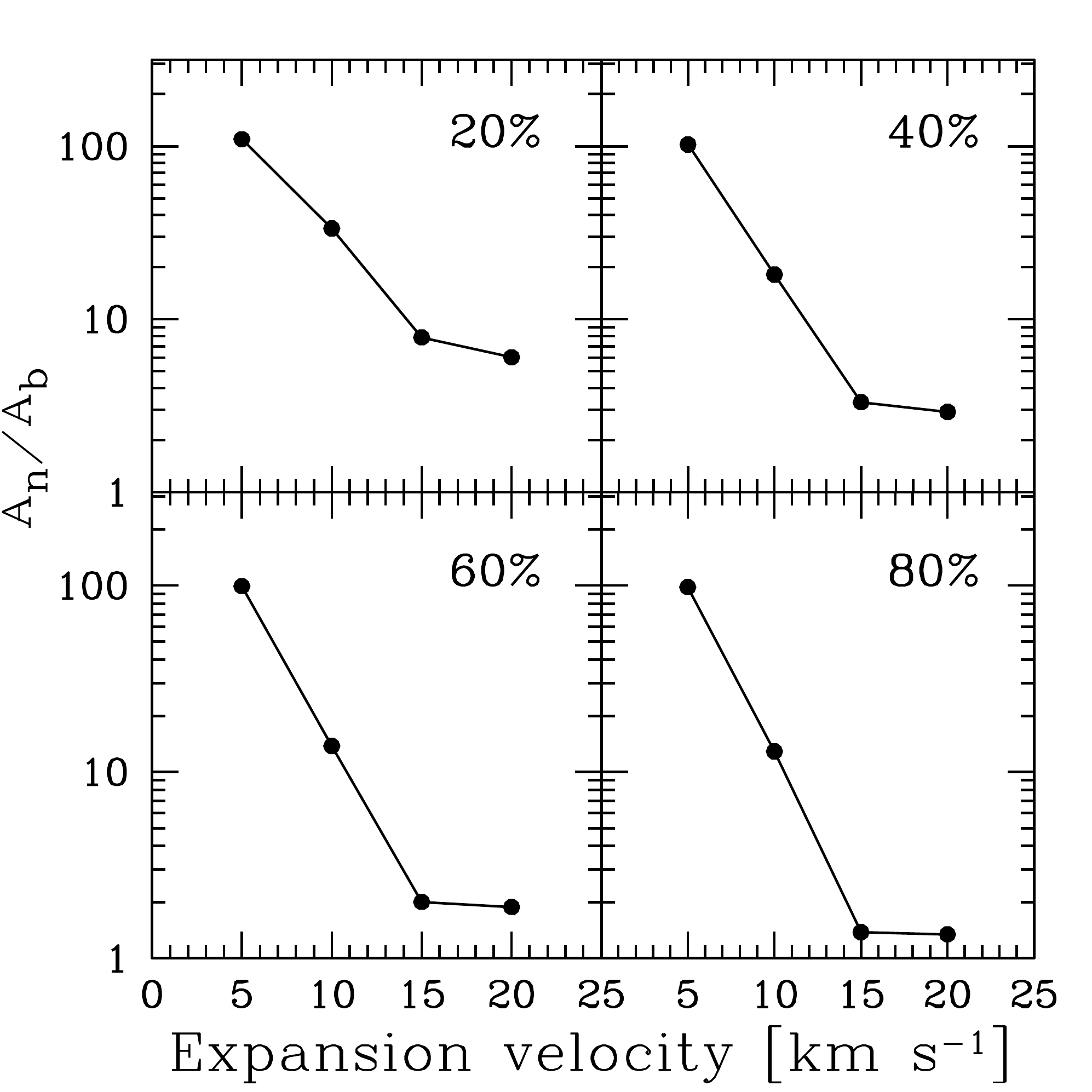}
\end{tabular}
\caption{\textit{Left panel}: single Gaussian velocity dispersions
  derived from the first set of model data cubes of NGC 2403 as a
  function of the input expansion velocities used to construct the
  models (see Sect.~\ref{sub:radialmotion}). \textit{Right panel:}
  $A_{n}/A_{b}$ ratio derived from the second set of model data cubes
  of NGC 2403 as a function of the expansion velocities (see
  Sect.~\ref{sub:radialmotion}). Percentage indicates the fraction of
  profiles having non-zero expansion velocities in the model data cubes.}
\label{fig:radial_motion}
\end{figure*}

Figure \ref{fig:radial_motion} shows the resulting velocity
dispersions (as measured with a single Gaussian) as a function of the
input expansion velocities. As expected, the velocity dispersions
increase with increasing expansion velocities. The increase is steeper
as we increase the fraction of gas with non-zero expansion velocities.
We see that a radial motion of 5 $\rm{km~s^{-1}}$ causes only minimal
broadening, independent of the area coverage. The most extreme model
with 20 $\rm{km~s^{-1}}$ radial motions and 80\% area coverage has a
superprofile that is broadened by $\sim$ 23\%. Large-scale non-circular motions
of this magnitude are rare in disk galaxies, and mostly only found in
the central parts of barred galaxies. We can compare this with the
work by \citet{trachternachetal08} who used harmonic decompositions of
the velocity fields of the THINGS galaxies to quantify the
non-circular motions there. They found a median amplitude of all
non-circular motions of $\tilde{A}_{r} = 6.7 \pm 5.9$ km s$^{-1}$.  Of
their 18 analyzed galaxies, 17 have $\tilde{A}_{r}$ $\lesssim$ 10 km
s$^{-1}$ and only one has $\tilde{A}_{r} > 20$ km s$^{-1}$.

Finally, we repeat the procedure described above, but rather than
assuming the same constant HI surface density disk for all models, we
assume that the profiles with non-circular motion have an amplitude
that is 30\% of that of the profiles without non-circular motions.
This choice will allow the non-circular motion profiles to mimic
low-level broad wings in the super profiles. In this case we fit all
resulting super profiles with double Gaussians. We show in Figure
\ref{fig:radial_motion} and Table \ref{tab:radial_motion} the fitted
$A_{n}/A_{b}$ ratios as a function of input expansion velocities. The
$A_{n}/A_{b}$ ratio decreases with increasing expansion velocities. If
we assume that the non-circular motions in the THINGS galaxies are of
order 10 km s$^{-1}$ or less \citep{trachternachetal08}, then these
motions are only able to introduce spurious broad wings containing at
most 7\% of the total flux, i.e., much smaller than observed. We
therefore conclude that radial motions cannot be a major cause of the
observed non-Gaussianity of the observed super profiles.

\begin{deluxetable}{c c c c c }
\centering
\tabletypesize{\scriptsize}
\tablecaption{Effect of the radial motions on the shapes of the 
	super profiles \label{tab:radial_motion}}
\tablewidth{0pt}
\tablehead{ 
	$V_{rad}$        & \multicolumn{4}{c}{$A_{n}/A_{b}$}\\
	(km s$^{-1}$)    & 20\%  & 40\%        & 60\%  & 80\%
}
\startdata
     5           & 109.8 & 102.1       & 99.3  & 98.2\\       
    10           & 33.5  & 18.1        & 13.8  & 12.9\\
    15           & 7.9   & 3.3         & 2.0   & 1.4\\
    20           & 6.0   & 2.9         & 1.9   & 1.3\\ 
\enddata
\tablecomments{$V_{rad}$: Input radial velocity used to 
construct the model data cubes. Percentage indicates the area 
coverage of gas with non-circular motions. $A_{n}/A_{b}$: 
Flux ratio of the narrow and broad components.}
\end{deluxetable}

\subsection{Thick disks and lagging halos}\label{sub:thickdisk}

Lastly, we investigate the possibility that our sample galaxies have a
thin disk embedded in a ``lagging" thick HI disk with a lower rotation
velocity, as was found for NGC 891 \citep{swatersetal97,
  oosterlooetal07} and NGC 2403 \citep{fraternalietal02}. The super
profiles then represent a ``cross-cut'' through the global dynamics of
the system, rather than a reflection of the state of the ISM. If this
were the case, the lagging disk would broaden the profiles
asymmetrically, depending on whether the approaching or the receding
side of the galaxy is studied. Adding these two halves would give
the impression of the super profile having broad wings.

\begin{figure*}\begin{tabular}{l}
\centerline{\rotatebox{0}{\resizebox{170mm}{!}{\includegraphics{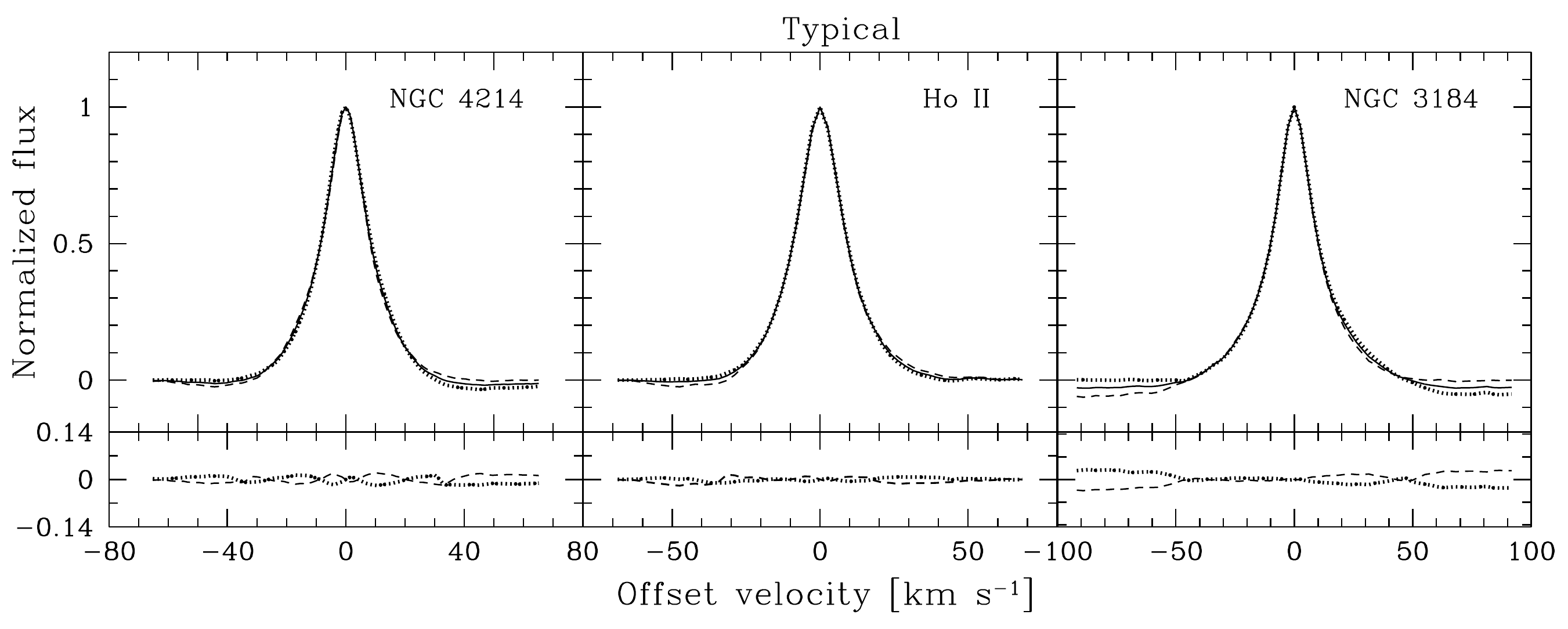}}}}\\\\
\centerline{\rotatebox{0}{\resizebox{170mm}{!}{\includegraphics{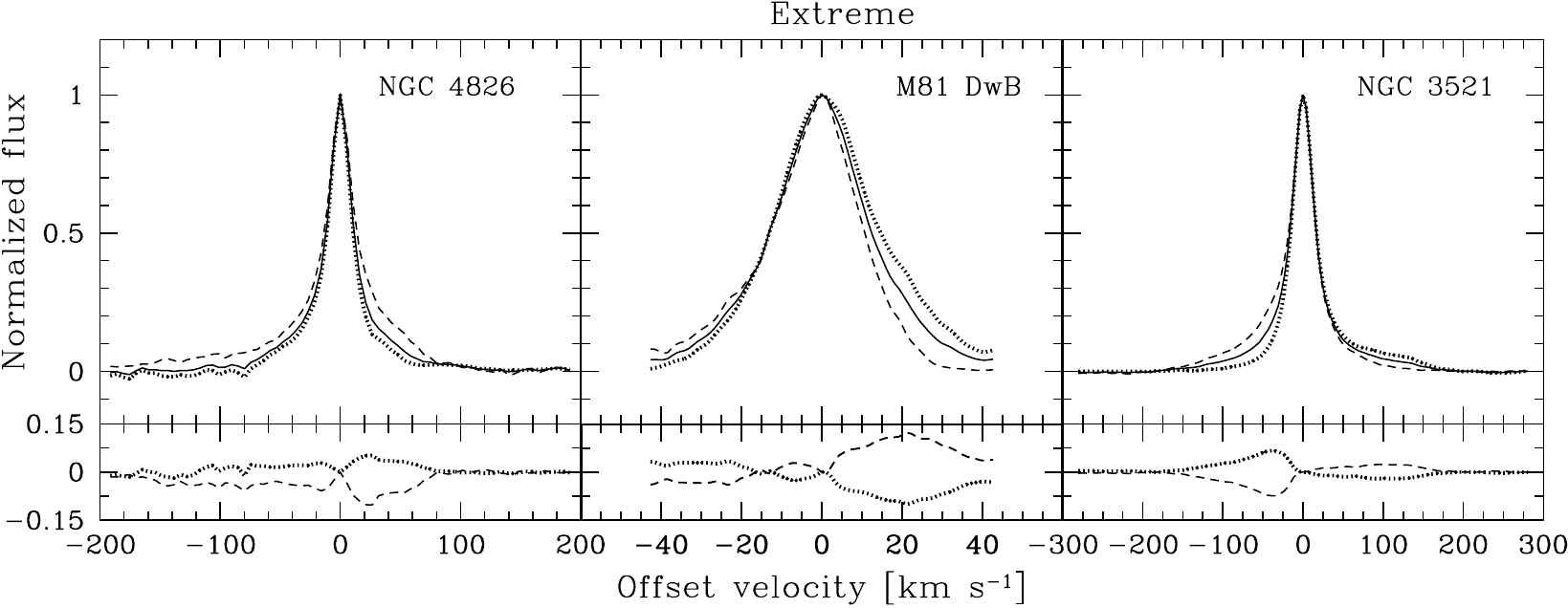}}}}
\end{tabular}
\caption{Super profiles extracted from the approaching and receding
  sides of the galaxies. The super profiles from the two halves of
 each galaxy are represented as dashed and dotted lines. The overall
  super profiles are represented by the solid lines. The bottom panels
  of each plots represent the difference between the super profiles
  from the two sides and the total super profiles. The super profiles
  are normalized to their peak values.}
\label{fig:approach_rec}
\end{figure*}

\begin{figure*}
\begin{tabular}{l l }
\includegraphics[width=3in,height=3in]{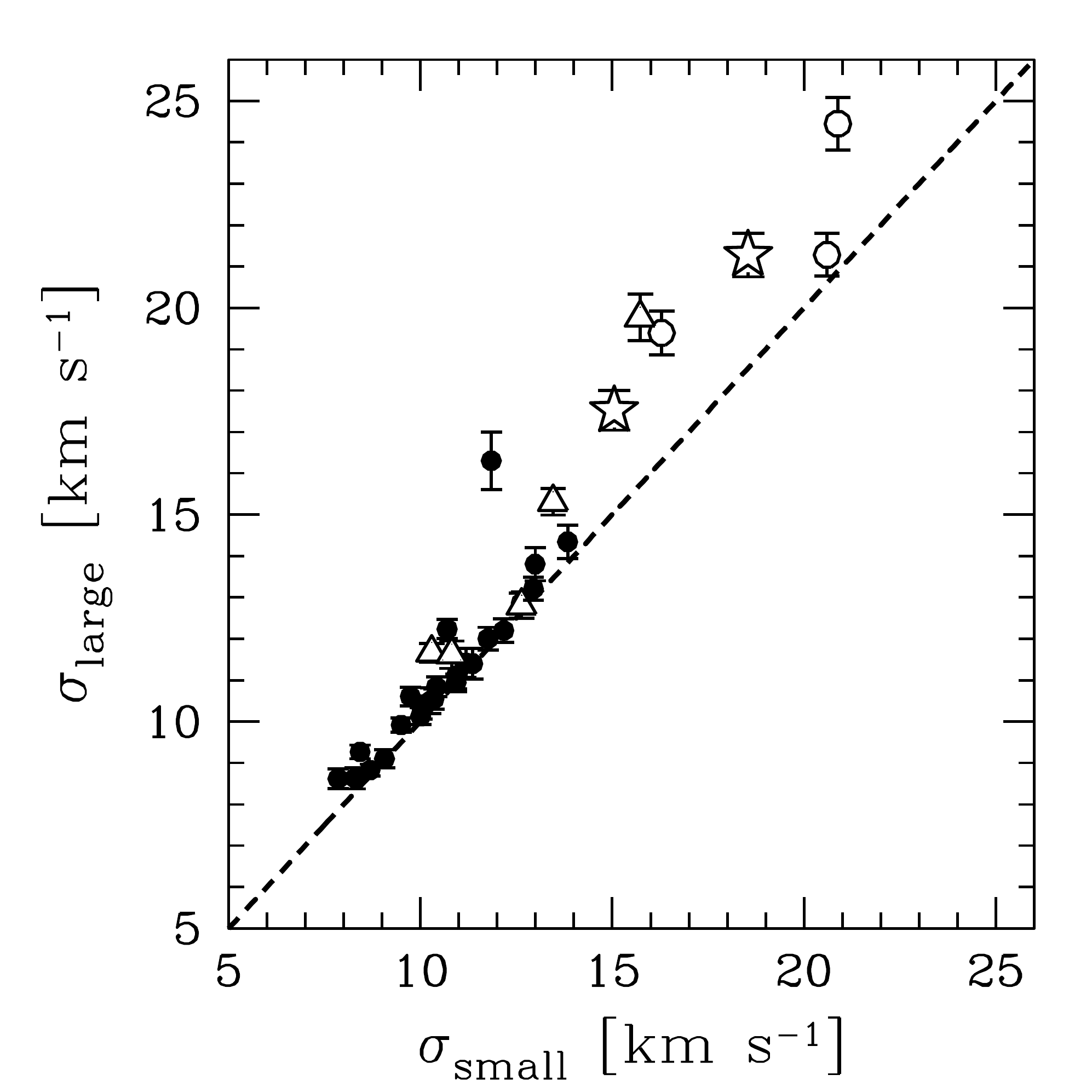}&%
\includegraphics[width=3in,height=3in]{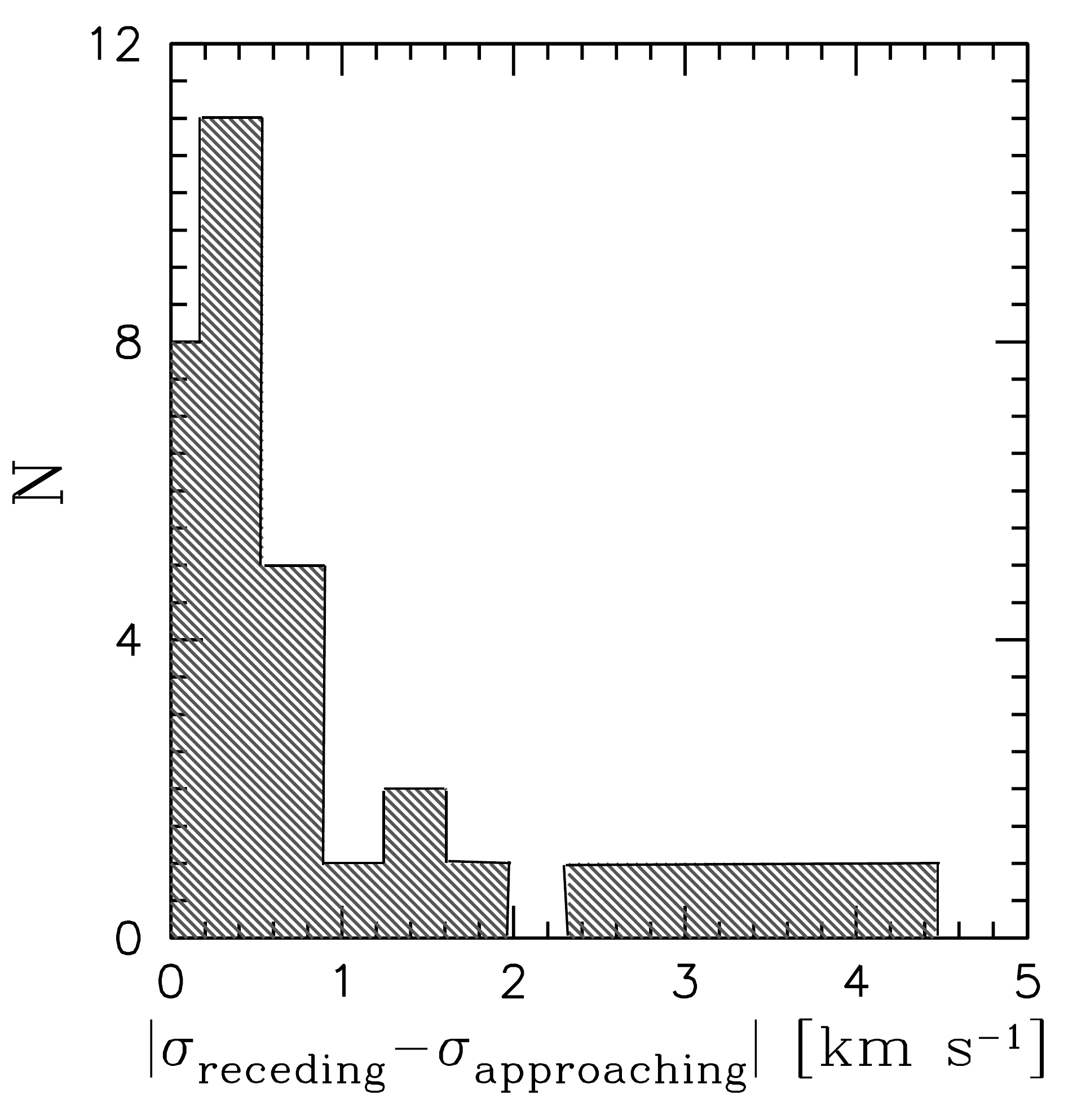} \\
\includegraphics[width=3in,height=3in]{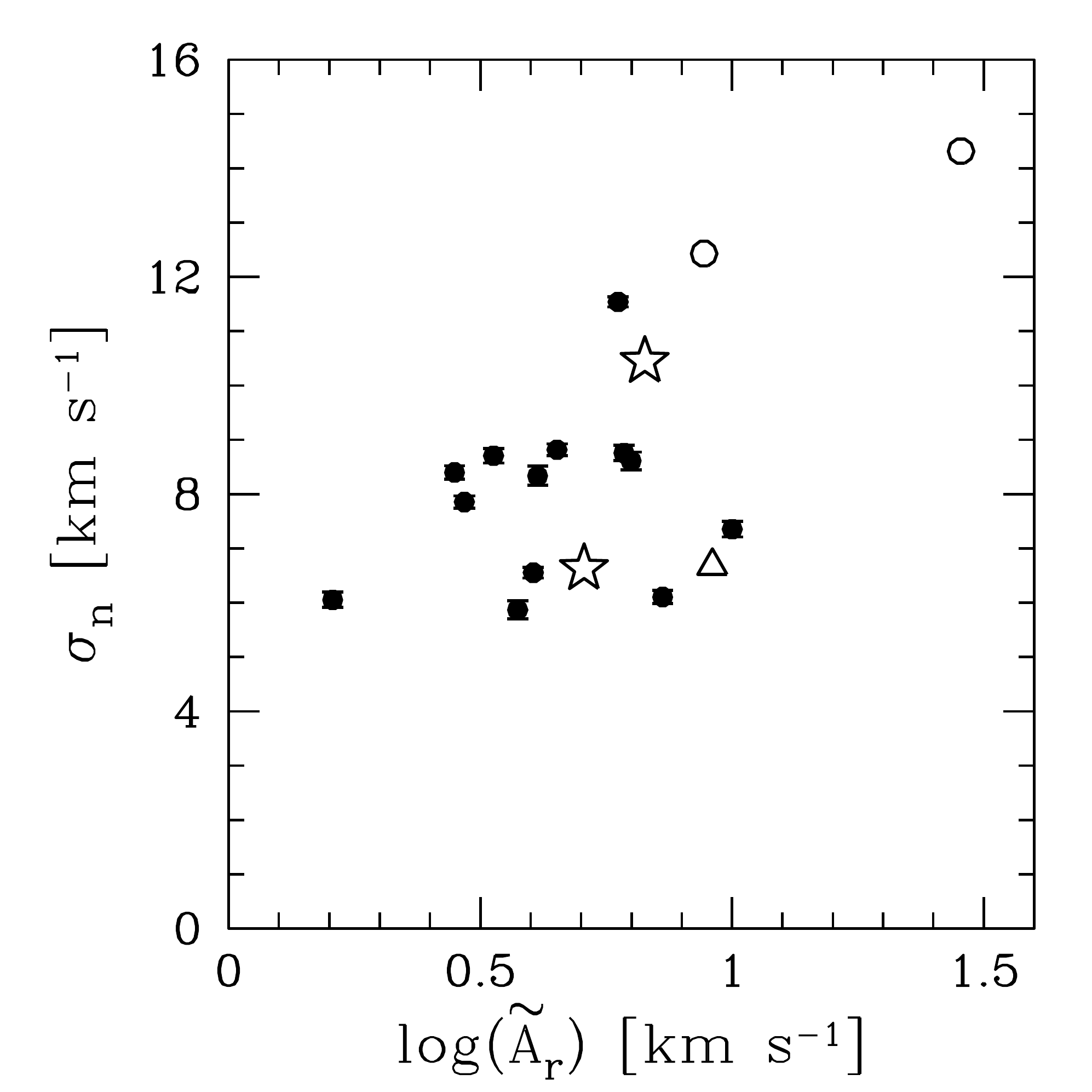}&%
\includegraphics[width=3in,height=3in]{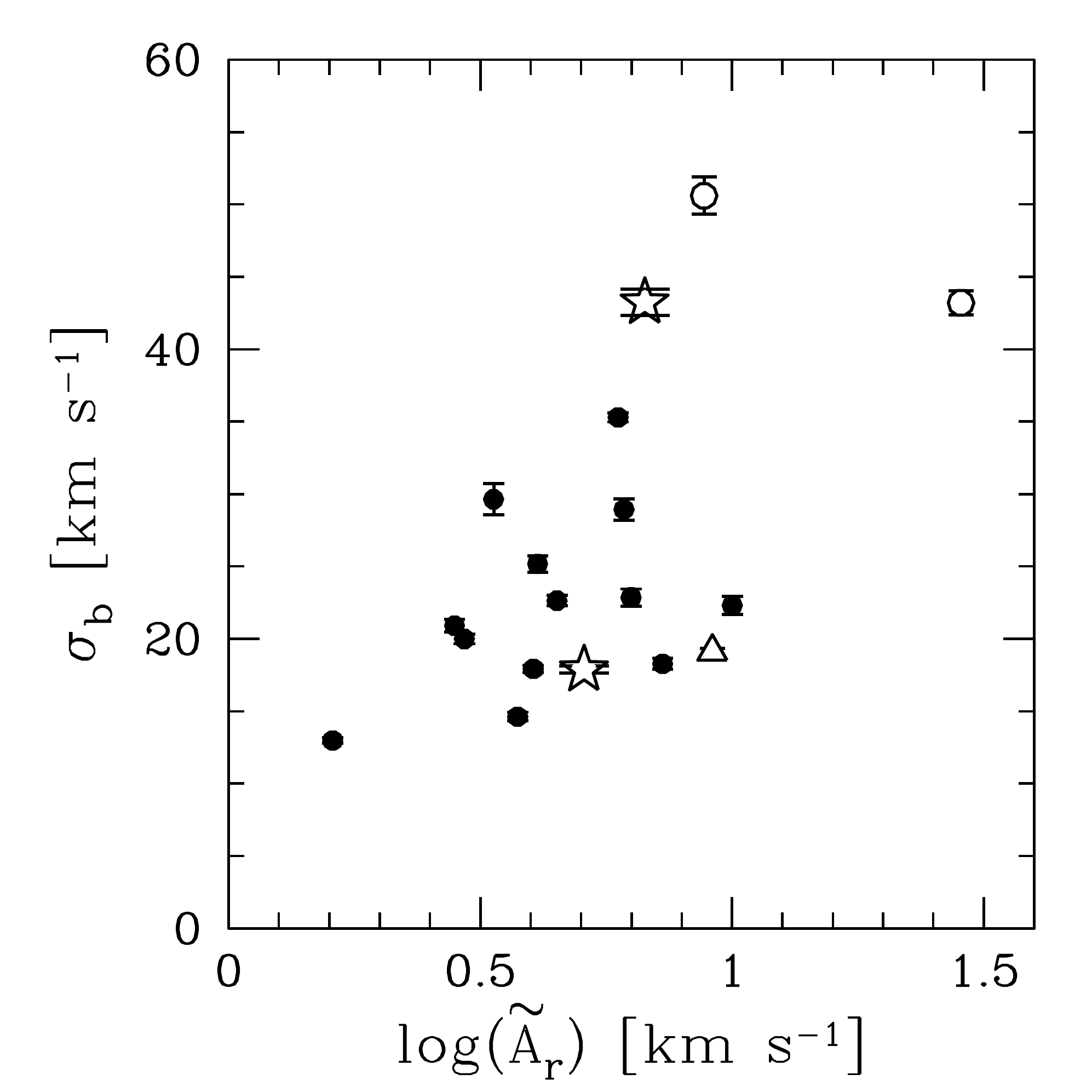} \\
\end{tabular}
\caption{\textit{Top left panel}: Comparison of the widths of the super
  profiles from the approaching and receding sides of the
  galaxies. The largest of each of these velocity dispersions is
  represented by $\rm{\sigma_{large}}$ and the smallest one by
  $\rm{\sigma_{small}}$. The dashed line represents the line of
  equality.\textit{Top right panel}: Histogram of the difference
  between the velocity dispersion derived from the approaching and
  receding halves of the galaxies. \textit{Bottom left panel:} Narrow
  component velocity dispersion as a function of the amplitudes of
  non-circular motions, $\tilde{A}$, derived by
  \citet{trachternachetal08}. \textit{Bottom right panel:} Broad
  component velocity dispersion as a function of $\tilde{A}$. Note
  that only the 18 galaxies that are part of
  \citet{trachternachetal08} sample are shown here.  Symbols are as in Fig.
  \ref{fig:faint_brigh_norm}. }
\label{fig:com_large_small}
\end{figure*}

\begin{figure}
    \begin{tabular}{l}
\includegraphics[width = 3in,height = 3in]{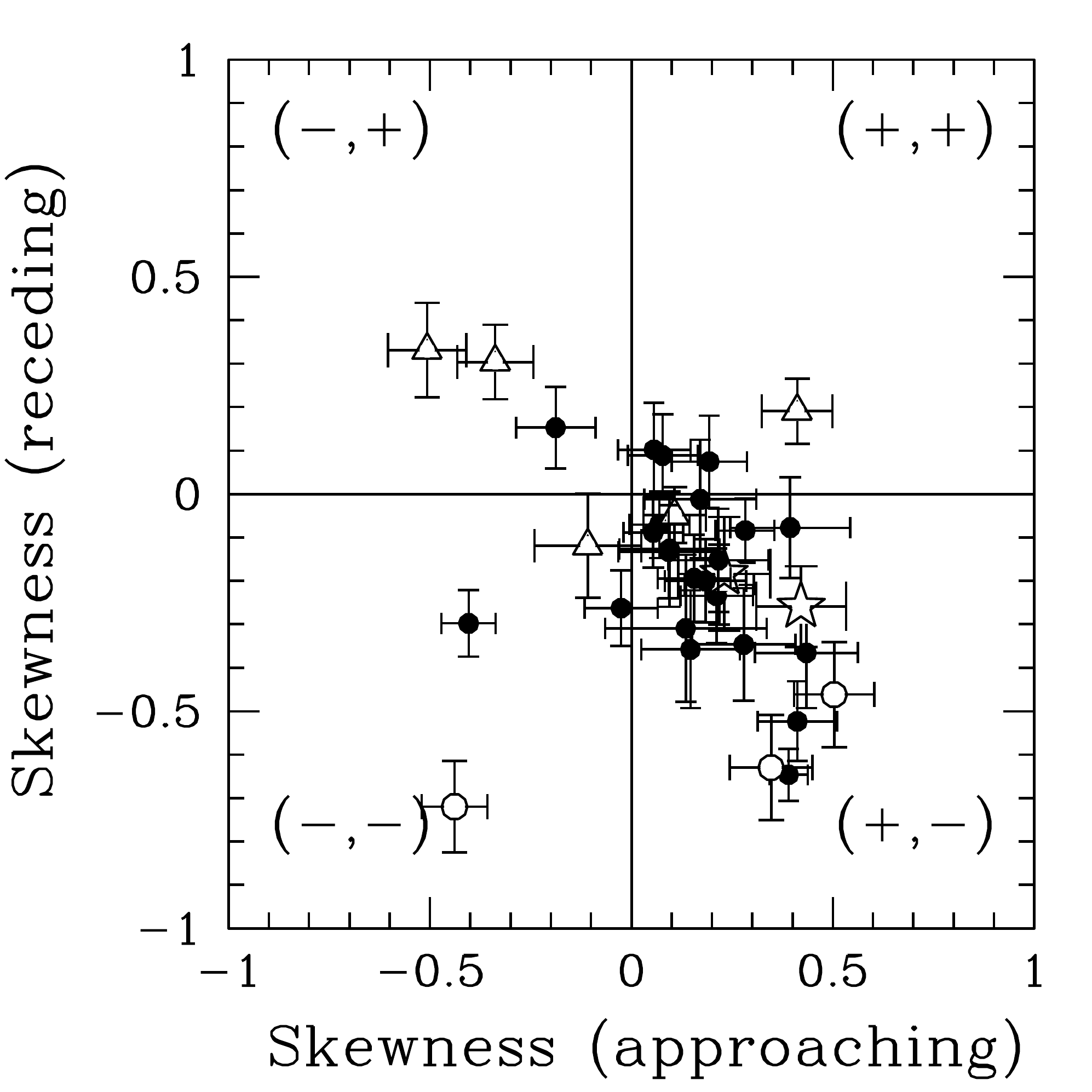}
    \end{tabular}
\caption{Skewness (defined as $4\sqrt{3}h_{3}$, where $h_{3}$ is the third Gauss-Hermite polynomial coefficient) 
of the super profiles derived from the receding 
and approaching side of each galaxy. The two outlying galaxies in the lower left 
quadrant are DDO 53 (top) and NGC 1569 (bottom).  The plus and minus signs 
represent the signs of the skewness parameters. Symbols are as in Fig.
  \ref{fig:faint_brigh_norm}.}
\label{fig:comp_skew_rec_ap}
\end{figure}

Figure \ref{fig:approach_rec} shows examples of super profiles from
the approaching and receding sides of a few typical sample galaxies
plotted on top of the super profiles derived from the entire
disk. Note that the super profiles plotted there are normalized to
their peak values. For reference we also show the most asymmetric
approaching and receding super profiles in the bottom panel of Fig.~\ref{fig:approach_rec}. 
We compare the velocity dispersions derived from the approaching and
receding sides in Figure \ref{fig:com_large_small}. For each galaxy,
the largest of the velocity dispersions derived from the two halves of
each galaxy is shown as $\sigma_{large}$, whereas the smallest one is
given by $\sigma_{small}$.  We also show a histogram of the difference
between the velocity dispersions derived from the two sides in Fig.~
\ref{fig:com_large_small}. Of the 34 galaxies, 24 have differences
less than 1 $\rm{km~s^{-1}}$; for the remaining 10 galaxies, the
difference is between 1 and 4.5 $\rm{km~s^{-1}}$. Of those 10
galaxies, 4 are interacting (NGC 3031, NGC 4449, NGC 5194, NGC 5457),
2 are kinematically disturbed (NGC 1569 and NGC 3521), one is known to
be a galaxy with a counter-rotating inner disk (NGC 4826). To complement 
our analysis, we also measure the coefficient of skewness of the super profiles 
in the approaching and receding sides. The result is presented in Figure 
\ref{fig:comp_skew_rec_ap}. Most of the points in Fig.~\ref{fig:comp_skew_rec_ap} 
are offsets towards the lower right quadrant. This means that super profiles in the receding sides 
tend to be negatively skewed, whereas those in the approaching sides tend to be 
positively skewed. These are consistent with but do not by themselves prove the existence of lagging 
thick HI disks. Most of the outliers in Fig.~ 
\ref{fig:comp_skew_rec_ap} will not be considered for further analysis as described in 
Sect.~\ref{sec:cleansample}. 

\section{Defining a clean sample}
\label{sec:cleansample}

\begin{figure*}
    \begin{tabular}{l l}
  \includegraphics[width = 3in,height = 3in]{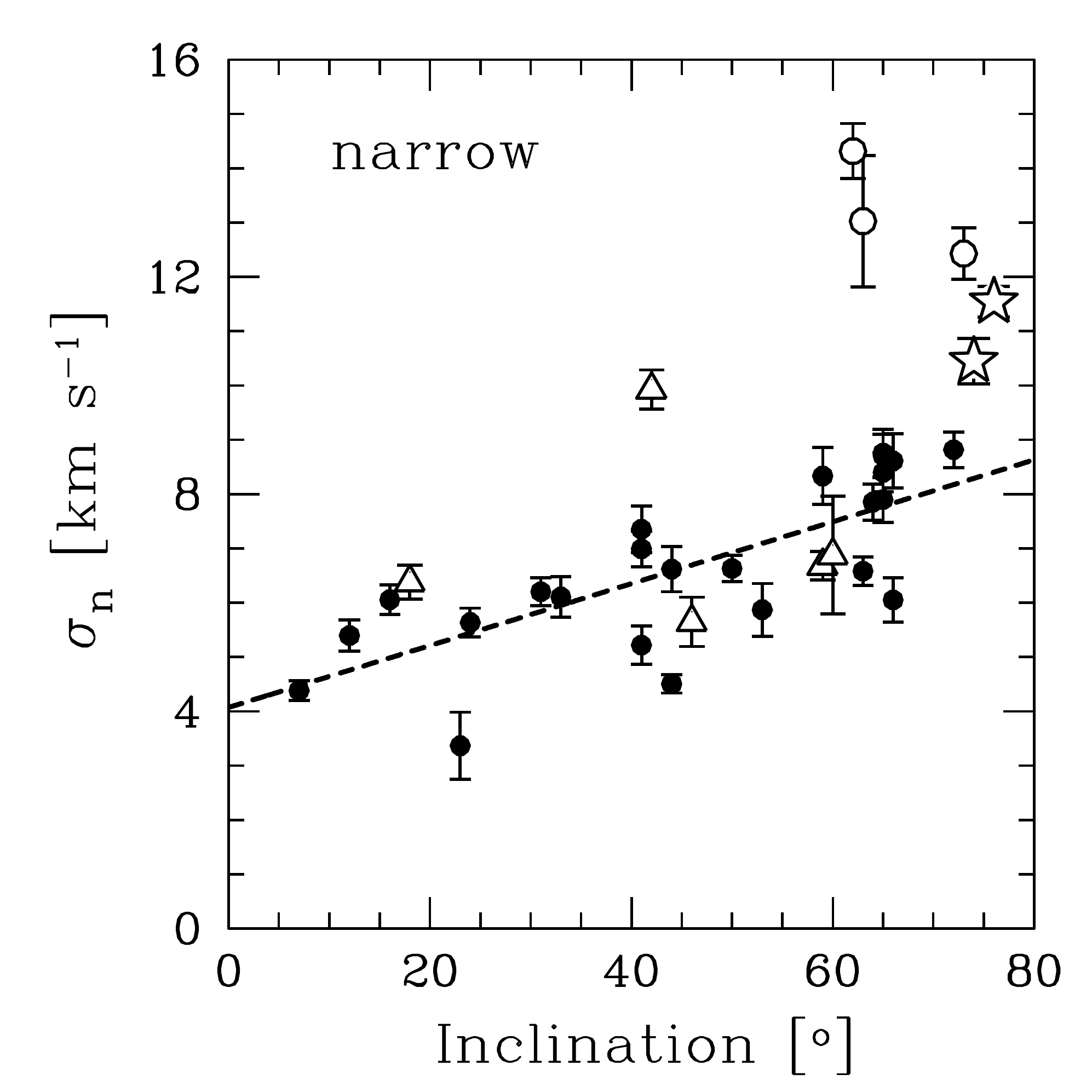}&
 \includegraphics[width = 3in,height = 3in]{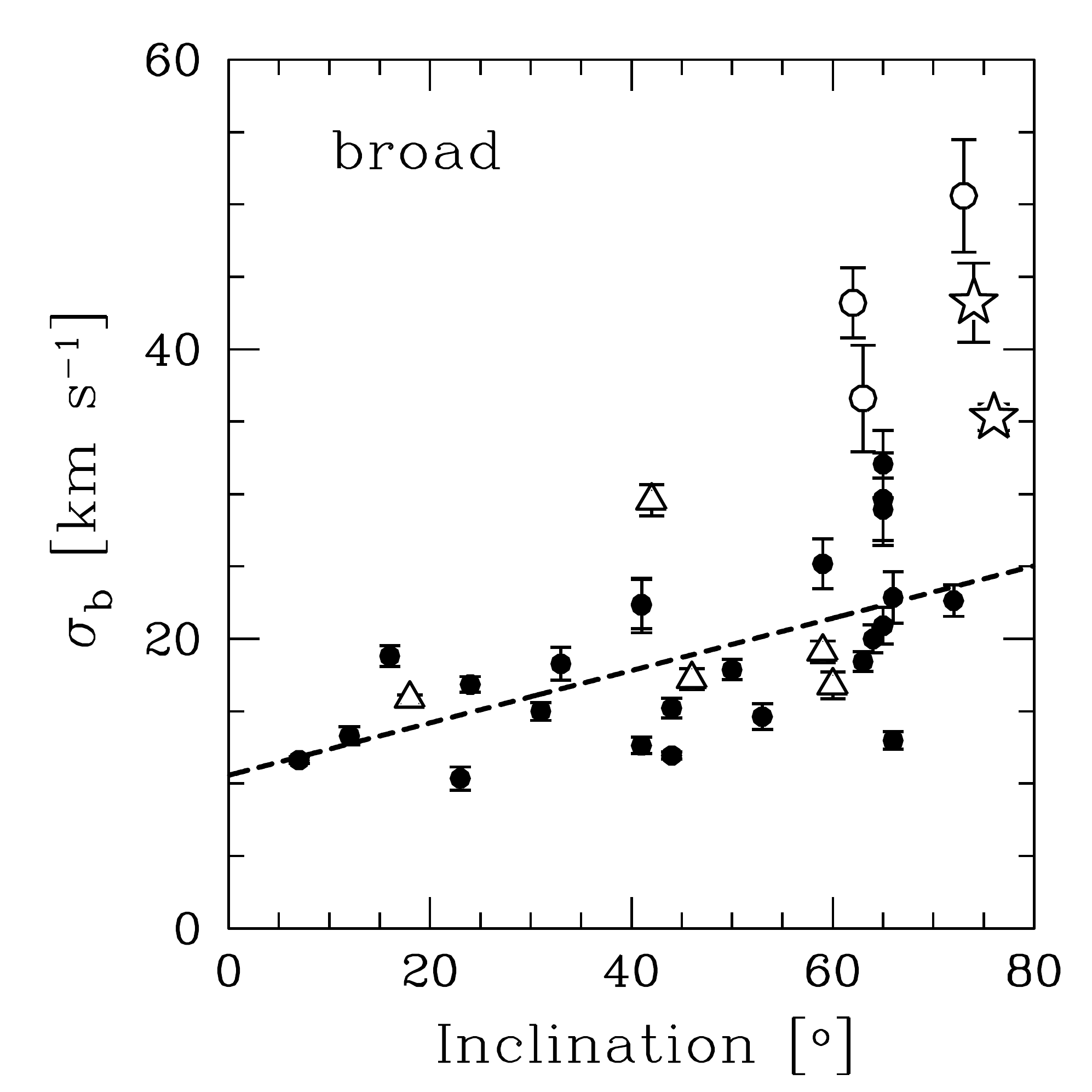}\\
 \includegraphics[width = 3in,height = 3in]{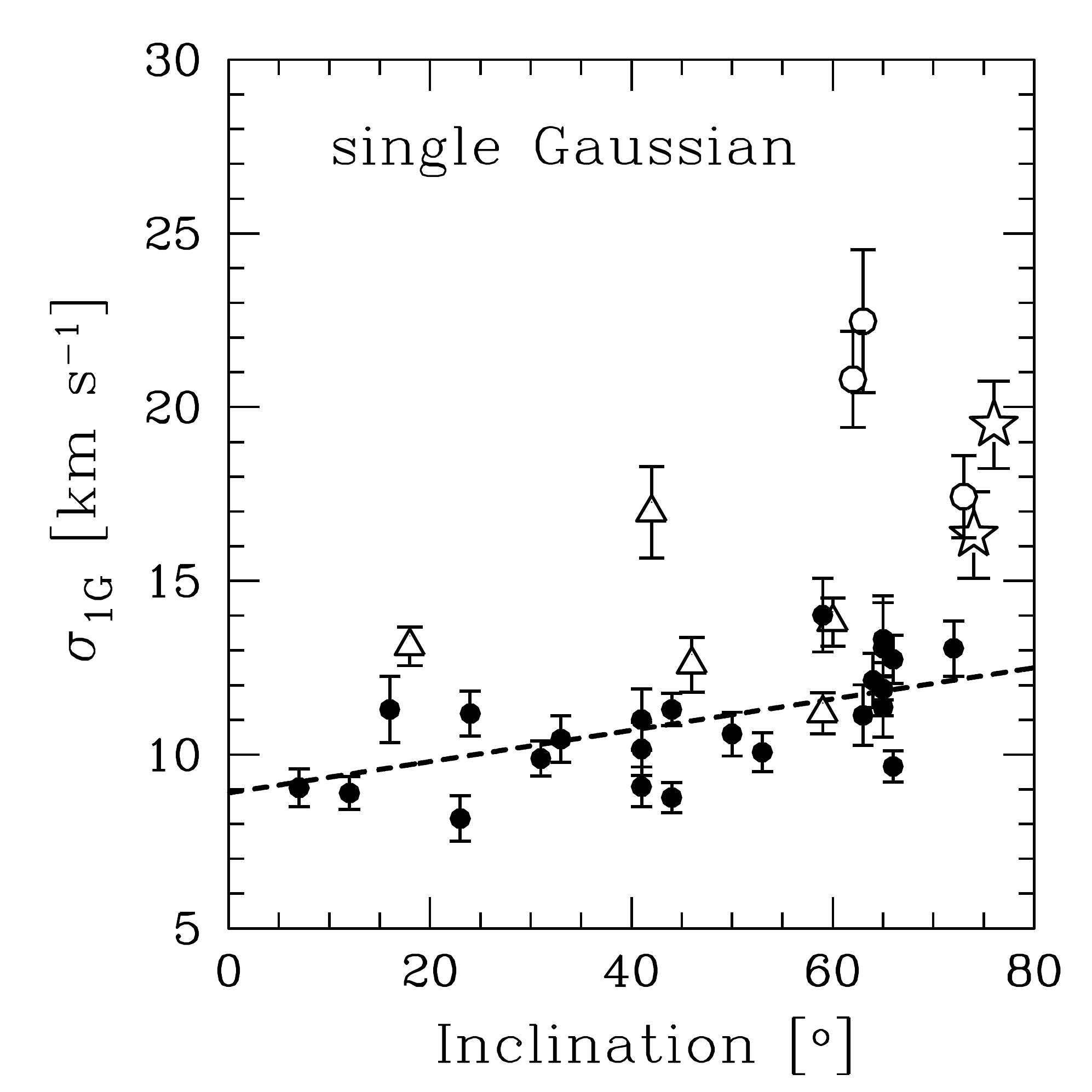}&
\includegraphics[width = 3in,height = 3in]{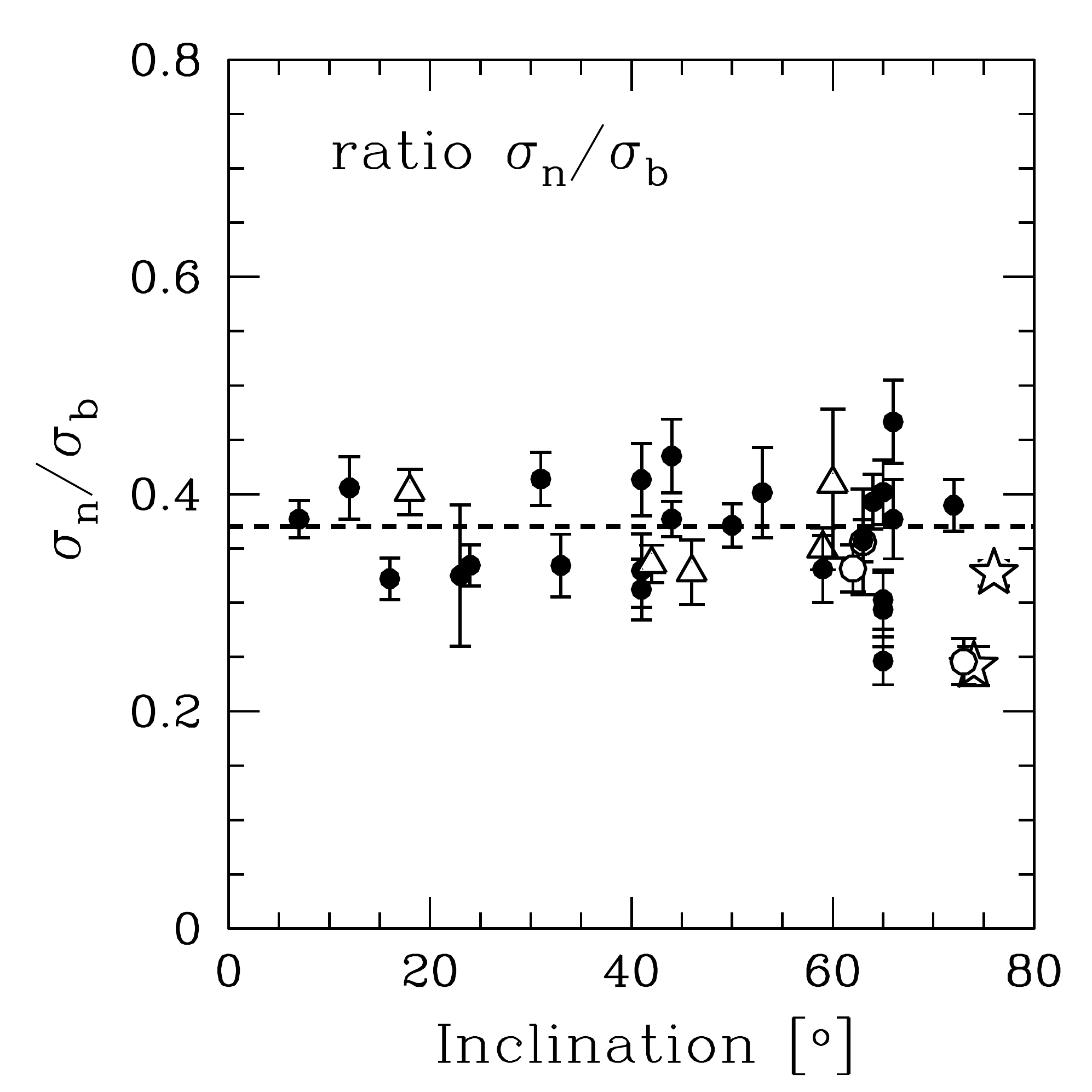}
    \end{tabular}
\caption{Relation between inclination and measured velocity
  dispersion. All error bars are 3 $\sigma$ error bars. The top left
  panel is for the narrow component. The top right panel is for the
  broad component and the bottom left panel is for the single Gaussian
  component. The dashed lines are linear least squares fit with slopes
  of 0.045$\pm$0.014 for the single Gaussian component,
  0.057$\pm$0.010 for the narrow component and 0.18$\pm$0.05 for the
  broad component. Note that only the filled data points were used in these
  fits. The bottom right panel is a plot of the ratio of the narrow
  and broad component velocity dispersions against inclination. The
  dashed line represents the weighted average value of
  $\sigma_{n}/\sigma_{b}$ $\backsimeq$ 0.37$\pm$0.04.  Symbols follow Fig.
  \ref{fig:faint_brigh_norm}.}
\label{fig:incl_disp}
 \end{figure*}

In this section we define a subsample of galaxies that do not suffer
from the effects of interaction or major bulk motions. Figure
\ref{fig:incl_disp} shows the measured super profile velocity
dispersions plotted against the inclinations of the THINGS
galaxies. We see a narrow relation with galaxies with higher
inclinations having broader profiles (as also shown for these galaxies
in \citealt{leroyetal08} and \citealt{tamburroetal09}).  Many of the
galaxies that do not follow the relation (NGC 1569, NGC 5194, NGC
2841, NGC 3627, NGC 3521, and NGC 7331) are known to be affected by
strong sources of turbulence or bulk motion (such as interaction with
other galaxies or starbursts). These galaxies are marked with
different symbols in Fig. \ref{fig:incl_disp}. One thing to note is
the remarkable constancy of the $\sigma_n/\sigma_b$ ratio. The
weighted average value of this ratio is $0.37\pm0.04$.

Based on inspection of Figures \ref{fig:com_large_small_MAJOR}, 
\ref{fig:rad_skew}, \ref{fig:com_large_small}, \ref{fig:comp_skew_rec_ap} and 
\ref{fig:incl_disp}, we exclude from the
rest of our analysis galaxies, that do not follow the observed trend
between inclination and velocity dispersion in Fig.
\ref{fig:incl_disp}. We also exclude galaxies where the difference in
velocity dispersion measured for the approaching and receding sides, or
PA+$90^{\circ}$ and PA+$270^{\circ}$ sides is more than 1.5 $\rm{km~s^{-1}}$.
Interacting galaxies will also not be considered. Based on these
criteria, we are left with 22 galaxies which we identify as our clean
sample. Their super profiles are expected to be less dominated by the various 
effects presented earlier. The clean sample is listed in Table \ref{tab:fitted_clean}.

\section{Effects of asymmetric profiles}\label{sub:asymm_effect}

Having identified a sample likely not to be suffering from large-scale
disturbances, we now take a detailed look at the shapes of the
observed input profiles in this sample.

If the individual profiles used in the stacking are themselves
asymmetric, this can also lead to the observed super profiles, as the
sum of a negatively skewed and a positively skewed profile will be one
with broader wings and a narrower peak than a Gaussian profile.

Here we check whether the broad components in the super profiles are
intrinsic or artifacts due to asymmetric input profiles. We do this by
selecting only symmetrical profiles, creating super profiles from
those symmetrical input profiles and checking whether broad and narrow
components are still present. 

To select symmetrical profiles, we create masks based on the
difference between the third order Hermite (HER3) velocity field
values and intensity weighted mean first moment values (IWM). The
former trace the velocity of the peak of the profile, the
latter the intensity weighted mean value. For asymmetric profiles
there will therefore be a difference between the HER3 and the IWM
values.  For symmetrical profiles the difference between HER3 and IWM
is small, and this can be used to efficiently select symmetrical input
profiles.  If super profiles created from only the symmetric profiles
still show broad components, then these are likely to be intrinsic.
We thus define three masks based on the difference in velocity
between HER3 and IWM, $\Delta_{H-I}$, as follows:\\ \textit{---
  Symmetrical profiles (SP)}: positions where $|\Delta_{H-I}| \leq
5\,\rm{km~s^{-1}}$.\\ \textit{--- Left-handed asymmetrical profiles
  (LHAP)}: positions where
$\Delta_{H-I}<-5\,\rm{km~s^{-1}}$.\\ \textit{--- Right-handed
  asymmetrical profiles (RHAP)}: positions where $\Delta_{H-I}
>5\,\rm{km~s^{-1}}$.

\begin{figure*}
    \begin{tabular}{l l l}
 \rotatebox{0}{\resizebox{56mm}{!}{\includegraphics[width = 0.6in, height = 0.6in]{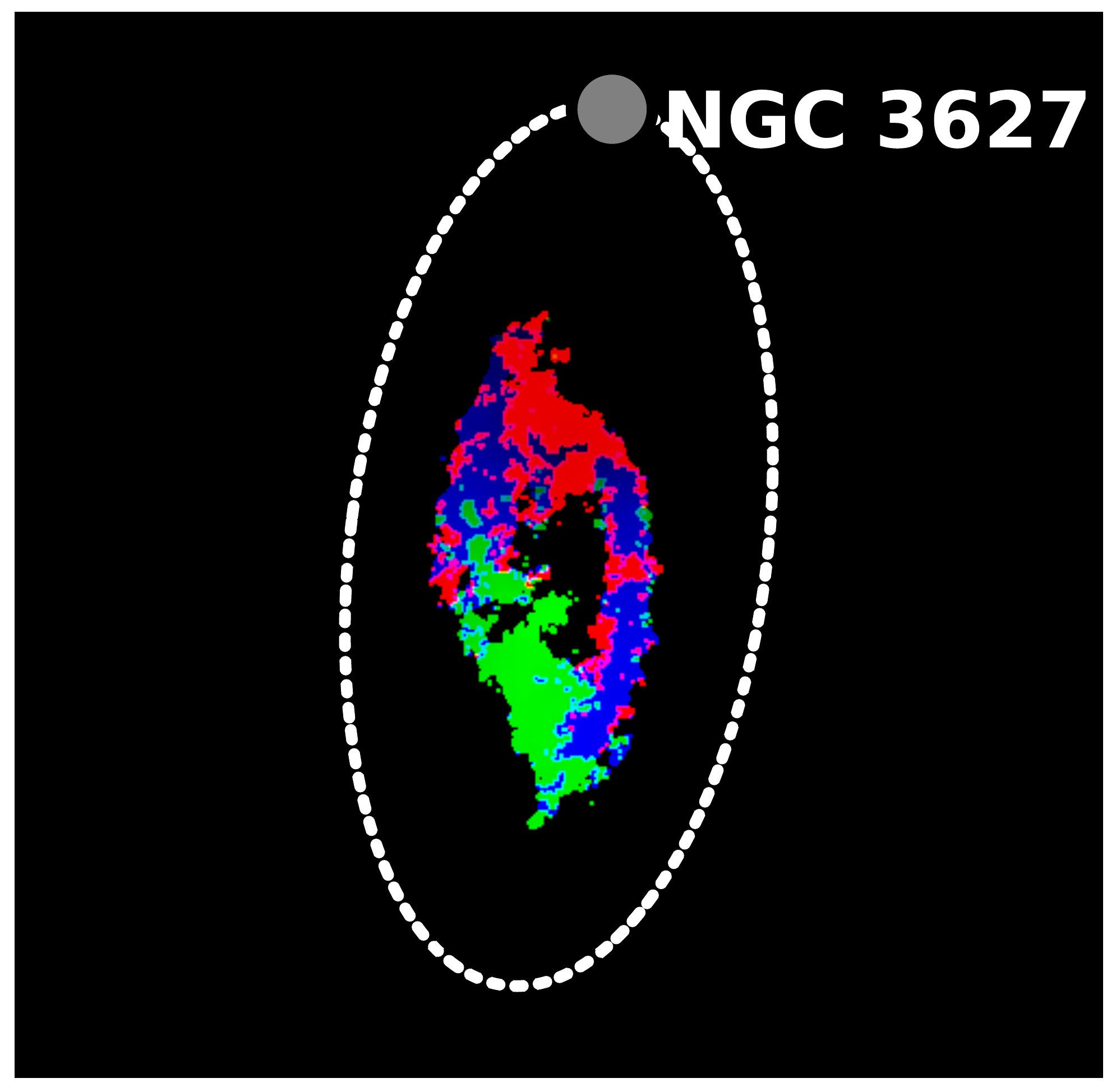}}}&
 \rotatebox{0}{\resizebox{56mm}{!}{\includegraphics[width = 0.6in, height = 0.6in]{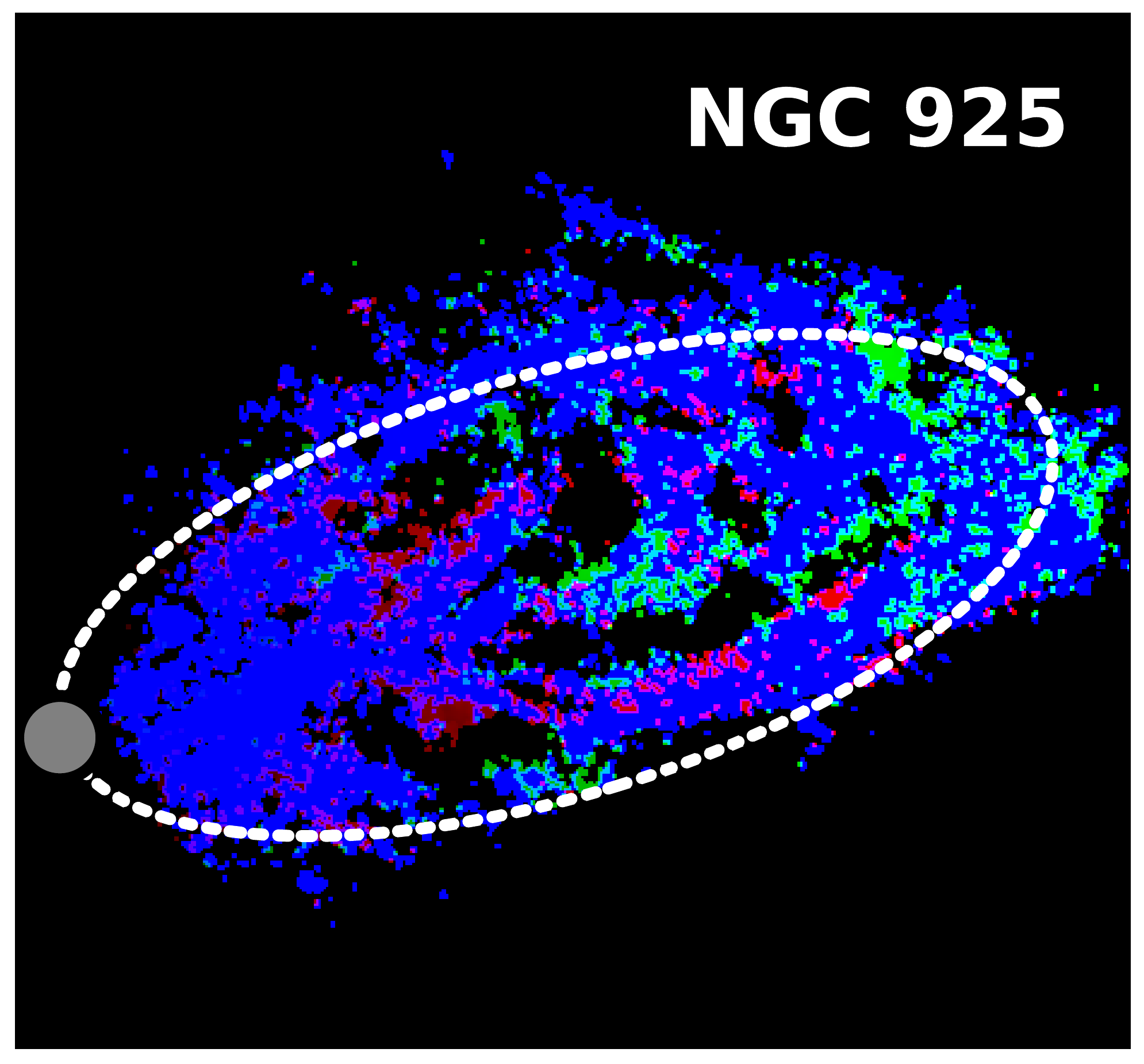}}}&
 \rotatebox{0}{\resizebox{56mm}{!}{\includegraphics[width = 0.6in, height = 0.6in]{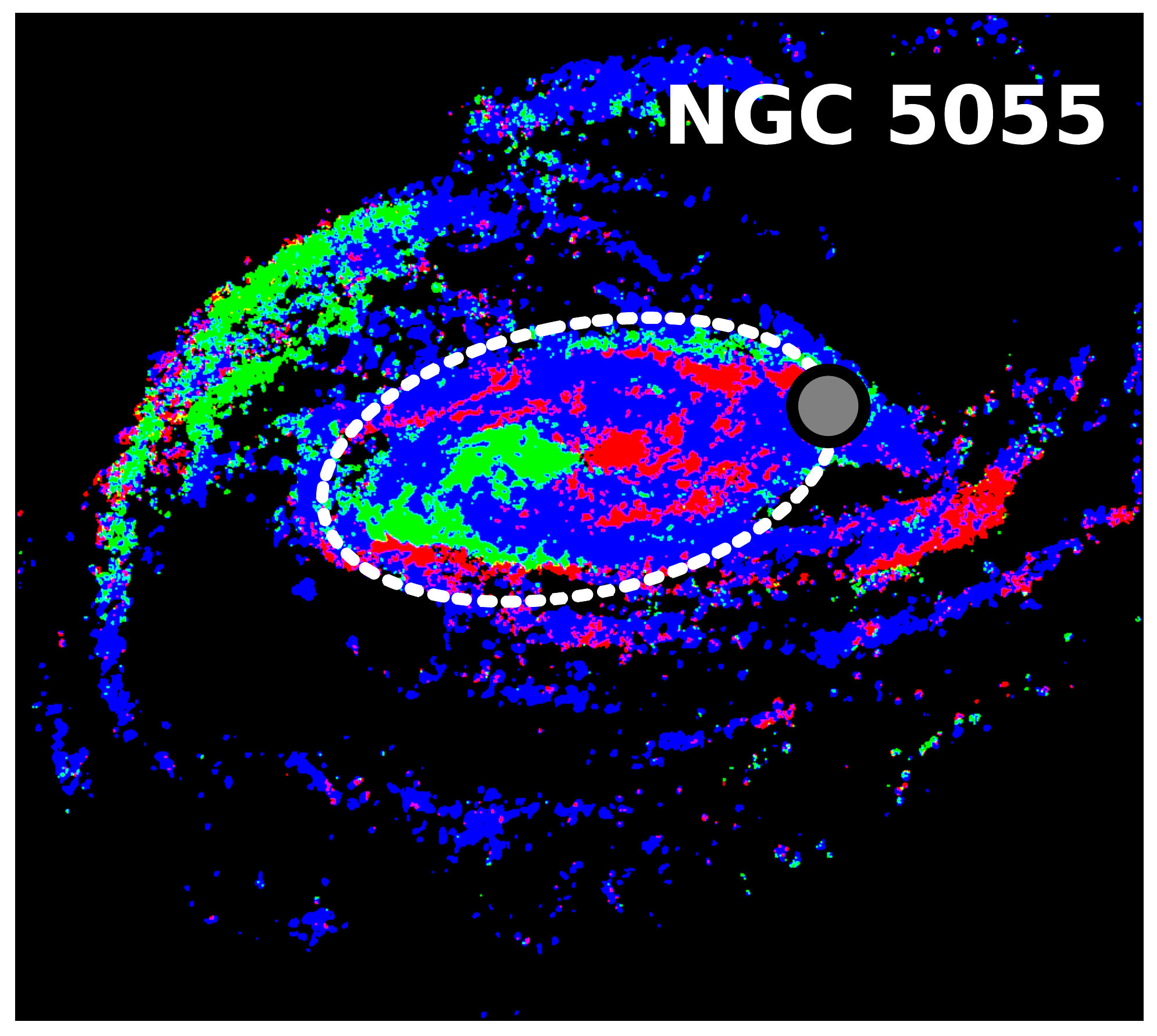}}}\\
\rotatebox{0}{\resizebox{56mm}{!}{\includegraphics[width = 0.6in, height = 0.6in]{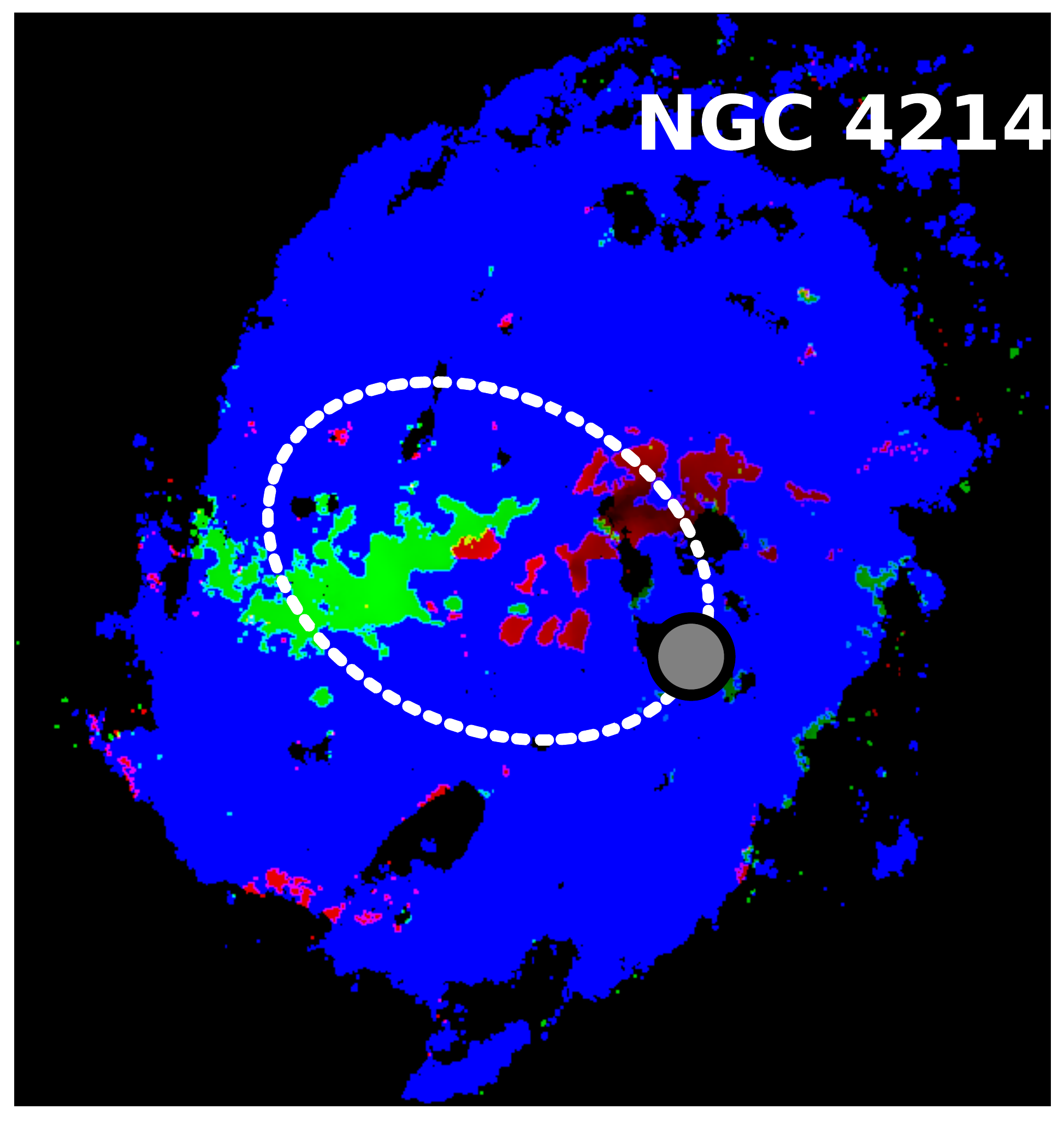}}}&
 \rotatebox{0}{\resizebox{56mm}{!}{\includegraphics[width = 0.6in, height = 0.6in]{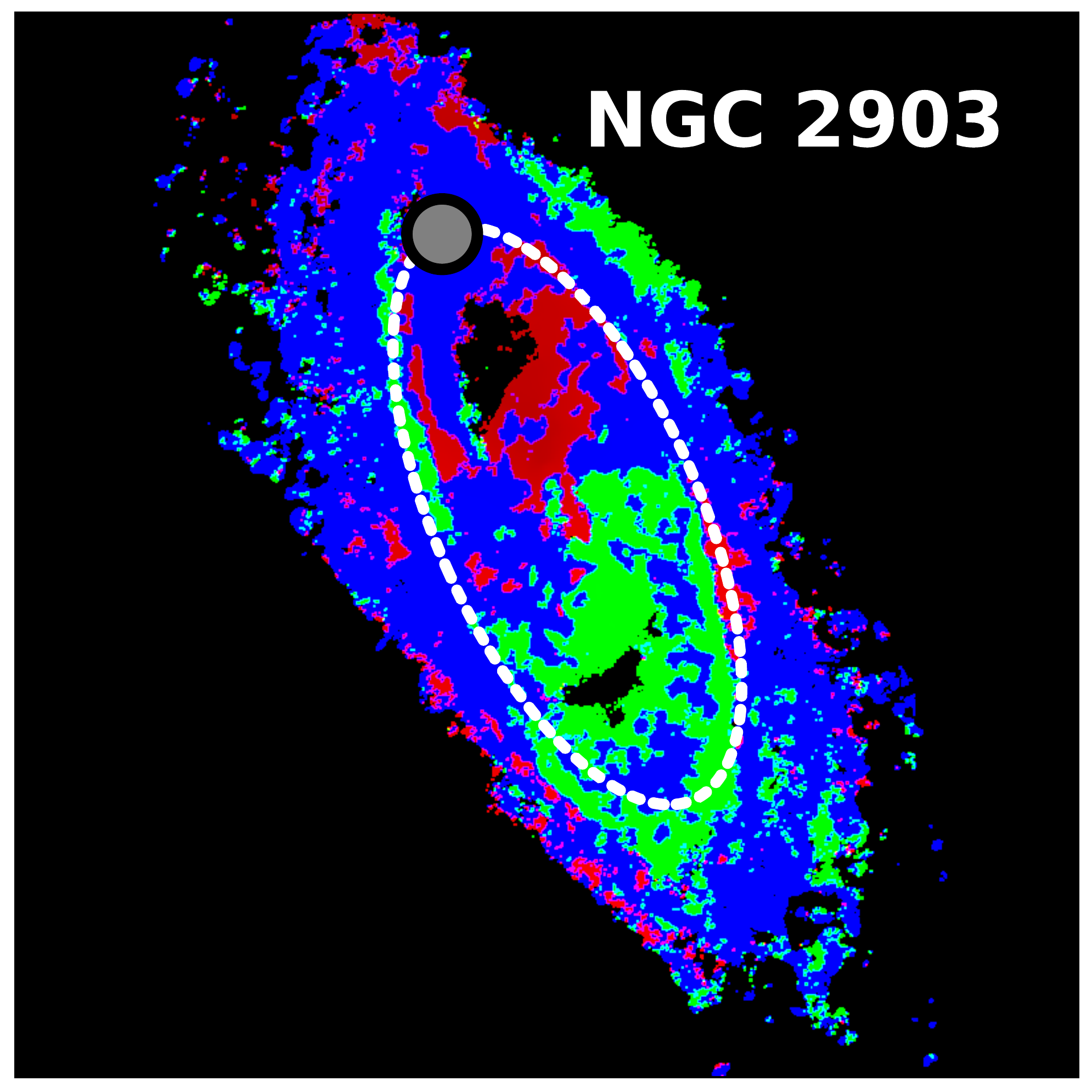}}}&
 \rotatebox{0}{\resizebox{56mm}{!}{\includegraphics[width = 0.6in, height = 0.6in]{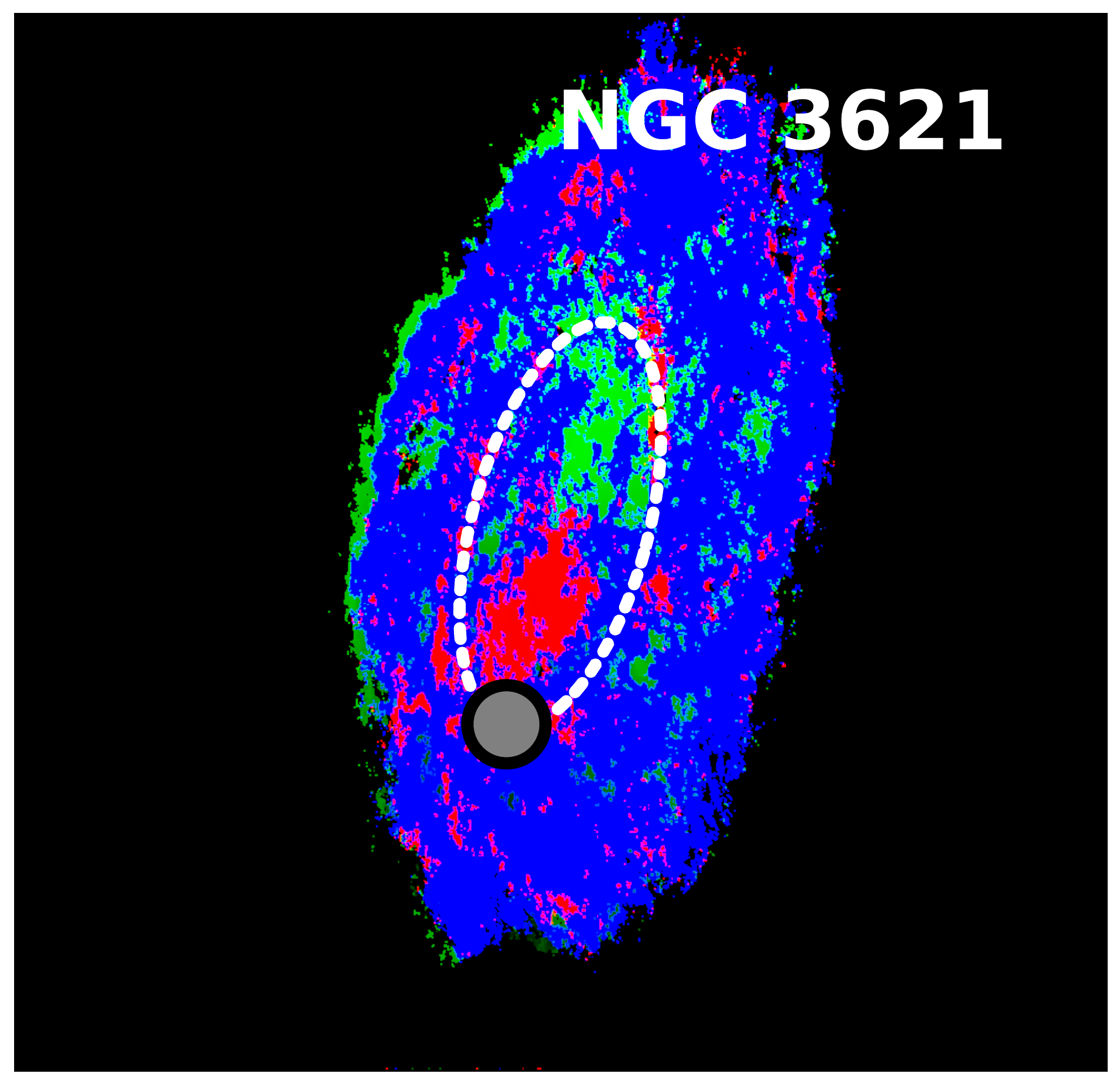}}}\\
 \rotatebox{0}{\resizebox{56mm}{!}{\includegraphics[width = 0.6in, height = 0.6in]{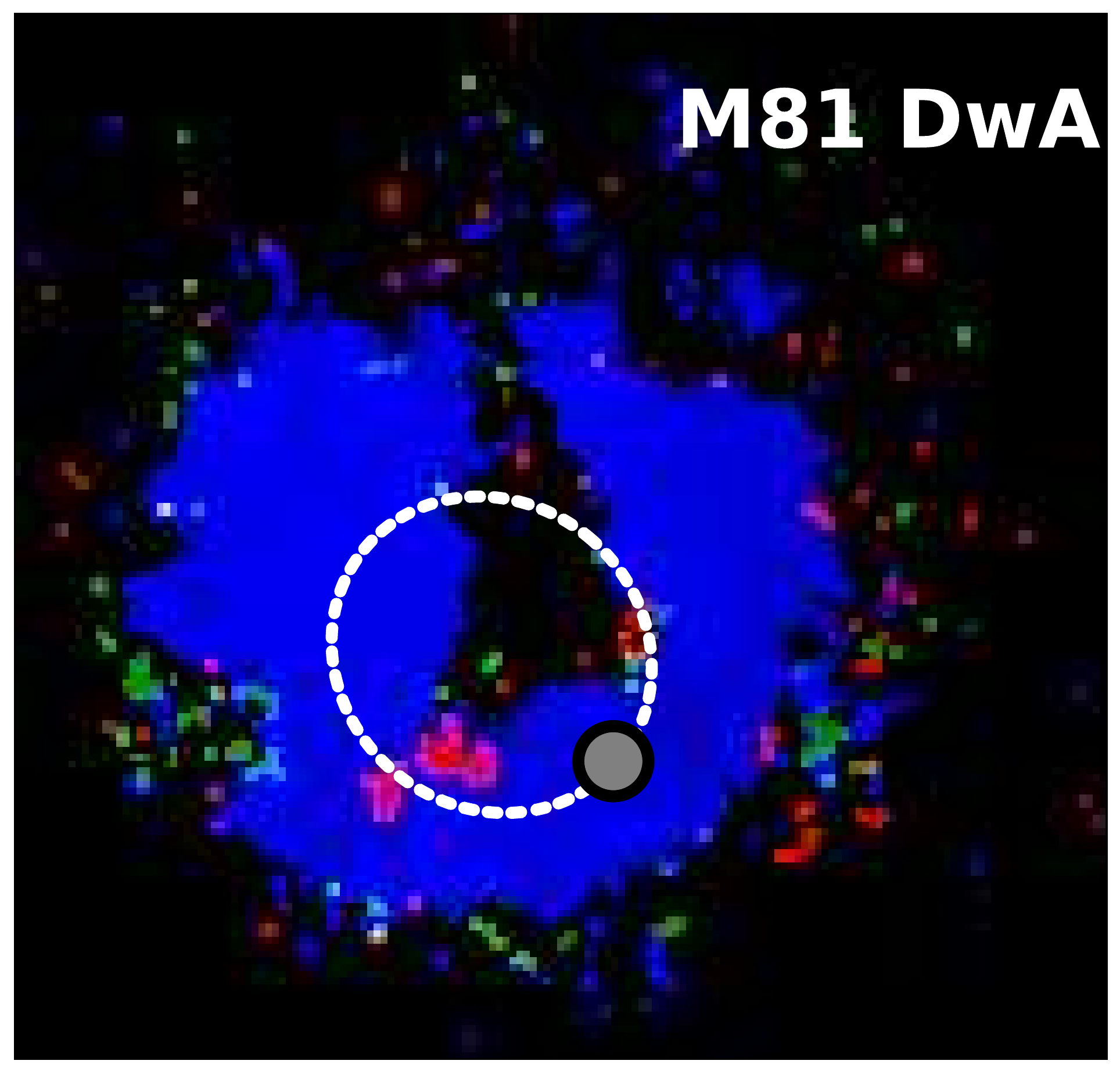}}}&
 \rotatebox{0}{\resizebox{56mm}{!}{\includegraphics[width = 0.6in, height = 0.6in]{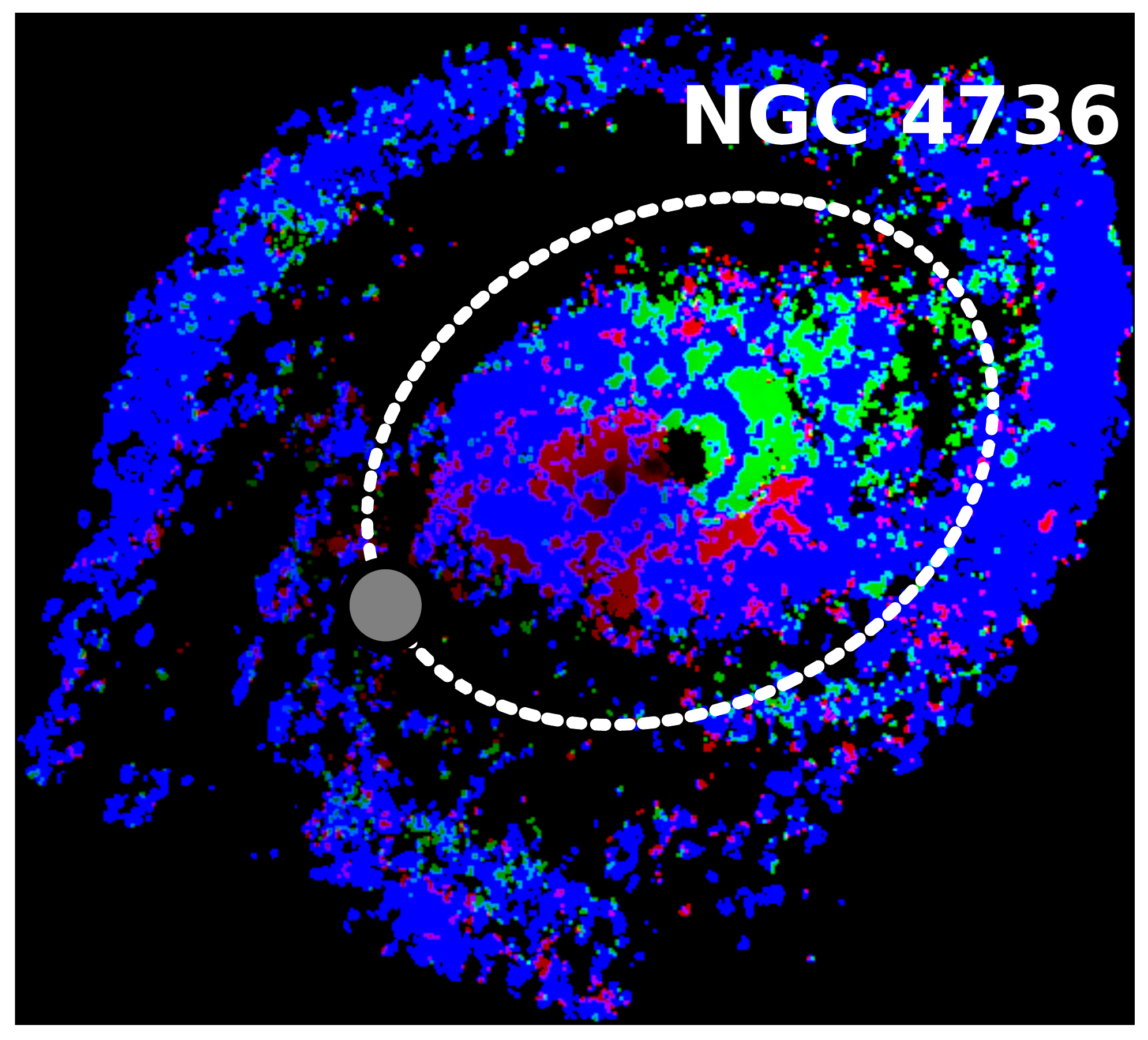}}}&
 \rotatebox{0}{\resizebox{56mm}{!}{\includegraphics[width = 0.6in, height = 0.6in]{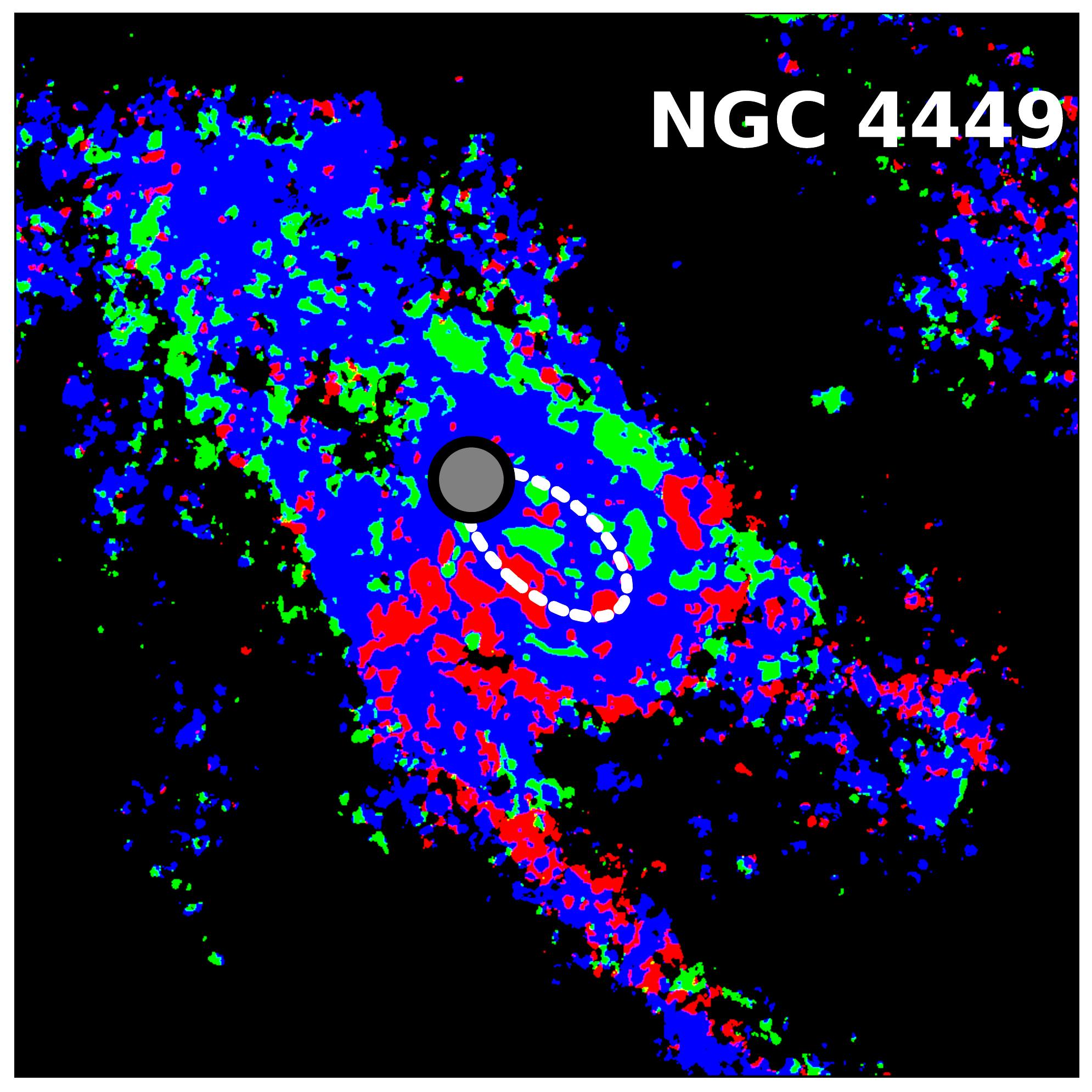}}}\\
 \rotatebox{0}{\resizebox{56mm}{!}{\includegraphics[width = 0.6in, height = 0.6in]{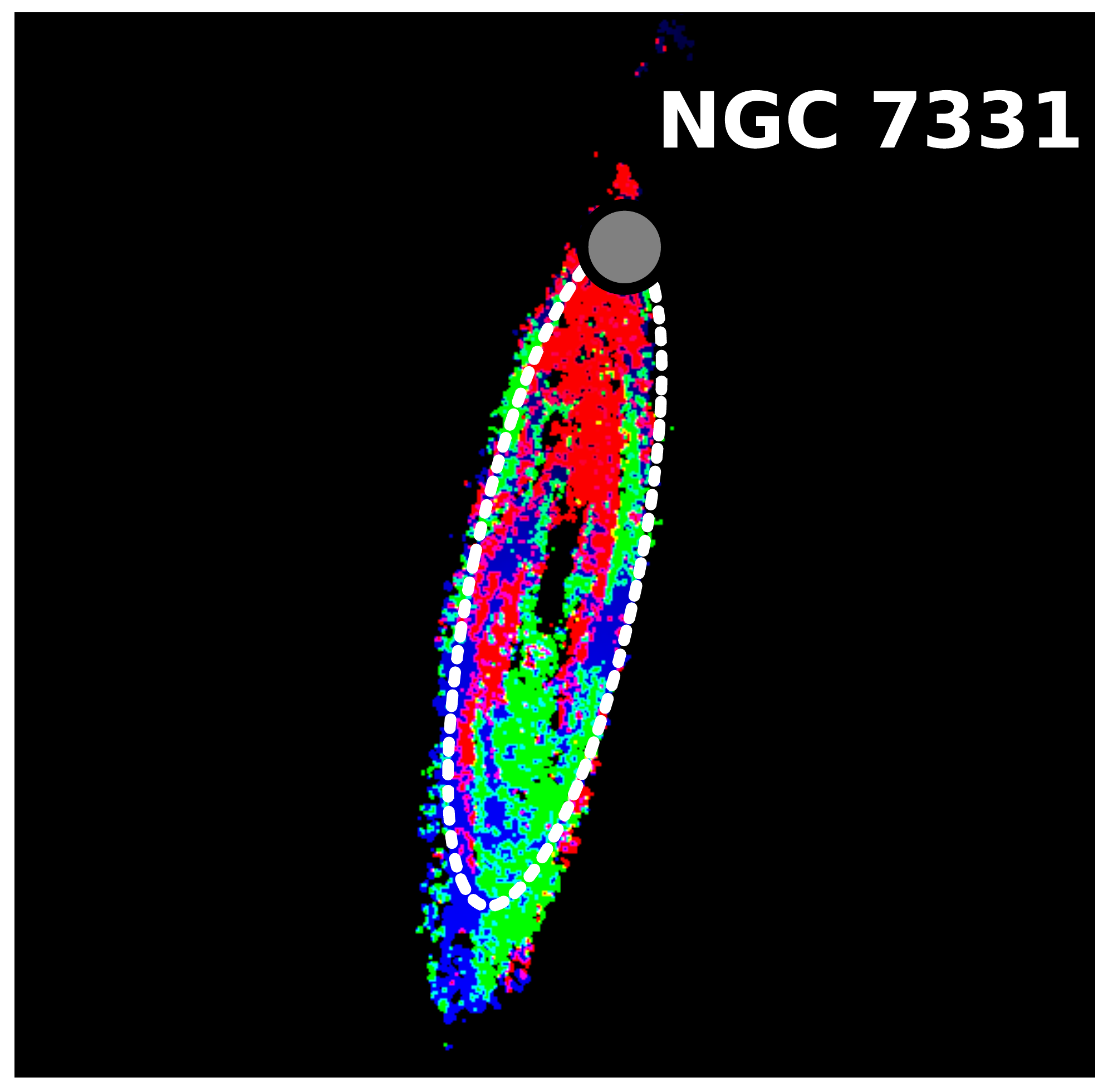}}}&
 \rotatebox{0}{\resizebox{56mm}{!}{\includegraphics[width = 0.6in, height = 0.6in]{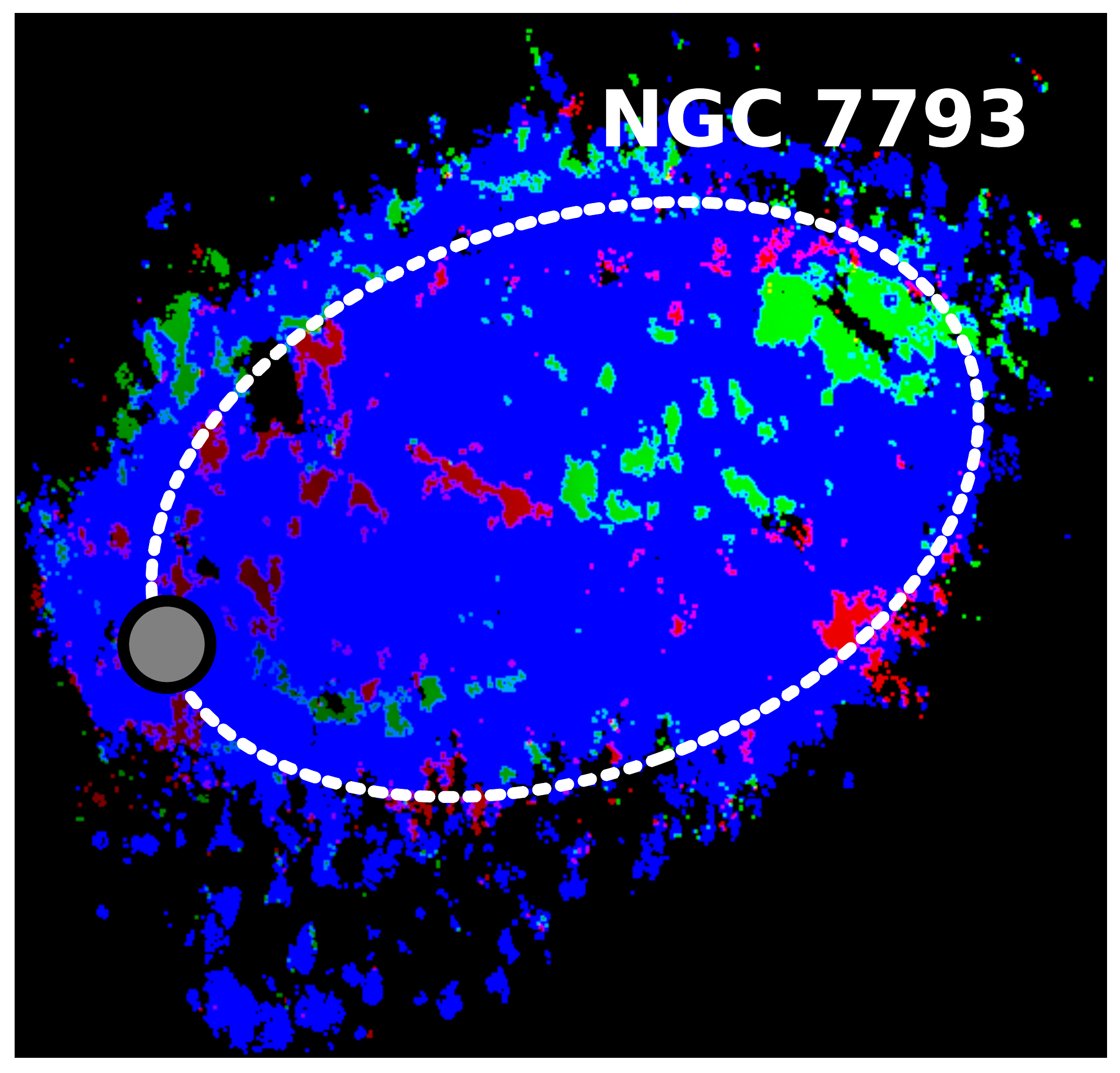}}}&
\rotatebox{0}{\resizebox{56mm}{!}{\includegraphics[width = 0.6in, height = 0.6in]{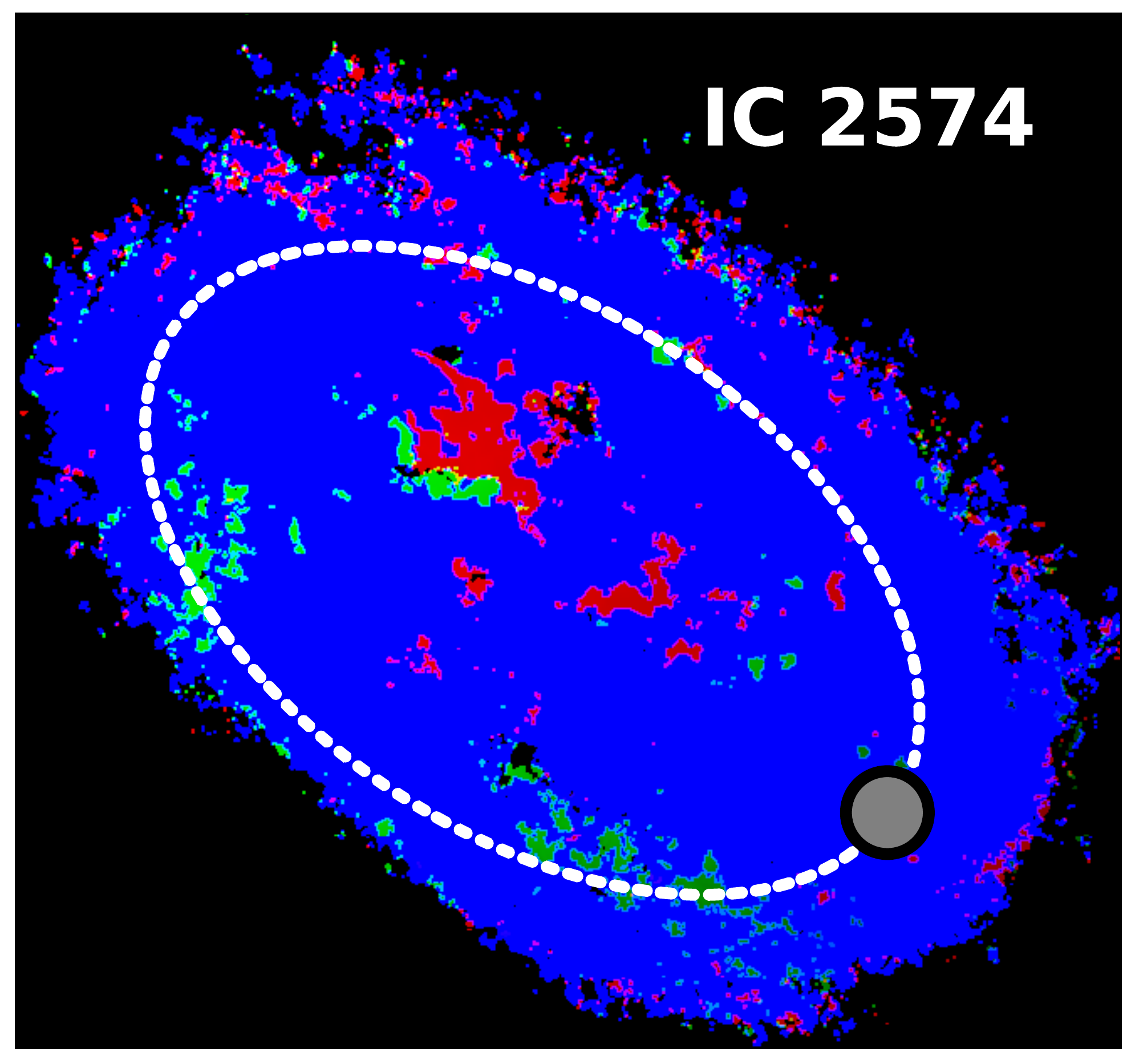}}}
    \end{tabular}
      \caption{Location of symmetric and asymmetric profiles for the
        THINGS galaxies. Blue pixels represent symmetric profiles
        (SP). Green and red pixels represent left-handed asymmetric
        profiles (LHAP) and right-handed asymmetric profiles (RHAP),
        respectively. Ellipses represent the optical
        radius $r_{25}$. The small grey circles indicate the approaching sides.}
\label{fig:location_asymm_and_sym}
 \end{figure*}

\begin{figure*}
    \begin{tabular}{l l l}
 \rotatebox{0}{\resizebox{55mm}{!}{\includegraphics[width = 0.55in, height = 0.55in]{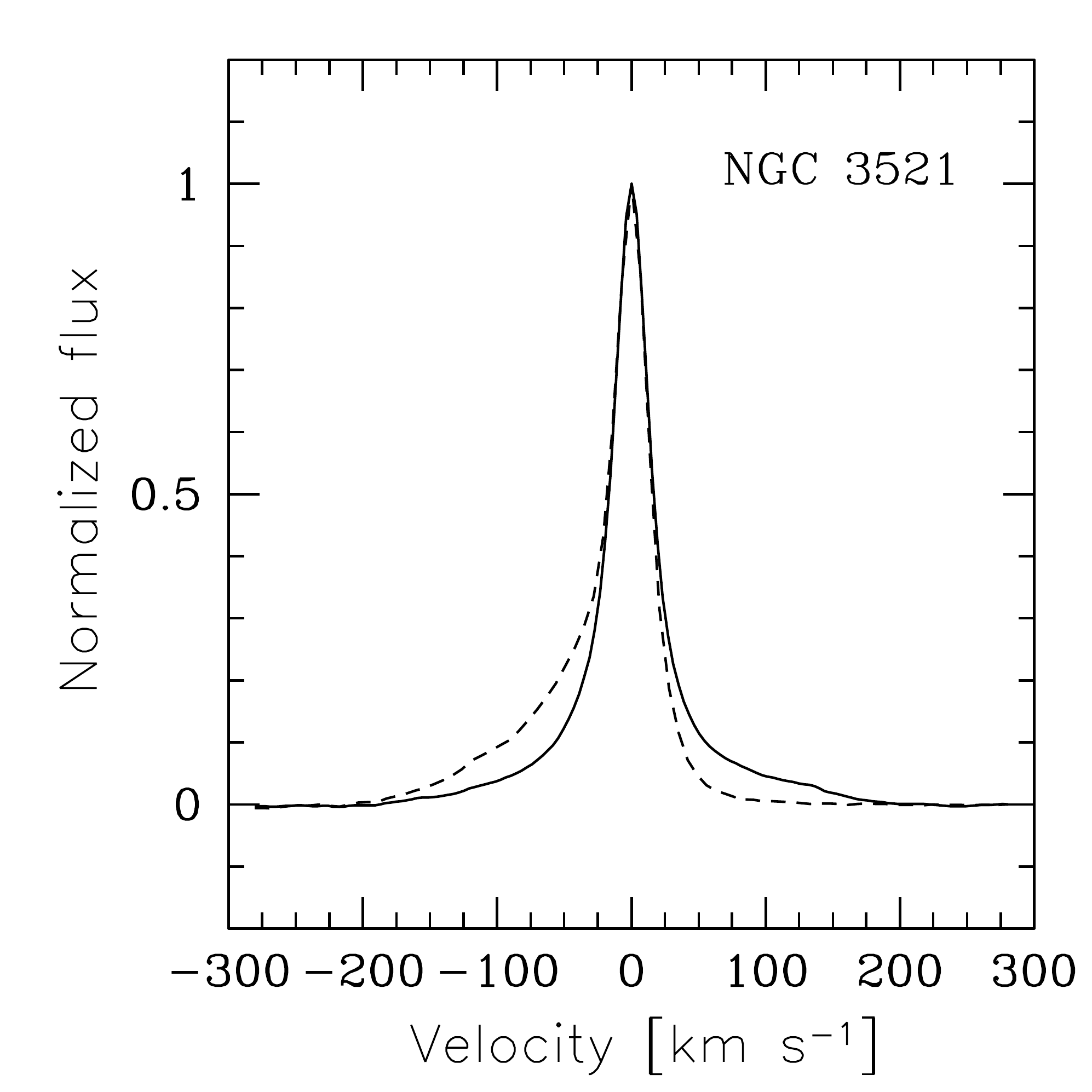}}}&
\rotatebox{0}{\resizebox{55mm}{!}{\includegraphics[width = 0.55in, height = 0.55in]{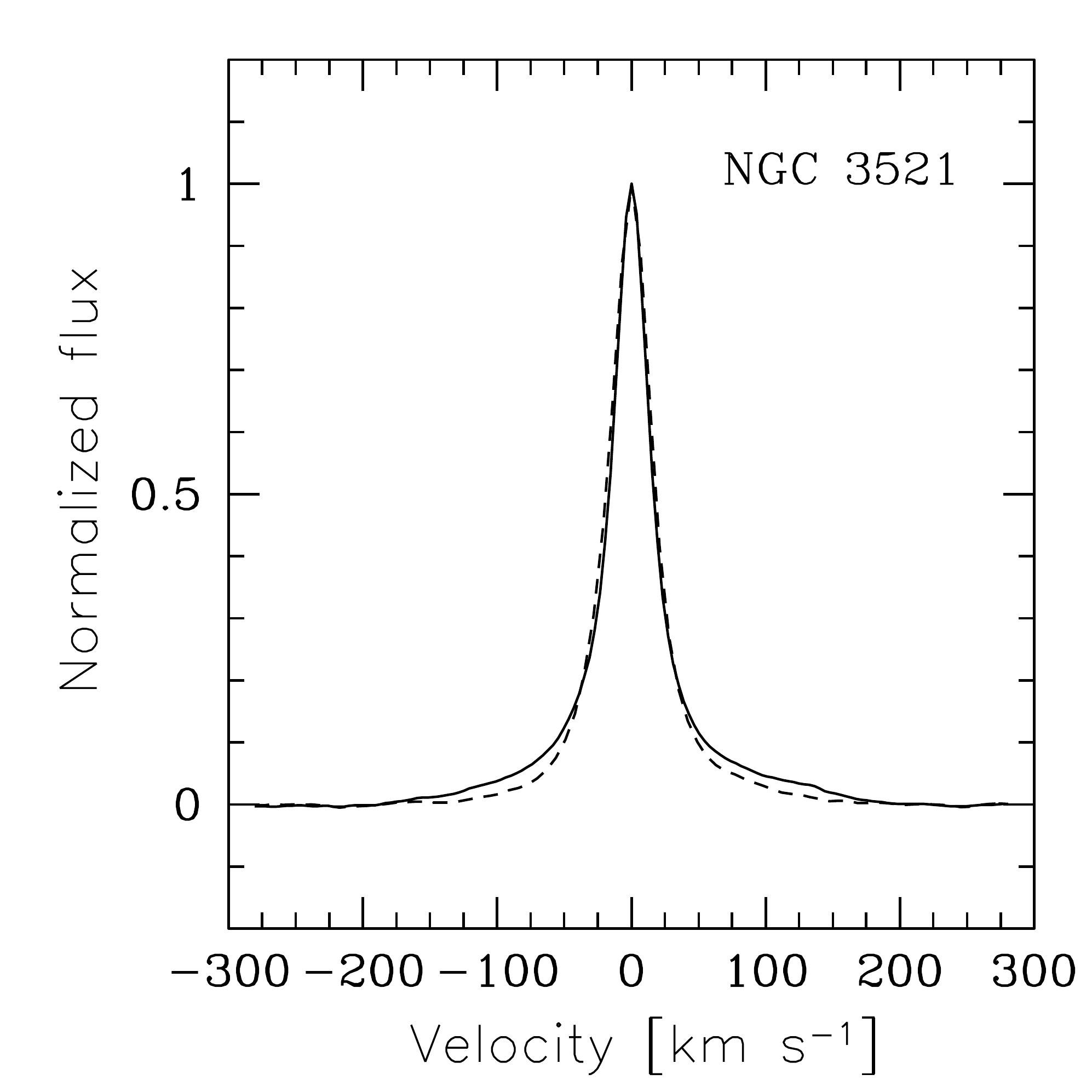}}}&
\rotatebox{0}{\resizebox{55mm}{!}{\includegraphics[width = 0.55in, height = 0.55in]{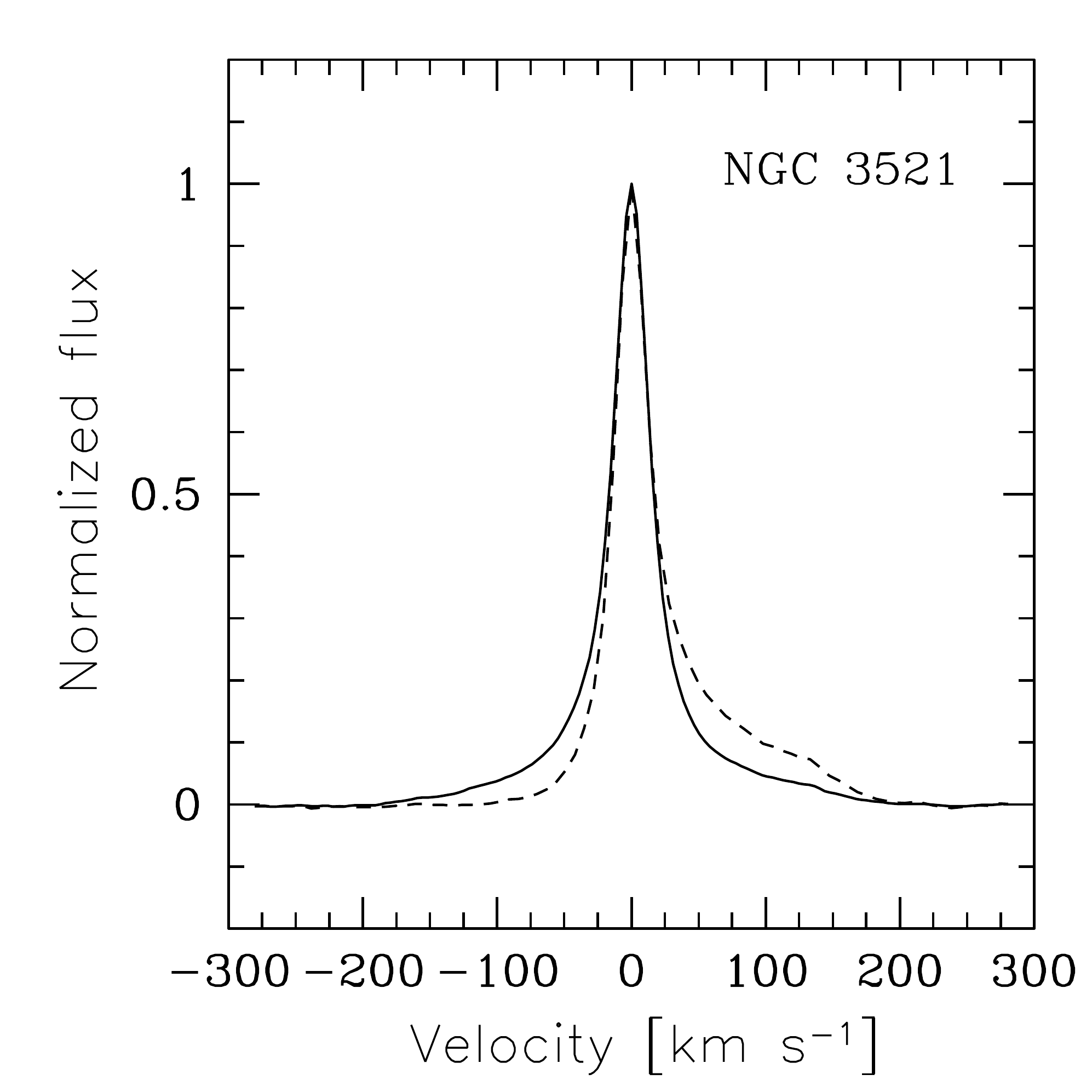}}}
  \end{tabular}
 \caption{Examples of super profiles of NGC 3521 using LHAP (left panel), 
	 SP (middle panel) and RHAP (right panel) 
 represented by the dashed lines, overplotted on top of the 
 total super profile (shown as solid lines).}%
\label{fig:msp_mlsp_super_profile}
 \end{figure*}

SP therefore selects symmetrical profiles; the other two masks LHAP
and RHAP select only asymmetrical profiles. We show for all galaxies
the location of the symmetrical and asymmetrical profiles in 
Fig.~\ref{fig:location_asymm_and_sym}. 
We create super profiles using the 
three masks and fit them with both single and double Gaussian
components. Examples of the super profiles of NGC 3521 created using
the three kinds of masks, overplotted on top of its total super profiles 
are shown in Fig.~\ref{fig:msp_mlsp_super_profile}. 
It is clear that the masks produce the expected results.

 \begin{figure*}
    \begin{tabular}{l l l}

 \rotatebox{0}{\resizebox{56mm}{!}{\includegraphics[width = 0.6in, height = 0.6in]{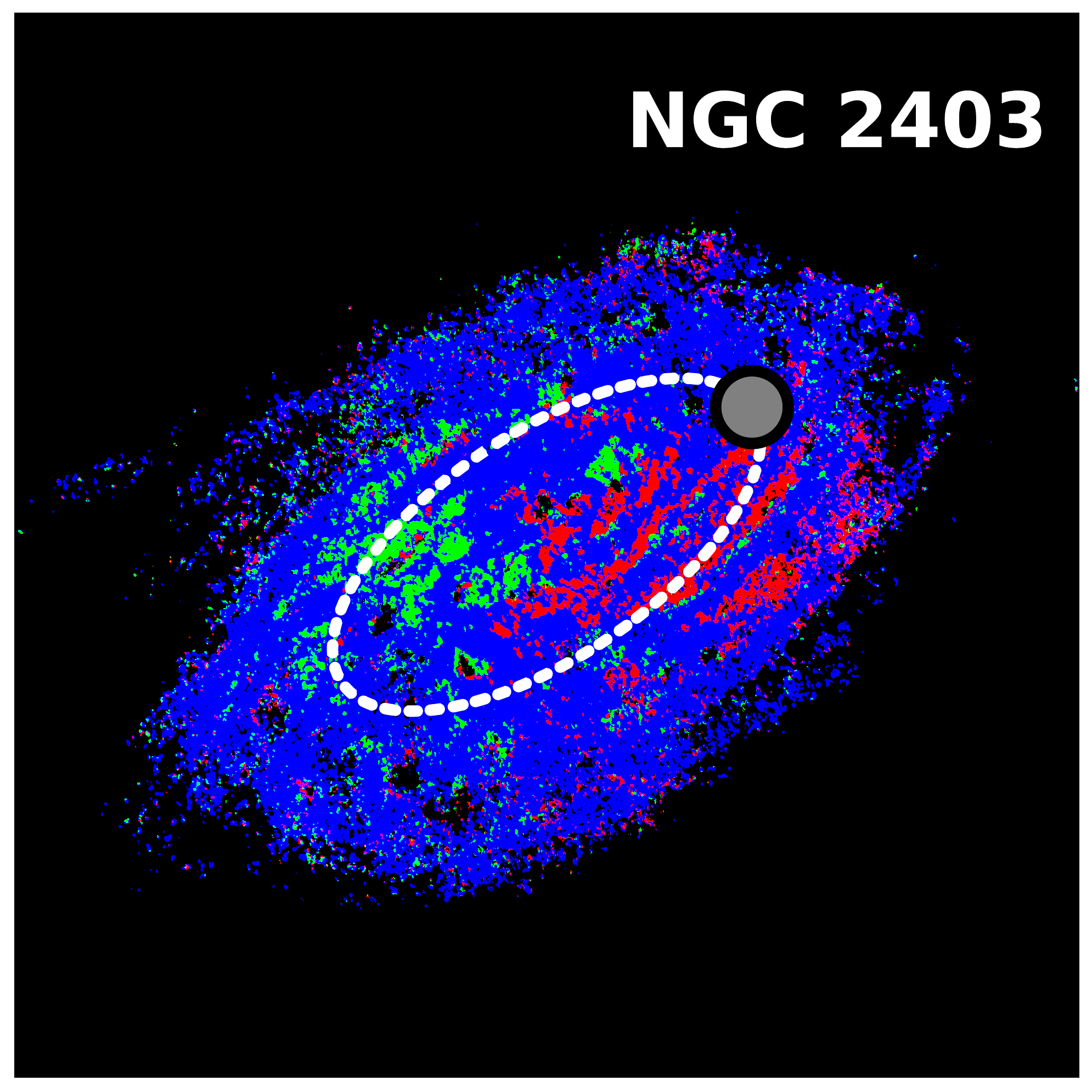}}}&
 \rotatebox{0}{\resizebox{56mm}{!}{\includegraphics[width = 0.6in, height = 0.6in]{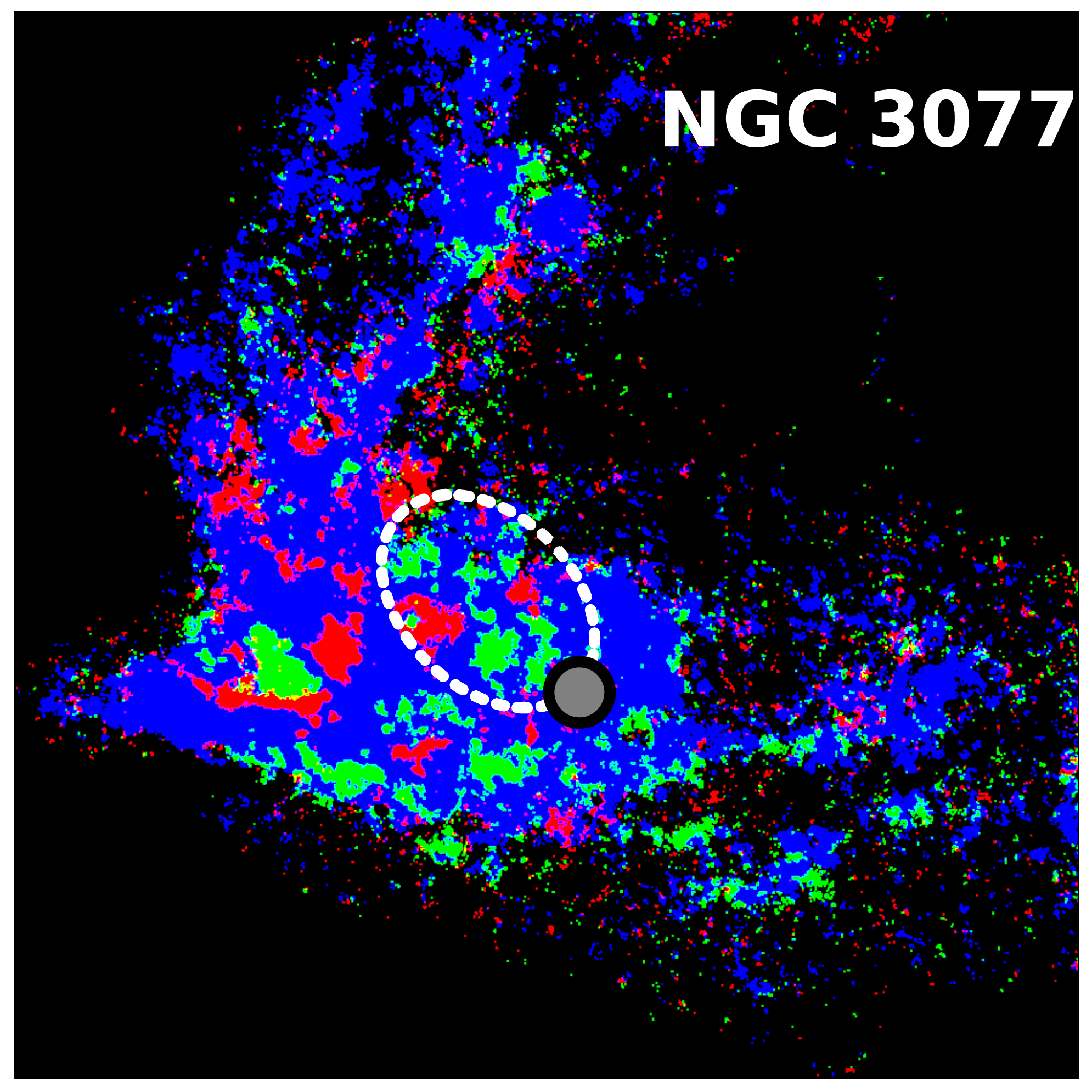}}}&
  \rotatebox{0}{\resizebox{56mm}{!}{\includegraphics[width = 0.6in, height = 0.6in]{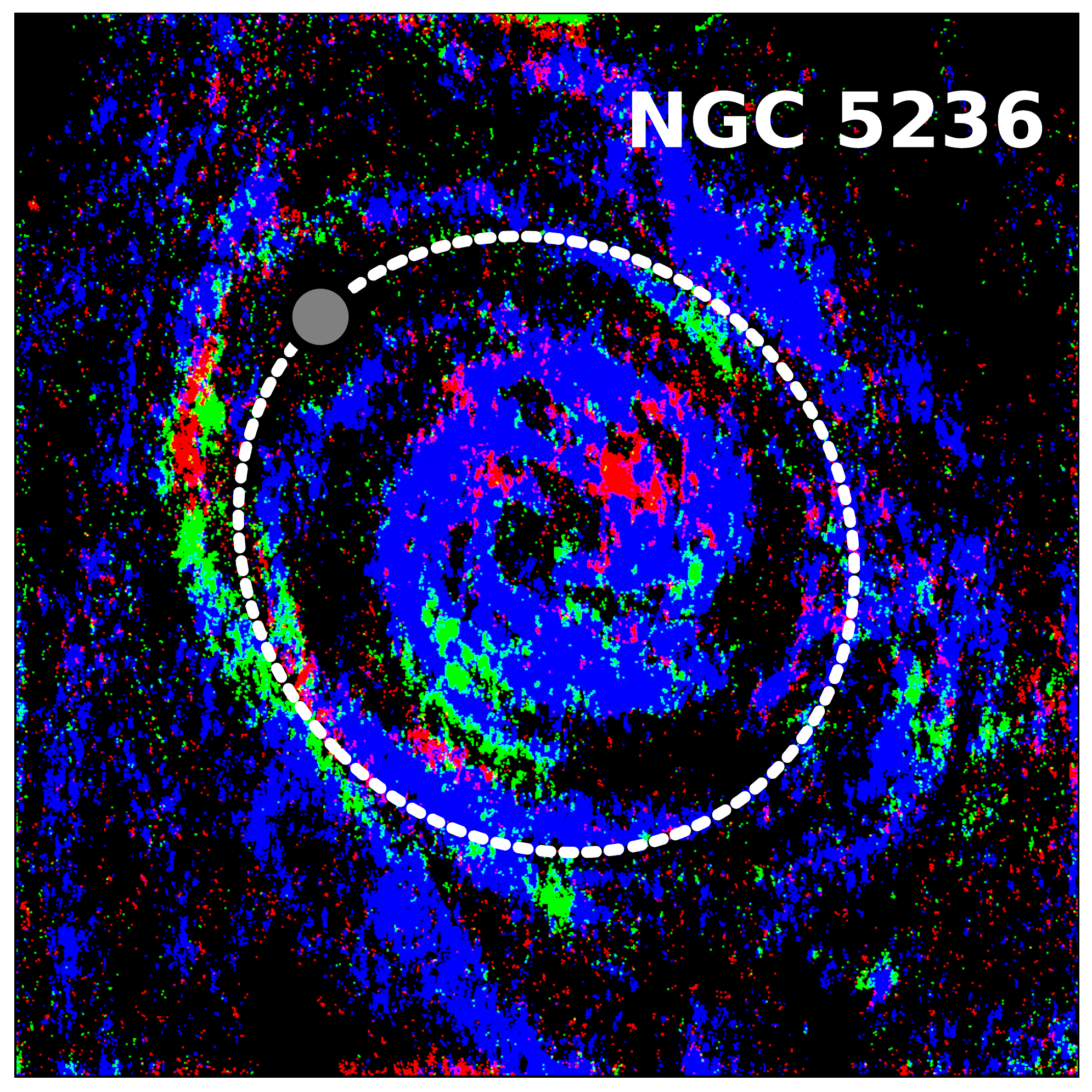}}}\\
   \rotatebox{0}{\resizebox{56mm}{!}{\includegraphics[width = 0.6in, height = 0.6in]{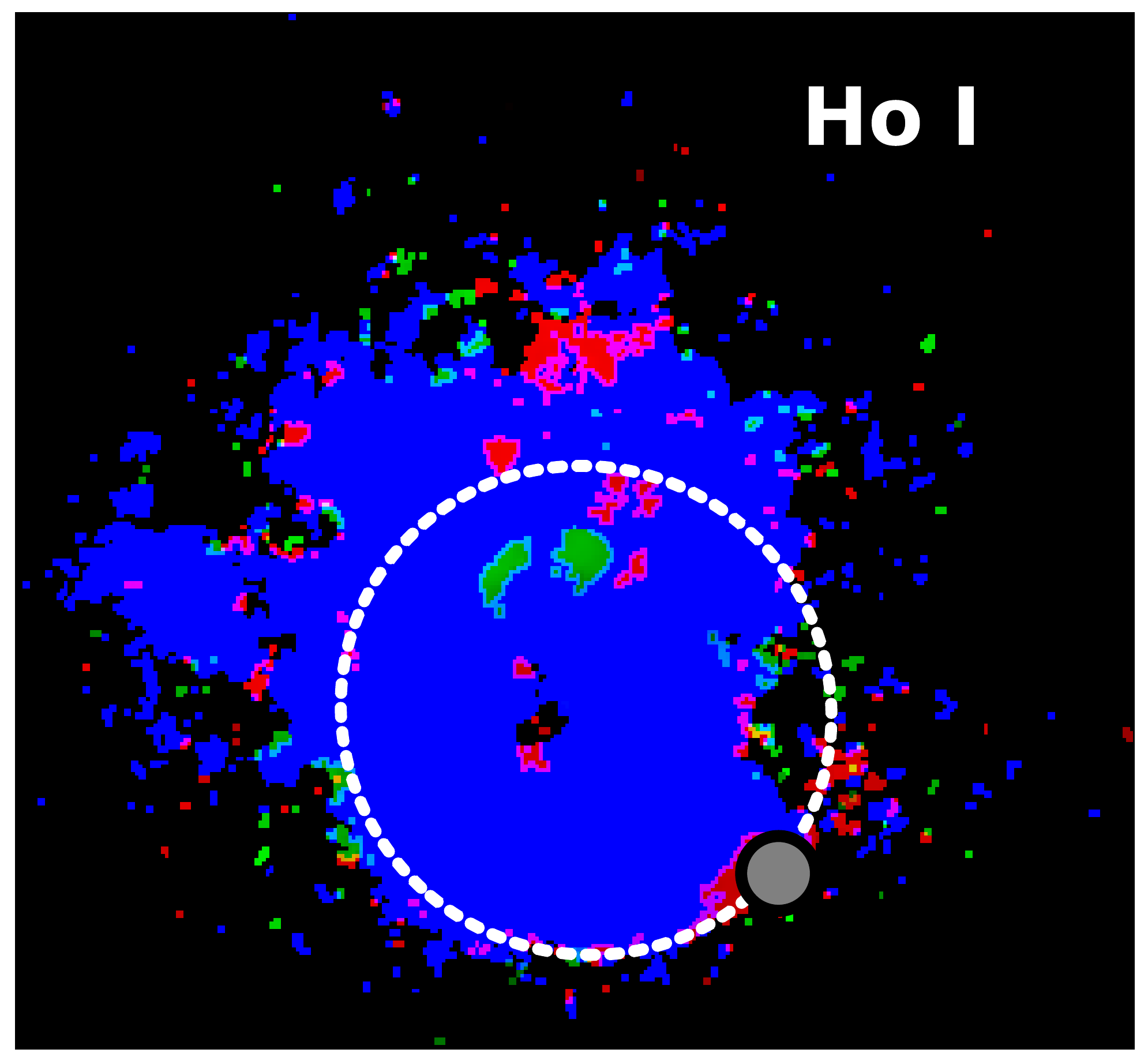}}}&
 \rotatebox{0}{\resizebox{56mm}{!}{\includegraphics[width = 0.6in, height = 0.6in]{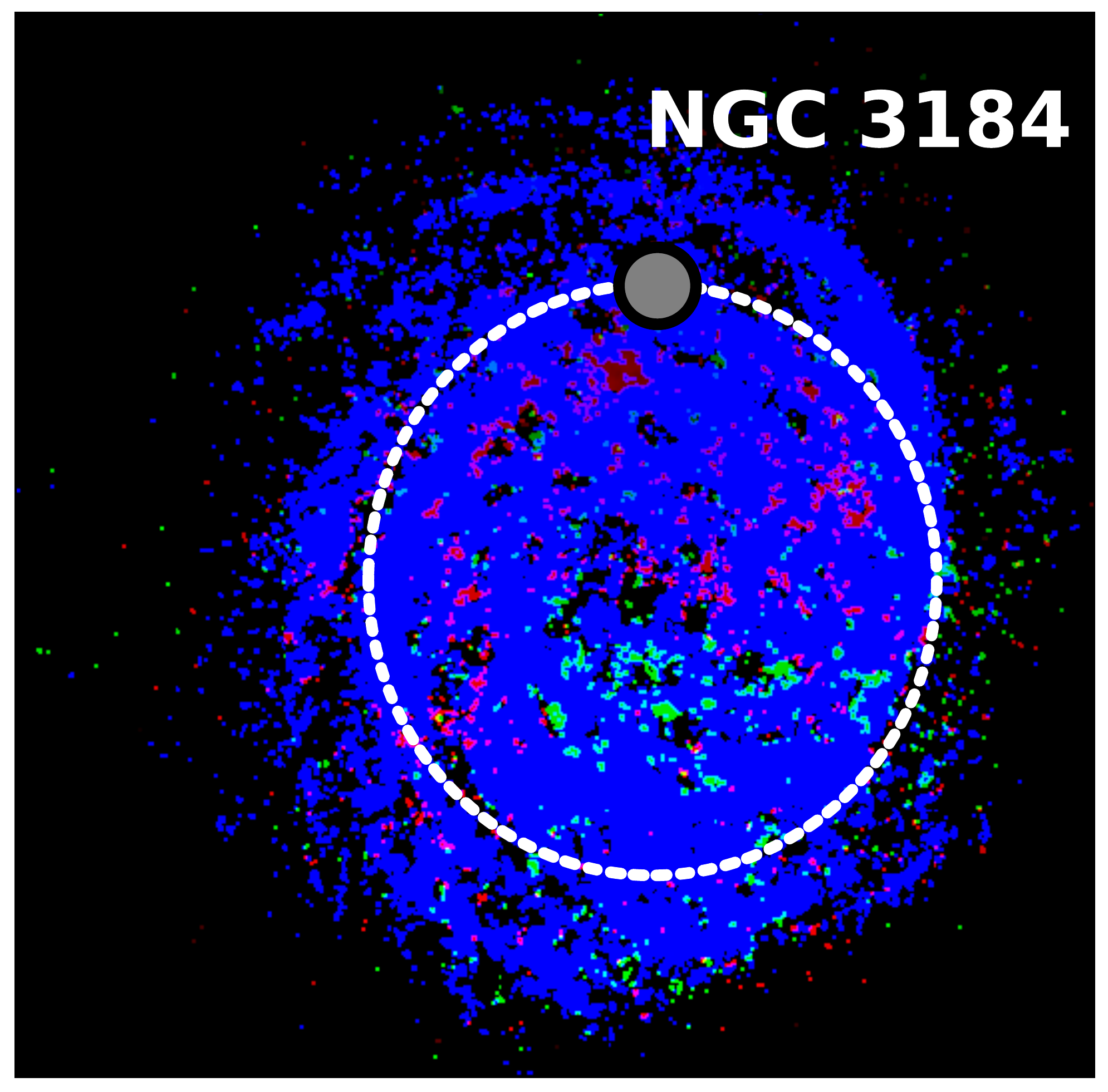}}}&
 \rotatebox{0}{\resizebox{56mm}{!}{\includegraphics[width = 0.6in, height = 0.6in]{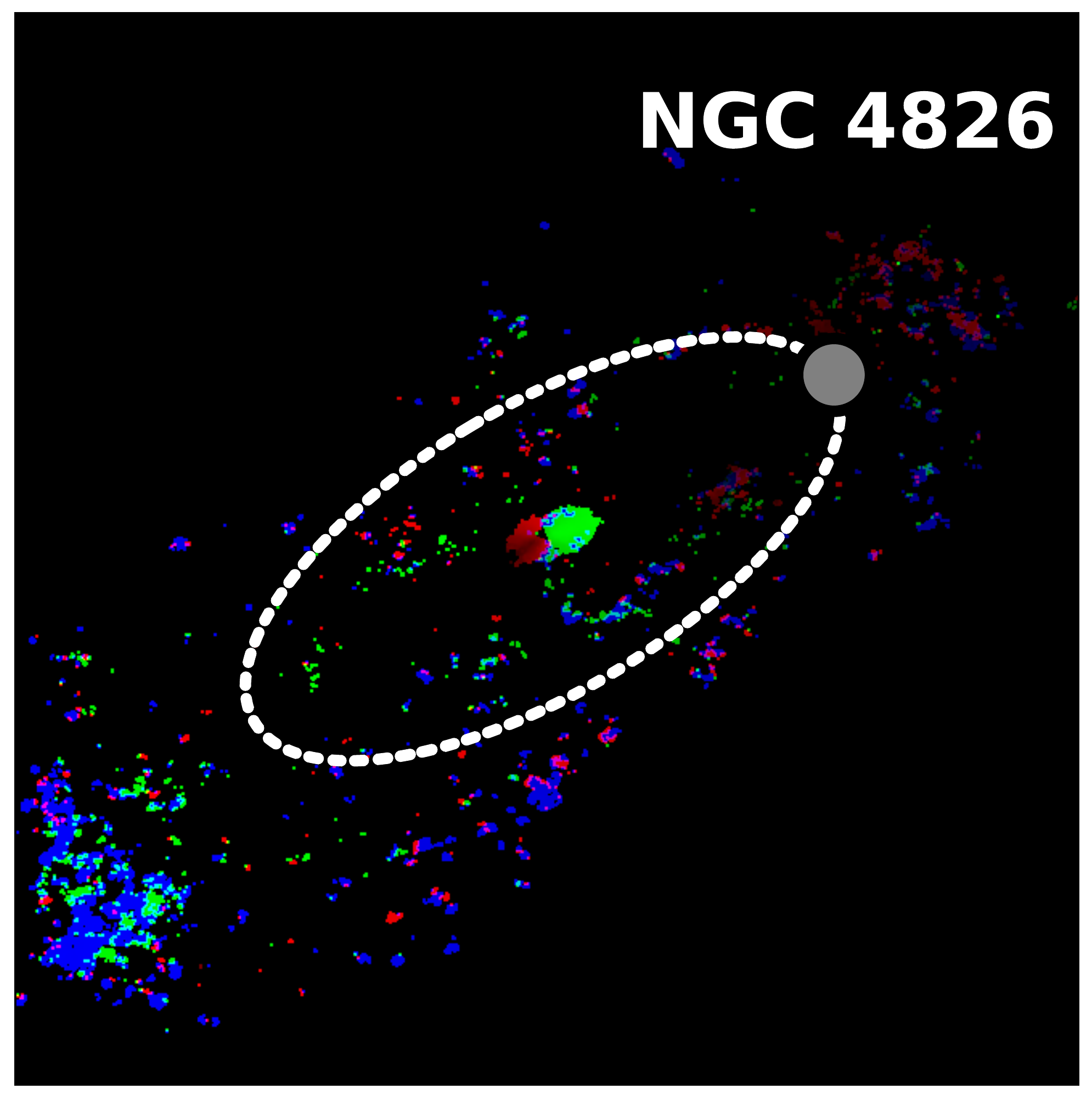}}}\\
 \rotatebox{0}{\resizebox{56mm}{!}{\includegraphics[width = 0.6in, height = 0.6in]{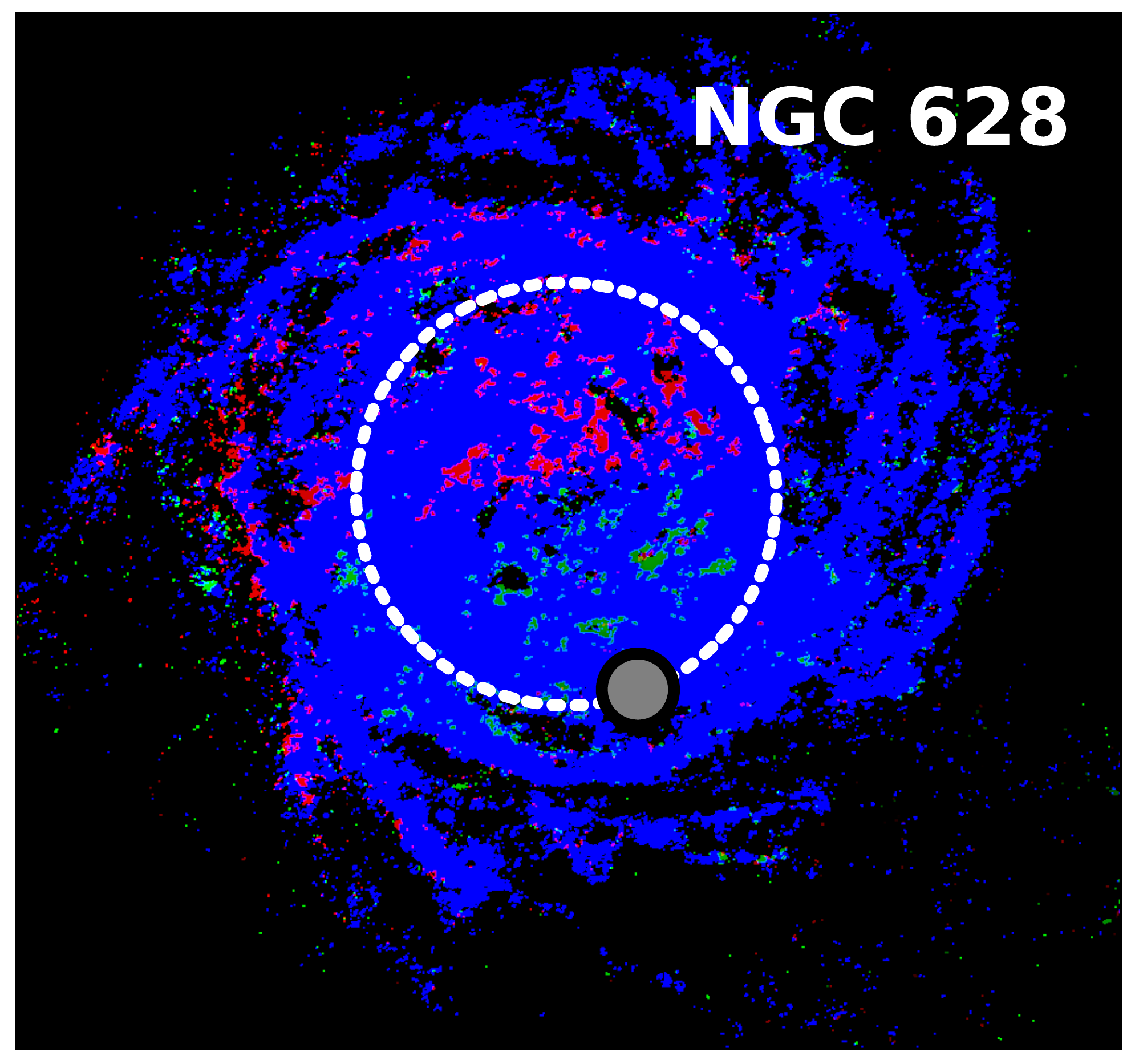}}}&
\rotatebox{0}{\resizebox{56mm}{!}{\includegraphics[width = 0.6in, height = 0.6in]{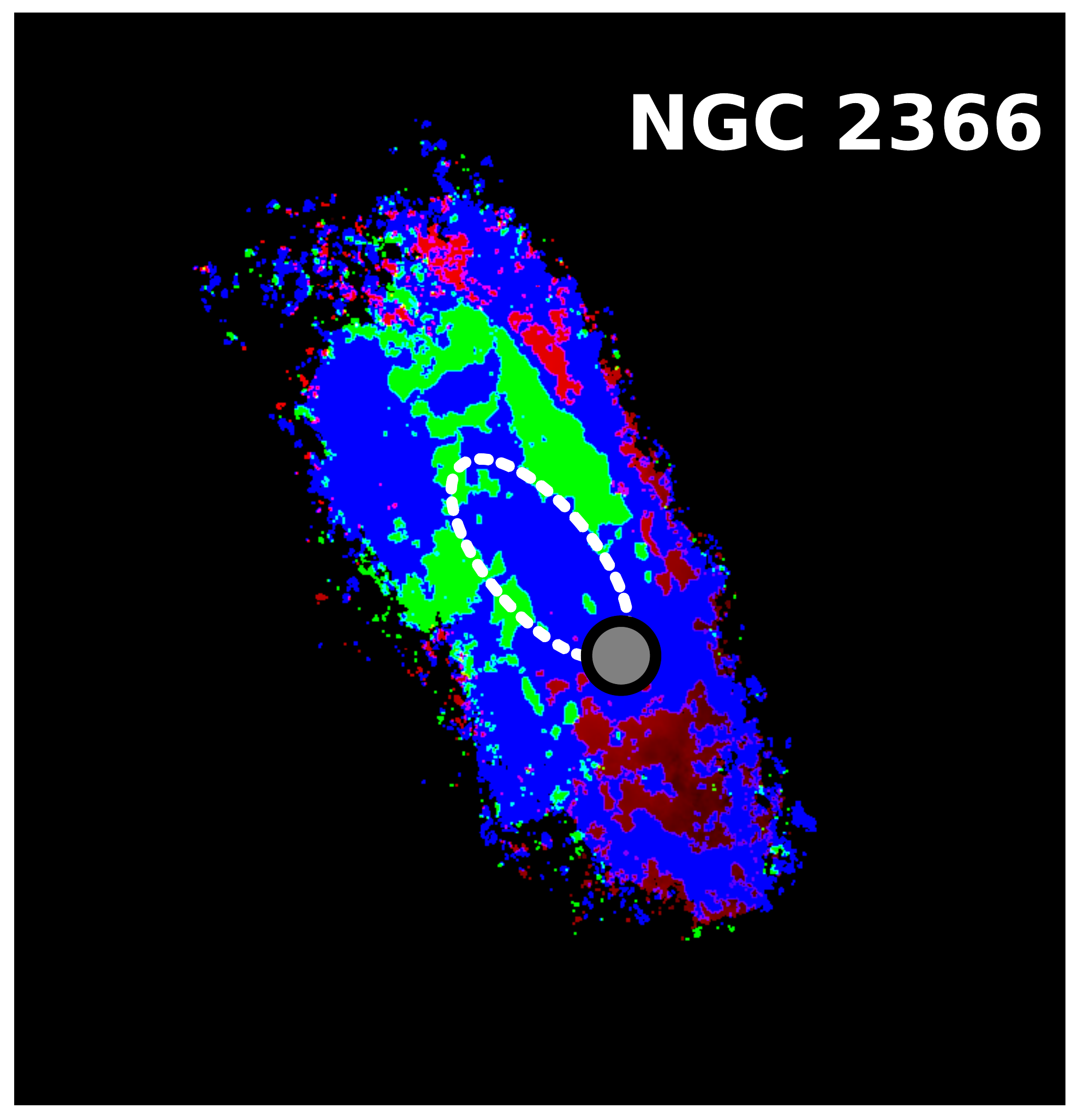}}}&
 \rotatebox{0}{\resizebox{56mm}{!}{\includegraphics[width = 0.6in, height = 0.6in]{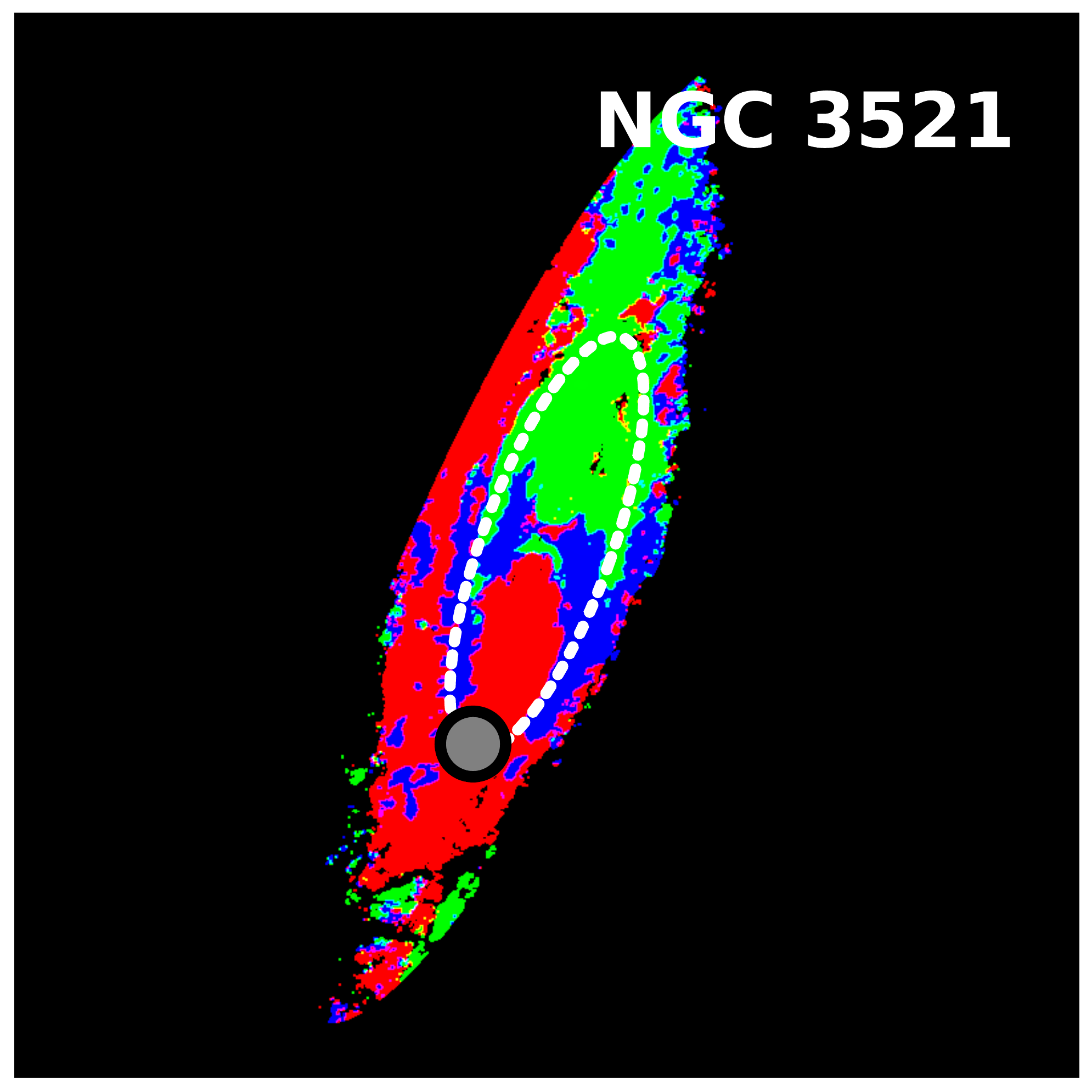}}}\\
 \rotatebox{0}{\resizebox{56mm}{!}{\includegraphics[width = 0.6in, height = 0.6in]{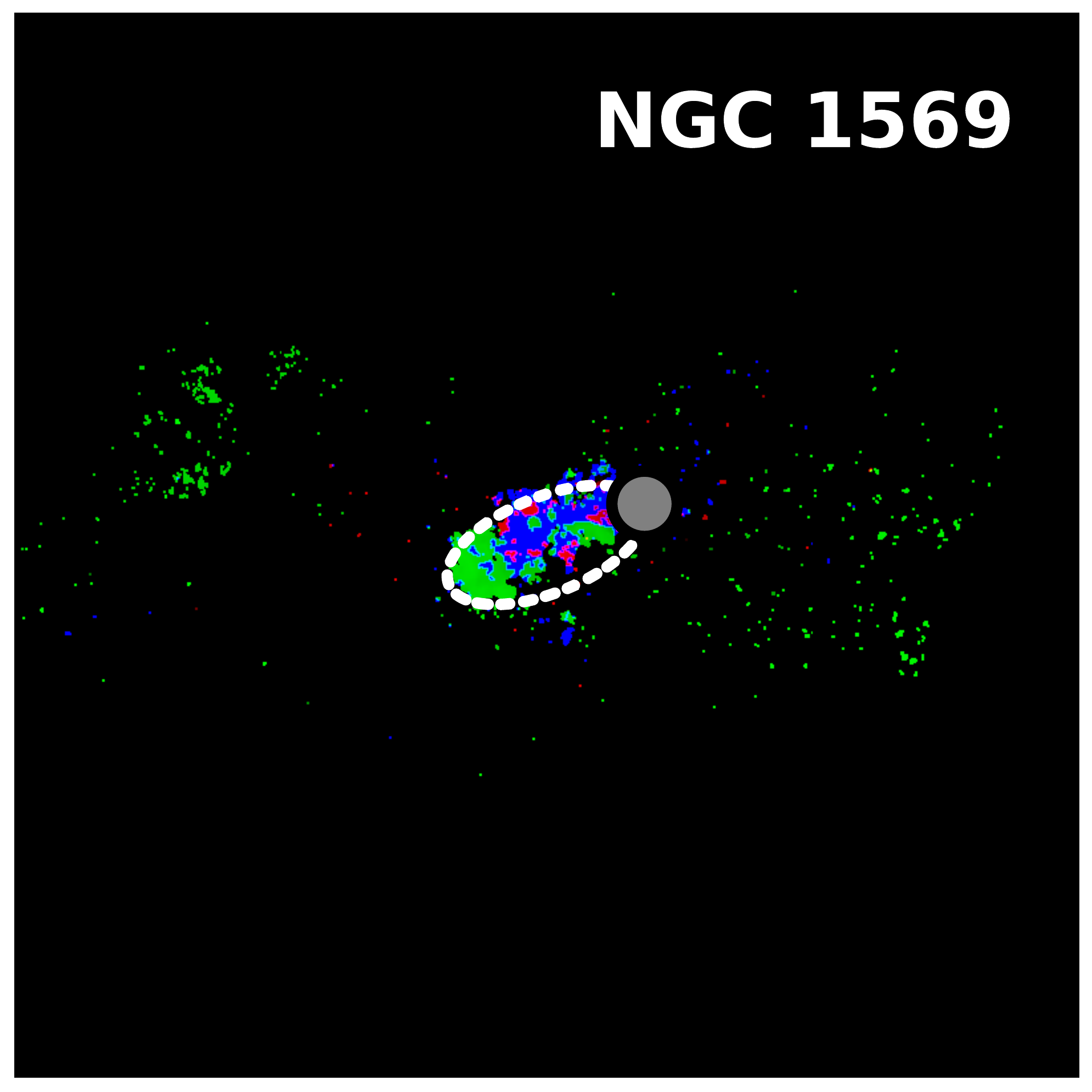}}}&
 \rotatebox{0}{\resizebox{56mm}{!}{\includegraphics[width = 0.6in, height = 0.6in]{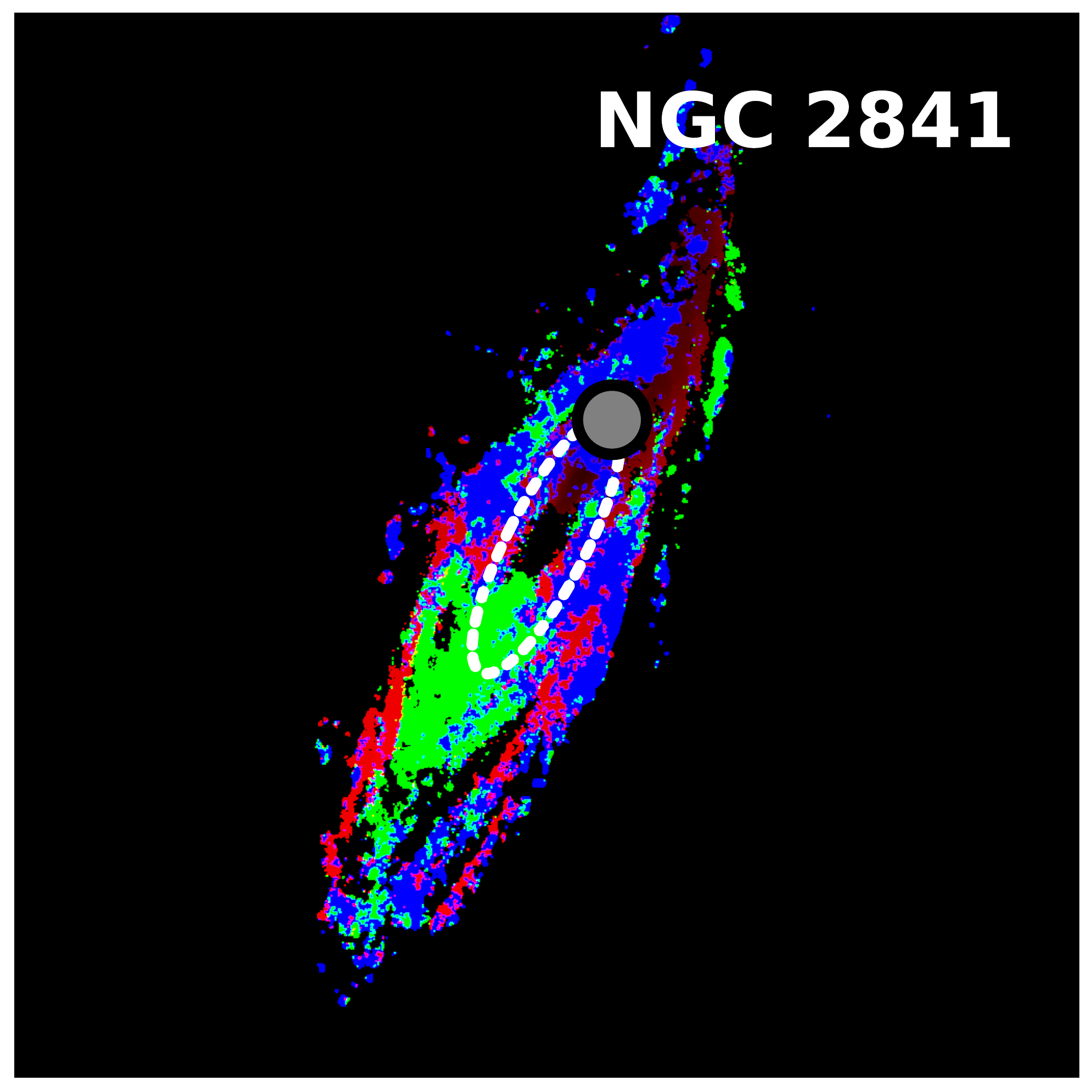}}}&
 \rotatebox{0}{\resizebox{56mm}{!}{\includegraphics[width = 0.6in, height = 0.6in]{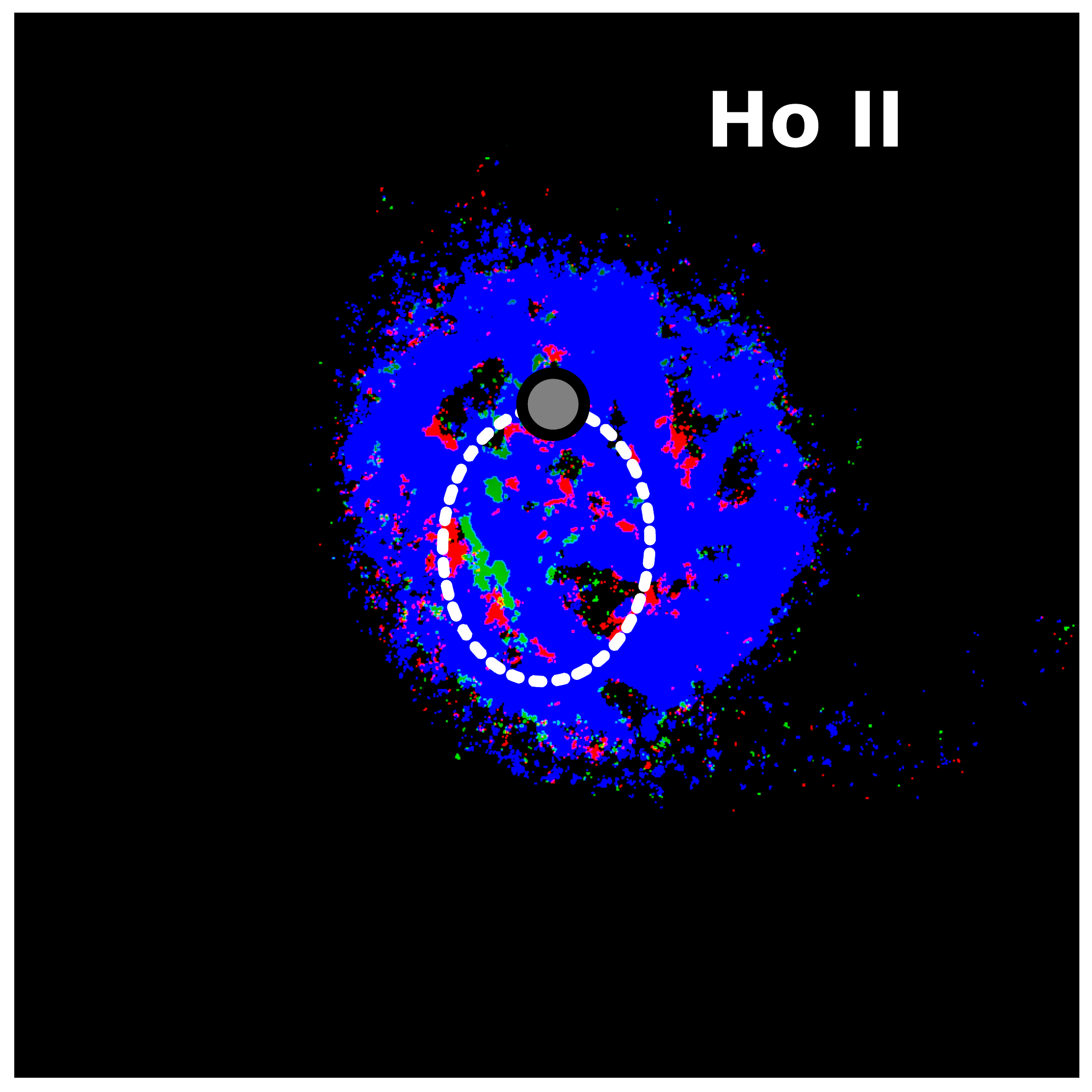}}}
\\\\ & & \hspace*{-12.12cm}\textbf{Figure \ref{fig:location_asymm_and_sym}}. \scriptsize{(continued).}
\end{tabular}
\end{figure*}

\begin{figure*}
    \begin{tabular}{l l l}
 \rotatebox{0}{\resizebox{56mm}{!}{\includegraphics[width = 0.6in, height = 0.6in]{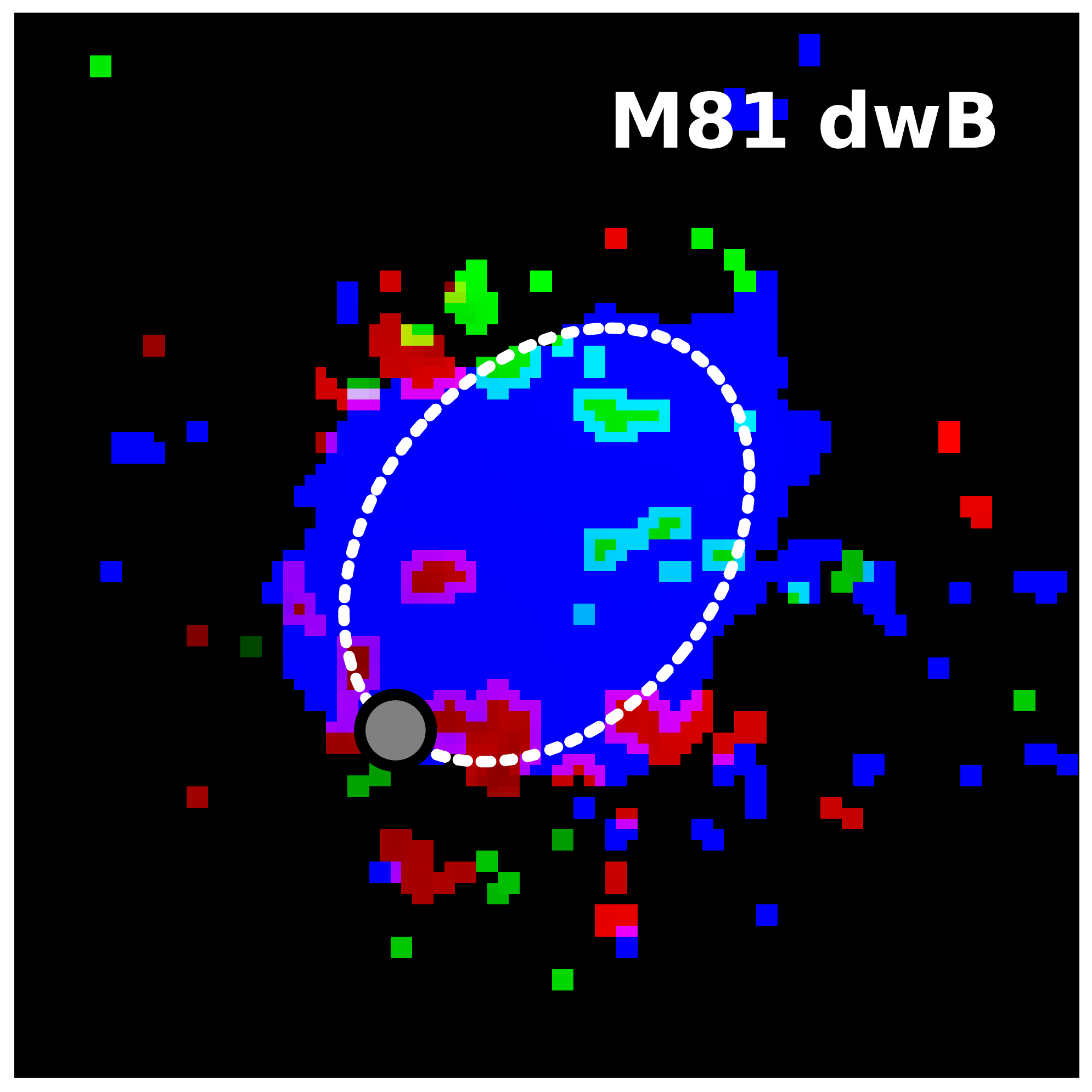}}}&
 \rotatebox{0}{\resizebox{56mm}{!}{\includegraphics[width = 0.6in, height = 0.6in]{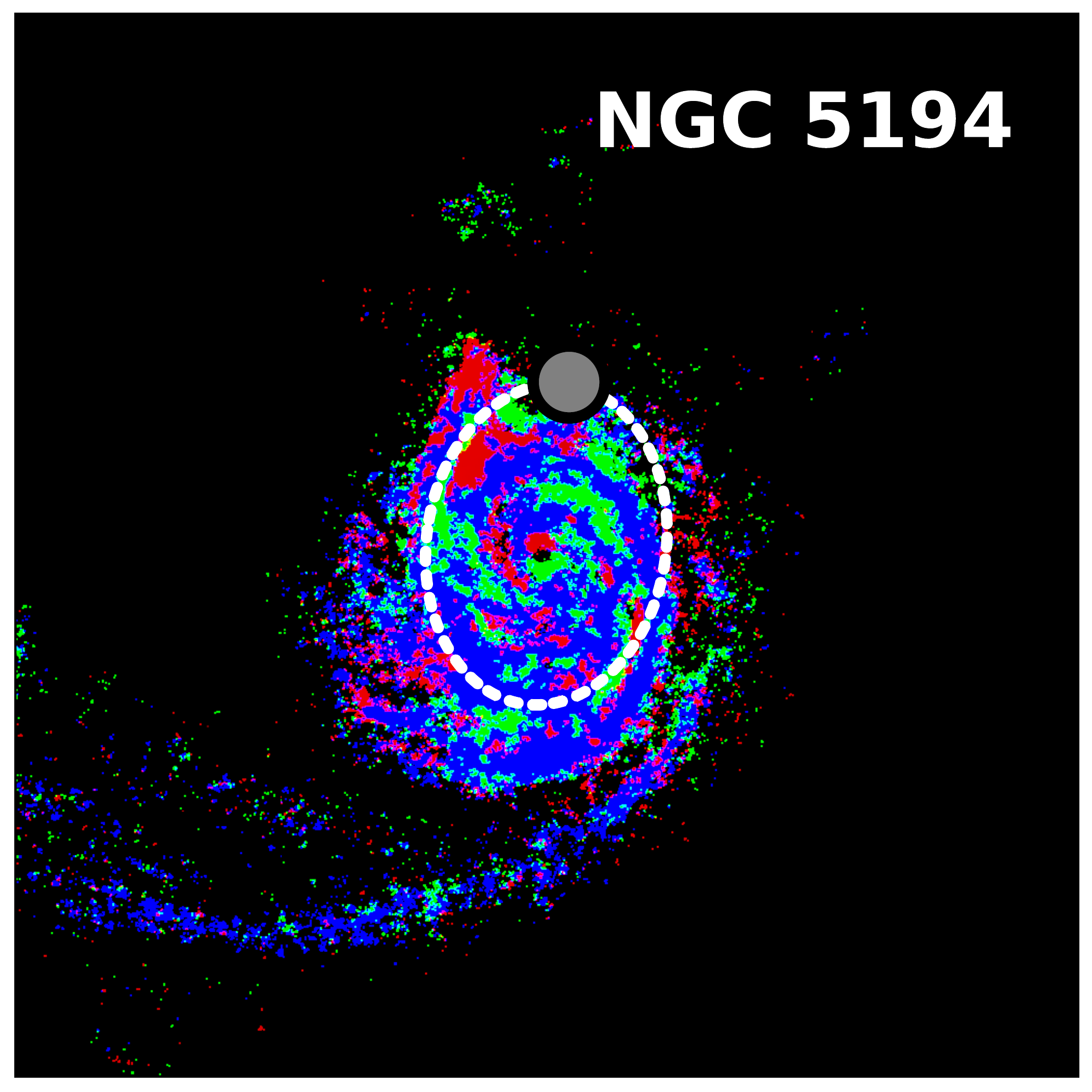}}}&
 \rotatebox{0}{\resizebox{56mm}{!}{\includegraphics[width = 0.6in, height = 0.6in]{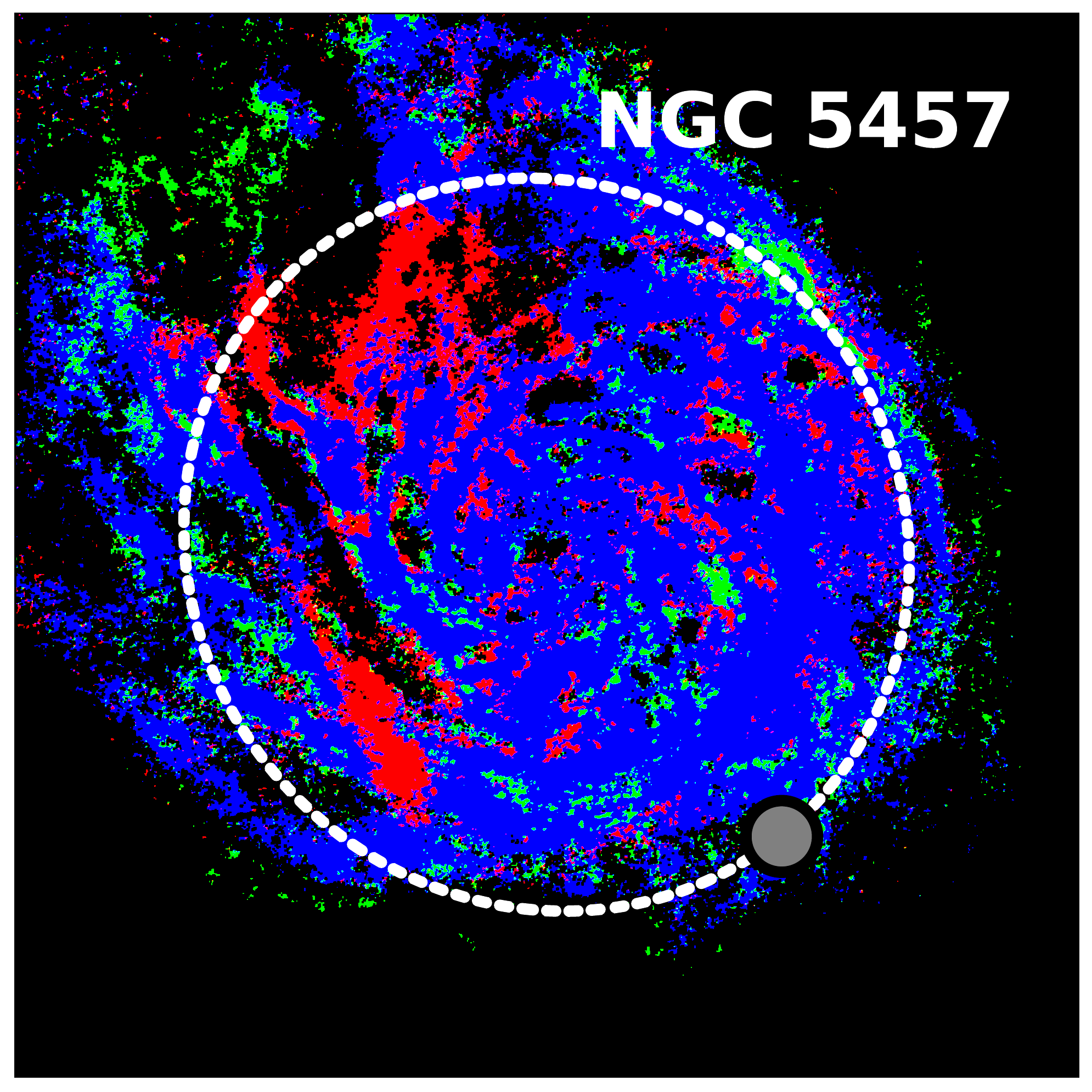}}}\\
  \rotatebox{0}{\resizebox{56mm}{!}{\includegraphics[width = 0.6in, height = 0.6in]{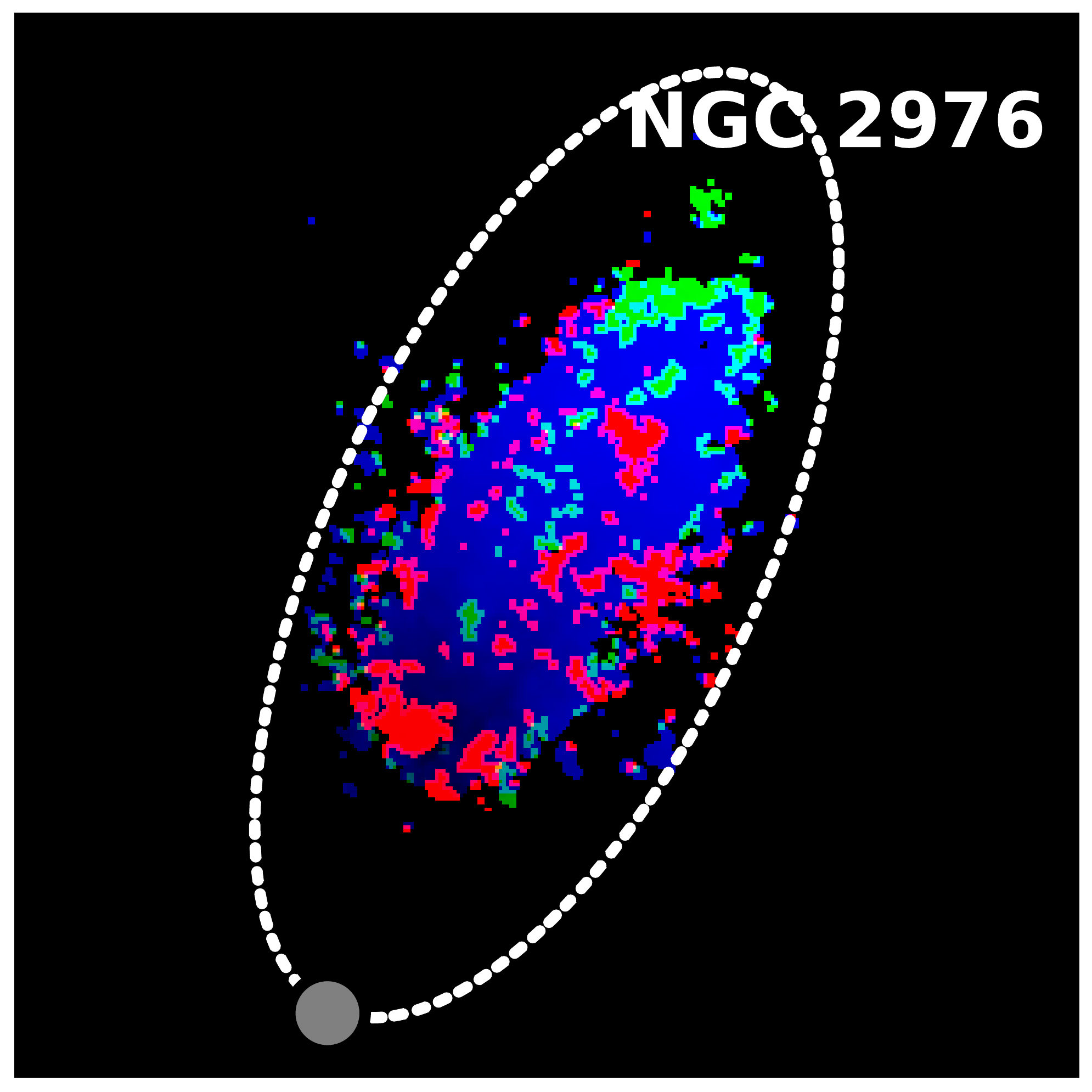}}}&
 \rotatebox{0}{\resizebox{56mm}{!}{\includegraphics[width = 0.6in, height = 0.6in]{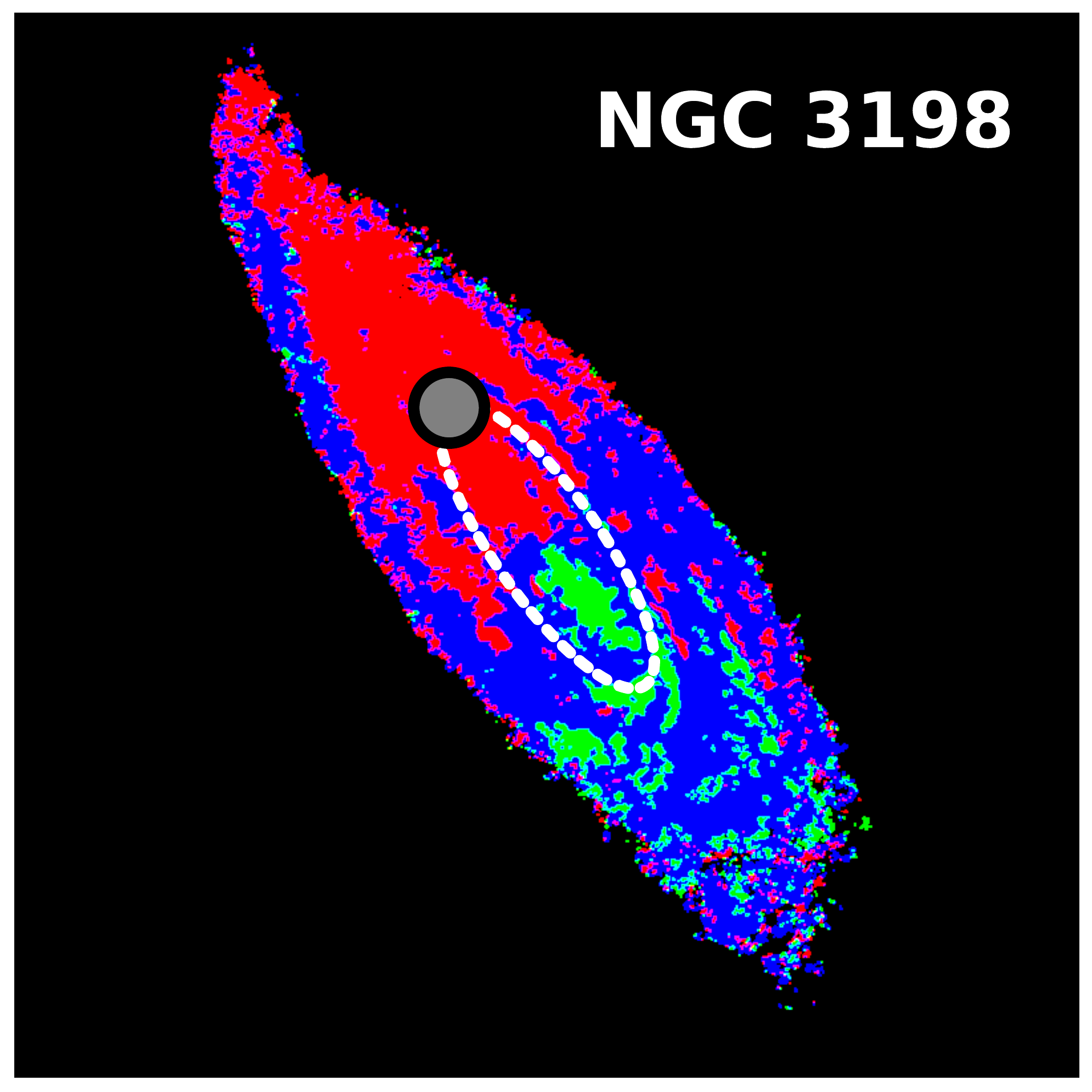}}}&
 \rotatebox{0}{\resizebox{56mm}{!}{\includegraphics[width = 0.6in, height = 0.6in]{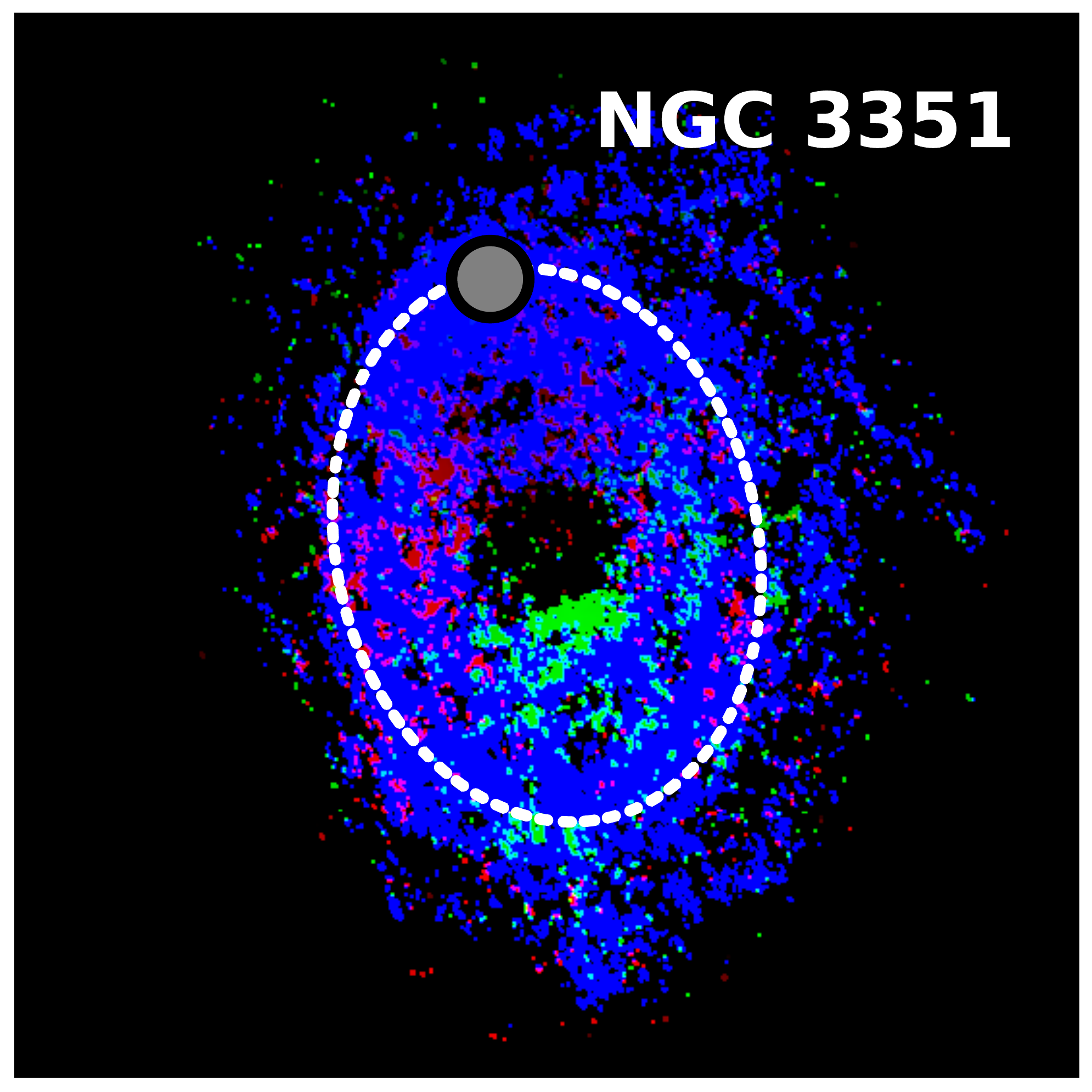}}}\\
  \rotatebox{0}{\resizebox{56mm}{!}{\includegraphics[width = 0.6in, height = 0.6in]{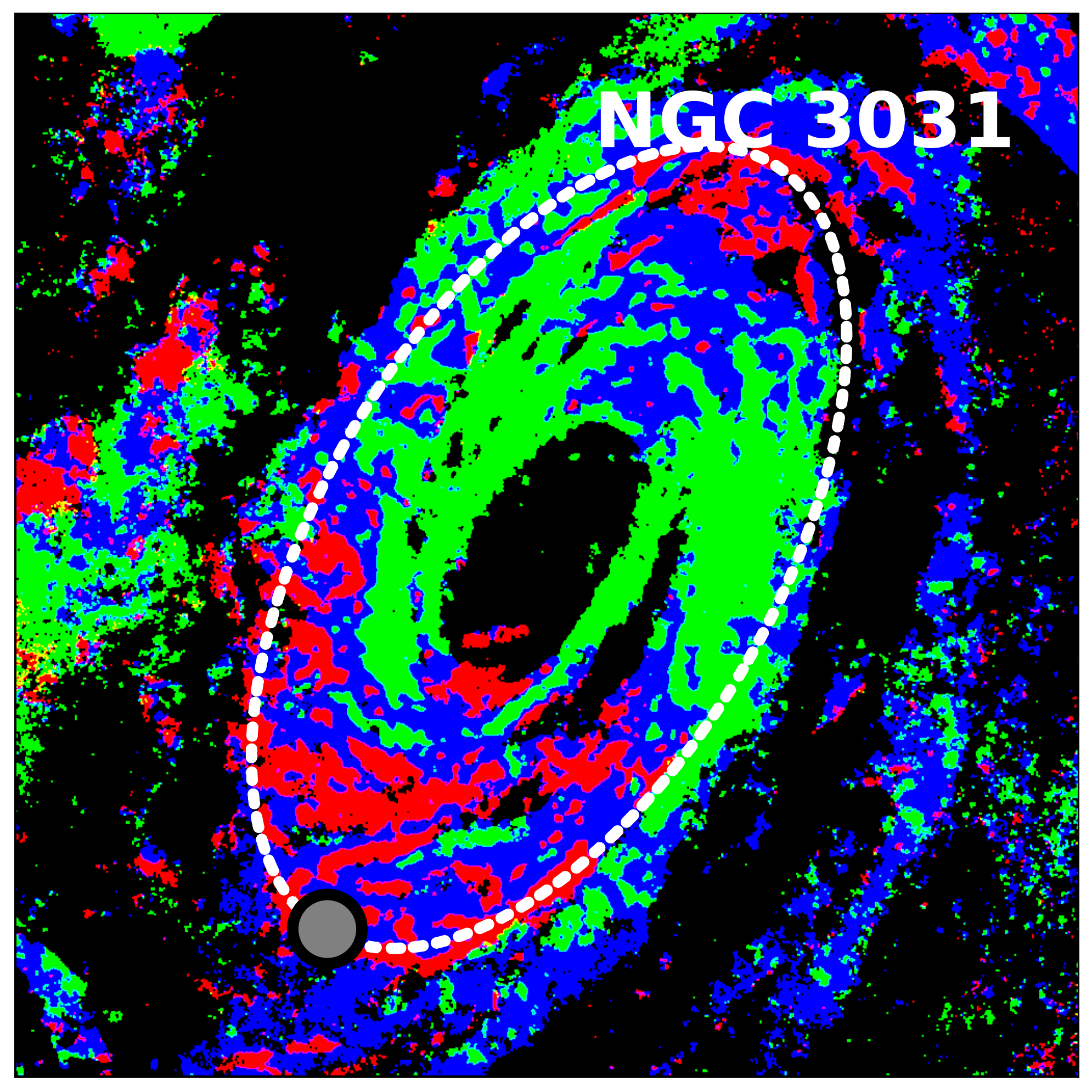}}}&
\rotatebox{0}{\resizebox{56mm}{!}{\includegraphics[width = 0.6in, height = 0.6in]{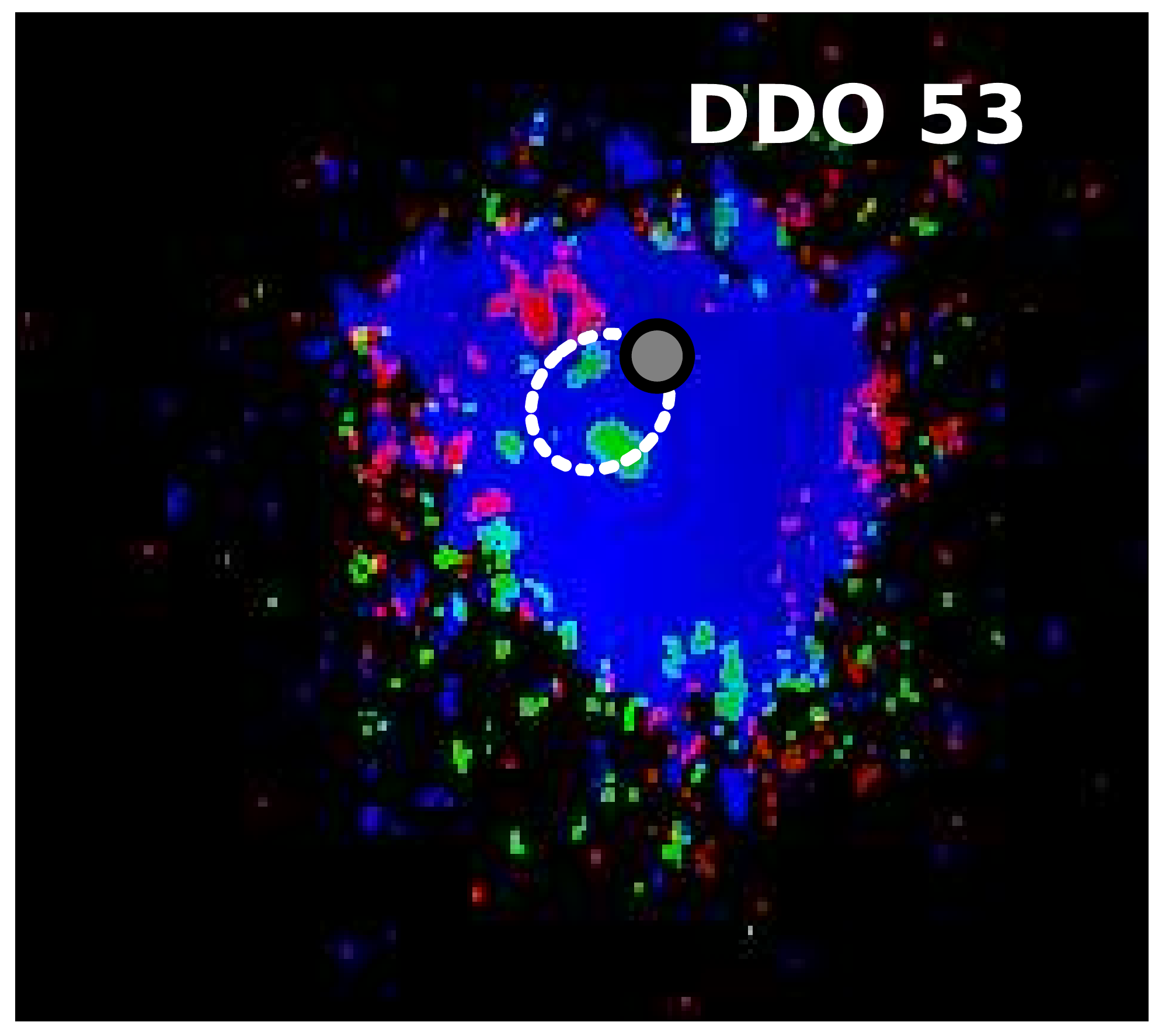}}}&
\rotatebox{0}{\resizebox{56mm}{!}{\includegraphics[width = 0.6in, height = 0.6in]{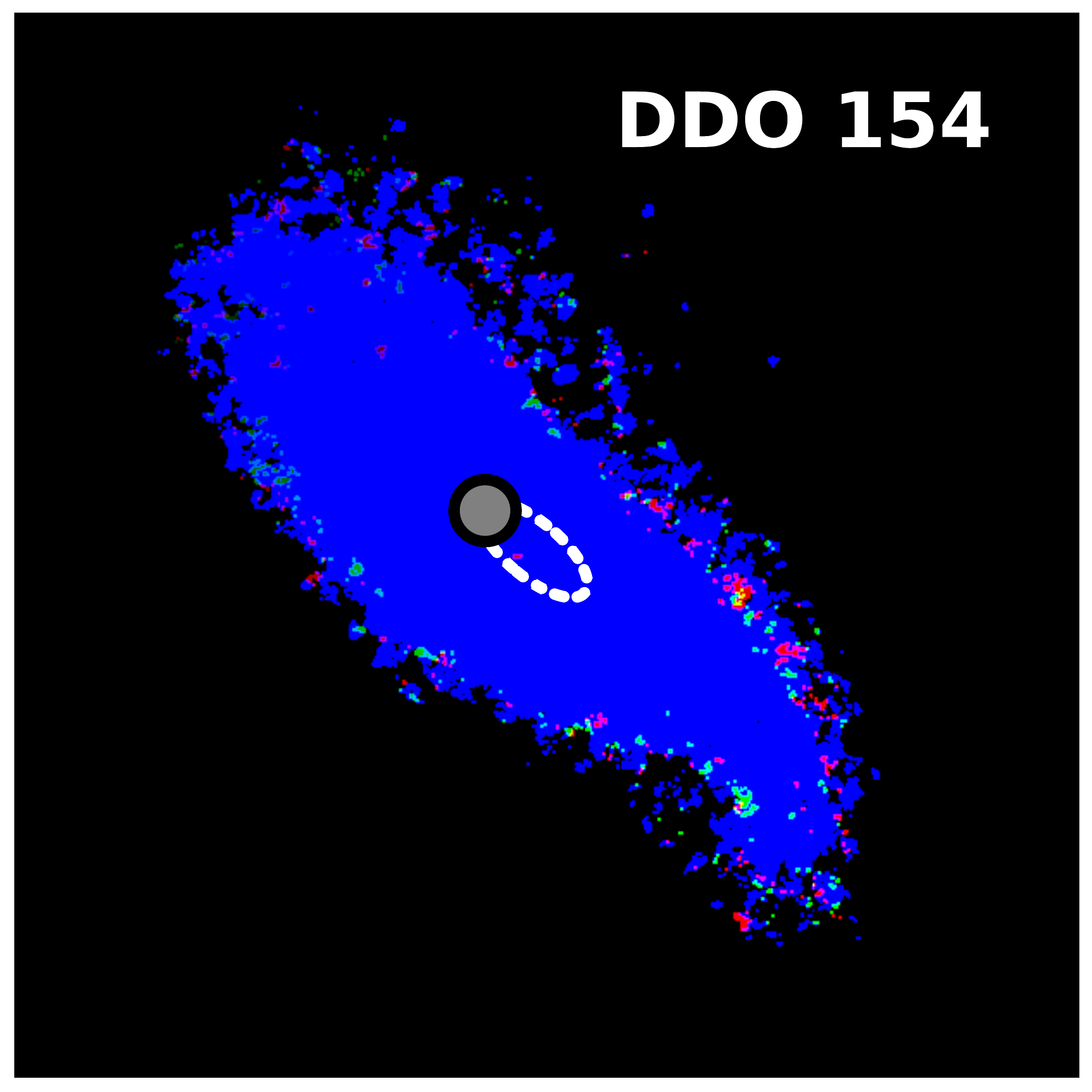}}}\\
\rotatebox{0}{\resizebox{56mm}{!}{\includegraphics[width = 0.6in, height = 0.6in]{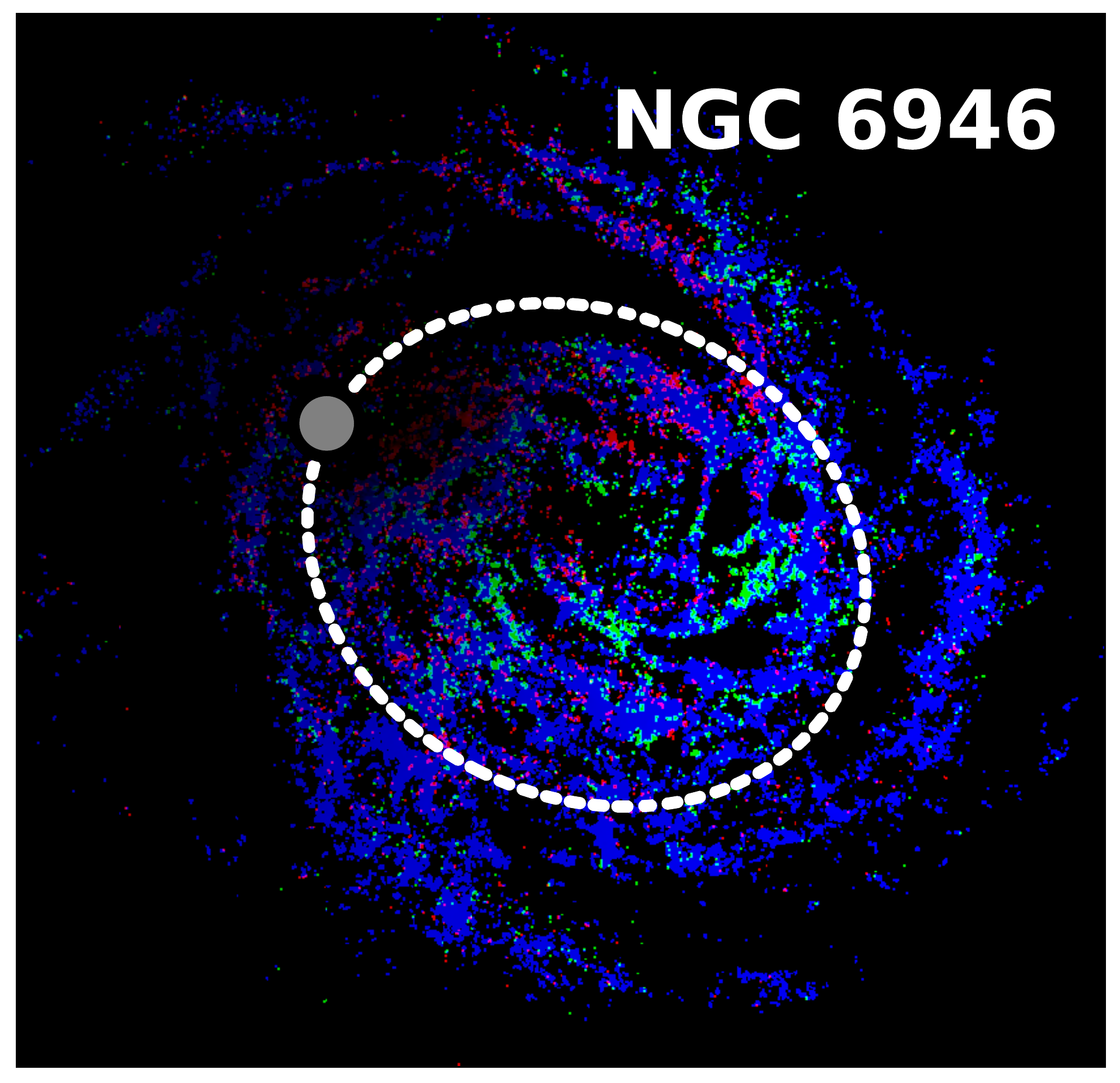}}}
\\\\ & & \hspace*{-12.2cm}\textbf{Figure \ref{fig:location_asymm_and_sym}}. \scriptsize{(continued).}
\end{tabular}
\end{figure*}

\begin{figure*}
    \begin{tabular}{l l}
  \includegraphics[width = 3in,height = 3in]{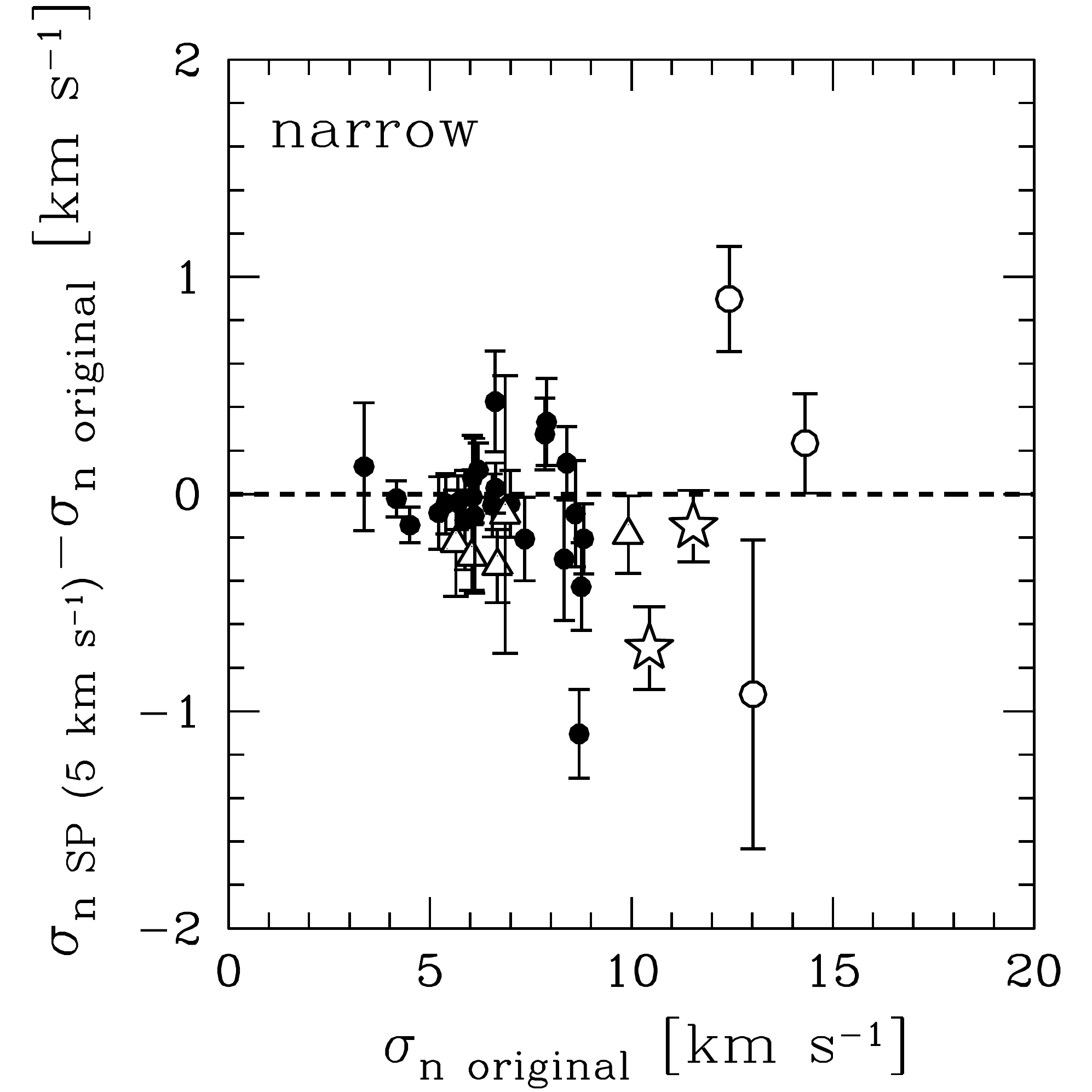}&
 \includegraphics[width = 3in,height = 3in]{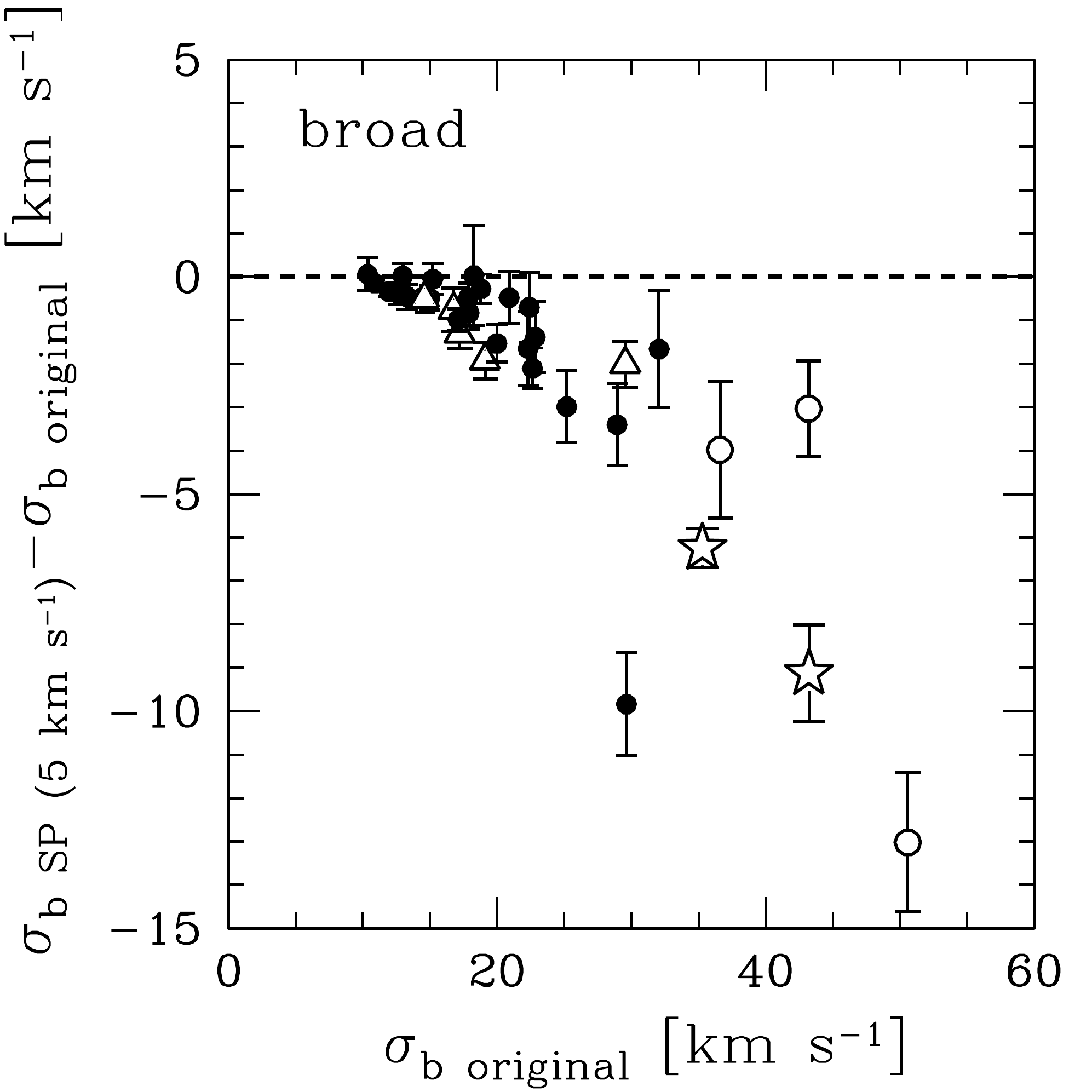}\\
 \includegraphics[width = 3in,height = 3in]{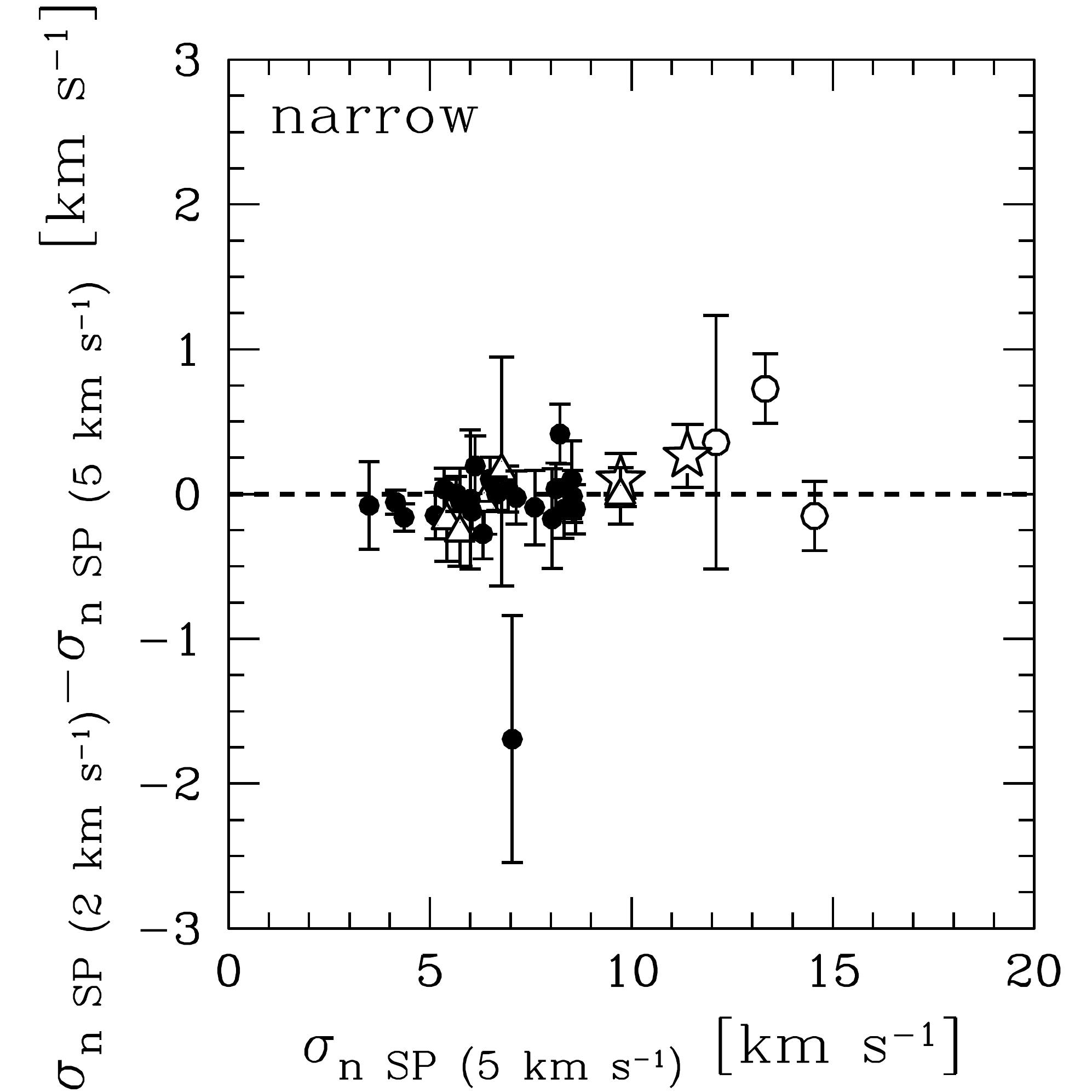}&
\includegraphics[width = 3in,height = 3in]{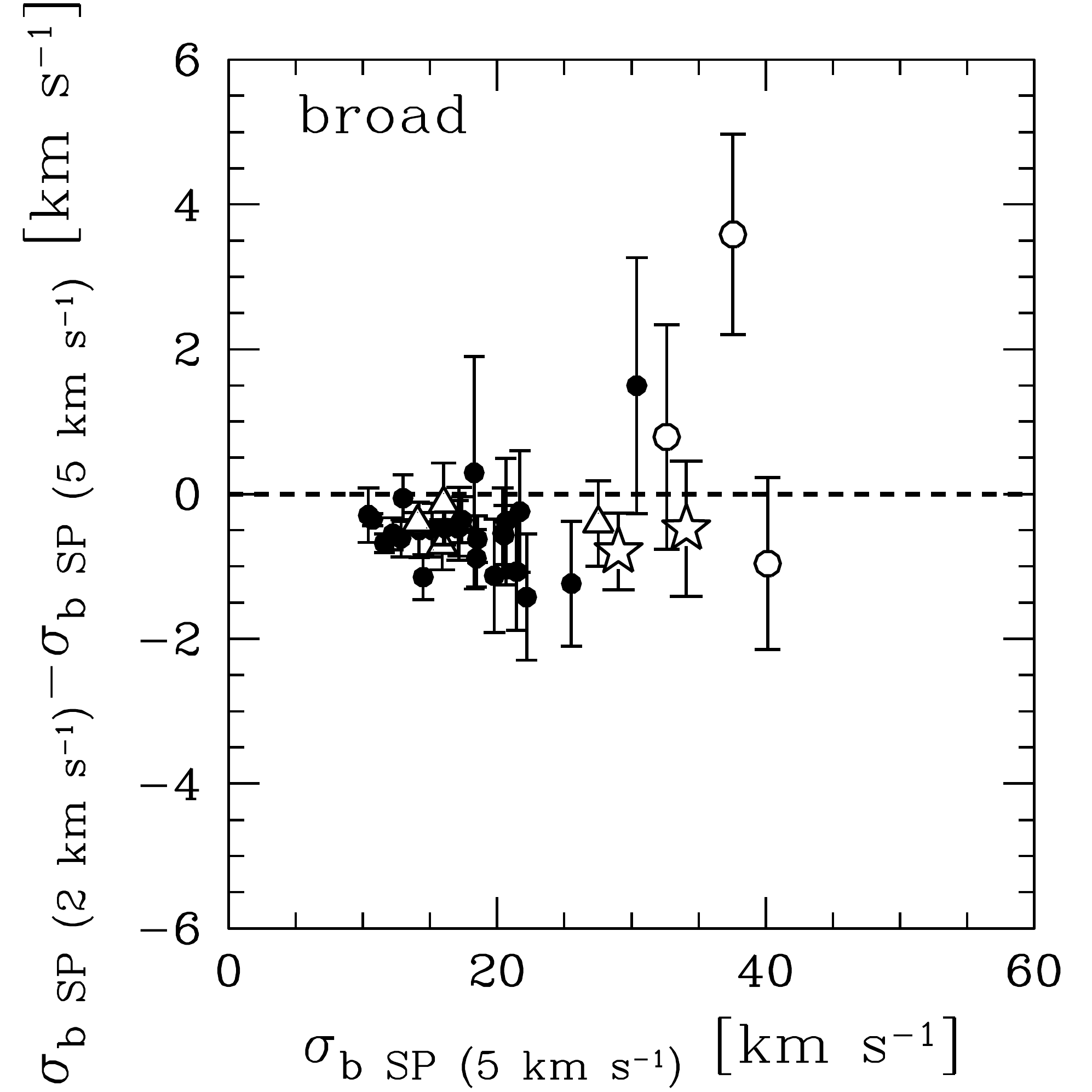}
    \end{tabular}
\caption{\textit{Top panels:} Comparison of the velocity dispersions
  of the original super profiles to those derived from SP profiles
  using $|\Delta_{H-I}| \leq 5$ km~s$^{-1}$. \textit{Bottom panel:}
  Comparison of the velocity dispersions derived from SP profiles
  using the $|\Delta_{H-I}| \leq 5$ km~s$^{-1}$ with those derived from
  SP profiles using $|\Delta_{H-I}| \leq 2$ km~s$^{-1}$. Dashed lines are
  lines of equality. Symbols are as in Fig. \ref{fig:faint_brigh_norm}.}
\label{fig:sym_asym_comp}
\end{figure*}

We compare the dispersions from the original super profiles (i.e.,
those derived from using the entire HI disk; see
Sect.~\ref{sec:method}) with those derived using the SP mask in
Fig.~\ref{fig:sym_asym_comp} (top panel). The dispersions of the
narrow component from the original super profiles are similar to those
from the SP super profiles. However, the dispersions of the broad
component from the original super profiles tend to be larger than
those from the SP super profiles. This is what we would expect if the
original profiles were broadened by asymmetry in the input
profiles. We can remove this effect by only selecting symmetric
profiles.

\begin{figure*}
    \begin{tabular}{l l}
 \includegraphics[width = 3in,height = 3in]{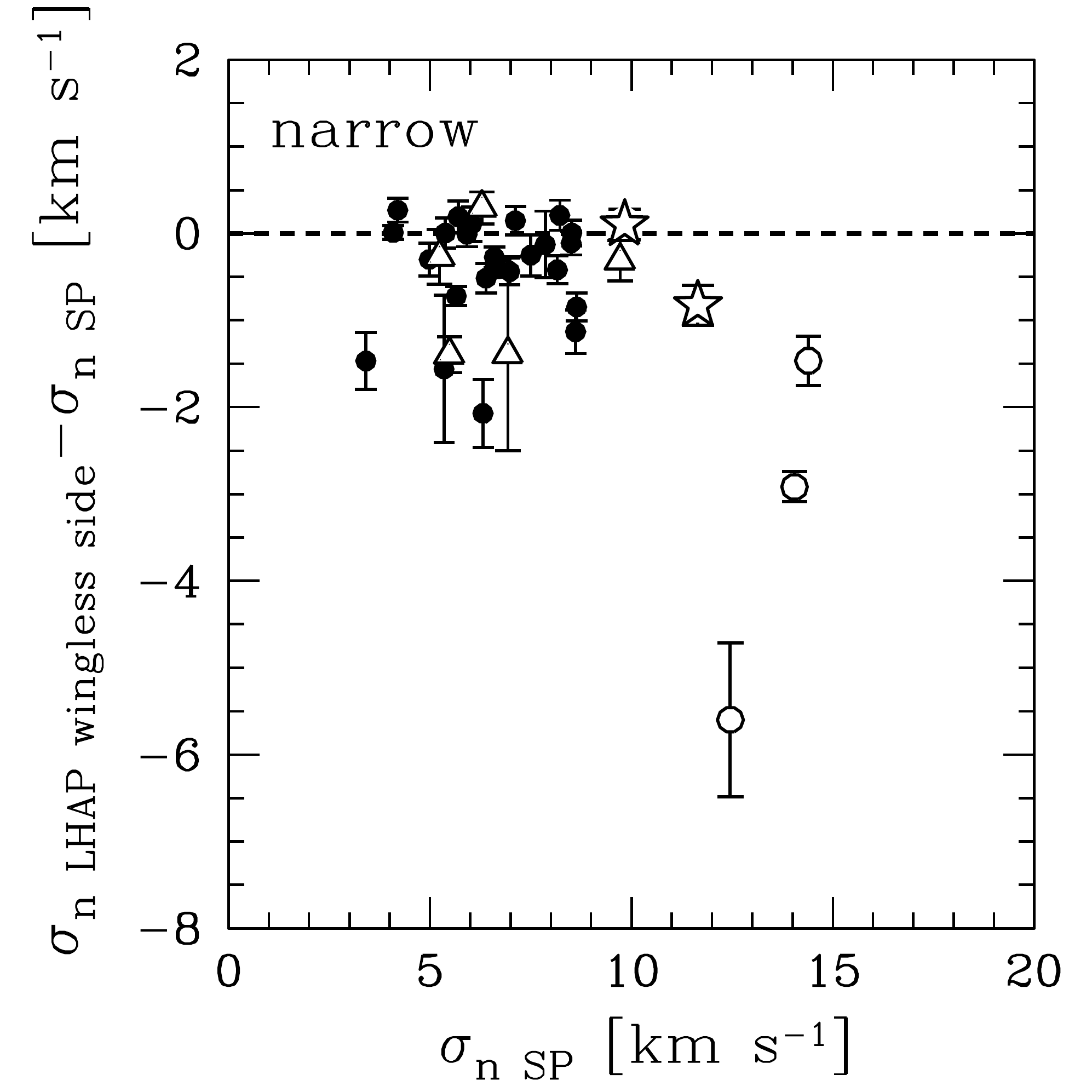}&
\includegraphics[width = 3in,height = 3in]{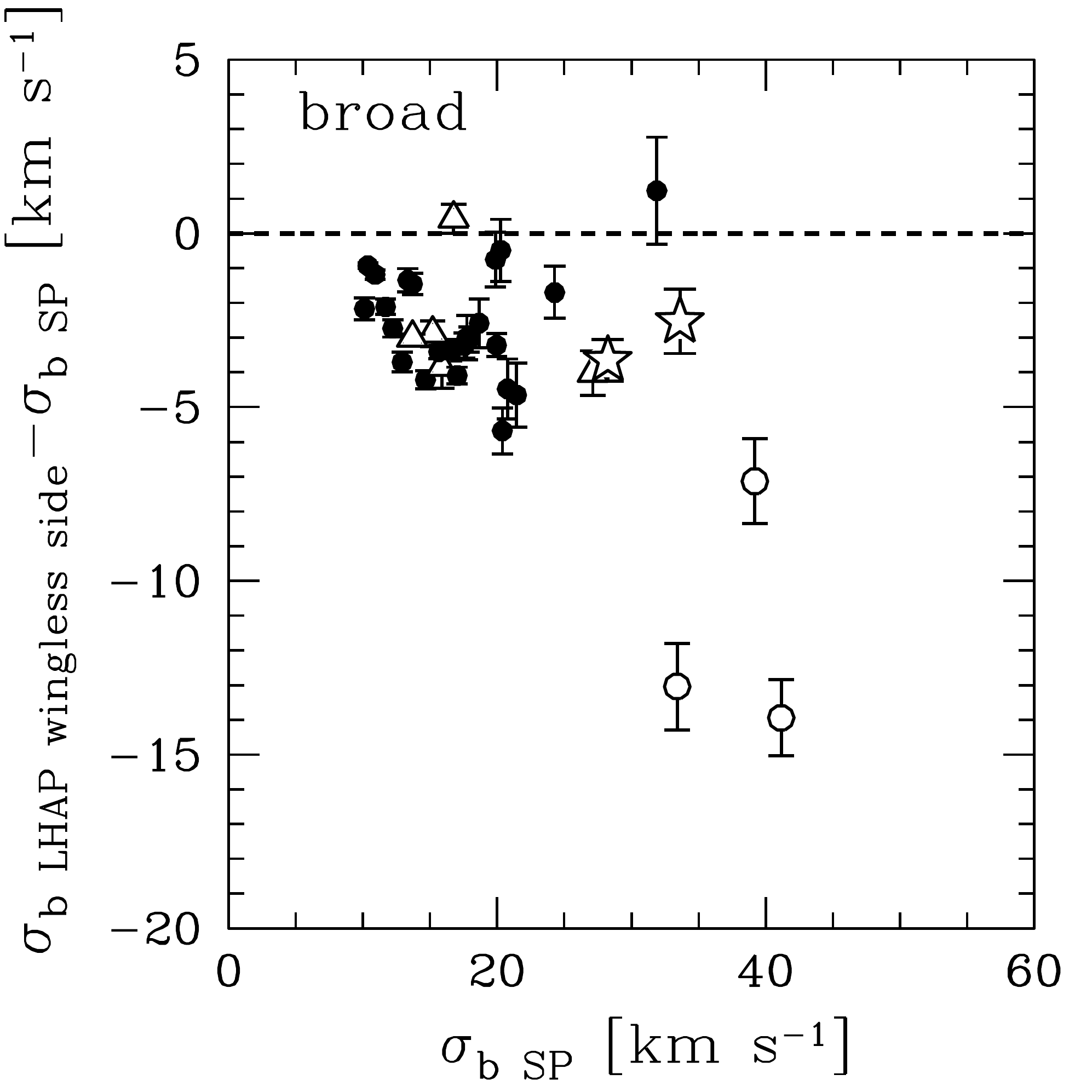}\\
\includegraphics[width = 3in,height = 3in]{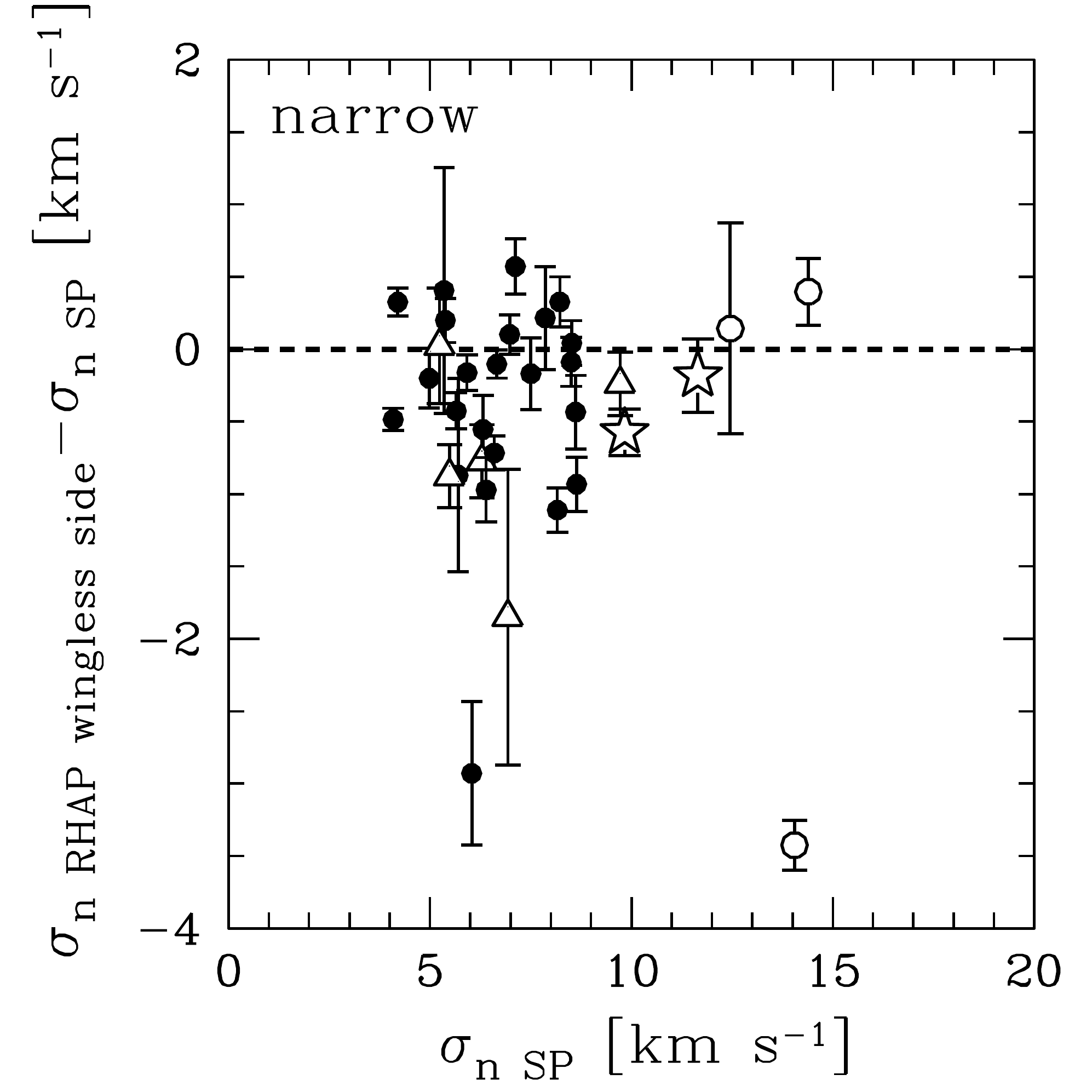}&
\includegraphics[width = 3in,height = 3in]{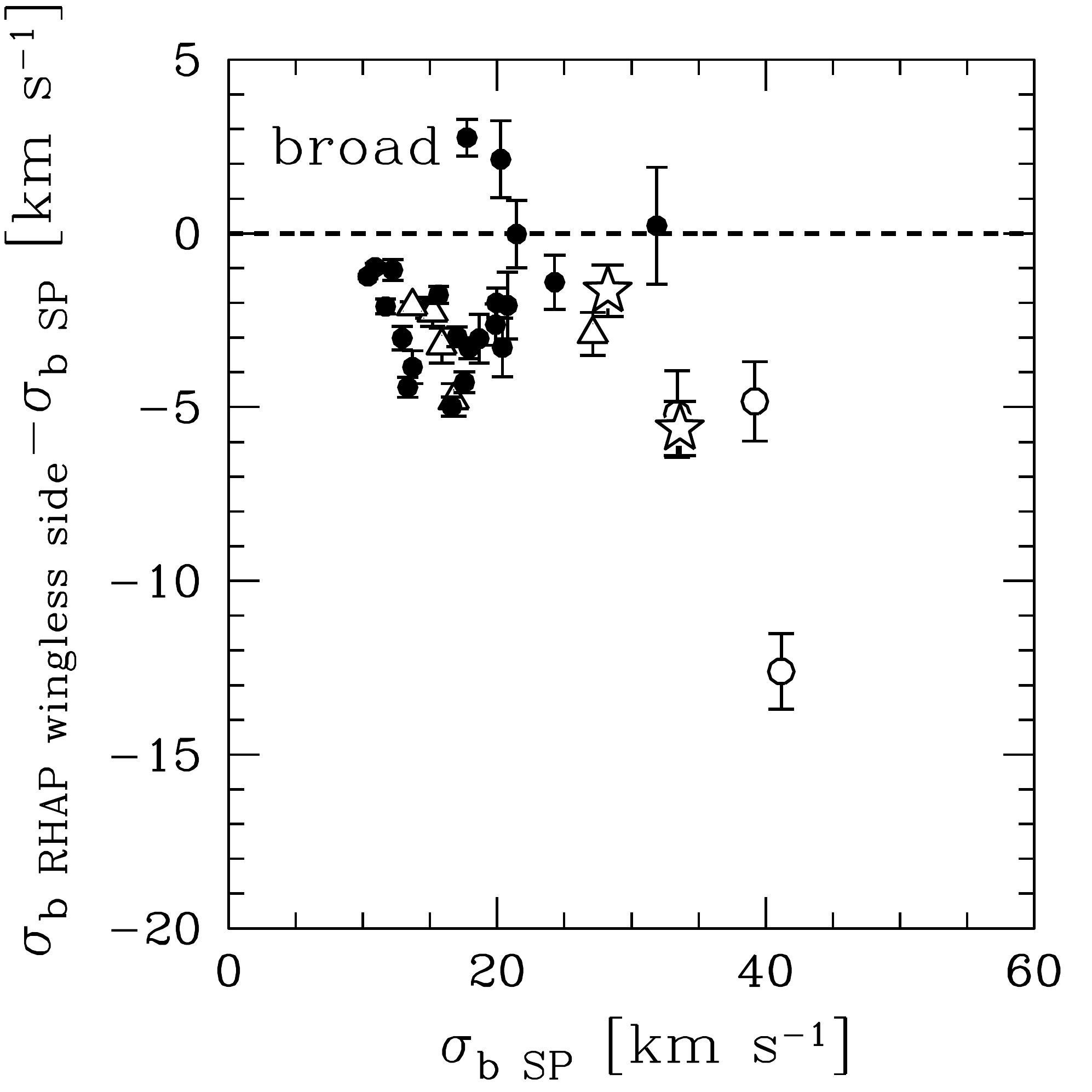}
\end{tabular}
\caption{Comparison of the velocity dispersions derived from the SP
  super profiles to those derived from the wingless sides of the LHAP
  (\textit{top panel}) and RHAP (\textit{bottom panel}) super
  profiles using $|\Delta_{H-I}| \leq 2$ km s$^{-1}$. The dashed
  lines are lines of equality. Symbols are as in Fig.
  \ref{fig:faint_brigh_norm}.}
\label{fig:wingless}
\end{figure*}
\begin{deluxetable*}{l r r r r l}
\centering
\tabletypesize{\scriptsize}
\tablecaption{Fitted parameters of the clean sample \label{tab:fitted_clean}}
\tablewidth{0pt}
\tablehead{
	\multicolumn{1}{c}{Galaxy}&\multicolumn{1}{c}{$\sigma_{1g}$}&
	\multicolumn{1}{c}{$\sigma_{n}$}&\multicolumn{1}{c}{$\sigma_{b}$} 
	&\multicolumn{1}{c}{$A_{n}/A{b}$}&\multicolumn{1}{c}{$\chi^{2}_{2G}/\chi^{2}_{1G}$}\\
	&\multicolumn{1}{c}{$(\rm{km s^{-1}})$}&$(\rm{km s^{-1})}$&$\rm{(km s^{-1})}$& & \\
	\multicolumn{1}{c}{1} &\multicolumn{1}{c}{2} &\multicolumn{1}{c}{3}&\multicolumn{1}{c}{4} 
	&\multicolumn{1}{c}{5}& \multicolumn{1}{c}{6}
}
\startdata
DDO 53 &9.7$\rm{\pm}$0.1&6.0$\rm{\pm}$0.1&13.3$\rm{\pm}$0.2& 0.41$\rm{\pm}$0.03 &$<0.01$ \\
DDO 154&9.6$\rm{\pm}$0.1&6.3$\rm{\pm}$0.1&12.9$\rm{\pm}$0.2& 0.50$\rm{\pm}$0.04& $<0.01$\\
Ho I   &8.9$\rm{\pm}$0.1&5.4$\rm{\pm}$0.1&12.2$\rm{\pm}$0.2& 0.39$\rm{\pm}$0.02&$~~~$0.02\\
Ho II  &8.9$\rm{\pm}$0.1&5.0$\rm{\pm}$0.1&11.7$\rm{\pm}$0.2& 0.33$\rm{\pm}$0.02 &$<0.01$\\
IC 2574&9.8$\rm{\pm}$0.2&5.7$\rm{\pm}$0.2&13.7$\rm{\pm}$0.3& 0.42$\rm{\pm}$0.03&$~~~$0.10\\
M81 dwA&8.2$\rm{\pm}$0.1&3.4$\rm{\pm}$0.1&10.1$\rm{\pm}$0.2& 0.14$\rm{\pm}$0.01&$~~~$0.01\\
\textbf{NGC 628}&8.8$\rm{\pm}$0.2&4.5$\rm{\pm}$0.1&11.6$\rm{\pm}$0.1& 0.27$\rm{\pm}$0.02&$~~~$0.03\\
NGC 925&12.6$\rm{\pm}$0.2&8.6$\rm{\pm}$0.2&20.4$\rm{\pm}$0.8& 0.74$\rm{\pm}$0.06&$~~~$0.02\\
NGC 2366&12.2$\rm{\pm}$0.2&8.2$\rm{\pm}$0.1&17.7$\rm{\pm}$0.3&0.58$\rm{\pm}$0.03 &$<0.01$\\
NGC 2403&10.7$\rm{\pm}$0.2&6.6$\rm{\pm}$0.1&16.8$\rm{\pm}$0.2& 0.60$\rm{\pm}$0.06&$~~~$0.02\\
\textbf{NGC 2903}&12.6$\rm{\pm}$0.3&8.2$\rm{\pm}$0.2&24.3$\rm{\pm}$0.7&0.78$\rm{\pm}$0.04&$<0.01$\\
NGC 2976&11.9$\rm{\pm}$0.2&8.5$\rm{\pm}$0.1&19.9$\rm{\pm}$0.4& 0.86$\rm{\pm}$0.05 &$<0.01$\\
\textbf{NGC 3184}&10.4$\rm{\pm}$0.3&5.9$\rm{\pm}$0.1&17.9$\rm{\pm}$0.2& 0.44$\rm{\pm}$0.01&$<0.01$\\
\textbf{NGC 3198}&13.1$\rm{\pm}$0.2&8.5$\rm{\pm}$0.1&20.0$\rm{\pm}$0.3& 0.64$\rm{\pm}$0.03&$~~~$0.07\\
\textbf{NGC 3351}&9.9$\rm{\pm}$0.2&7.0$\rm{\pm}$0.1&21.5$\rm{\pm}$0.7& 0.95$\rm{\pm}$0.06 &$~~~$0.02\\
\textbf{NGC 3621}&11.3$\rm{\pm}$0.2&7.5$\rm{\pm}$0.2&18.7$\rm{\pm}$0.7& 0.71$\rm{\pm}$0.06&$~~~$0.16\\
NGC 4214&8.3$\rm{\pm}$0.1&4.2$\rm{\pm}$0.0&10.9$\rm{\pm}$0.1&0.29$\rm{\pm}$0.01&$~~~$0.25\\
\textbf{NGC 4736}&10.0$\rm{\pm}$0.2&7.1$\rm{\pm}$0.1&20.3$\rm{\pm}$0.7&0.99$\rm{\pm}$0.07&$~~~$0.03\\
\textbf{NGC 5055}&13.3$\rm{\pm}$0.3&7.9$\rm{\pm}$0.3&20.8$\rm{\pm}$0.7&0.55$\rm{\pm}$0.05&$~~~$0.17\\
\textbf{NGC 5236}&10.6$\rm{\pm}$0.2&5.5$\rm{\pm}$0.1&15.1$\rm{\pm}$0.1& 0.36$\rm{\pm}$0.03&$~~~$0.03\\
\textbf{NGC 6946}&10.1$\rm{\pm}$0.2&6.4$\rm{\pm}$0.3&17.5$\rm{\pm}$0.1&0.65$\rm{\pm}$0.09&$~~~$0.02\\
NGC 7793&10.4$\rm{\pm}$0.2&6.7$\rm{\pm}$0.1&17.0$\rm{\pm}$0.2& 0.63$\rm{\pm}$0.02&$~~~$0.21\\
\enddata
\tablecomments{Column 1: Name of galaxy; Column 2: 
Velocity dispersions derived from the single Gaussian fit; Column 3: 
Velocity dispersions of the narrow component; Column 4: Velocity 
dispersions of the broad component; Column 5: $A_{n}$/$A_{b}$ 
ratio. Column 6: Ratio of the $\chi^{2}$ values from the single and 
double Gaussian fitting. Spiral galaxies \citep[adopting the 
definition of][]{leroyetal08} are marked in bold face; the rest 
are dwarf galaxies (using the same definition).}
\end{deluxetable*}
\begin{figure}
\begin{tabular}{l}
\includegraphics[width = 3in,height = 3in]{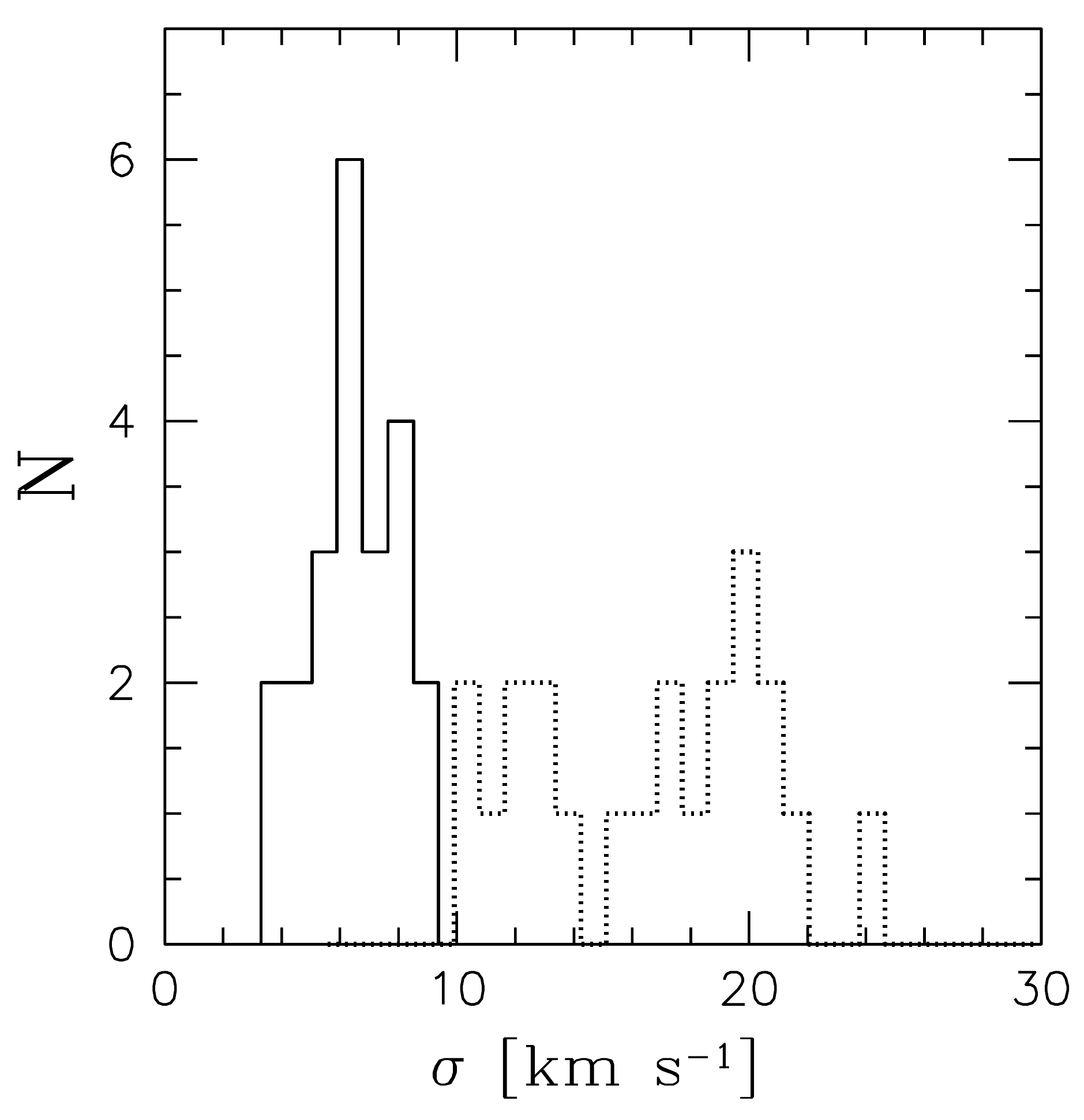}
\end{tabular}
\caption{Histograms of the velocity dispersion of the clean
  sample. The solid line histogram represents the narrow component. The
  dotted line histogram represents the broad component.}
\label{fig:hist_clean}
\end{figure}

To test whether the $|\Delta_{H-I}| \leq 5\, \rm{km~s^{-1}}$
criterion produces symmetrical profiles, we create new masks with
$|\Delta_{H-I}| \leq 2\, \rm{km~s^{-1}}$. This selects only very
symmetrical profiles but at the cost of a reduced number of input
profiles. Using the more strict condition results in a loss of about
60\% of the profiles. Figure \ref{fig:sym_asym_comp} (bottom panel)
compares the $|\Delta_{H-I}| \leq 5\, \rm{km~s^{-1}}$ SP mask velocity
dispersions with the $|\Delta_{H-I}| \leq 2\, \rm{km~s^{-1}}$ equivalent.
It is clear that the dispersions of the broad components of the latter
tend to be smaller by $\sim 1\, \rm{km~s^{-1}}$.
Despite the smaller number of input profiles in the $|\Delta_{H-I}| \leq
2\, \rm{km~s^{-1}}$ case, Fig.~\ref{fig:sym_asym_comp} shows that the
uncertainties have not increased dramatically.  In the following, we
will therefore use the 2 $\rm{km~s^{-1}}$ condition unless stated otherwise.

Results from the single and double Gaussian fitting are summarized in
Table \ref{tab:fitted_clean}. This convincingly shows that both narrow
and broad components are needed to fit the profiles. We also fit the
wingless sides of LHAP and RHAP super profiles (i.e., the ``Gaussian''
half of the profile with single and double Gaussian components. Here
we also find that the two-components fit perform better. 

The broad component velocity dispersions
derived from the wingless side of the LHAP and RHAP super profiles
tend to be smaller than those derived from the SP super
profiles. However, there is no major difference between the narrow
component velocity dispersions derived from the wingless side of LHAP
and RHAP super profiles and those from SP super profiles (see
Fig.~\ref{fig:wingless}). The
measurements of the narrow component velocity dispersion are thus more
robust than those of the broad component.

\section{Super profiles and global properties of galaxies}\label{sec:globaltrend}

Figure \ref{fig:hist_clean} shows the distribution of the narrow and
broad component velocity dispersions derived using the SP profiles in 
the clean sample galaxies. The narrow component has a much
tighter distribution than the broad component, with mean values of
$6.5\pm 1.5~ \rm{km~s^{-1}}$ and $16.8\pm 4.3 ~\rm{km~s^{-1}}$,
respectively. We adopt the definition of spirals and dwarfs given in
\citet{leroyetal08}. They define dwarfs as galaxies with rotation
velocities $V_{rot}\leq125~\rm{km~s^{-1}}$, stellar masses
$M_\ast\lesssim10^{10}~M_\odot$, and absolute $B$ magnitude
$M_B\gtrsim-20~\rm{mag}$. Galaxies more massive than this are defined
to be spirals. Dividing our clean sample into dwarfs and
  spirals, we find their velocity dispersions are identical within the
  uncertainties. For the dwarfs, we find a mean narrow component
velocity dispersion of $6.5 \pm 1.5 ~\rm{km~s^{-1}}$ and a mean broad
component velocity dispersion of $15.1 \pm 3.6~\rm{km~s^{-1}}$. For
the spirals, we find values of $6.7\pm~1.4~\rm{km~s^{-1}}$ and
$18.7\pm4.3~ \rm{km~s^{-1}}$, respectively.

The $A_{n}/A_{b}$ ratios differ though. For the spirals we find a
mean of $0.63\pm0.24$; for the dwarfs a value of $0.49\pm0.20$,
indicating a larger importance of the narrow component in spirals. We
summarize the fitted parameters of the clean sample super profiles in
Table \ref{tab:fitted_clean}.

In the following, we explore possible correlation between global galaxy
properties and super profile shapes.  If the narrow and broad
components identified here are related to the CNM and the WNM phases
of the ISM, we expect some correlation between these two components
and tracers of star formation.  For example, \citet{schaye04} suggests
that the transition from the warm to the cold phase of the ISM
triggers disk instabilities which eventually leads to star formation.
We use FUV-NUV colors, H$\alpha$ luminosities and
metallicities as star formation indicators. The FUV-NUV colors are
drawn from \citet{leeetal11} or else \citet{muaozetal09}. The
(FUV-NUV) colors have been corrected for galactic extinction. Emission
in the UV arises from O and later-type B stars and therefore traces
star formation over a timescale of $\lesssim$ 100-200 Myrs. The
H$\alpha$ luminosities are taken from \citet{kennicutetal08}. 
Emission in H$\alpha$ is mainly caused by hot stars such as O and
early-type B stars which are massive and short lived. Thus,
H$\alpha$ emission is a tracer of recent star formation over a
timecale of $\lesssim$ 10 Myrs. This timescale is roughly comparable
to the time it takes for the WNM to cool \citep[e.g.][]{wolfireetal03}. 
The metal abundance is taken from
\citet{moustakasetal10} or alternatively from Table 1 of
\citet{walteretal08}. \citet{moustakasetal10} used both empirical
\citep{pt05} and theoretical \citep{kk04} calibrations to compute
gas-phase abundances. Here we use the average values of the
characteristic (globally averaged) metallicities of the two
calibrations listed in Table 9 of \citet{moustakasetal10}.  As
discussed in \citet{moustakasetal10}, the abundance is likely to be
underestimated by up to $\sim$0.2-0.3 dex in the empirical
calibrations and overestimated by a similar amount in the theoretical
calibrations. Thus, taking the average values of the two calibrations
is expected to yield an abundance estimate that is close to the
``true" value. Metallicity plays an important role in cold phase
formation \citep{schaye04, krumholzeta09, walchetal11} as gas can be
cooled, for example, via collisional excitation of heavy elements such
as oxygen. 

The relationship between the super profile parameters and these
different star formation indicators are presented below. We plot in
Figure \ref{fig:globaltrendone} the velocity dispersion ratio
($\sigma_{n}/\sigma_{b}$) and the $A_{n}/A_{b}$ ratio of the narrow
and broad components as a function of metallicities, (FUV-NUV) colors
and $\rm{H{\alpha}}$ luminosities. 

We find that the velocity dispersion ratio decreases with increasing
metallicity with a correlation coefficient $R~\sim-0.56$. In principle this 
could reflect more efficient cooling, but as turbulence must also play 
a large role in determining the value of the velocity dispersion, it is more 
likely that this trend is a reflection of a metallicity-star formation rate relation. 
High metallicity galaxies tend to have higher star formation rates, and for these 
high star formation 
rate (high metallicity) galaxies, the broad component velocity dispersion 
is likely to be larger which will give a lower $\sigma_{n}/\sigma_{b}$.

The $\sigma_{n}/\sigma_{b}$ ratio is higher for bluer (FUV-NUV)
colors; the correlation coefficient is $R~\sim-0.61$. As low
metallicity galaxies are on average also bluer in (FUV-NUV), this relation probably
reveals similar information as the metallicity \textit{vs}
$\sigma_{n}/\sigma_{b}$ relation.
There is only a weak correlation between dispersion ratio and 
$\rm{H{\alpha}}$ luminosities ($R~\sim-0.20$). This weak
correlation may be due to the fact that the super profile
parameters and the $\rm{H{\alpha}}$ luminosities have not been derived
from the same regions within the galaxies. This will be an interesting topic 
for follow-up studies.

In terms of the $A_{n}/A_{b}$ ratio, which should measure the amount
of cold gas (CNM) relative to warm gas (WNM), we find that the
fraction of the narrow components relative to the broad components
tends to be high for high metallicity galaxies ($R~\sim0.53$).  We
find similar relations in terms of (FUV-NUV) colors and
$\rm{H{\alpha}}$ luminosities with correlation coefficients
$R~\sim0.53$ and $R~\sim0.47$, respectively. These correlations are 
all preliminary indications that the presence of cold gas is associated with 
star formation.

\section{Does the narrow component trace molecular gas?} \label{sec:moleculargas}

Star formation and molecular hydrogen are intimately connected.
Unfortunately, molecular hydrogen is not directly observable and CO
emission is usually used as a tracer. Usually a constant
CO-to-$\rm{H_{2}}$ conversion factor is assumed to derive the amount
of H$_2$ present in a galaxy.  However, the strength of the CO
emission line depends on the metal abundance of galaxies and CO is
difficult to detect in low metallicity environments. CO may also not
trace all the $\rm{H_{2}}$ content of galaxies because unlike
$\rm{H_{2}}$, CO cannot self-shield \citep[e.g.,][]{krumholzeta11} and
therefore, in regions with low metal and dust abundances where both CO
and $\rm{H_{2}}$ are present, UV photons from young stars will easily
dissociate CO while leaving $\rm{H_{2}}$ intact.

As H$_2$ forms as the cold phase of the neutral medium cools, we test
here whether the narrow component identified in the super profiles can
be used to infer the amount of $\rm{H_{2}}$ in galaxies.  This is
motivated by early observations made by \citet{younglo96,younglo97}
and \citet{deblokwalter06}. They found that the narrow component tends
to be located near regions of star formation. The narrow component
could thus be associated with the star fomation in a manner similar to
that of H$_2$.  Also, from a theoretical point of view, HI must pass
through a cold phase before turning molecular.

We consider the 11 galaxies in the clean sample that have also been
detected in CO in the HERA CO-Line Extragalactic Survey
\citep[HERACLES,][]{heracles}. HERACLES provides maps of CO $J$ = 2
$\rightarrow$ 1 emission in 18 THINGS galaxies (though the full
HERACLES sample is larger) at 13$^{\prime\prime}$ angular resolution
and 2.6 km s$^{-1}$ velocity resolution using observations with the
IRAM 30 m telescope. The CO-detected galaxies have metallicities
ranging from 8.25 to 9.12 in units of [12 + log(O/H)].

If the narrow component is indeed associated with molecular
  gas then we expect correlations between the $A_{n}/A_{b}$ ratio and
the $\rm{H_{2}}$ to $\rm{HI}$ mass ratio.  When masses are measured in
identical regions, this translates in correlations between the
respective surface densities as well. We also test for correlations
with the narrow and broad components velocity dispersion ratios.

We derive super profiles and average HI and H$_2$ mass surface
densities in the overlap regions in the sample galaxies where
\emph{both} CO and HI are detected. Restricting
ourselves to the overlap regions means we are testing only the regime
where both gas phases are present. We do not investigate the regime
where all gas has turned molecular and no HI is present anymore (such
as the central parts of some galaxies where central HI deficiencies
are filled with molecular gas).

The values for 
$\Sigma_{\rm{HI}}$ and $\Sigma_{\rm{H_{2}}}$ are calculated using

\begin{align}
\Sigma_{\rm{HI}}  &= 0.02~\cos i ~I_{\rm{HI}} \\
\Sigma_{\rm{H_{2}}}  &= 4.4~ \frac{X}{X_{\rm{CO}}} ~\frac{1}{R_{21}} ~\cos i~ I_{CO}(2 \rightarrow 1)
\end{align}
where $\Sigma_{\rm{HI}}$ and $\Sigma_{\rm{H_{2}}}$ are in units of
$M_{\odot}~\rm{pc^{-2}}$, and $I_{\rm{HI}}$ and $I_{\rm{CO}}$ are both
in K km s$^{-1}$. The mass surface densities are corrected for
inclination and include a factor 1.36 to correct for the presence of
helium. In the above Equation, $R_{21}$ is a CO $J$ = 2 $\rightarrow$ 1 to $J$ = 1
$\rightarrow$ 0 line ratio which we assume to be equal to 0.7 \citep{schrubaetal12}, 
$X$ is a conversion factor normalized to the Galactic CO ($J$ =
1 $\rightarrow$ 0)-to-H$_{2}$ conversion factor $X_{\rm{CO}}$ = 2.0
$\times$ 10$^{20}$ cm$^{-2}$ $(\rm{K~km~s^{-1}})^{-1}$. Here we adopt
$X = X_{\rm{CO}}$. 

\begin{figure*}[htb]
\centering
    \begin{tabular}{l l l}
\rotatebox{0}{\resizebox{58mm}{!}{\includegraphics[width = 0.6in,height = 0.6in]{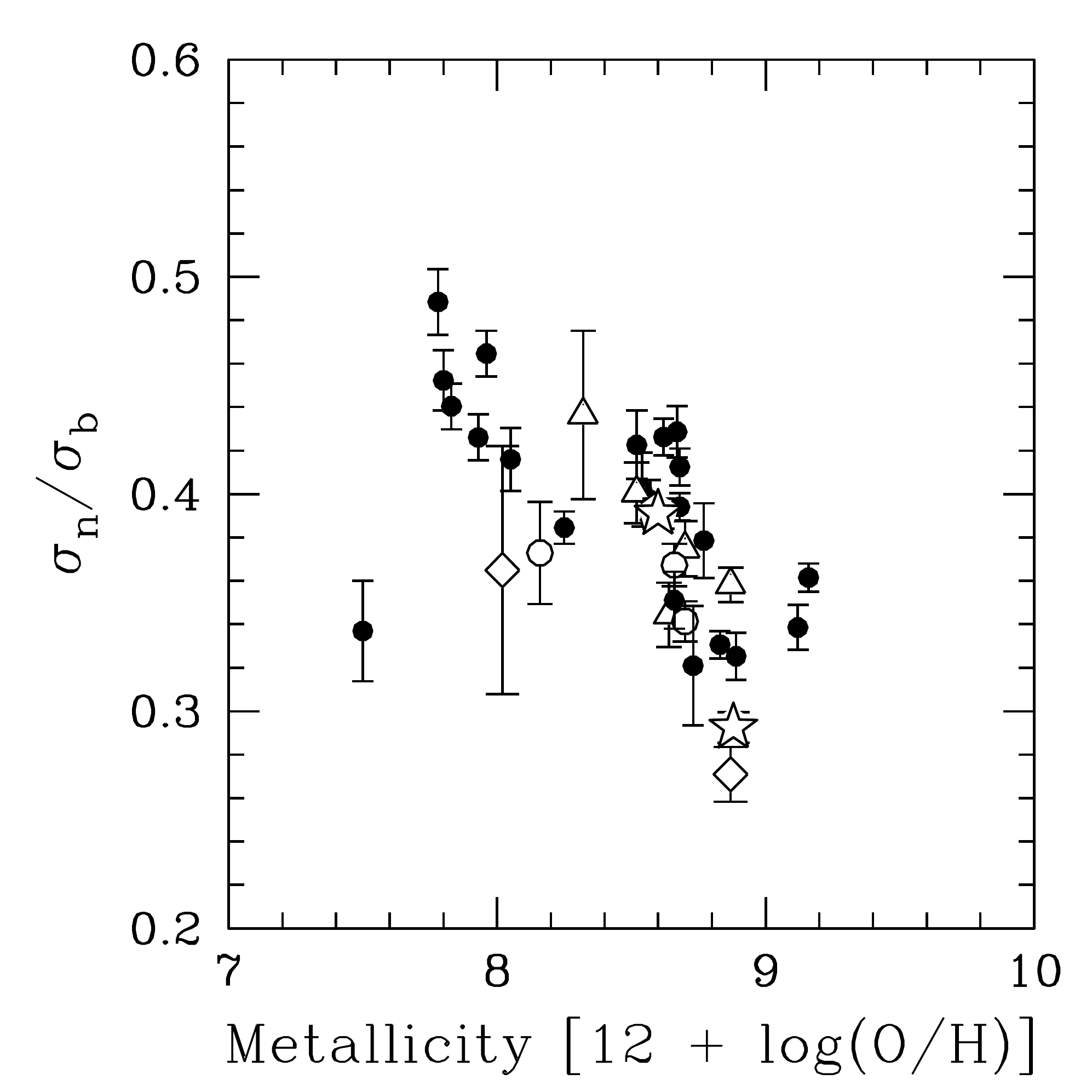}}}&
\rotatebox{0}{\resizebox{58mm}{!}{\includegraphics[width = 0.6in,height = 0.6in]{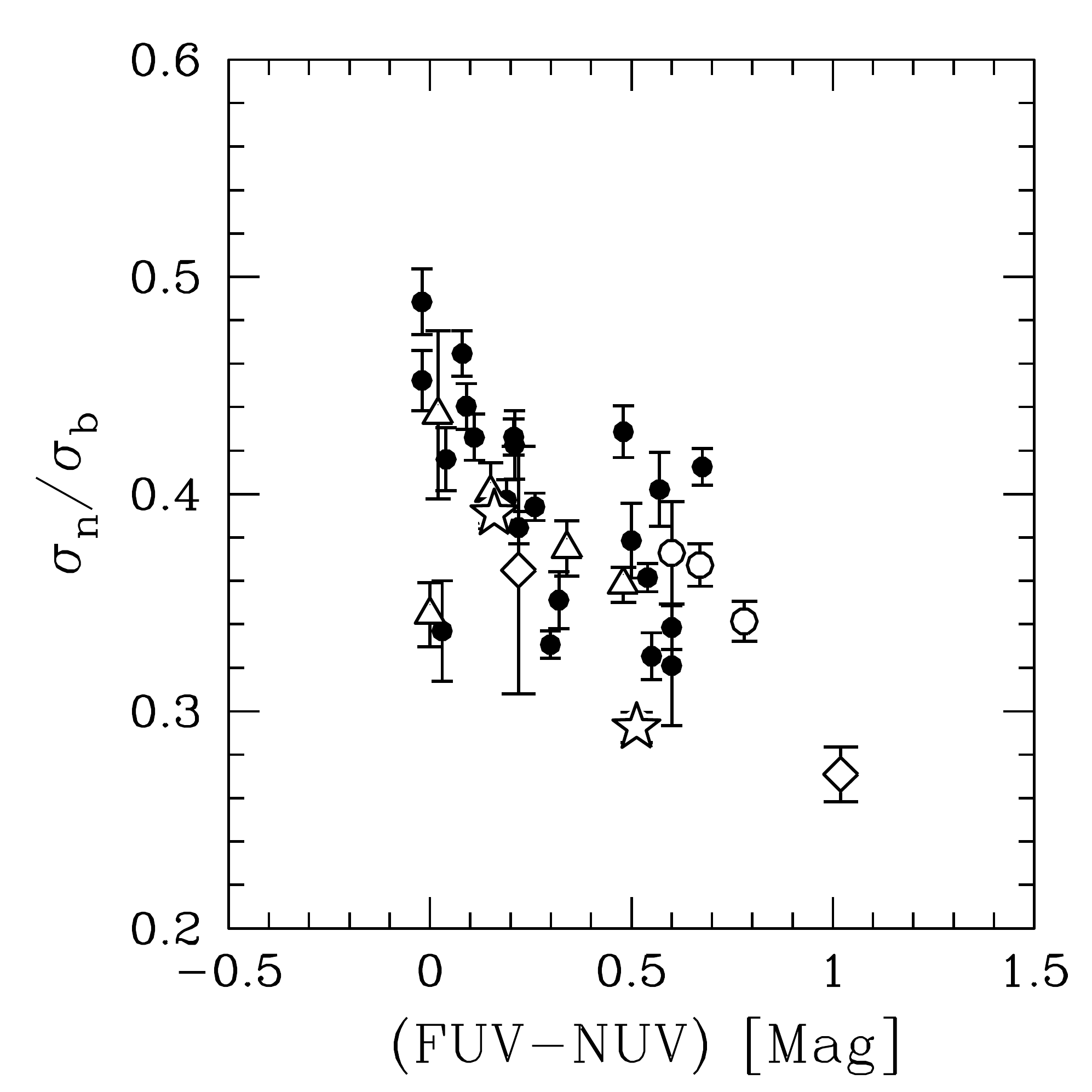}}}&
\rotatebox{0}{\resizebox{58mm}{!}{\includegraphics[width = 0.6in,height = 0.6in]{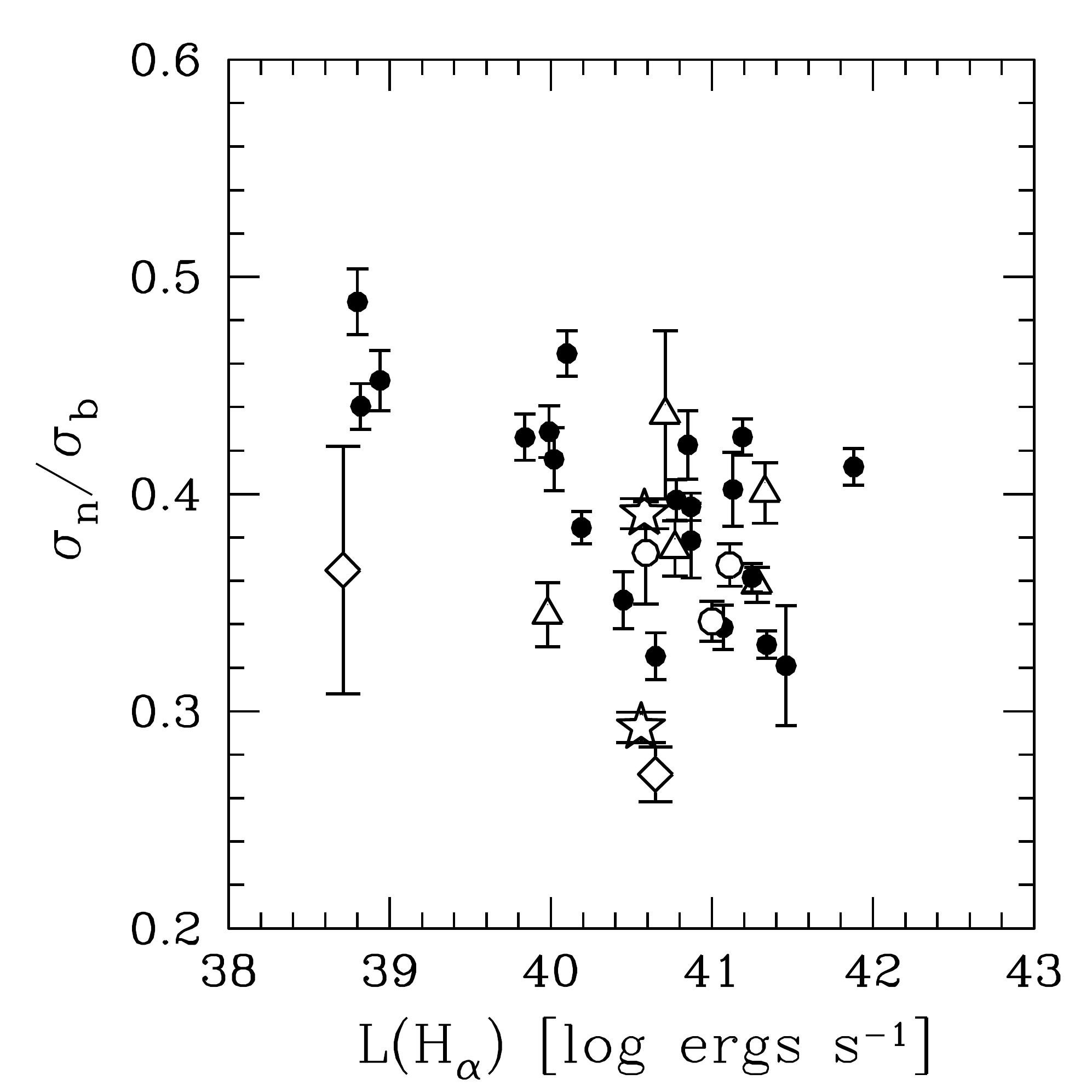}}}\\
\rotatebox{0}{\resizebox{58mm}{!}{\includegraphics[width = 0.6in,height = 0.6in]{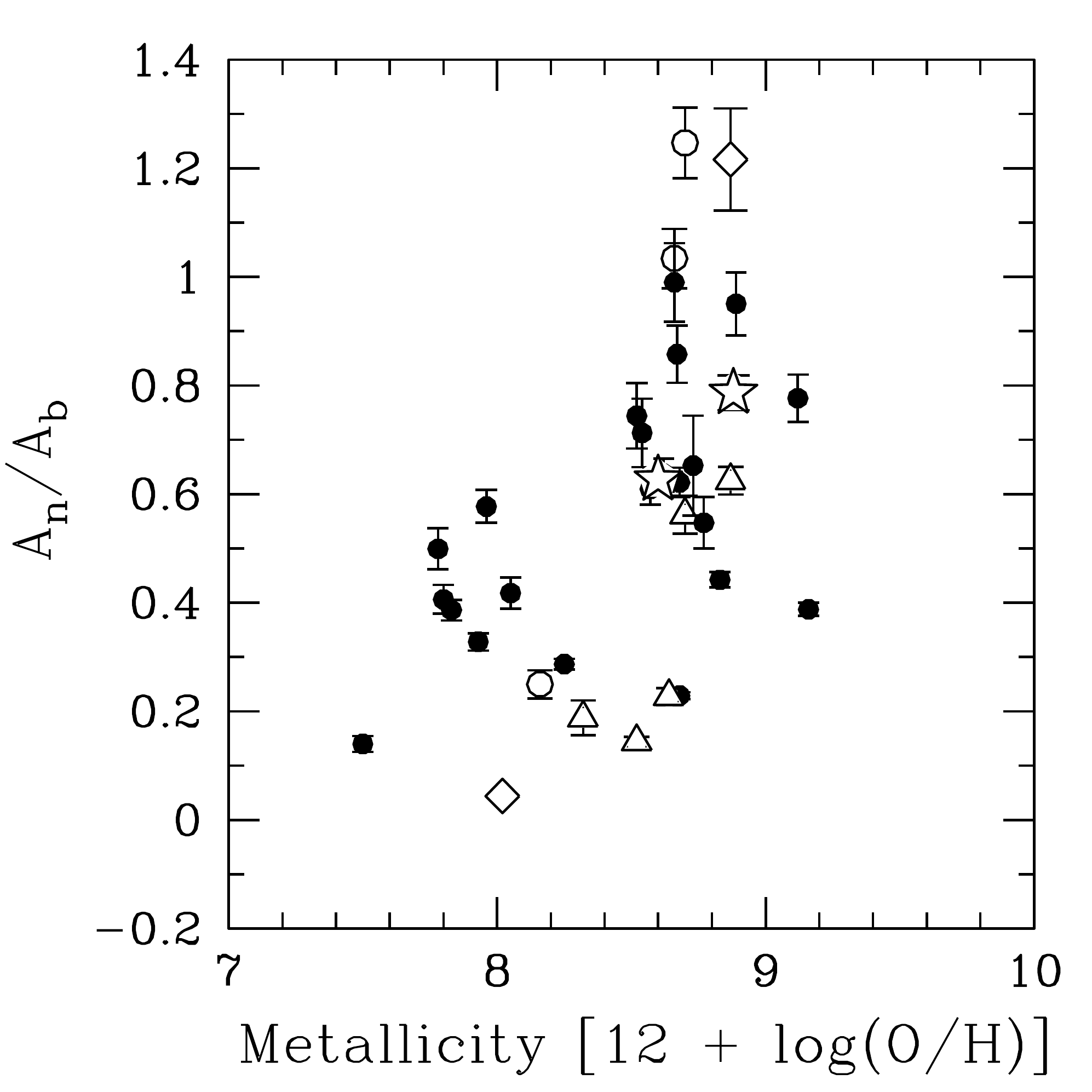}}}&
\rotatebox{0}{\resizebox{58mm}{!}{\includegraphics[width = 0.6in,height = 0.6in]{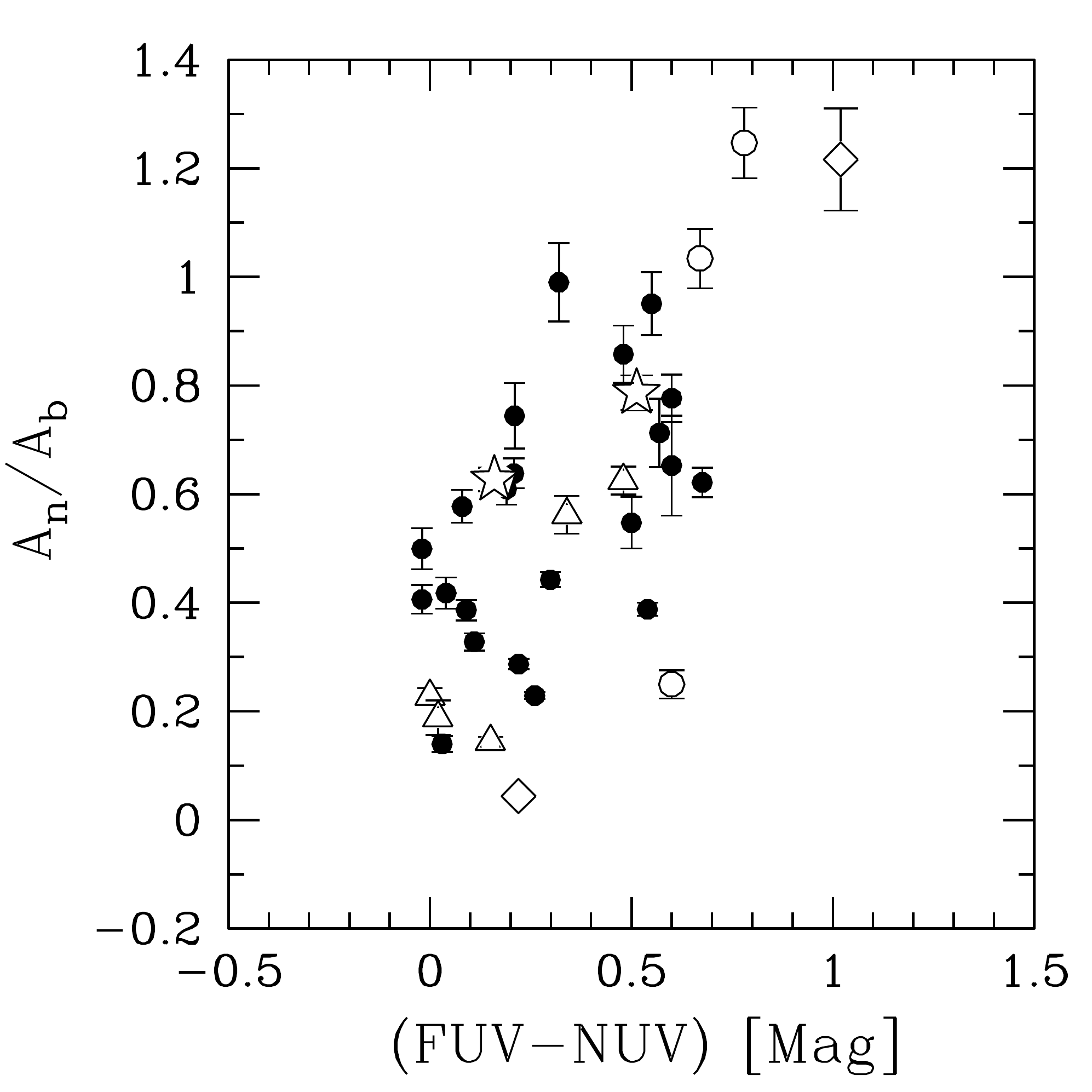}}}&
\rotatebox{0}{\resizebox{58mm}{!}{\includegraphics[width = 0.6in,height = 0.6in]{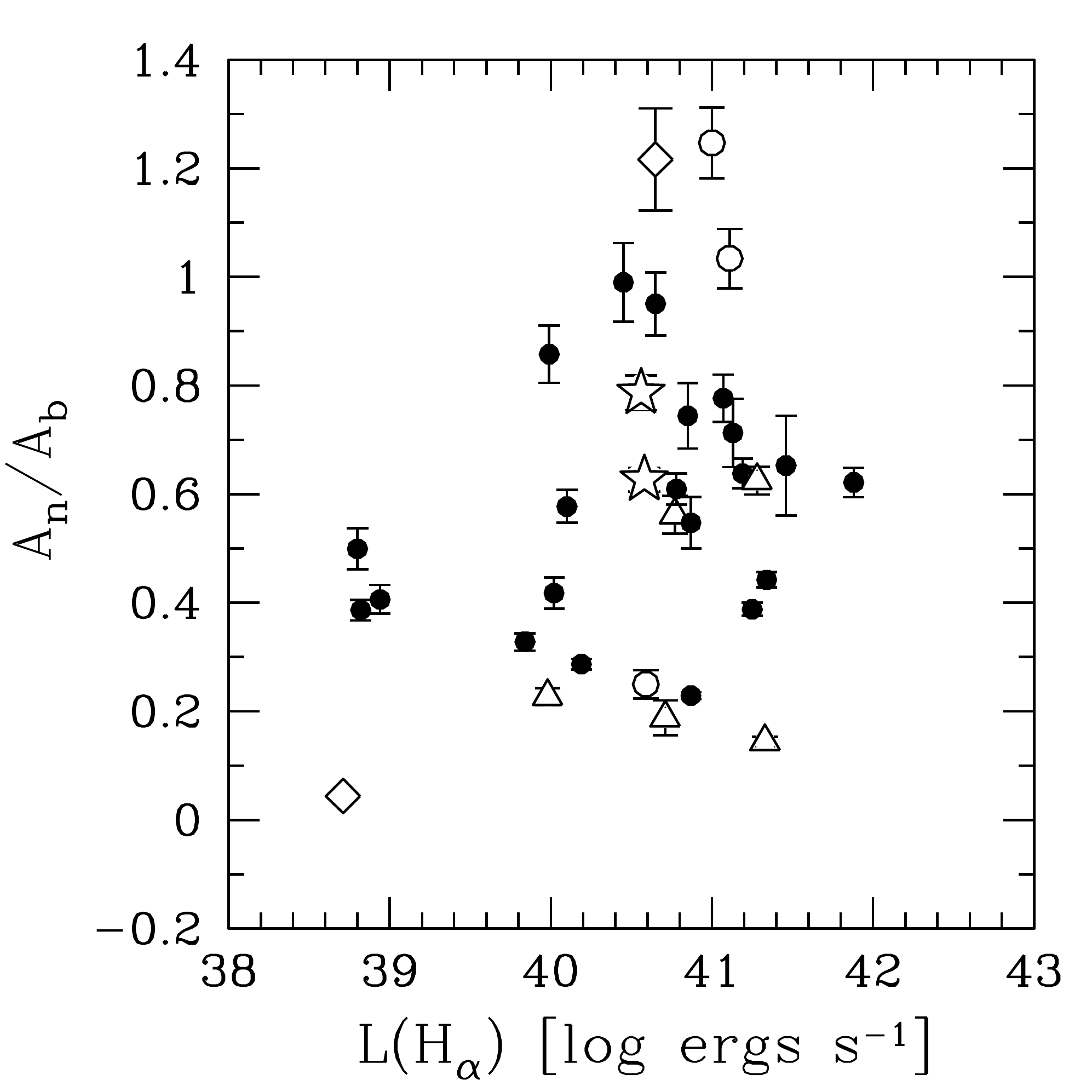}}}
\end{tabular}
\caption{\textit{Top panel}: velocity dispersion ratio as a function
  of metallicity, (FUV-NUV) colors and $\rm{H{\alpha}}$
  luminosities. \textit{Bottom panel:} Flux ratio of the narrow and
  broad components as a function of metallicity, (FUV-NUV) colors and
  $\rm{H{\alpha}}$ luminosities. The triangle symbols represent
  interacting galaxies (those that are tidally interacting with a
  nearby companion or are being affected by their environment; these
  are NGC 3031, NGC 4449, NGC 5194, NGC 5457, NGC 3077). The open
  circle symbols represent galaxies that are disturbed kinematically
  due to the effect of e.g., star formation (NGC 1569, NGC 3521, NGC
  3627). The star symbols indicate non-interacting galaxies that have
  an anomalously high velocity dispersion (NGC 7331, NGC 2841). The
  diamond shaped symbols represent non-interacting and non-disturbed galaxies
   where the dispersions measured on the opposite sides of the galaxies differ by more than 1.5 
   $\rm{km s^{-1}}$.
   (NGC 4826 and M81
  DwB; see Section \ref{sec:cleansample} for the choice of this value). Filled circle
  symbols represent our clean sample galaxies.}
\label{fig:globaltrendone}
 \begin{tabular}{l l}
\centering
\includegraphics[width = 3in,height = 3in]{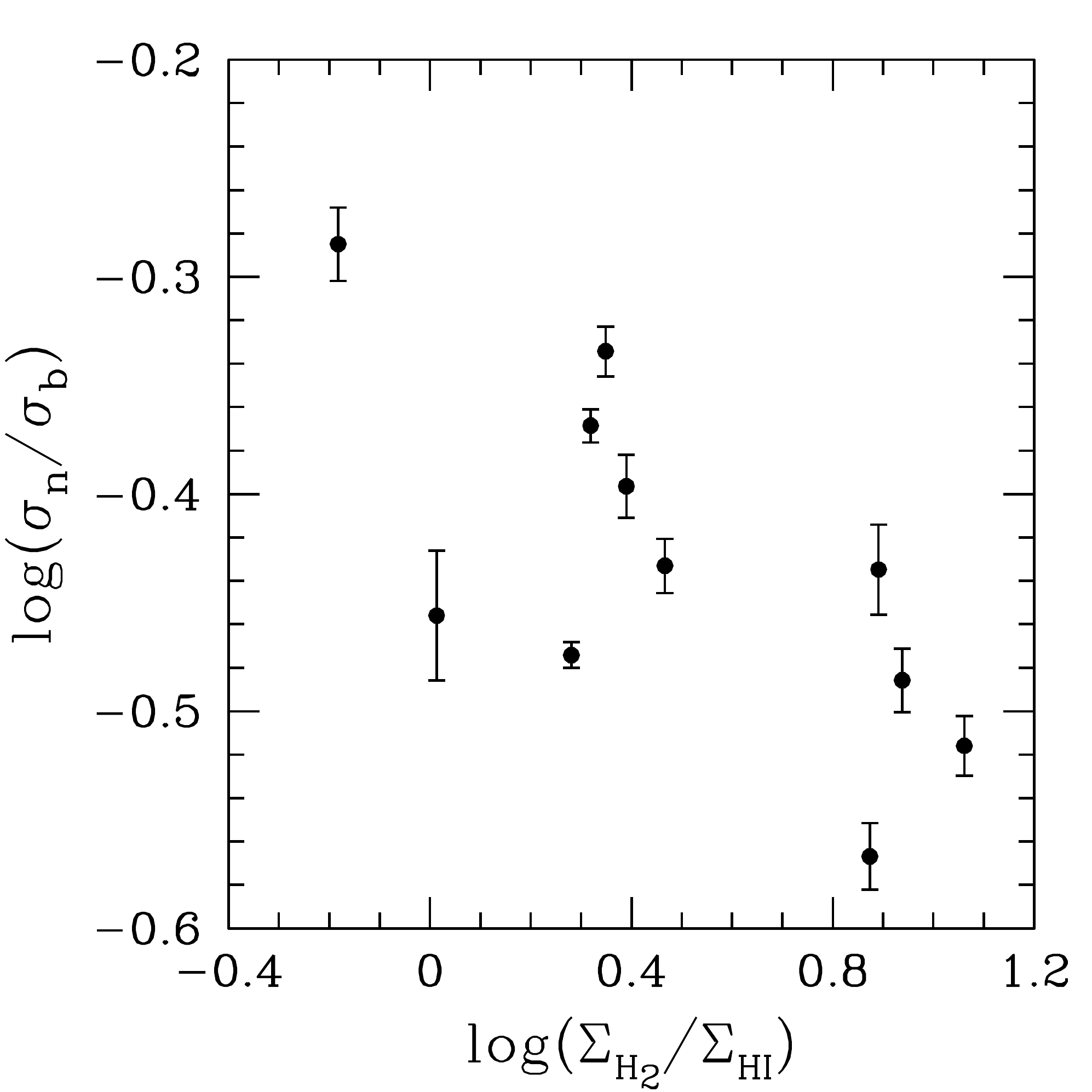}&
\includegraphics[width = 3in,height = 3in]{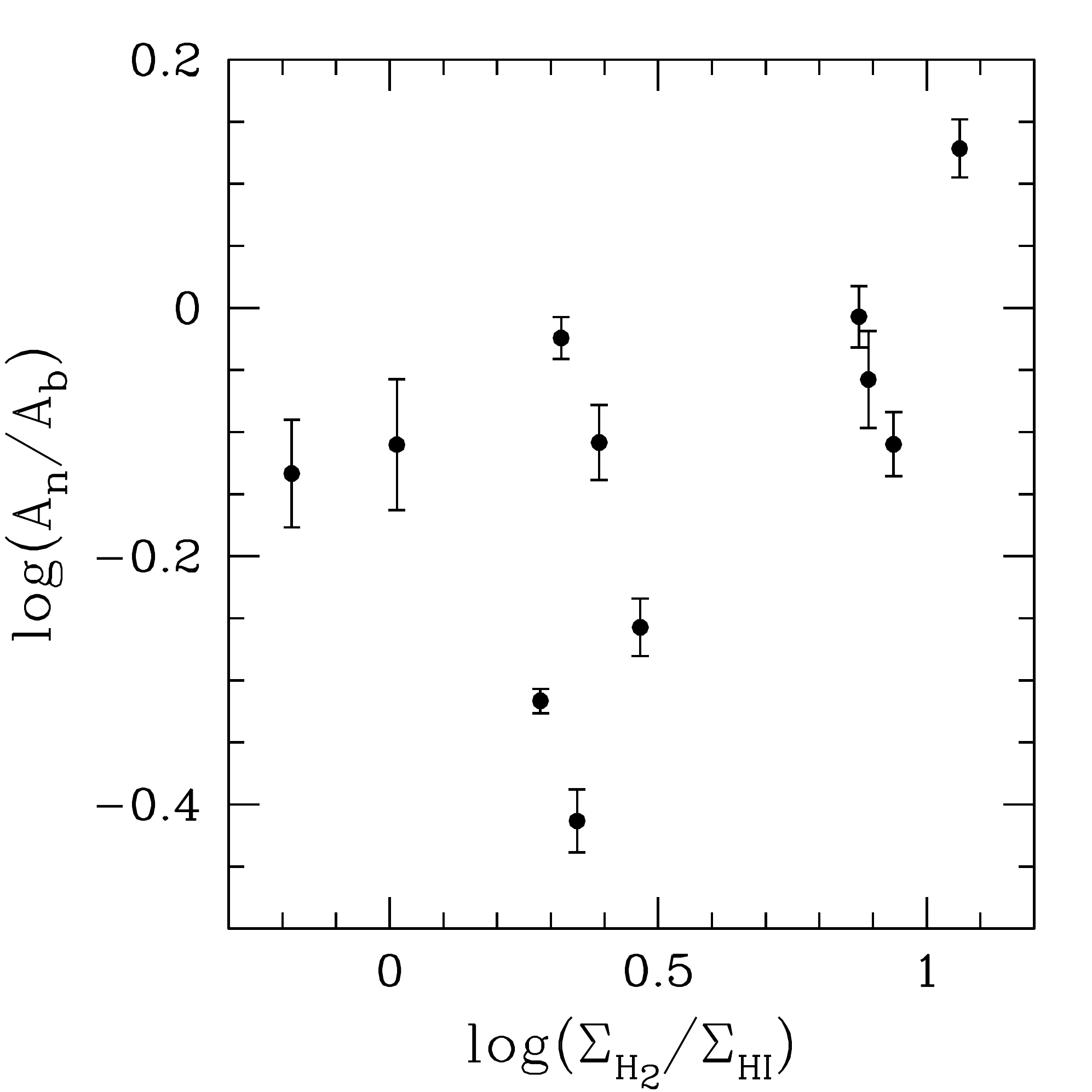}
\end{tabular}
\caption{Velocity dispersion (left panel) and
  $A_{n}/A_{b}$ ratio (right panel) as a function of $\rm{H_{2}}$ and HI mass
  surface density ratio.}
\label{anab_h1h2}
\end{figure*}

In Figure \ref{anab_h1h2} we compare $\log(A_{n}/A_{b})$ and
$\log(\sigma_{n}/\sigma_{b})$ with the values for
$\log(\Sigma_{\rm{H_{2}}}/\Sigma_{\rm{HI}})$ as derived for the same
areas in each galaxy. The observed upper limit on
  $\sigma_{n}/\sigma_{b}$ and $A_{n}/A_{b}$ seems to vary
  systematically with $\Sigma_{\rm{H_{2}}}/\Sigma_{\rm{HI}}$.  The
respective correlation coefficients are $R = 0.47$, and $R =
-0.70$. A possible interpretation of the observed trend between $\log(\sigma_{n}/\sigma_{b})$ 
and $\log(\Sigma_{\rm{H_{2}}}/\Sigma_{\rm{HI}})$ is that galaxies having larger 
fractions of molecular gas are expected to have more abundant star formation, which 
drives turbulence to the ISM and results in higher $\sigma_{b}$ values and thus lowers 
the $\log(\sigma_{n}/\sigma_{b})$ ratio. The trend in $A_{n}/A_{b}$ ratios might reflect 
the fact that atomic gas passes though the CNM phase before turning into molecular. Thus, 
galaxies having larger $A_{n}/A_{b}$ (more cold atomic gas) ratio are expected 
to have larger molecular gas fractions.
This result serves as early indication that the narrow 
component may be associated with molecular gas. It will be interesting 
to study the spatial distribution of these two gas components in more detail and 
investigate whether the locations of the narrow components correlate with 
those of the molecular gas.
\section{Summary} \label{sec:summary}

We have conducted in-depth analyses of the shapes of the HI velocity
profiles of galaxies from The HI Nearby Galaxy Survey (THINGS)
sample. To minimize the effect of noise on the velocity profiles, we
have constructed high signal-to-noise (S/N) profiles by aligning
individual profiles in velocity and stacking them. We call these
 \textit{super profiles}. We have quantified the
relevant systematic effects that may change the intrinsic shapes of
the super profiles and defined a clean sample where the observed
shapes of the super profiles are less affected by these effects.

We have fitted the super profiles with single, double and triple
Gaussian models, as well as Lorentzian models. Based on a $\chi^{2}$
analysis, we found that the shapes of the super profiles are
optimally described by the sum of a narrow and a broad Gaussian
component. The narrow component velocity dispersions of the clean
sample range from 3.4 to 8.6 $\rm{km~s^{-1}}$ with a mean of $6.5\pm
1.5~ \rm{km~s^{-1}}$. The broad component velocity dispersions range
from 10.1 to 24.3 $\rm{km~s^{-1}}$ with a mean of $16.8\pm 4.3
~\rm{km~s^{-1}}$. Note that due to the limitation of the data, 
the derived velocity dispersions 
are overestimated by at most 20\%. The combined effects of 
the finite channel spacing, 
the limited spatial resolution (see Figure \ref{fig:galmod_param}) 
and the inclination (see Figure \ref{fig:incl_disp}) 
contribute about 15\% while the effects of radial motions 
(whose magnitude is expected to be less than 
$\sim$10 $\rm{km~s^{-1}}$, see Section \ref{sub:radialmotion}) 
contribute about 5\%. 

We have investigated possible correlations between the shapes of the
super profiles and global properties of galaxies. We found that the
flux ratio of the narrow and broad components tends to be high for high
metallicity, high star formation rate galaxies. The flux ratio
also increases with increasing bluer FUV-NUV colors. We interprete these
correlations as evidence that cold HI gas is associated with star formation. 
In addition, the velocity
dispersion ratio of the narrow and broad components decreases with
increasing metallicity, FUV-NUV colors and $\rm{H{\alpha}}$
luminosities. 

We present tentative evidence that the narrow component is associated
with molecular gas. We find that upper limits on
$\sigma_{n}/\sigma_{b}$ and $A_{n}/A_{b}$ change with
$\rm{\Sigma_{H2}/\Sigma_{HI}}$, suggesting that the cold neutral phase 
and the molecular phase are related.  
The observed trend between the $A_{n}/A_{b}$ ratio and the 
$\rm{\Sigma_{H2}/\Sigma_{HI}}$ reflects the fact 
that atomic gas passes through the CNM phase before turning into molecular. 
Based on this preliminary analysis, it is expected that the location of 
the narrow component (CNM) correlates with the location of molecular gas. Higher 
resolution pixel-by-pixel studies of the HI line profiles should be able to confirm this. 

\acknowledgments
We thank the anonymous referee for helpful and constructive
comments that have contributed to this paper.
R.I. acknowledges financial support from the South African 
Research Chairs Initiative of the Department of Science and Technology 
and National Research Foundation.

\clearpage
\clearpage
\appendix
\section{Super profiles: Gaussian and Lorentzian}
\begin{figure*}[htb]
\centering
    \begin{tabular}{l}    
    \includegraphics[scale=.36]{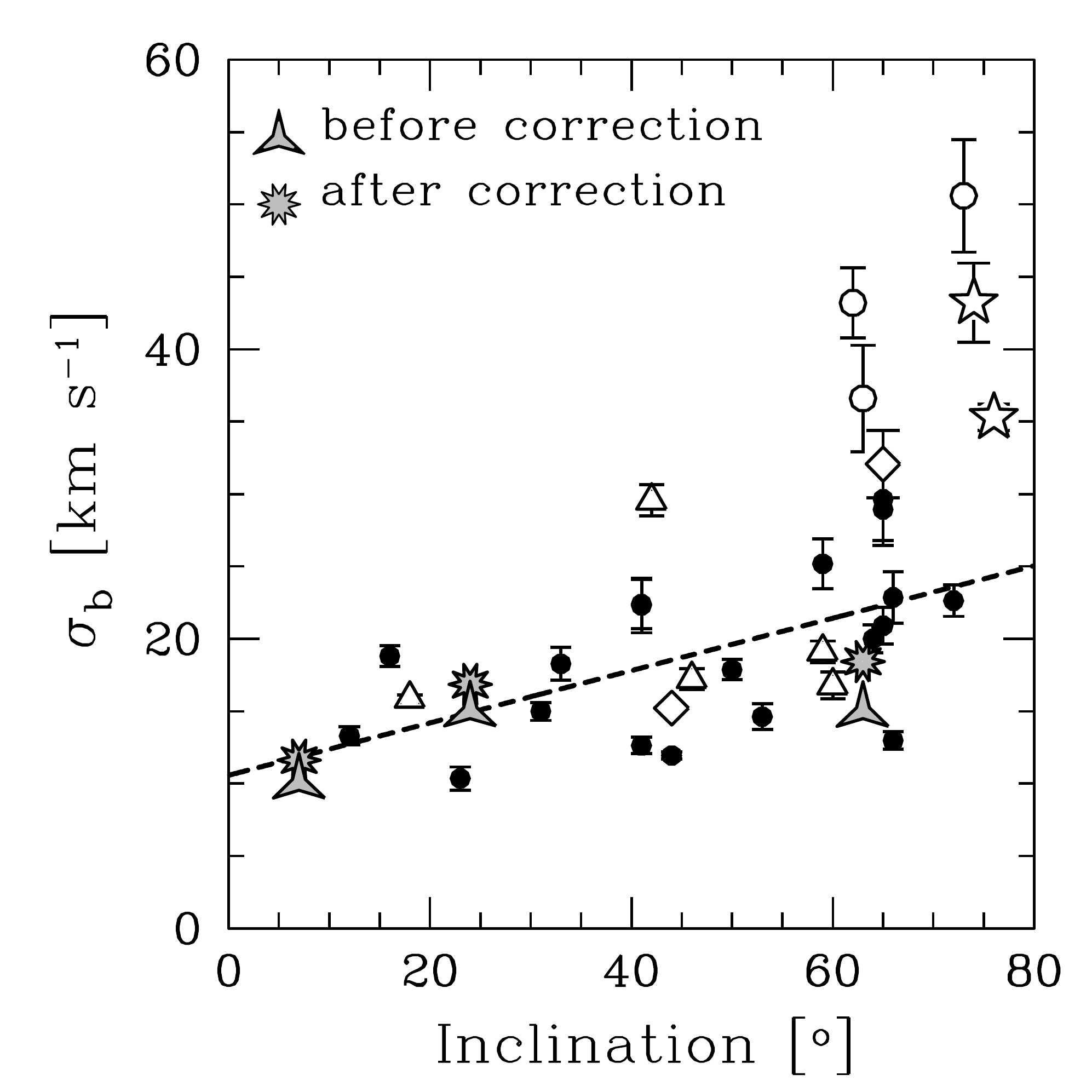}
\end{tabular}
\caption{Broad component velocity dispersion as a function of inclination. 
The solid symbols represent our clean sample. The grey symbols represent the 
velocity dispersions of (from left to right) NGC 628, NGC 5236 and NGC 2403 
before and after the negative bowl correction.}
\label{fig:trust_broad}
\end{figure*}

In this appendix, we present the super profiles of the entire THINGS sample
(see Section \ref{sec:method} for an explanation on how these profiles
were generated). Note that we correct for the presence of
a negative bowl in the super profiles of NGC 628, NGC 2403, NGC 5236
and NGC 5457 by fitting a polynomial to the baseline and substract
the polynomial from the super profiles. Also note that NGC 628, NGC 2403, and
NGC 5236 are among the clean sample galaxies 
(see Section \ref{sec:cleansample}). In general, the negative bowl 
correction is small  for these galaxies (of the order of 10\% 
for the velocity dispersions and 20\% for the fitted area) and their broad component 
parameters do not appear 
as outliers when plotted against other quantities such as inclination 
(see Figure \ref{fig:trust_broad} for illustration). 
We summarise in Table \ref{tab:trust_broad} and Table \ref{tab:trust_broad_area} 
the super profiles' parameters before and after the negative bowl correction. 

Following \citet{braun97}, we have also fitted the super profiles with a
Lorentzian function. We compare the reduced $\chi^{2}$ values from the
double and Lorentzian fitting in Table \ref{tab:table_lorentz}. Of the
34 fitted super profiles, 4 can be well fitted by either a double
Gaussian or a Lorentzian function (i.e., their reduced $\chi^{2}$
values agree within 10$\%$). For 25 super profiles, a double Gaussian
function is clearly preferred. Finally, for 5 super
profiles, a Lorentzian function best describe the shapes of the
profiles. Overall, a double Gaussian function seems to be an optimal
description of the profiles. We do not attempt to make any physical
interpretation of the Lorentzian fitting results, however we summarize
the fitted parameters in Table \ref{tab:table_lorentz}.

\begin{figure*}
    \begin{tabular}{l l l}
  \rotatebox{0}{\resizebox{58mm}{!}{\includegraphics[width = 0.6in,height = 0.6in]{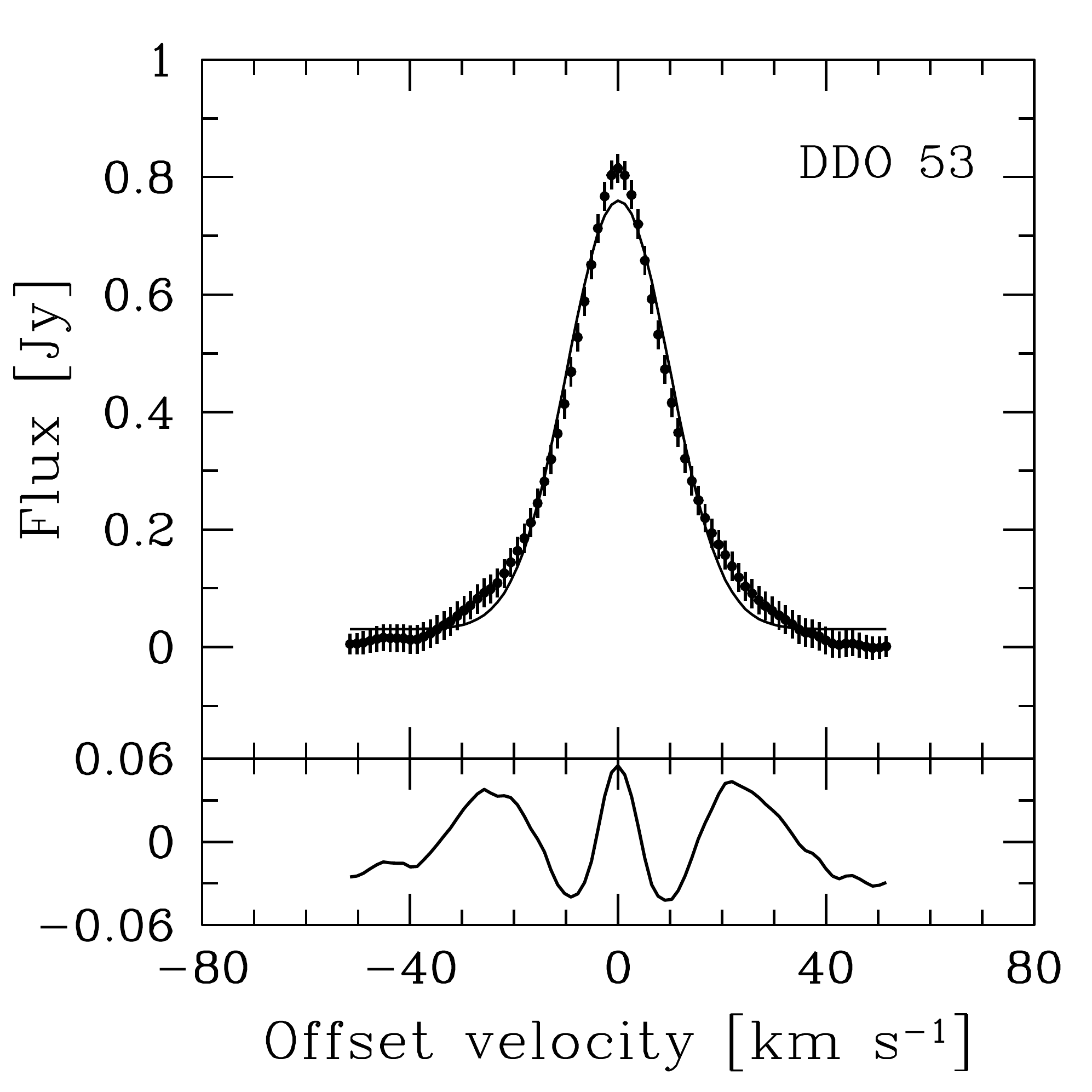}}}&
\rotatebox{0}{\resizebox{58mm}{!}{\includegraphics[width = 0.6in,height = 0.6in]{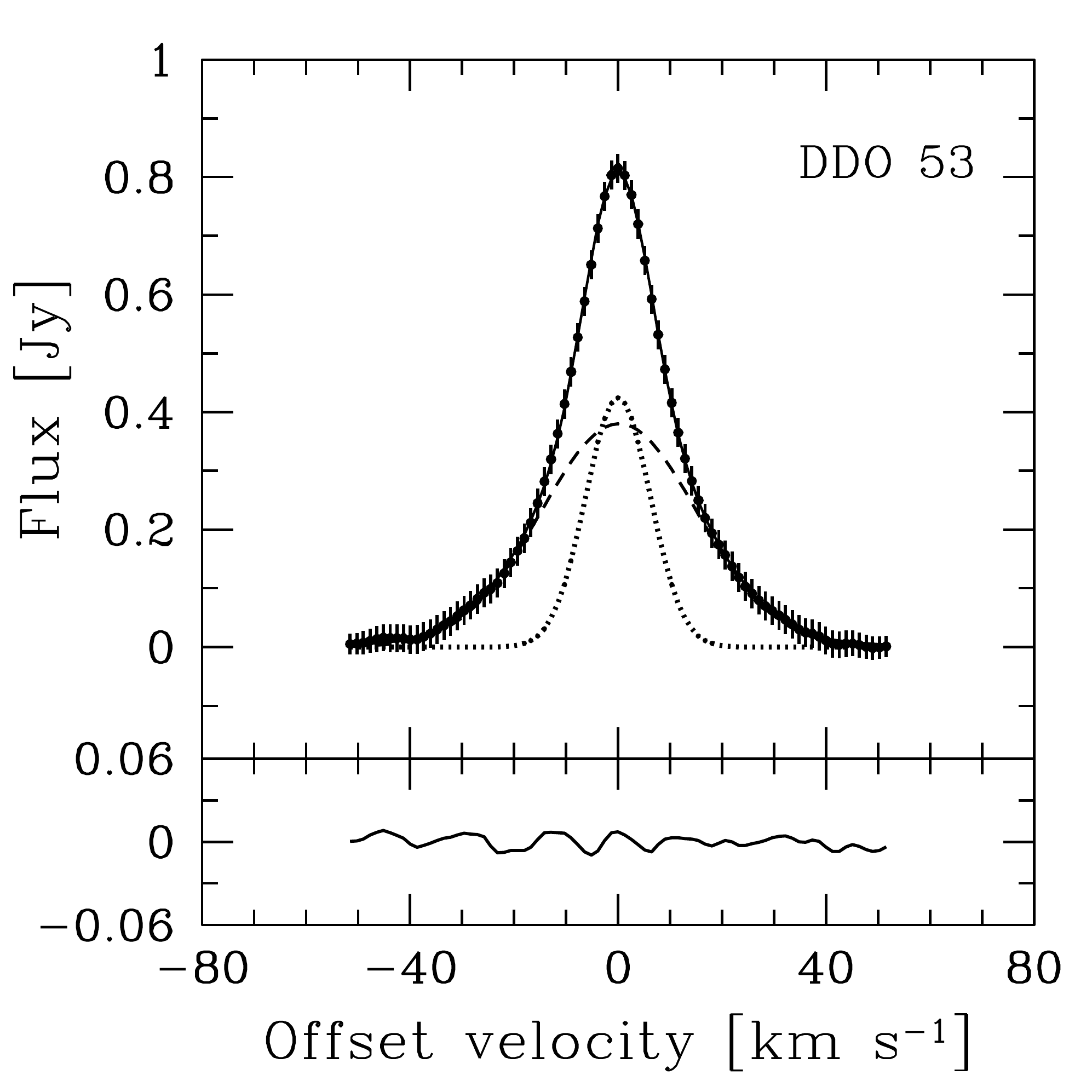}}}&
 \rotatebox{0}{\resizebox{58mm}{!}{\includegraphics[width = 0.6in,height = 0.6in]{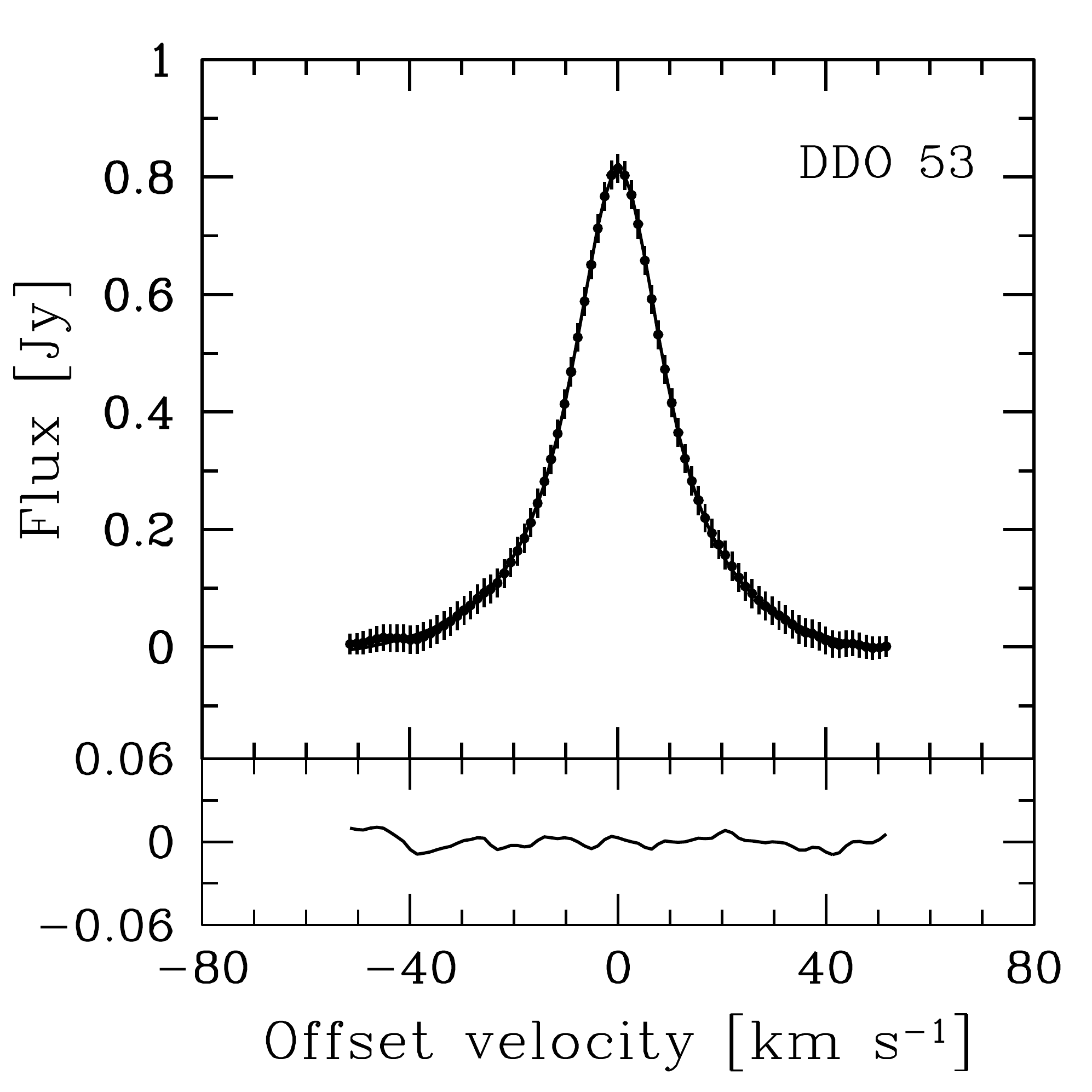}}}\\
   \rotatebox{0}{\resizebox{58mm}{!}{\includegraphics[width = 0.6in,height = 0.6in]{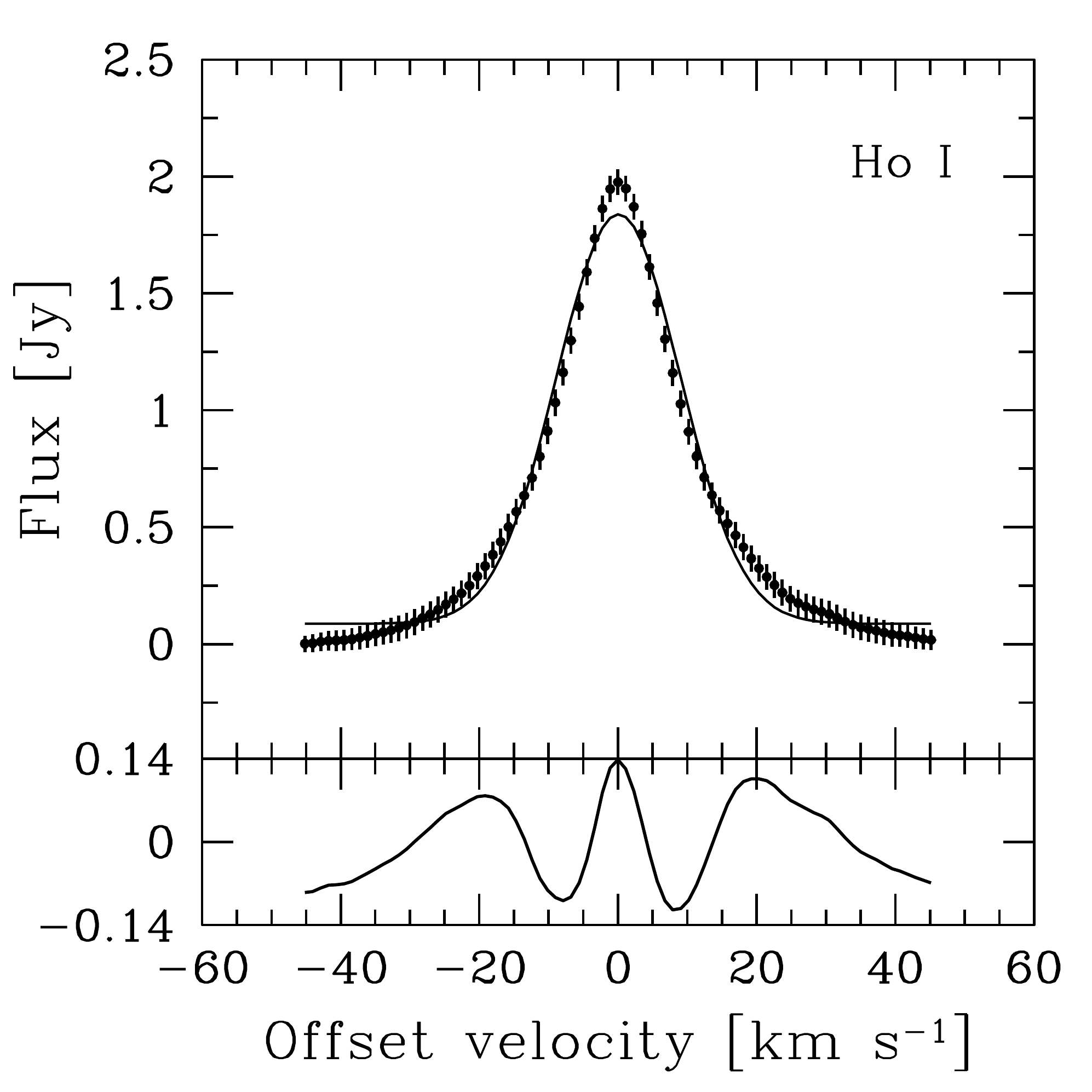}}}&
\rotatebox{0}{\resizebox{58mm}{!}{\includegraphics[width = 0.6in,height = 0.6in]{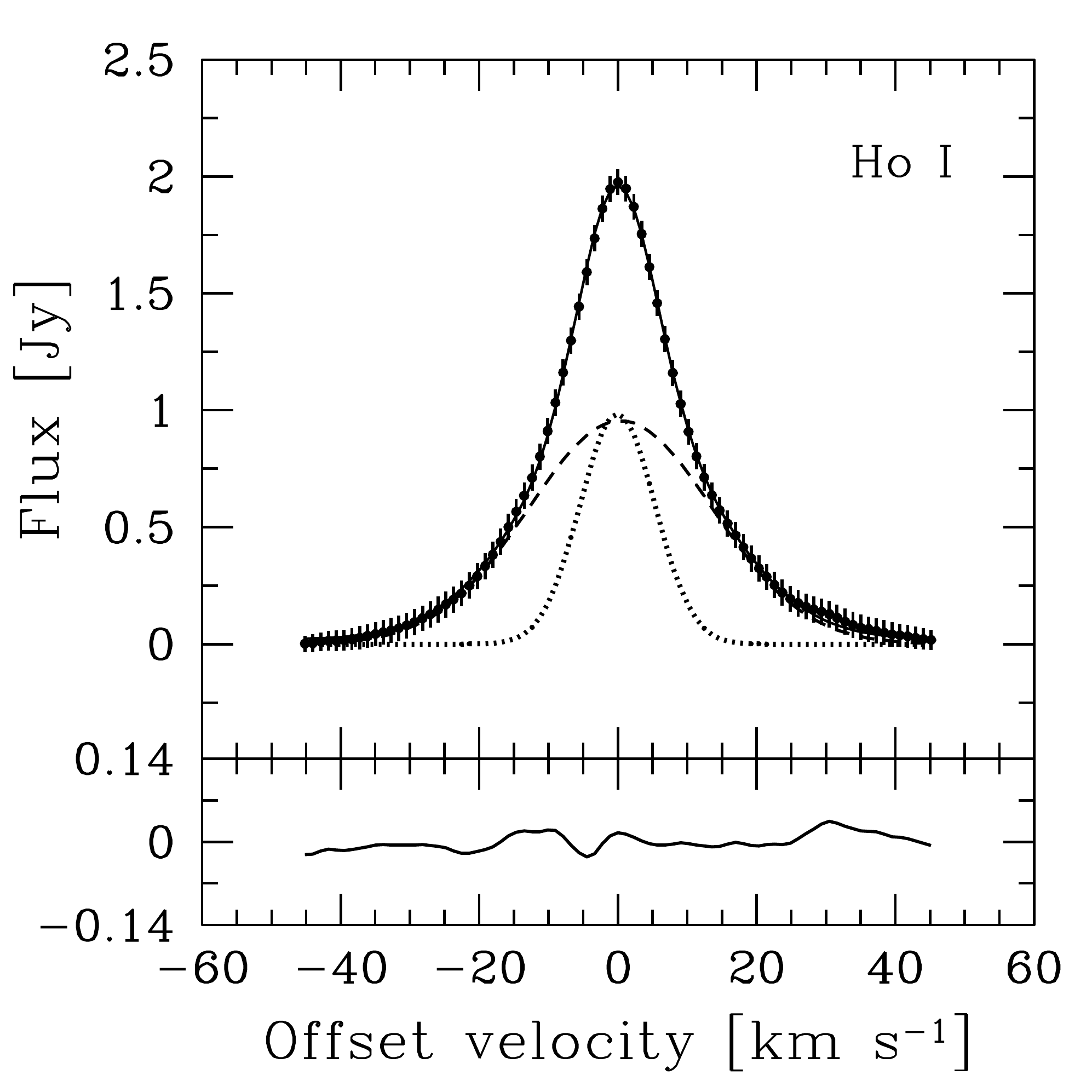}}}&
 \rotatebox{0}{\resizebox{58mm}{!}{\includegraphics[width = 0.6in,height = 0.6in]{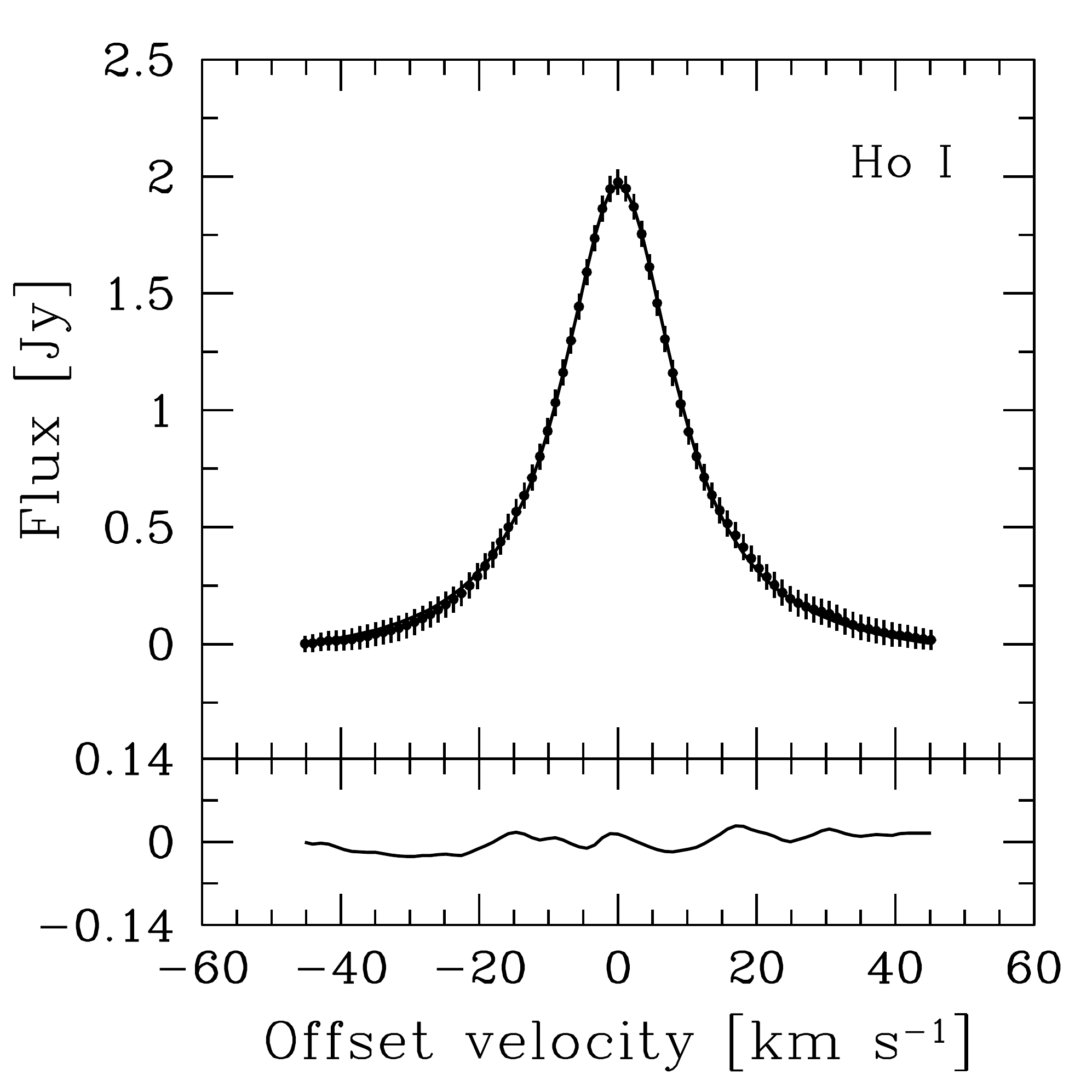}}}\\
   \rotatebox{0}{\resizebox{58mm}{!}{\includegraphics[width = 0.6in,height = 0.6in]{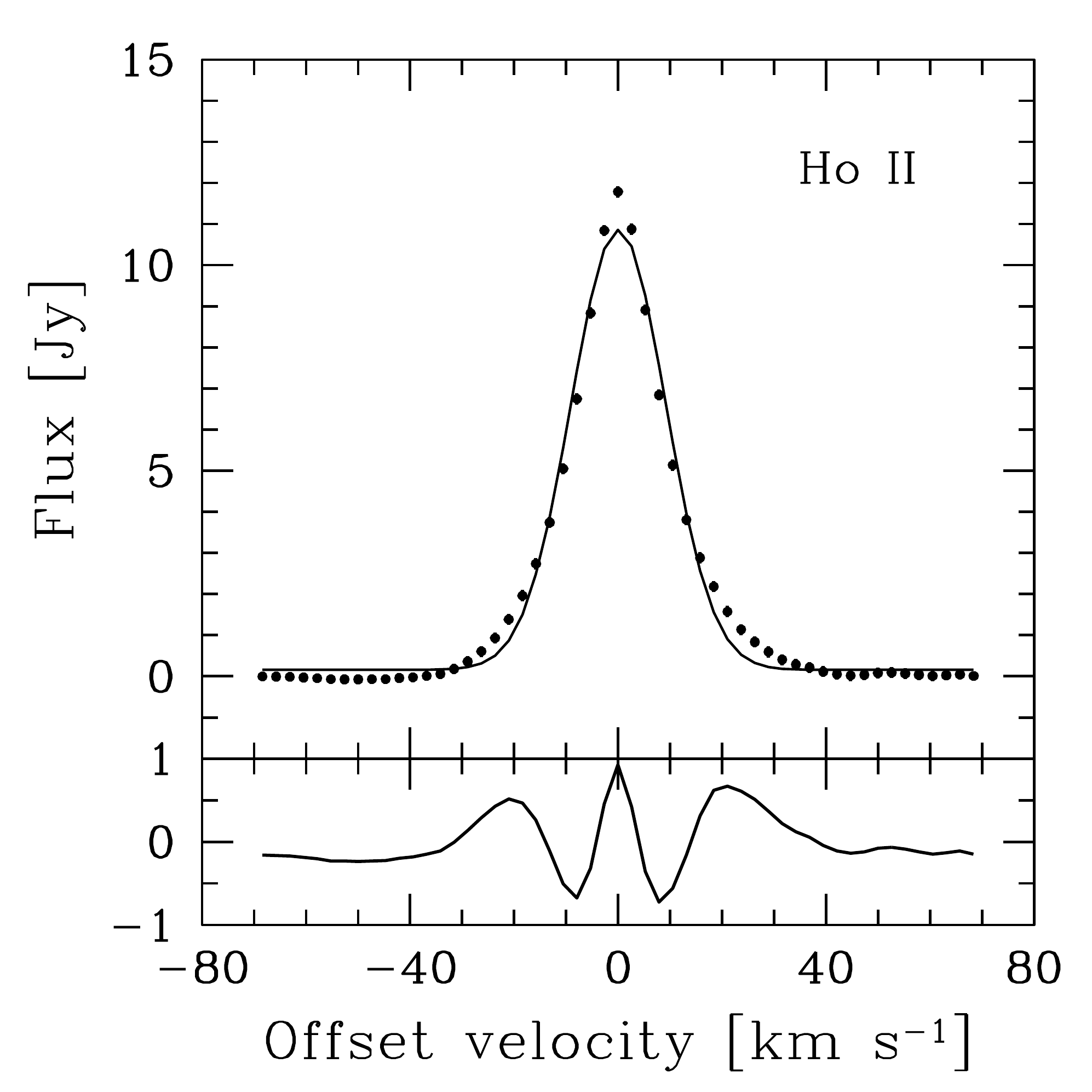}}}&
\rotatebox{0}{\resizebox{58mm}{!}{\includegraphics[width = 0.6in,height = 0.6in]{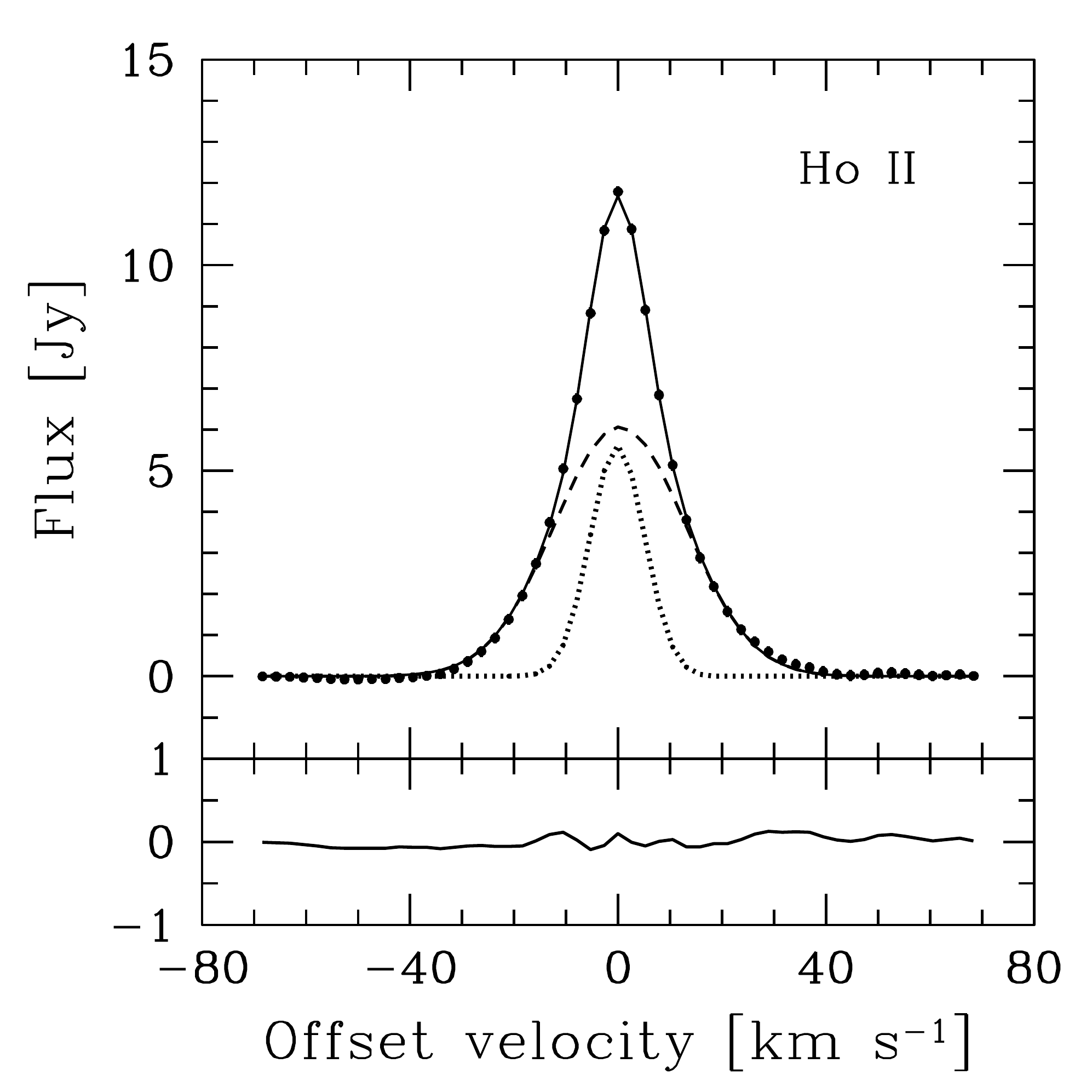}}}&
 \rotatebox{0}{\resizebox{58mm}{!}{\includegraphics[width = 0.6in,height = 0.6in]{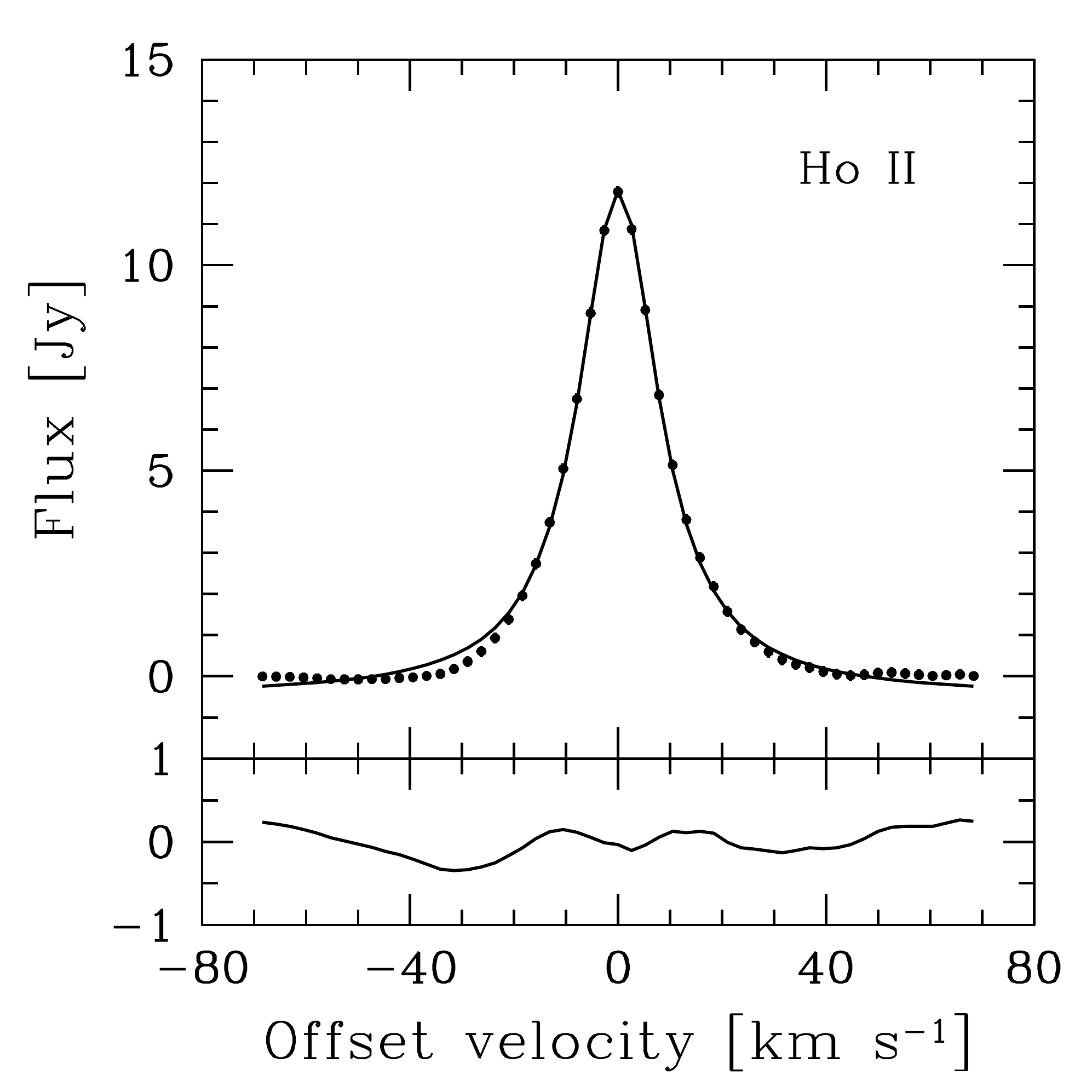}}}
\end{tabular}
\caption{Super profiles of the THINGS galaxies. For each galaxy, we show the
  results from the single (left panel) and double (middle panel)
  Gaussian fitting. The right panels show a fit with a Lorentzian
  function. The bottom panels of each set of plots show the residuals
  from the fits. The filled circles indicate the data. The solid black
  lines represent the results from the single, the sum of the double
  Gaussian, and the Lorentzian fitting. The dotted and the dashed
  lines in the middle panel represent the narrow and broad components required in the
  double Gaussian fitting. We plot error bars as 3$\sigma$ error bars,
  though in most cases they are smaller than the symbols plotted.}
\label{fig:app1}

\end{figure*}

The super profiles of the THINGS galaxies are shown in
Fig.~\ref{fig:app1}.  For each galaxy, the left panel
represents a plot of the super profile fitted with a single Gaussian
component whereas the middle and right panels represent super profiles
fitted with a double Gaussian and a Lorentzian functions,
respectively. The filled circles indicate the data points. The solid
black lines represent the results from fitting with a single and double Gaussian,
as well as Lorentzians. The dashed and the dotted lines represent the
narrow and broad components required in the double Gaussian
fitting. We show the residual from the three kinds of fits at the
bottom panel of each figure.  For the four galaxies  with negative bowls 
mentioned earlier, the open and the 
filled circle symbols represent the data points before and after the correction, 
respectively. The dashed lines represent the polynomial fits to the negative bowls.

\clearpage
\begin{deluxetable*}{l c c c }[!hbp]
\centering
\tabletypesize{\scriptsize}
\tablecaption{Velocity dispersion of super profiles before and after 
the negative bowl correction \label{tab:trust_broad}}
\tablewidth{0pt}
\tablehead{
	GALAXY & $\sigma_{1G}$ & $\sigma_{n}$ &$\sigma_{b}$\\ 
	&$(\rm{km s^{-1}})$& $(\rm{km s^{-1}})$& $(\rm{km s^{-1}})$\\
	&&&\\
       &\multicolumn{2}{c}{Before negative bowl correction}&
}
\startdata
	NGC 628  &8.4&4.0&10.0\\
	NGC 2403 &10.4&6.0& 15.0\\
	NGC 5236 &10.7& 5.1 &15.0\\
	\cutinhead{After negative bowl correction}
	NGC 628 &9.0 &4.4&11.6\\
	NGC 2403&11.1&6.6&18.4\\
	NGC 5236&11.2&5.6&16.8\\
\enddata
\tablecomments{$\sigma_{1G}$: Velocity dispersion
   derived from the single component Gaussian fit. 
   $\sigma_{n}$: Velocity
   dispersion of the narrow component. $\sigma_{b}$: Velocity
   dispersion of the broad component.}
\end{deluxetable*}
\begin{deluxetable}{l c c c}[!hbp]
\centering
\tabletypesize{\scriptsize}
\tablecaption{Comparison between the fitted area of super profiles before and after 
the negative bowl correction \label{tab:trust_broad_area}}
\tablewidth{0pt}
\tablehead{
	GALAXY& $A_{1G, b. c.}/A_{1G, a. c.}$ & $A_{n, b. c.}$/$A_{n, a. c.}$ 
	& $A_{b, b. c.}$/$A_{b, a.c.}$
}
\startdata
	NGC 628 &0.89&0.67&0.93\\
	NGC 2403&0.91&0.76&0.94\\ 
	NGC 5236&0.95&0.81&0.97\\
\enddata
\tablecomments{b. c. = Before correction (i.e. before negative bowl correction). 
a. c = After correction (i.e. after correcting for the negative bowl).}
\end{deluxetable}

\begin{deluxetable*}{l r r r c}
\centering
\tabletypesize{\scriptsize}
\tablecaption{Parameters of Lorentzian fit to the THINGS super profiles \label{tab:table_lorentz}}
\tablewidth{0pt}
\tablehead{
Galaxy&\multicolumn{1}{c}{$\rm{FWHM}$}&\multicolumn{1}{c}{$\rm{Area}$}&
\multicolumn{1}{c}{$\rm{mean}$}&\multicolumn{1}{c}{$\chi^{2}_{2G}/\chi^{2}_{Lor}$}\\ 
          &\multicolumn{1}{c}{$\rm{(km~ s^{-1})}$}
	  &\multicolumn{1}{c}{$\rm{(Jy~km~ s^{-1})}$}&\multicolumn{1}{c}{$\rm{(km~ s^{-1})}$}&\\
\multicolumn{1}{c}{1}&\multicolumn{1}{c}{2}&\multicolumn{1}{c}{3}&
\multicolumn{1}{c}{14}&\multicolumn{1}{c}{5}
}
\startdata
DDO 53	&22.0$\rm{\pm}0.1$&29.6$\rm{\pm}0.1$&0.06$\rm{\pm}0.02$&0.88\\
DDO 154	&20.1$\rm{\pm}0.3$&130.6$\rm{\pm}1.9$&0.01$\rm{\pm}0.09$&0.08\\
Ho I	&19.9$\rm{\pm}0.1$&64.2$\rm{\pm}0.4$&0.07$\rm{\pm}0.03$&0.93\\
Ho II	&18.7$\rm{\pm}0.3$&360.5$\rm{\pm}4.6$&0.09$\rm{\pm}0.08$&0.16\\
IC 2574	&19.9$\rm{\pm}0.3$&622.2$\rm{\pm}7.9$&0.14$\rm{\pm}0.09$&0.26\\
M81 dwB	&27.7$\rm{\pm}0.2$&6.1$\rm{\pm}0.0$&-0.10$\rm{\pm}0.03$&0.53\\
M81 dwA	&20.0$\rm{\pm}0.4$&7.1$\rm{\pm}0.2$&0.05$\rm{\pm}0.07$&0.54\\
NGC 3031&21.0$\rm{\pm}0.3$&1721.0$\rm{\pm}15.9$&0.38$\rm{\pm}0.08$&0.22\\
NGC 628	&17.4$\rm{\pm}0.3$&612.0$\rm{\pm}10.1$&-0.02$\rm{\pm}0.10$&0.00\\
NGC 925	&25.7$\rm{\pm}0.1$&338.0$\rm{\pm}1.3$&-0.04$\rm{\pm}0.03$&3.59\\
NGC 1569&51.5$\rm{\pm}1.8$&101.7$\rm{\pm}3.8$&-1.60$\rm{\pm}0.40$&0.19\\
NGC 2366&25.6$\rm{\pm}0.2$&351.9$\rm{\pm}2.0$&-0.16$\rm{\pm}0.04$&0.72\\
NGC 2403&21.0$\rm{\pm}0.2$&1650$\rm{\pm}14.8$&0.10$\rm{\pm}0.08$&0.01\\
NGC 2841&31.6$\rm{\pm}0.3$&225.4$\rm{\pm}1.9$&0.10$\rm{\pm}0.10$&0.70\\
NGC 2903&25.9$\rm{\pm}0.2$&344.1$\rm{\pm}1.7$&-0.10$\rm{\pm}0.05$&2.24\\
NGC 2976&24.4$\rm{\pm}0.3$&48.0$\rm{\pm}0.5$&0.20$\rm{\pm}0.09$&0.25\\
NGC 3077&24.2$\rm{\pm}0.4$&374.2$\rm{\pm}5.3$&0.03$\rm{\pm}0.13$&0.21\\
NGC 3184&22.7$\rm{\pm}0.3$&227.9$\rm{\pm}2.3$&0.10$\rm{\pm}0.08$&0.30\\
NGC 3198&25.6$\rm{\pm}0.2$&254.5$\rm{\pm}1.9$&-0.03$\rm{\pm}0.08$&0.26\\
NGC 3351&19.8$\rm{\pm}0.1$&85.9$\rm{\pm}0.4$&-0.00$\rm{\pm}0.04$&2.27\\
NGC 3521&34.6$\rm{\pm}0.3$&355.0$\rm{\pm}2.4$&-0.04$\rm{\pm}0.09$&1.13\\
NGC 3621&22.4$\rm{\pm}0.2$&837.8$\rm{\pm}7.2$&0.18$\rm{\pm}0.08$&0.92\\
NGC 3627&41.9$\rm{\pm}0.2$&47.8$\rm{\pm}0.0$&0.03$\rm{\pm}0.06$&1.87\\
NGC 4214&17.9$\rm{\pm}0.2$&324.9$\rm{\pm}2.9$&-0.02$\rm{\pm}0.05$&0.11\\
NGC 4449&27.9$\rm{\pm}0.7$&444.4$\rm{\pm}9.5$&-0.28$\rm{\pm}0.20$&0.14\\
NGC 4736&21.5$\rm{\pm}0.1$&106.9$\rm{\pm}0.4$&-0.05$\rm{\pm}0.03$&4.03\\
NGC 4826&25.7$\rm{\pm}$0.5&13.6$\rm{\pm}0.2$&-0.24$\rm{\pm}0.16$&0.35\\
NGC 5055&26.8$\rm{\pm}0.3$&537.0$\rm{\pm}5.3$&0.01$\rm{\pm}0.11$&0.52\\
NGC 5194&33.2$\rm{\pm}0.3$&274.2$\rm{\pm}2.2$&-0.21$\rm{\pm}0.10$&0.26\\
NGC 5236&21.2$\rm{\pm}0.2$&473.6$\rm{\pm}3.4$&-0.09$\rm{\pm}0.06$&0.02\\
NGC 5457&23.6$\rm{\pm}0.6$&2629.0$\rm{\pm}51.7$&-0.08$\rm{\pm}0.18$&0.00\\
NGC 6946&20.1$\rm{\pm}0.2$&425.3$\rm{\pm}3.5$&-0.06$\rm{\pm}0.07$&0.65\\
NGC 7331&37.5$\rm{\pm}0.3$&277.6$\rm{\pm}1.6$&0.04$\rm{\pm}0.09$& 0.22\\
NGC 7793&21.0$\rm{\pm}0.1$&353.1$\rm{\pm}1.6$&0.01$\rm{\pm}0.04$&0.72\\
\enddata
\tablecomments{Column 1: Name of galaxy; Column 2: Full Width at half maximum ; 
	Column 3: Integrated area; Column 4: Fitted central velocity values; 
	Column 5: Ratio of the reduced $\rm{\chi^{2}}$ of the double Gaussian and Lorentzian fitting.}
\end{deluxetable*}

\begin{figure*}
    \begin{tabular}{l l l} 
   \rotatebox{0}{\resizebox{58mm}{!}{\includegraphics[width = 0.6in,height = 0.6in]{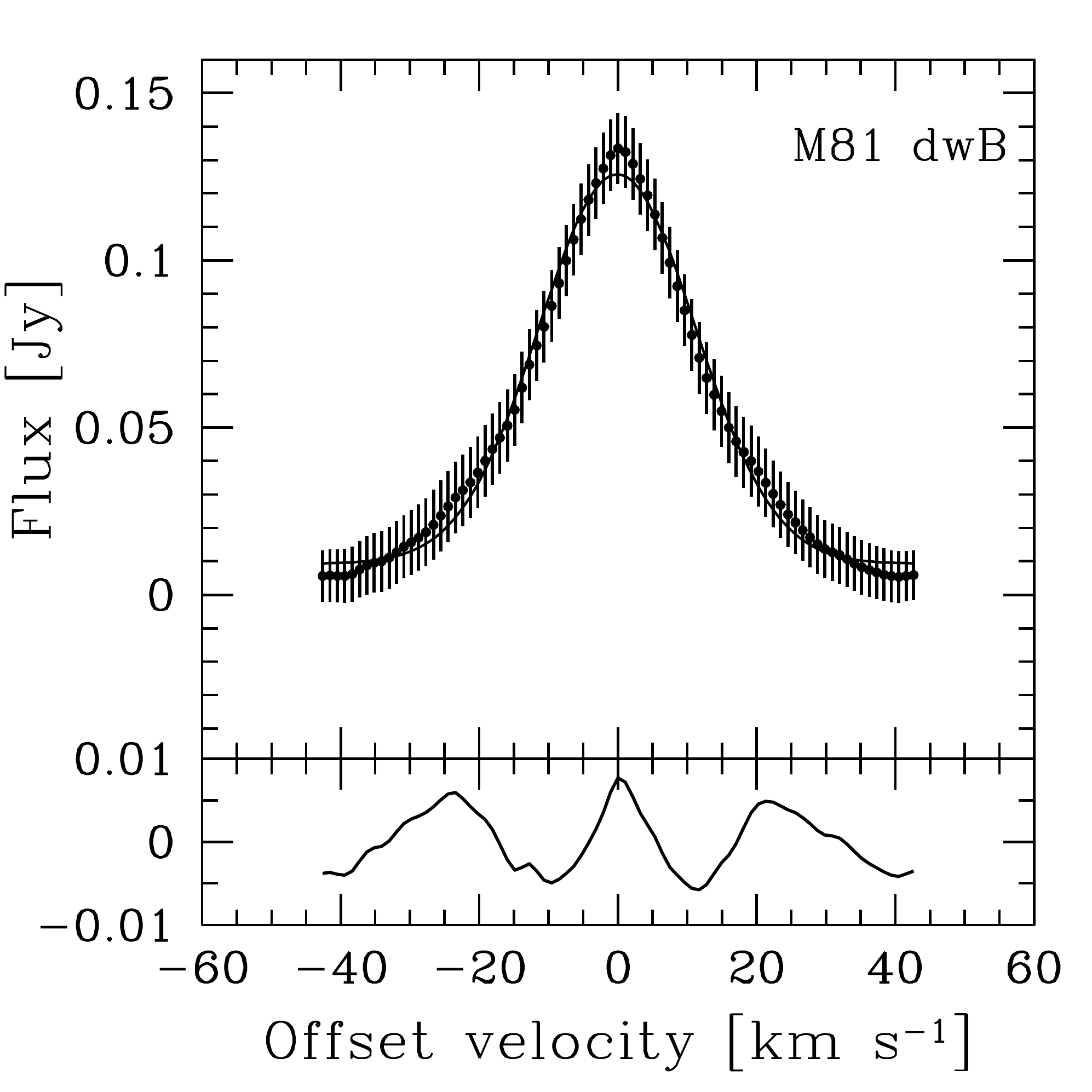}}}&
\rotatebox{0}{\resizebox{58mm}{!}{\includegraphics[width = 0.6in,height = 0.6in]{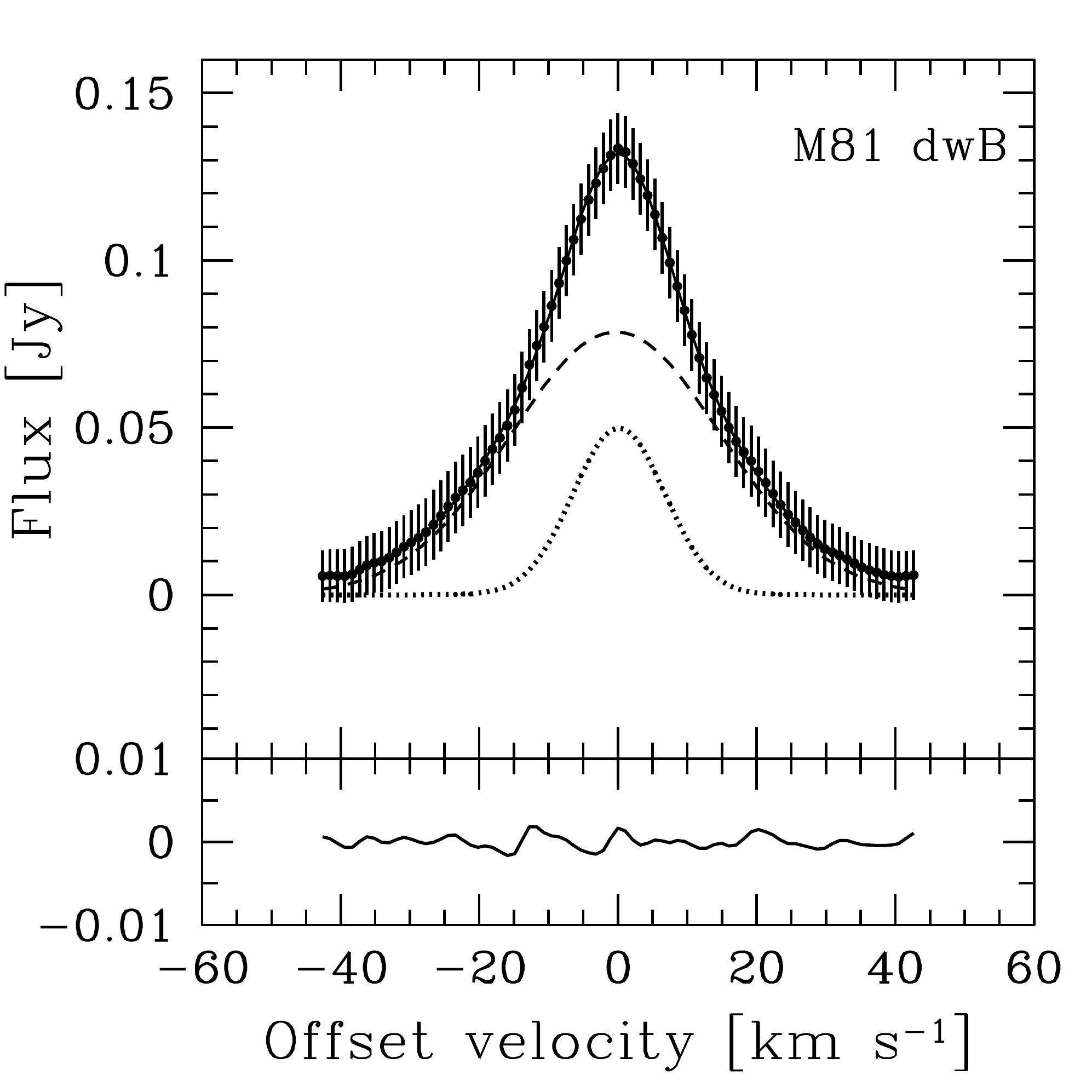}}}&
 \rotatebox{0}{\resizebox{58mm}{!}{\includegraphics[width = 0.6in,height = 0.6in]{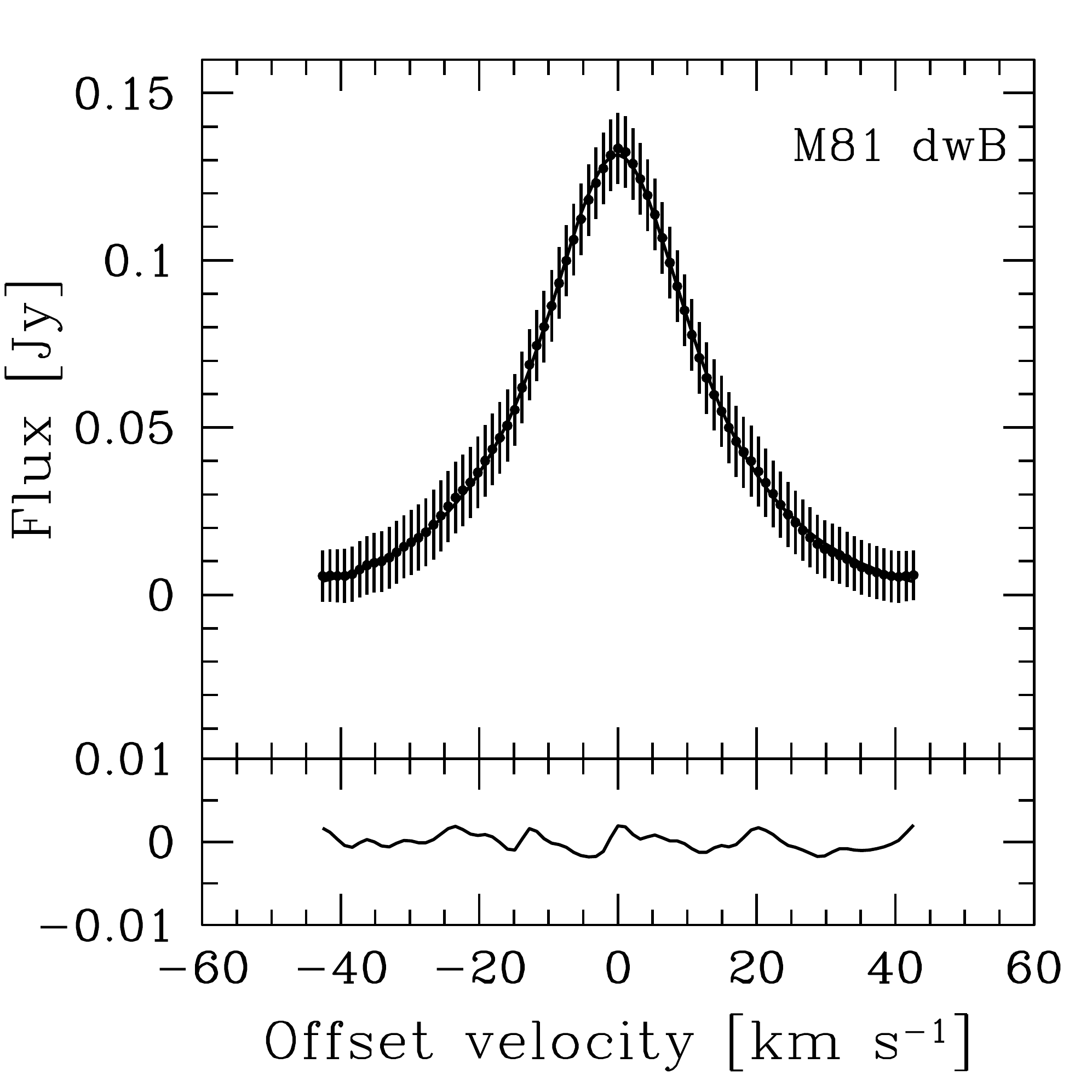}}}\\
   \rotatebox{0}{\resizebox{58mm}{!}{\includegraphics[width = 0.6in,height = 0.6in]{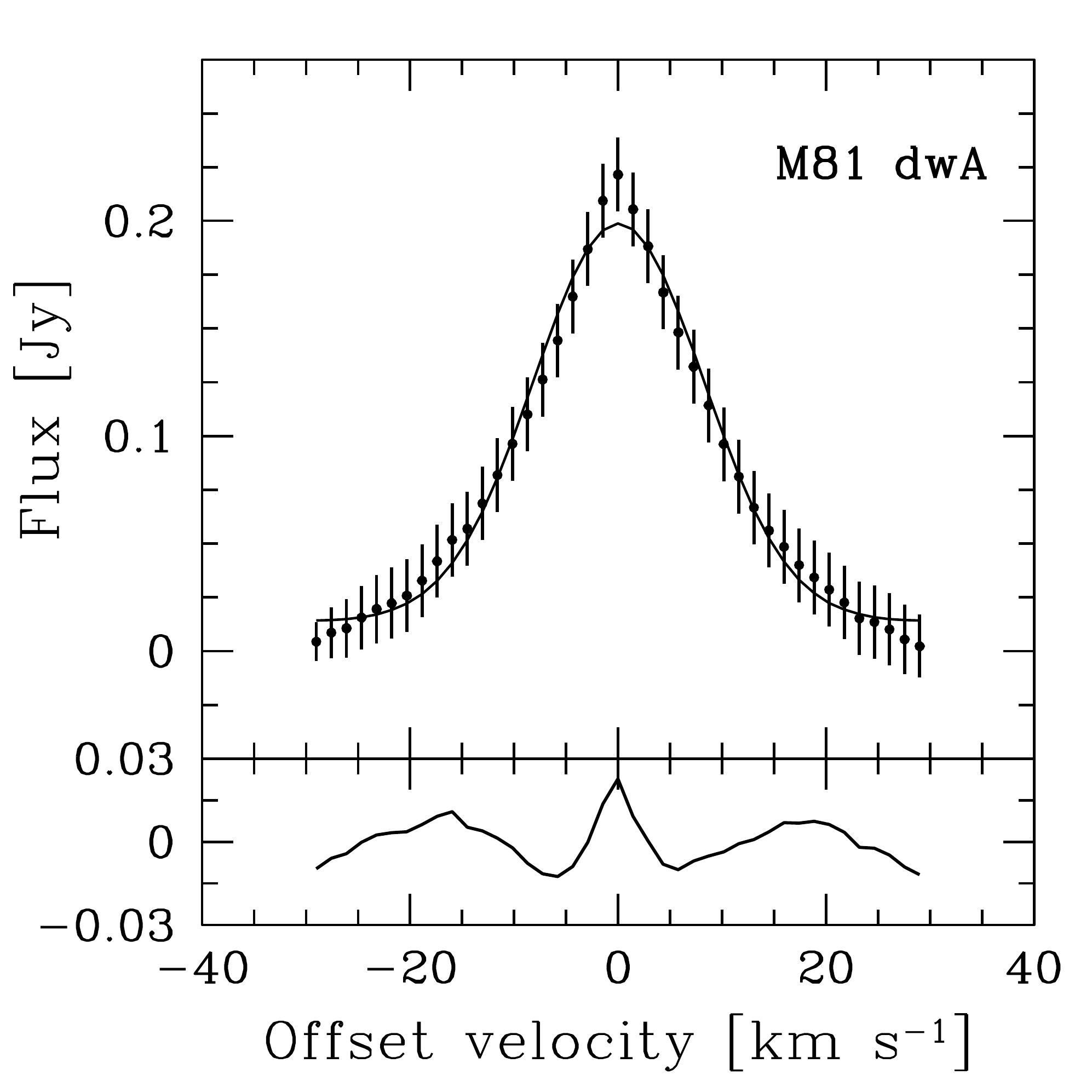}}}&
\rotatebox{0}{\resizebox{58mm}{!}{\includegraphics[width = 0.6in,height = 0.6in]{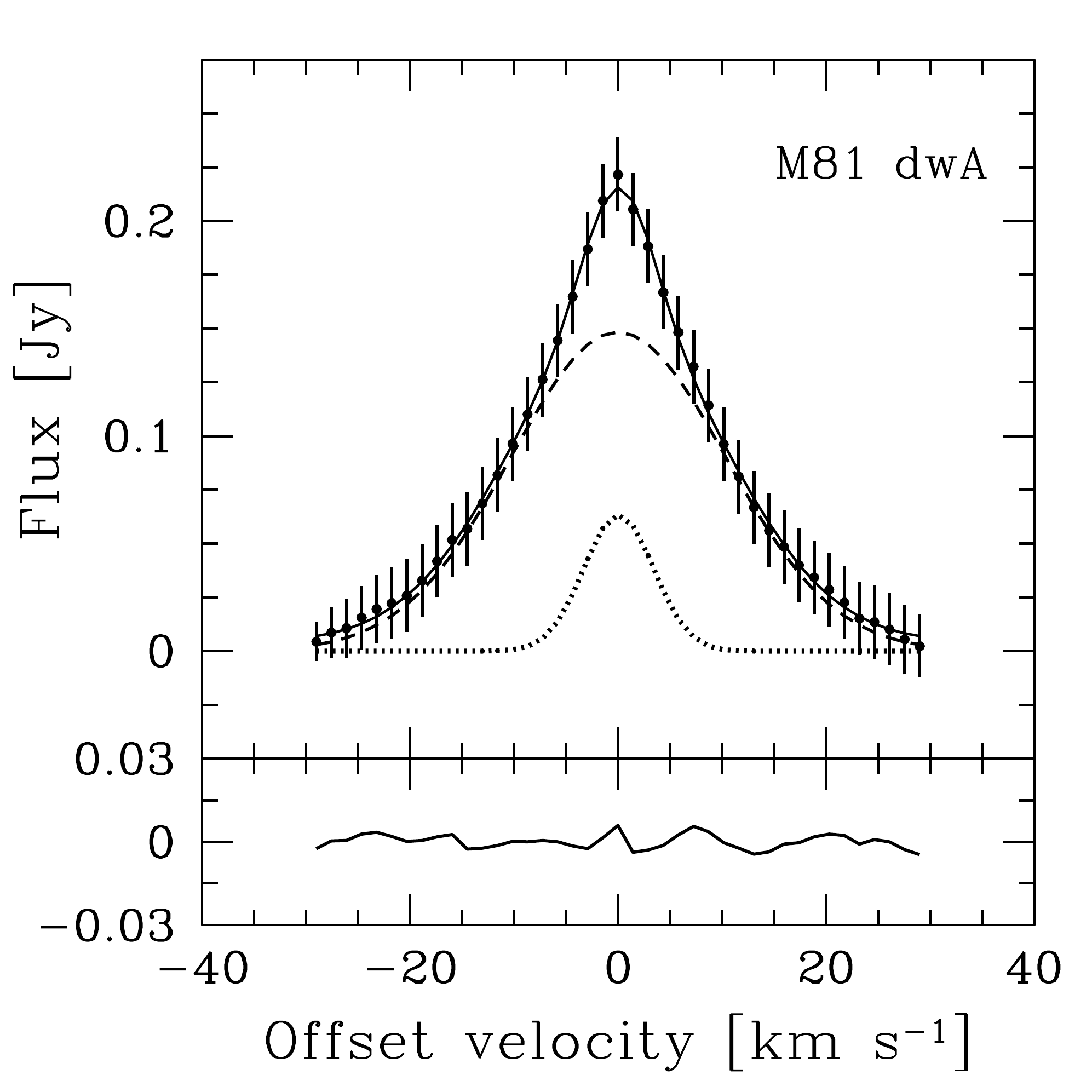}}}&
 \rotatebox{0}{\resizebox{58mm}{!}{\includegraphics[width = 0.6in,height = 0.6in]{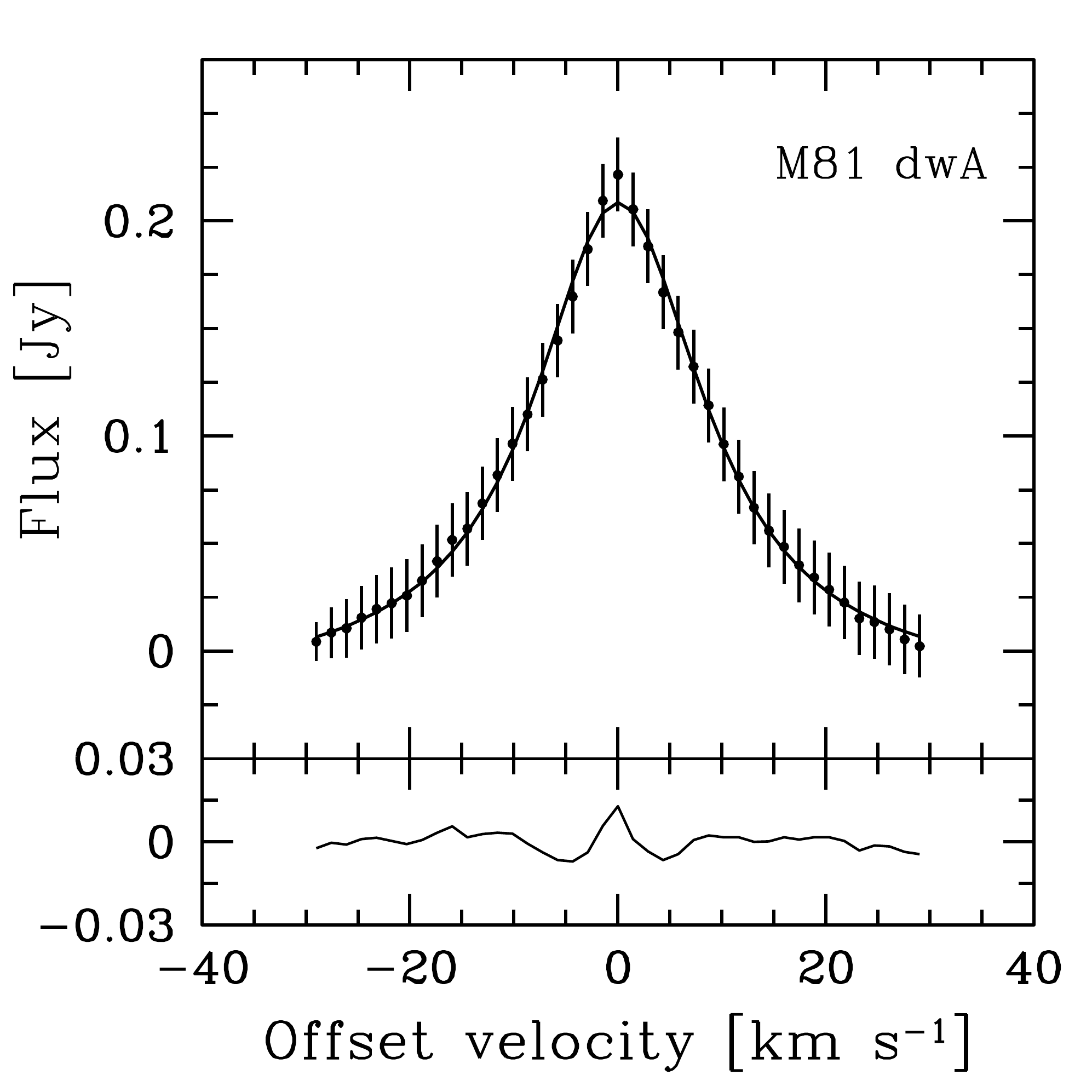}}}\\
   \rotatebox{0}{\resizebox{58mm}{!}{\includegraphics[width = 0.6in,height = 0.6in]{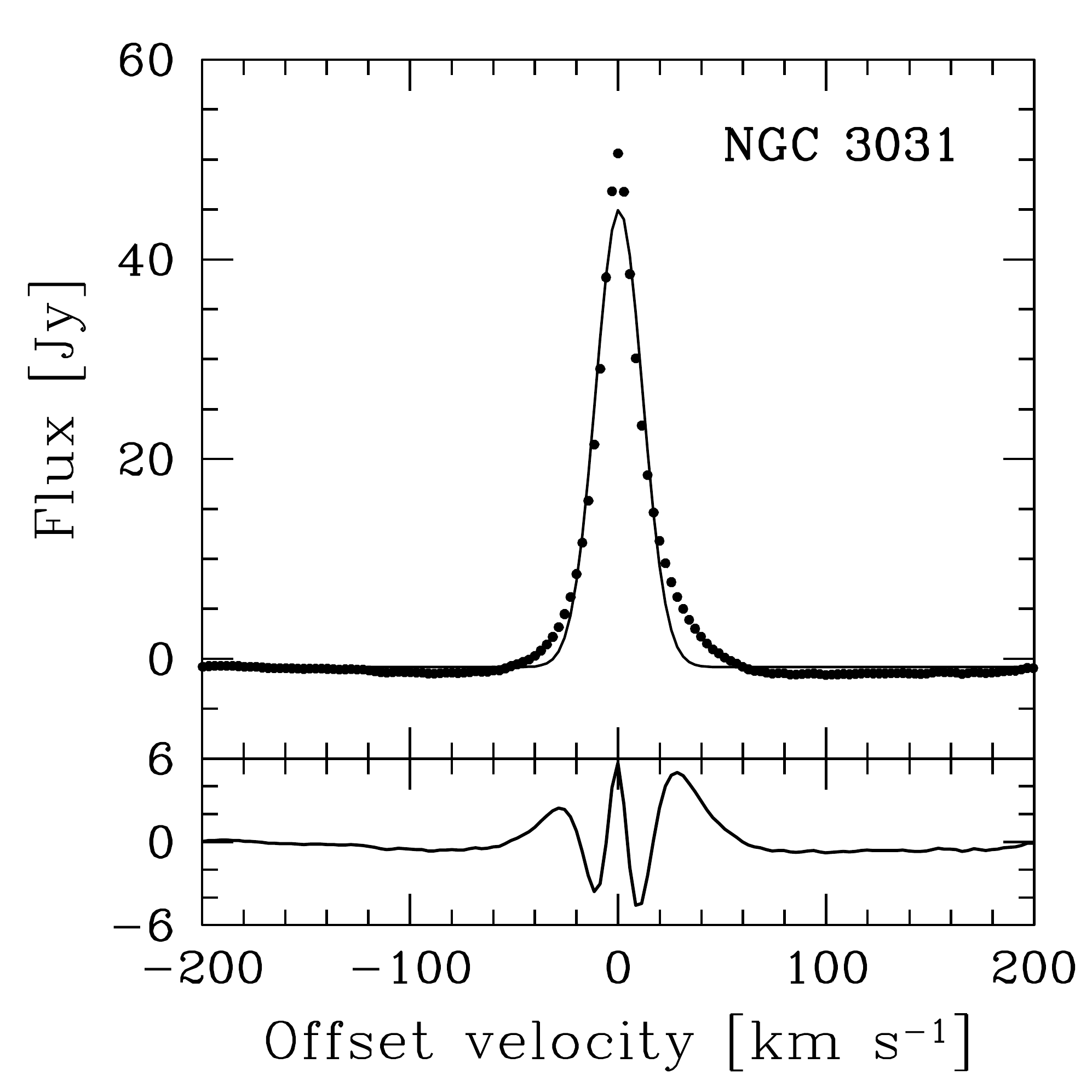}}}&
\rotatebox{0}{\resizebox{58mm}{!}{\includegraphics[width = 0.6in,height = 0.6in]{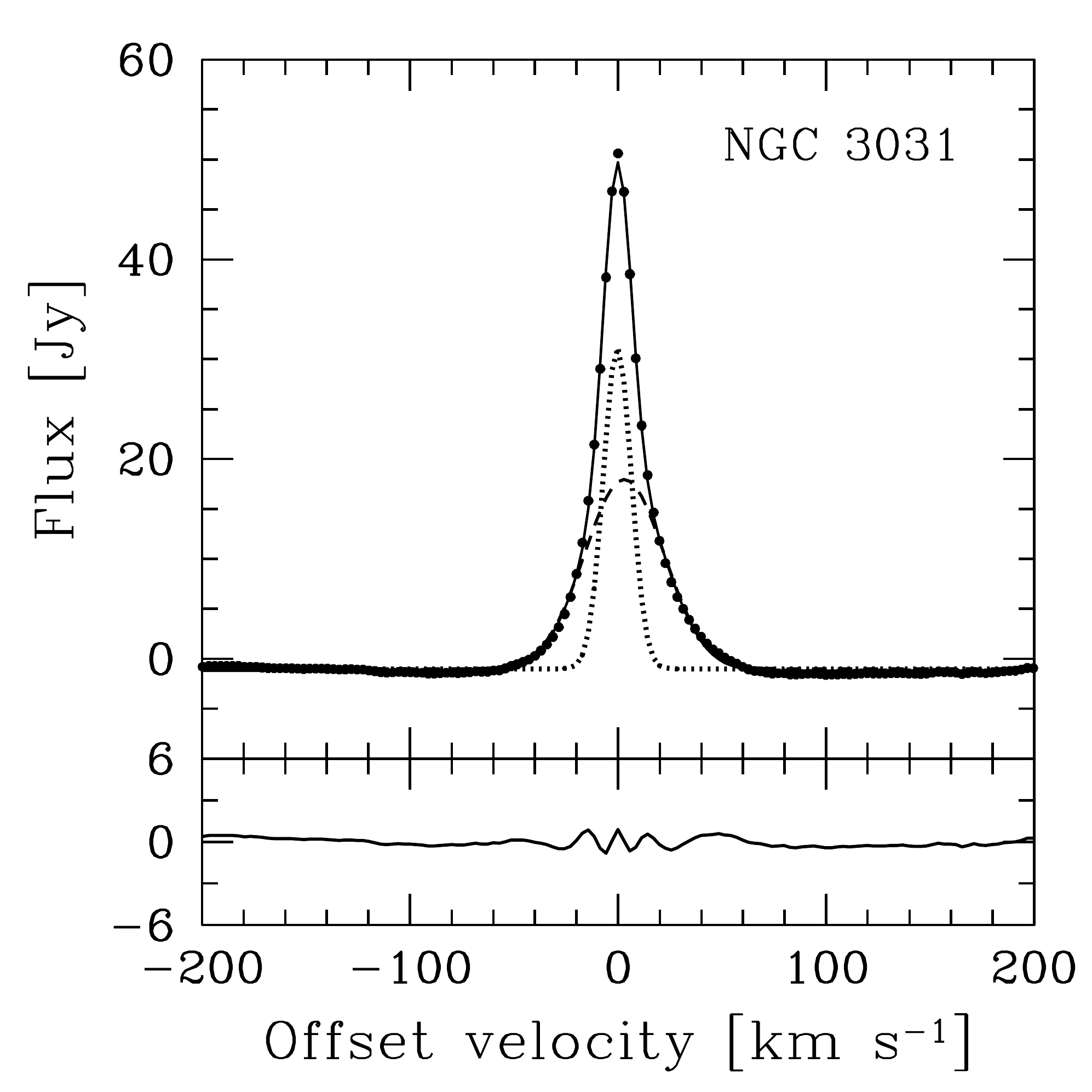}}}&
 \rotatebox{0}{\resizebox{58mm}{!}{\includegraphics[width = 0.6in,height = 0.6in]{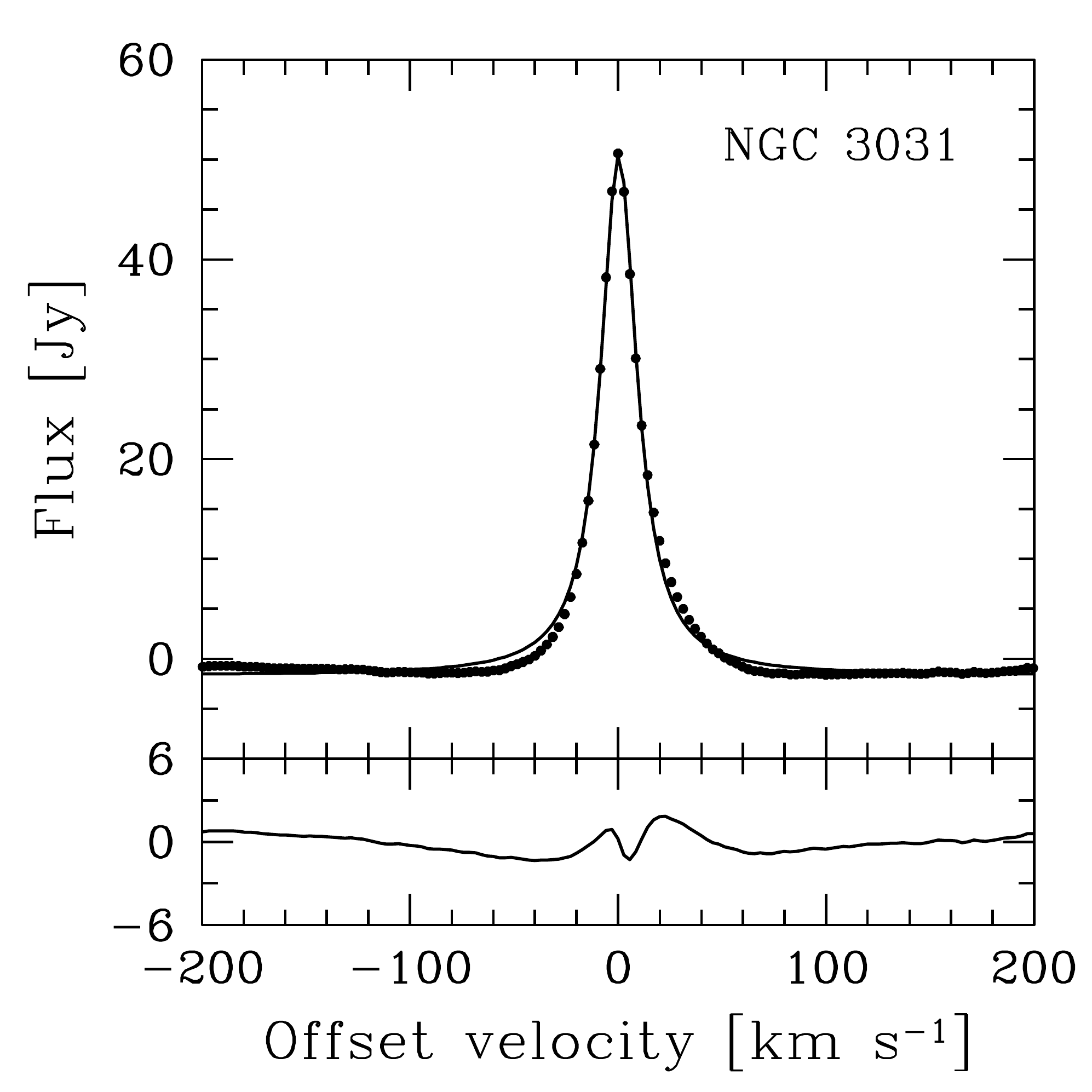}}}\\
     \rotatebox{0}{\resizebox{58mm}{!}{\includegraphics[width = 0.6in,height = 0.6in]{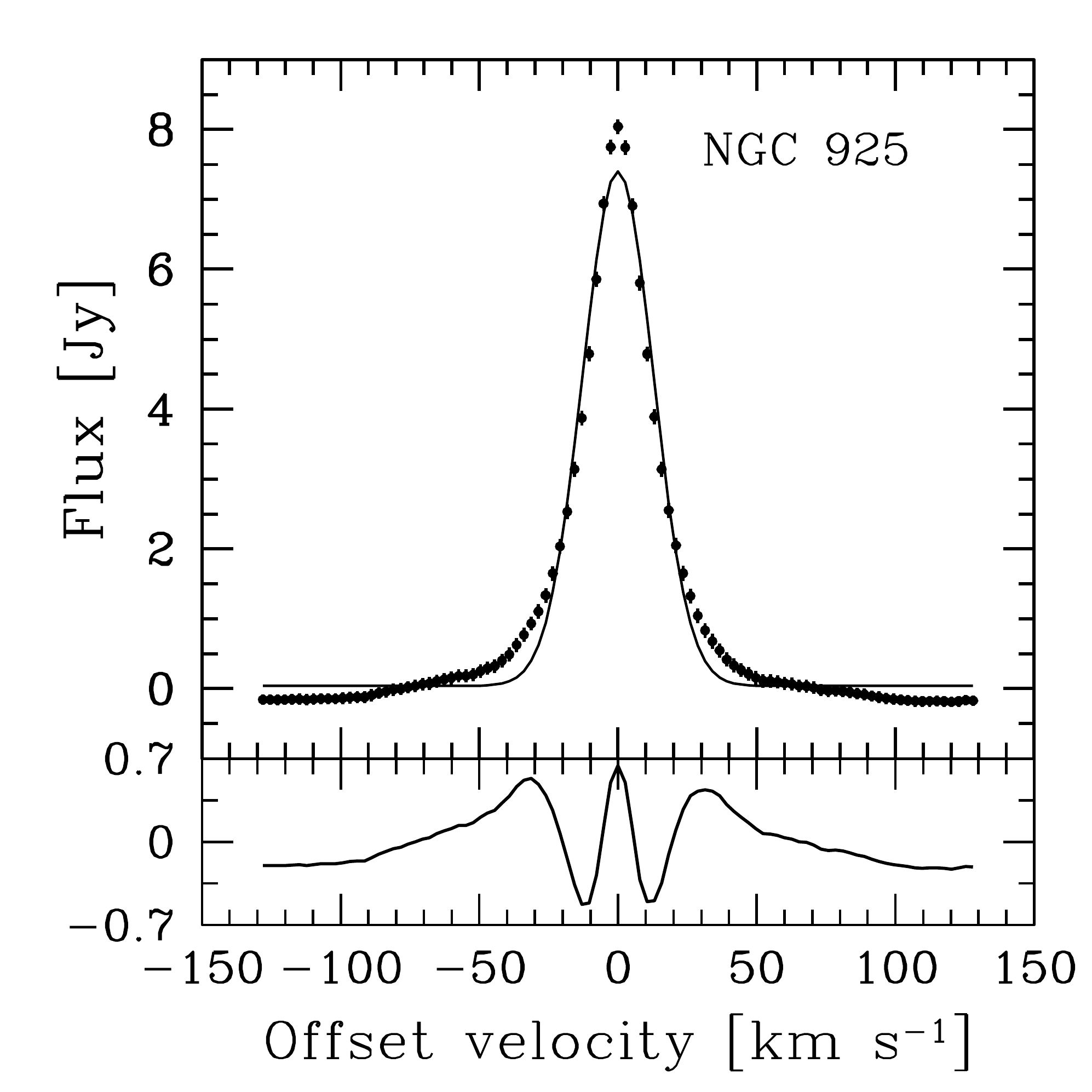}}}&
\rotatebox{0}{\resizebox{58mm}{!}{\includegraphics[width = 0.6in,height = 0.6in]{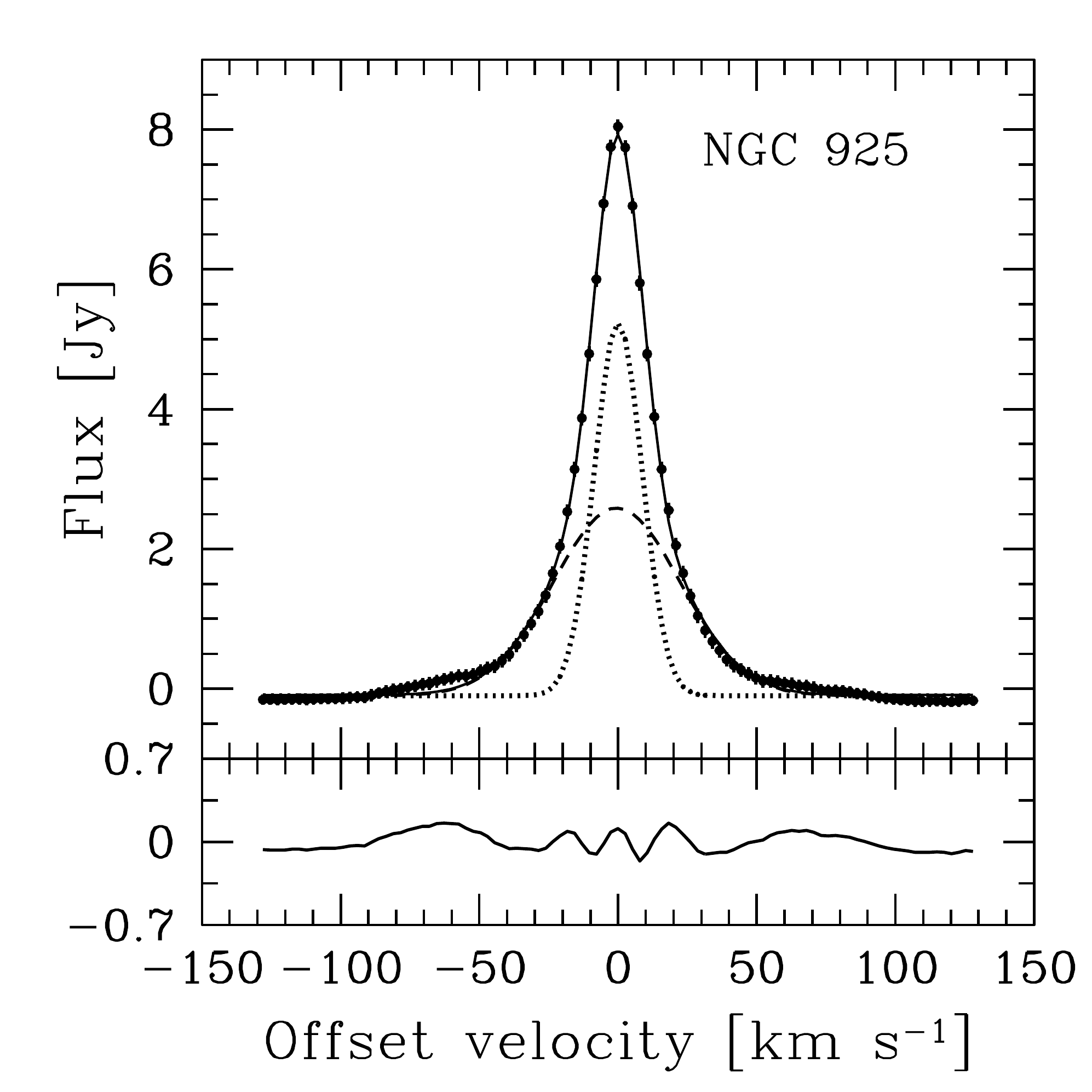}}}&
 \rotatebox{0}{\resizebox{58mm}{!}{\includegraphics[width = 0.6in,height = 0.6in]{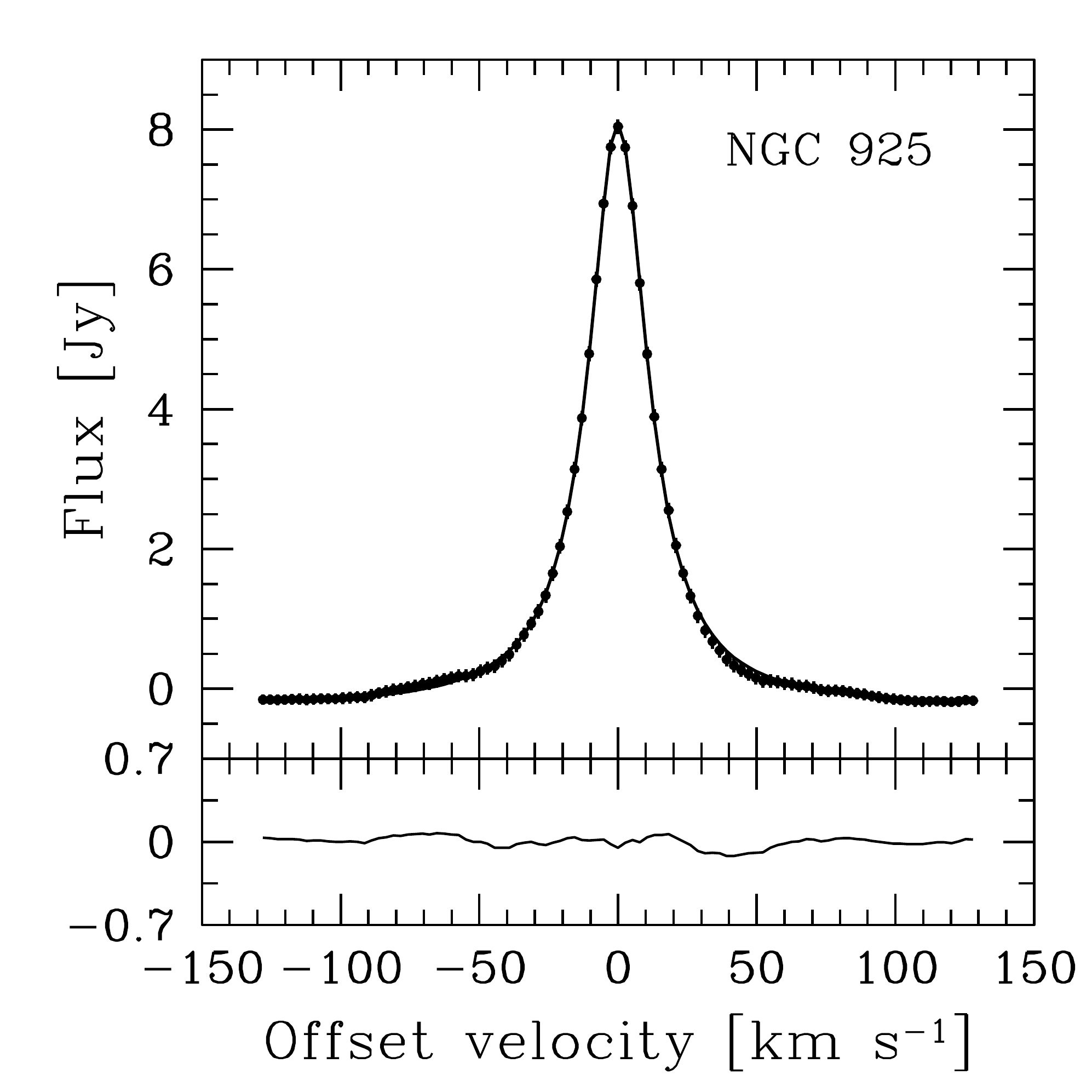}}}
\\\\ & & \hspace*{-12.12cm}\textbf{Figure \ref{fig:app1}}. \scriptsize{(continued).}
\end{tabular}
\end{figure*}

\begin{figure*}
    \begin{tabular}{l l l} 
   \rotatebox{0}{\resizebox{58mm}{!}{\includegraphics[width = 0.6in,height = 0.6in]{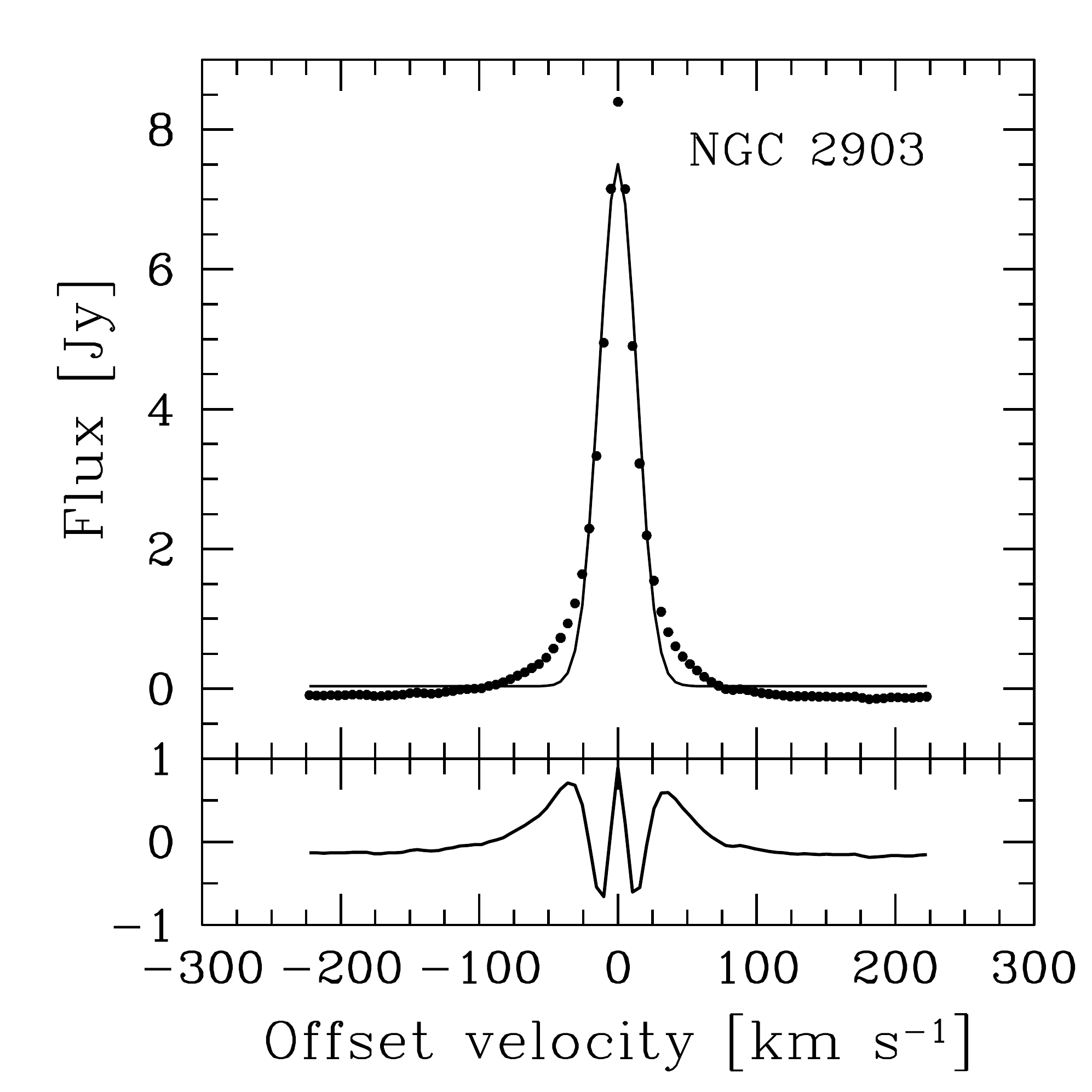}}}&
\rotatebox{0}{\resizebox{58mm}{!}{\includegraphics[width = 0.6in,height = 0.6in]{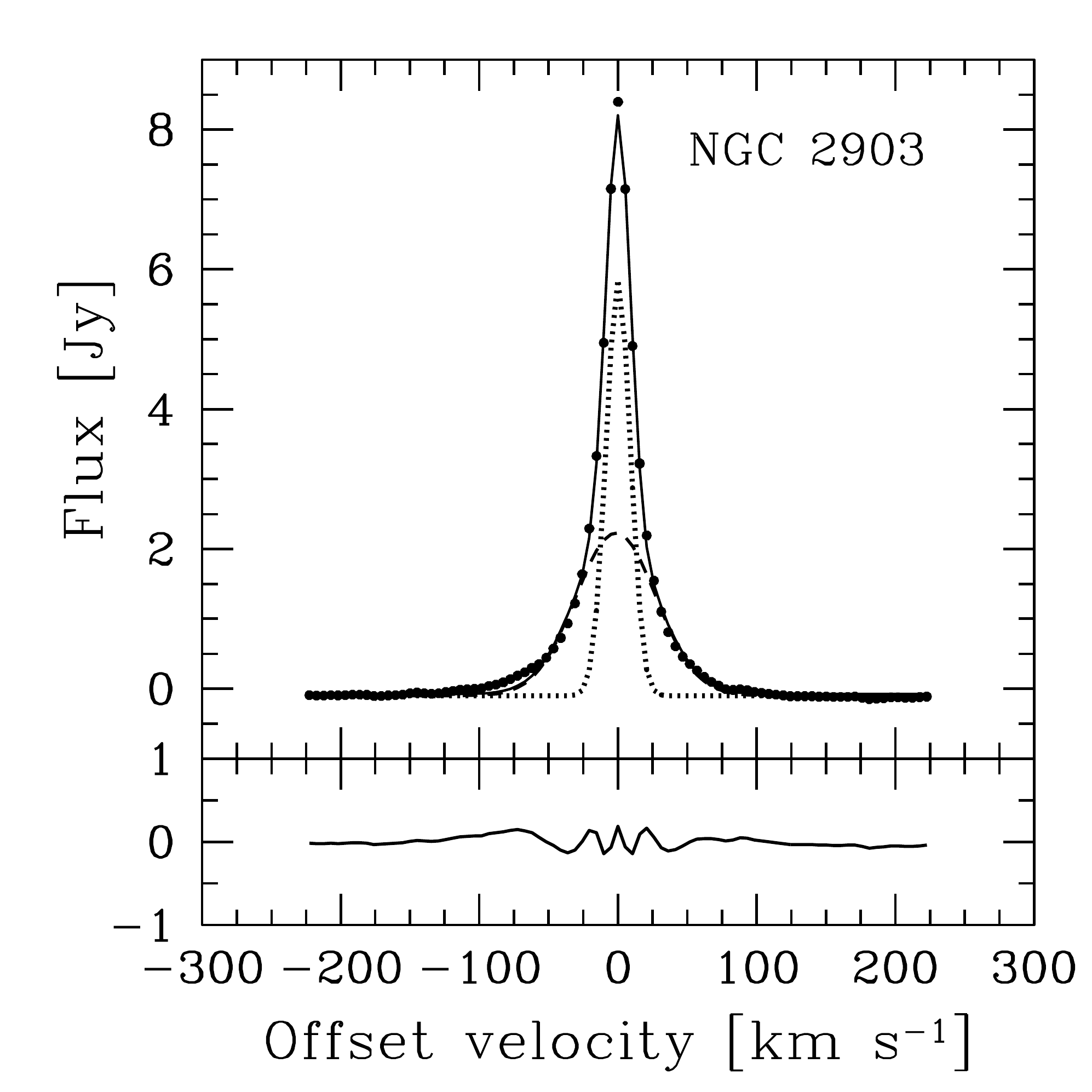}}}&
 \rotatebox{0}{\resizebox{58mm}{!}{\includegraphics[width = 0.6in,height = 0.6in]{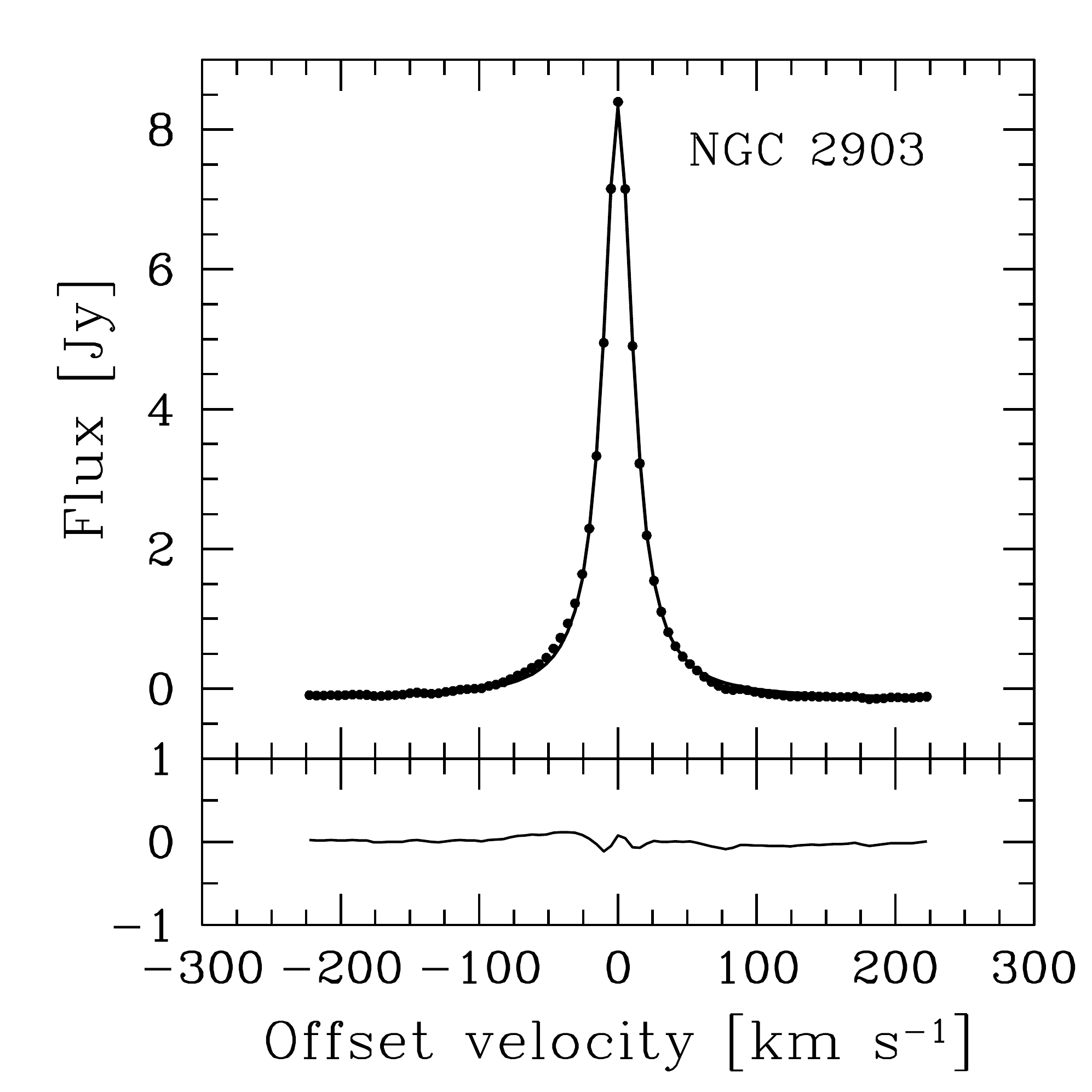}}}\\
   \rotatebox{0}{\resizebox{58mm}{!}{\includegraphics[width = 0.6in,height = 0.6in]{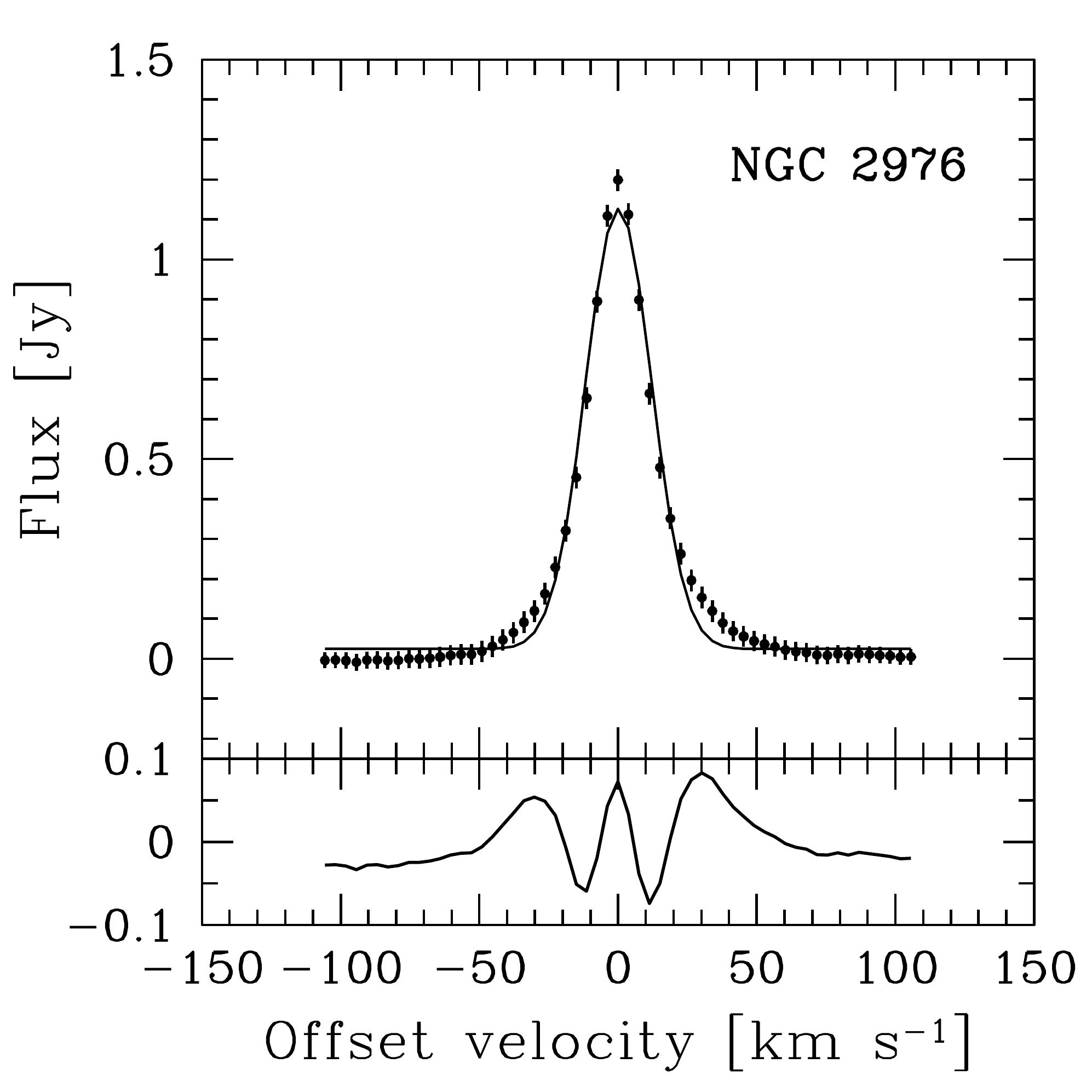}}}&
\rotatebox{0}{\resizebox{58mm}{!}{\includegraphics[width = 0.6in,height = 0.6in]{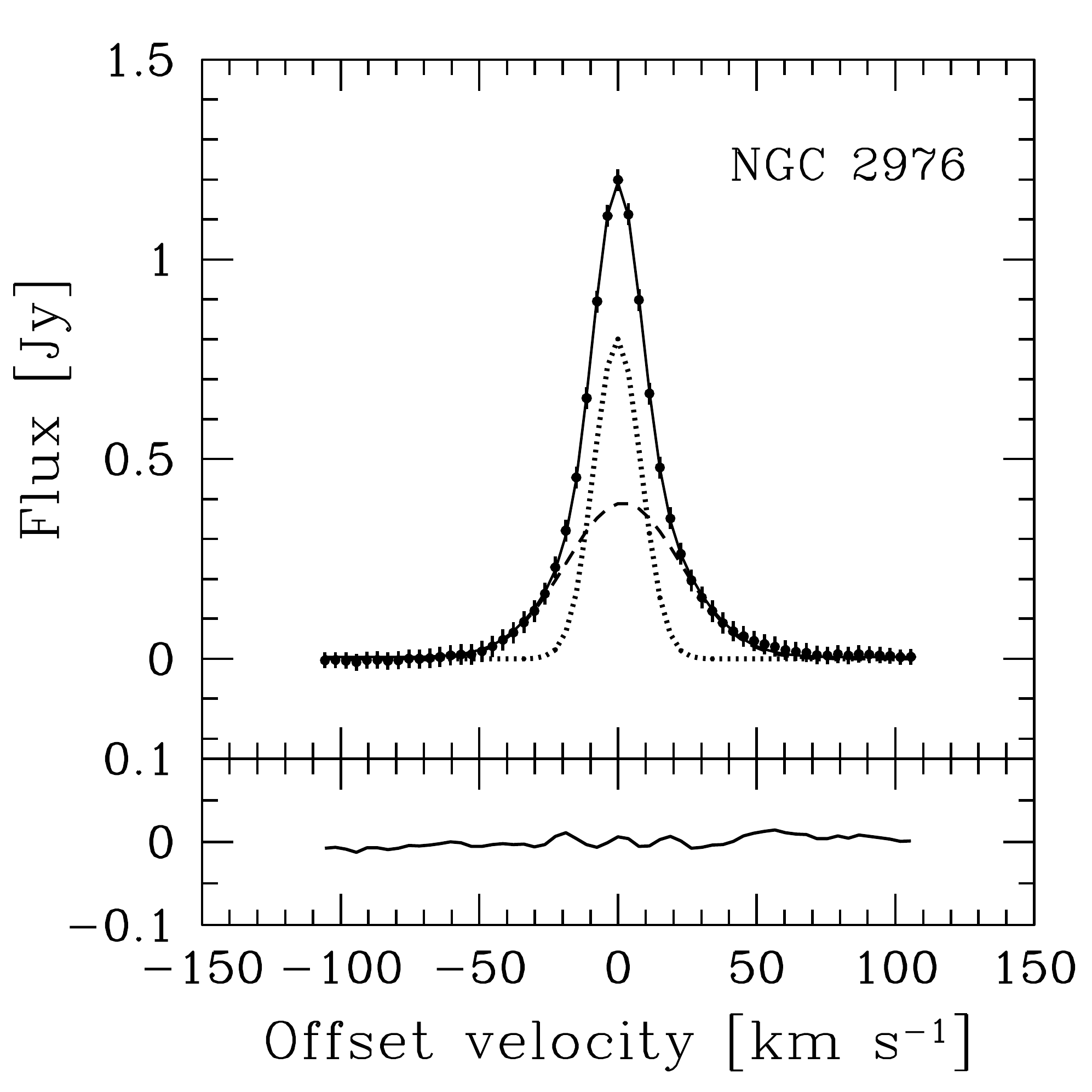}}}&
 \rotatebox{0}{\resizebox{58mm}{!}{\includegraphics[width = 0.6in,height = 0.6in]{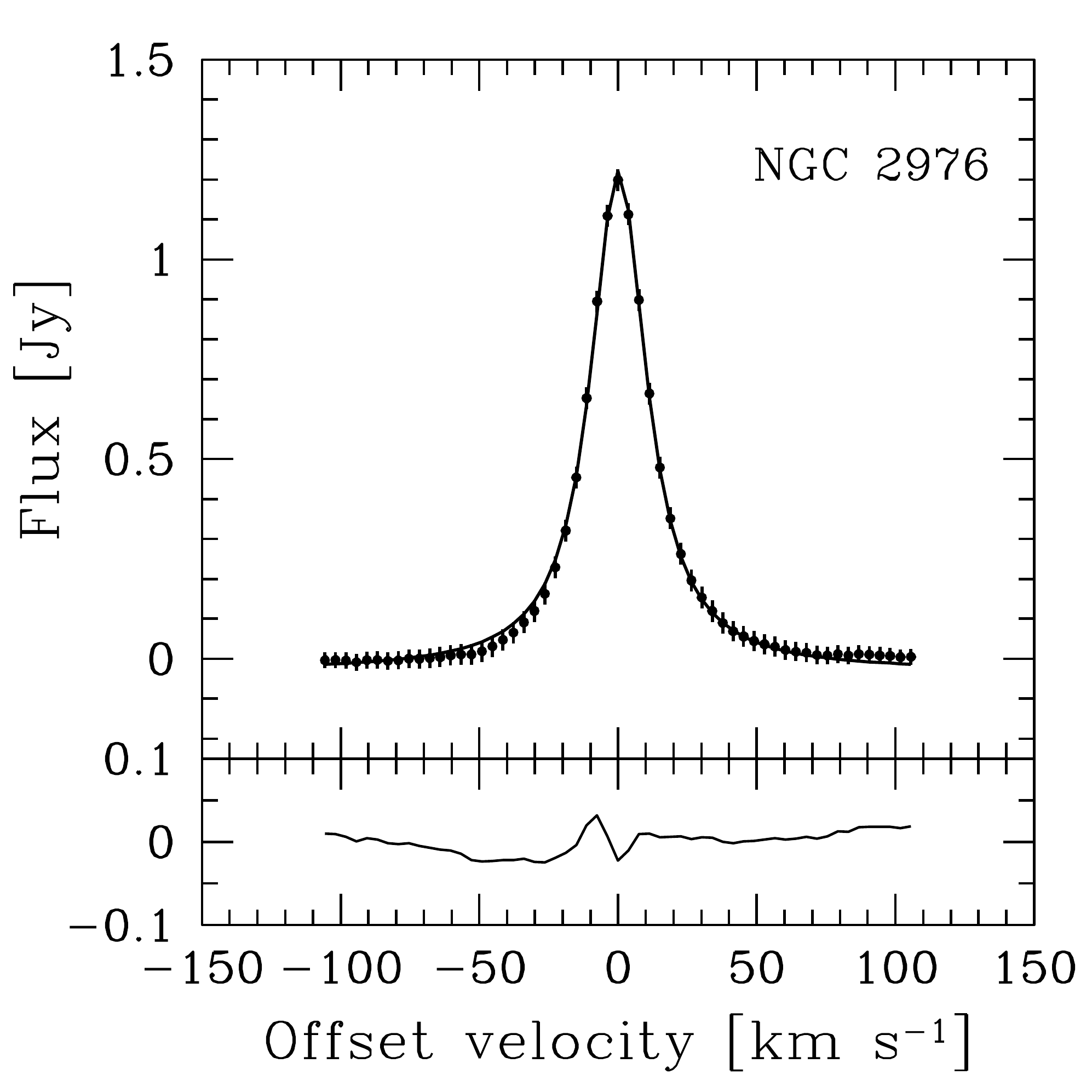}}}\\
   \rotatebox{0}{\resizebox{58mm}{!}{\includegraphics[width = 0.6in,height = 0.6in]{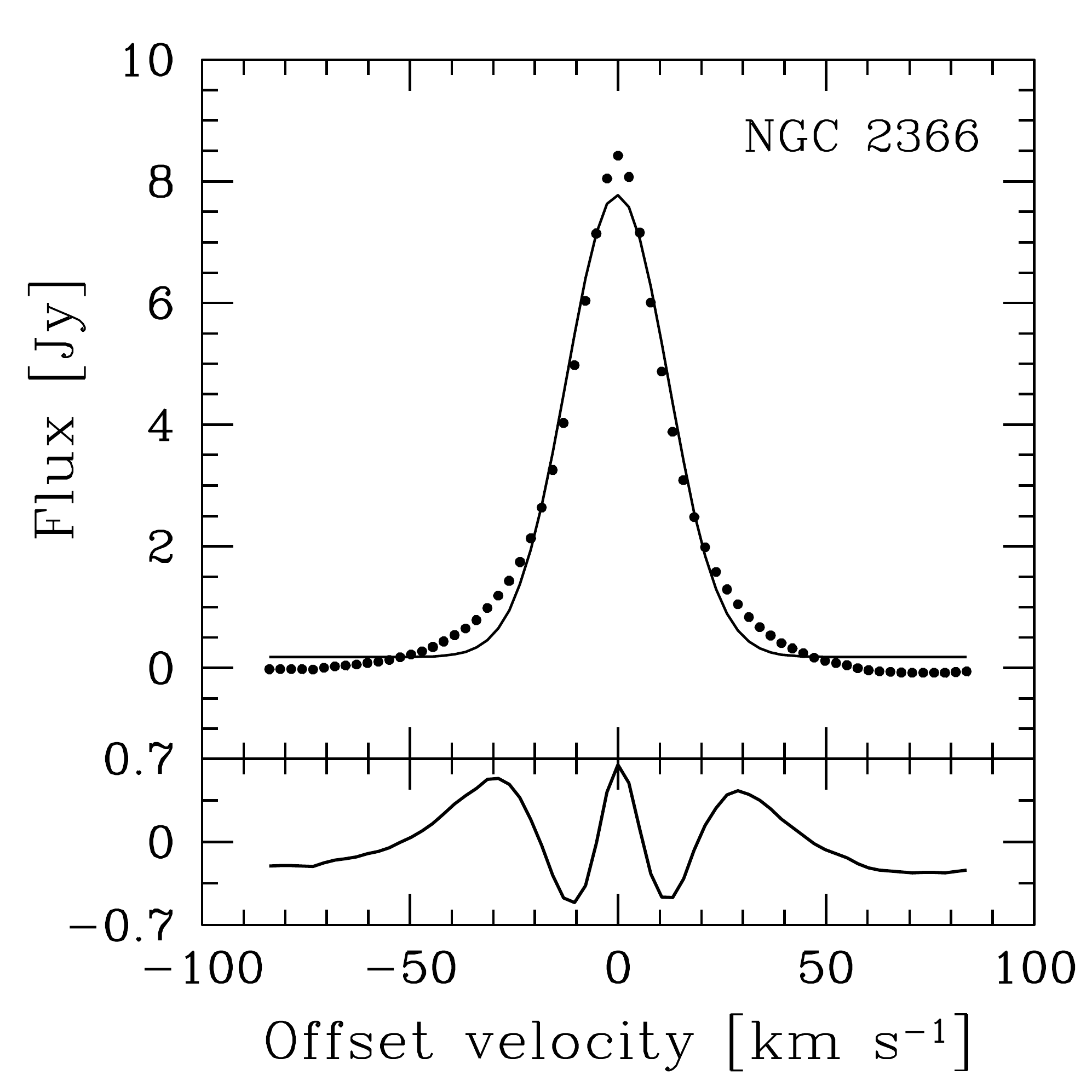}}}&
\rotatebox{0}{\resizebox{58mm}{!}{\includegraphics[width = 0.6in,height = 0.6in]{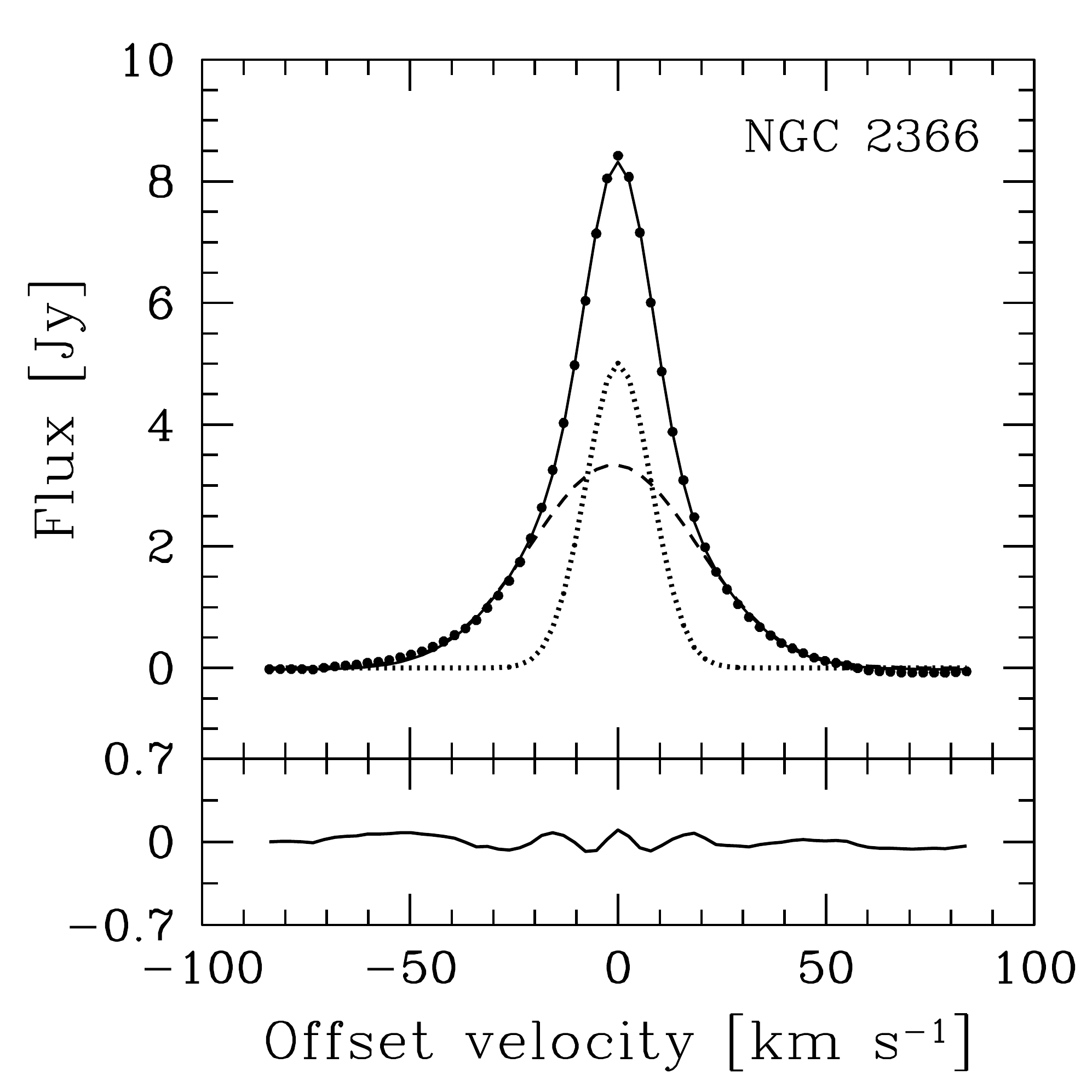}}}&
 \rotatebox{0}{\resizebox{58mm}{!}{\includegraphics[width = 0.6in,height = 0.6in]{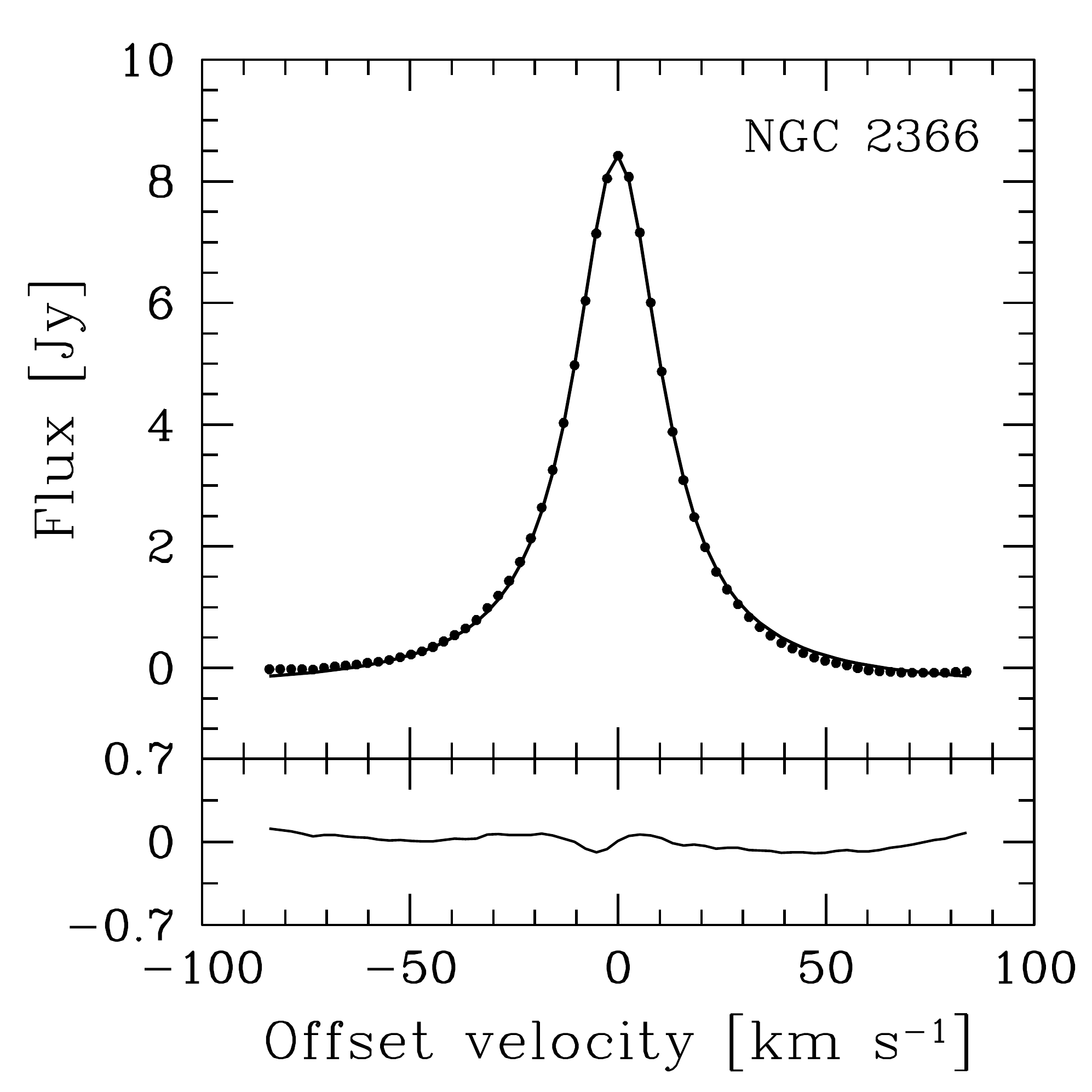}}}\\
   \rotatebox{0}{\resizebox{58mm}{!}{\includegraphics[width = 0.6in,height = 0.6in]{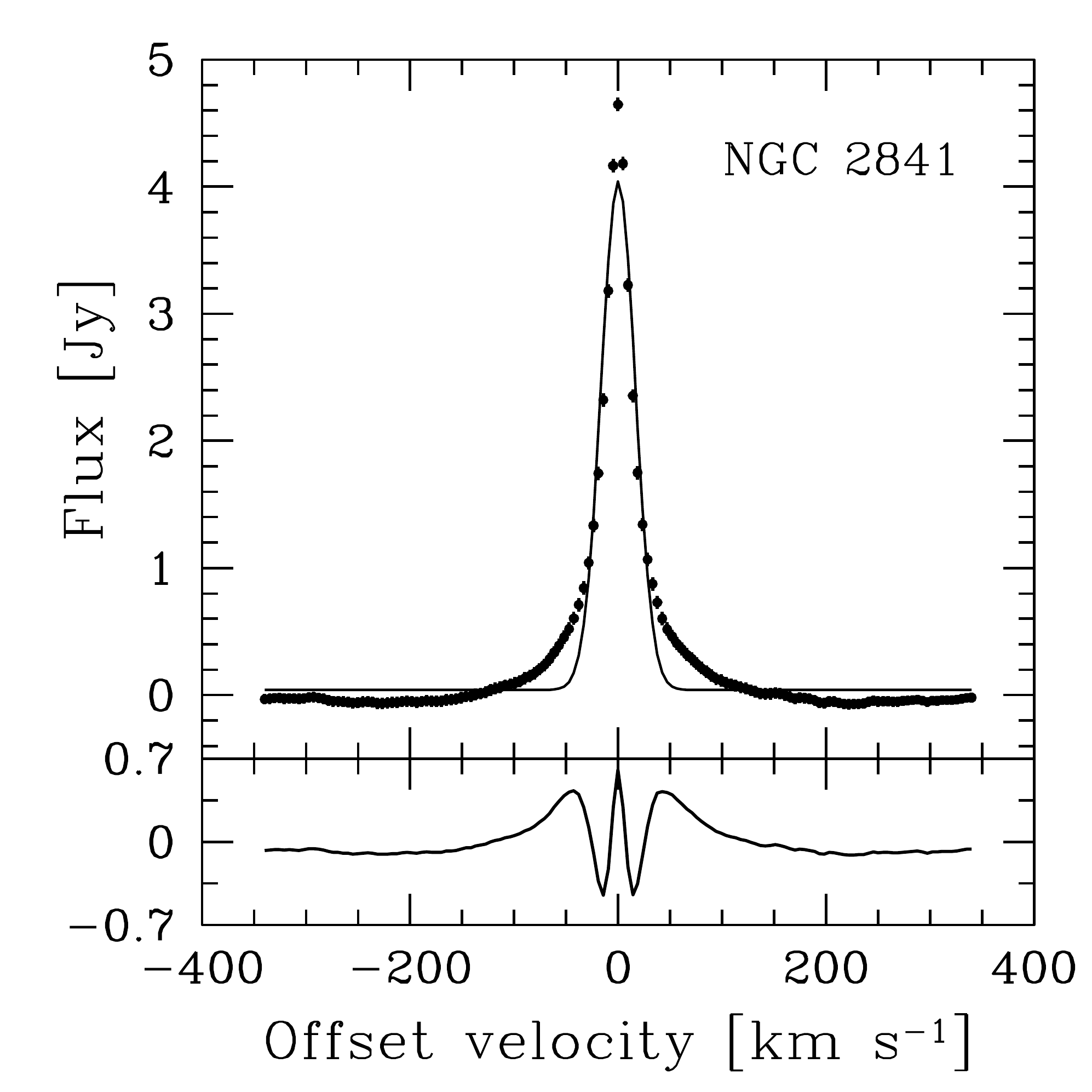}}}&
\rotatebox{0}{\resizebox{58mm}{!}{\includegraphics[width = 0.6in,height = 0.6in]{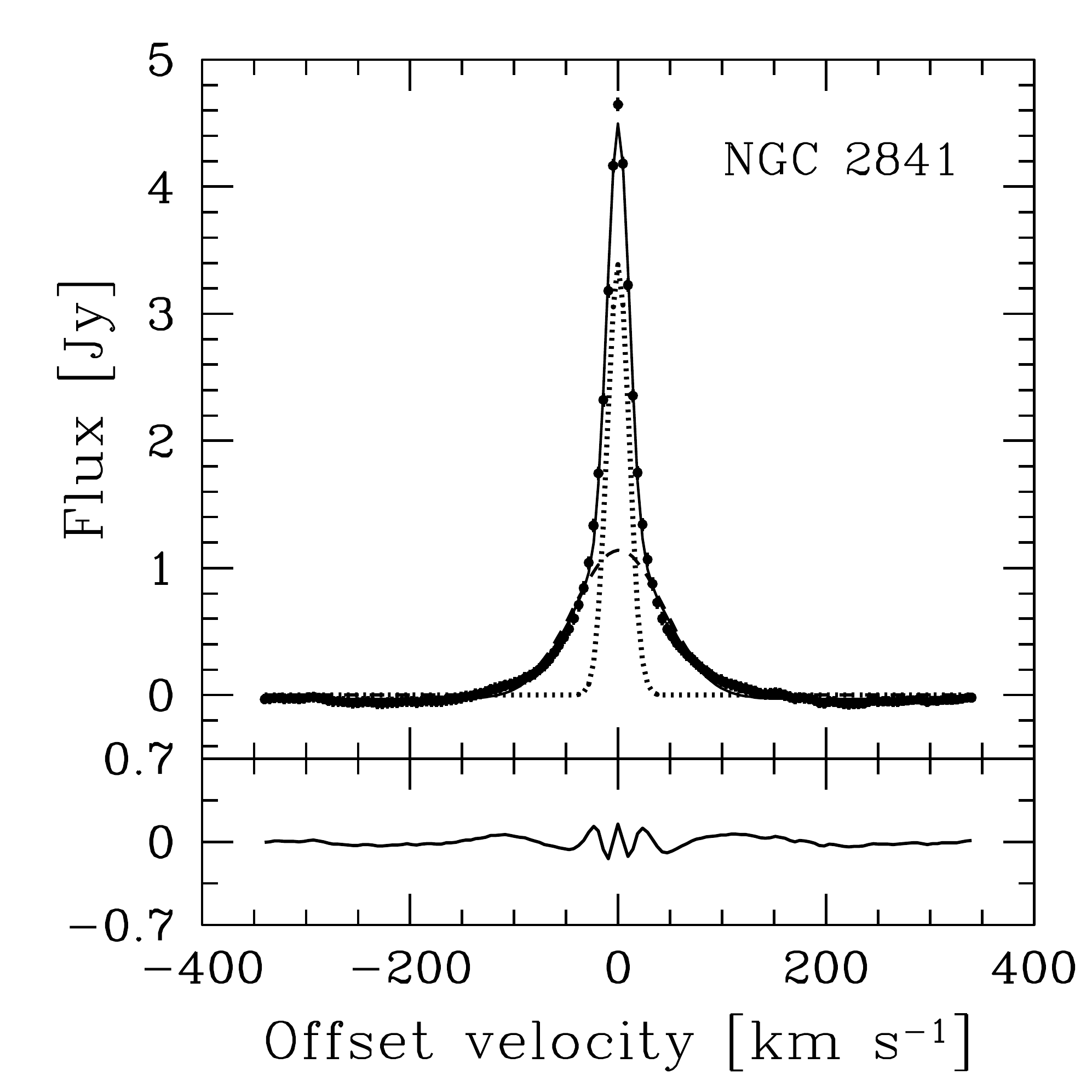}}}&
 \rotatebox{0}{\resizebox{58mm}{!}{\includegraphics[width = 0.6in,height = 0.6in]{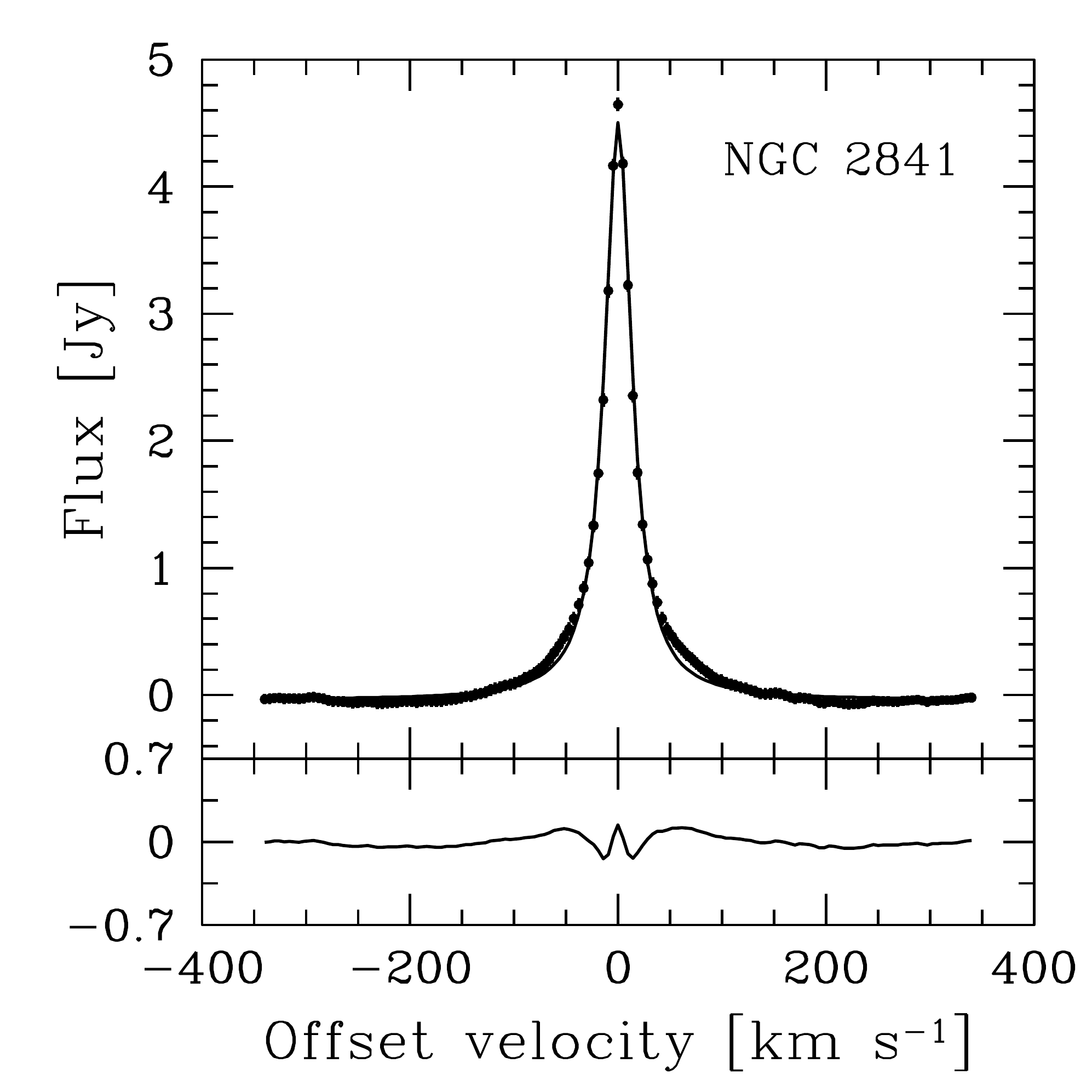}}}
 \\\\ & & \hspace*{-12.12cm}\textbf{Figure \ref{fig:app1}}. \scriptsize{(continued).}
\end{tabular}

\end{figure*}

\begin{figure*}
    \begin{tabular}{l l l}
   \rotatebox{0}{\resizebox{58mm}{!}{\includegraphics[width = 0.6in,height = 0.6in]{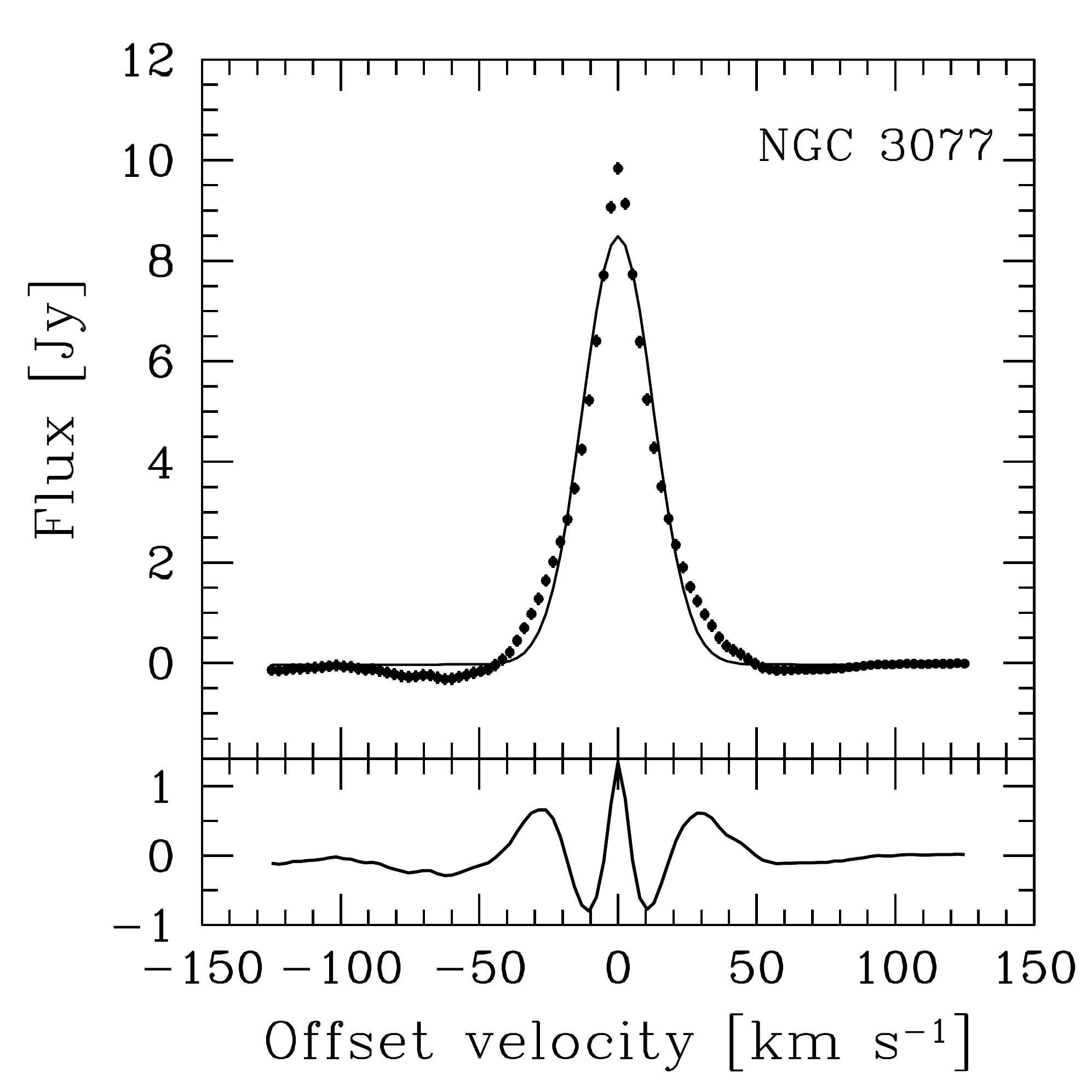}}}&
\rotatebox{0}{\resizebox{58mm}{!}{\includegraphics[width = 0.6in,height = 0.6in]{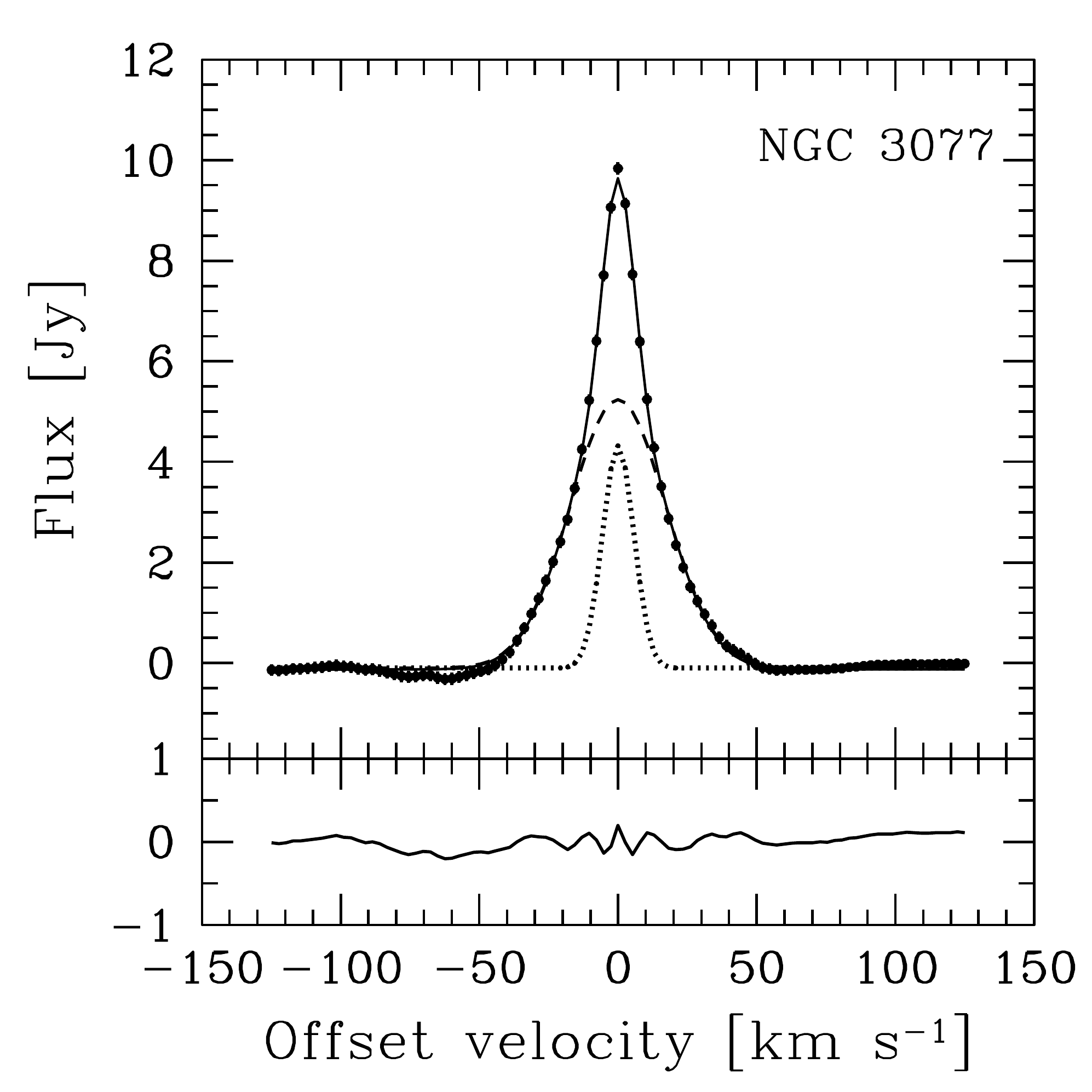}}}&
 \rotatebox{0}{\resizebox{58mm}{!}{\includegraphics[width = 0.6in,height = 0.6in]{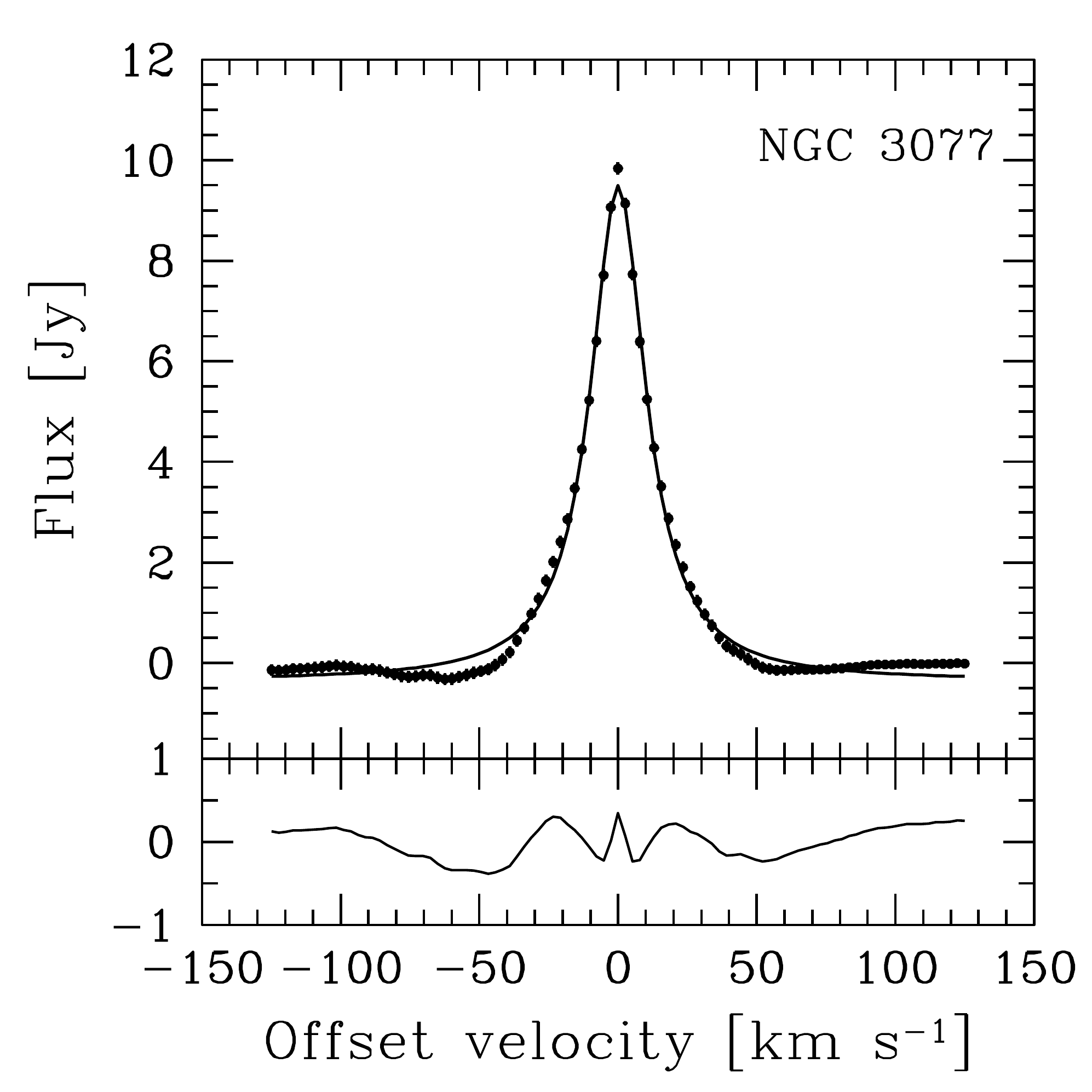}}}\\
   \rotatebox{0}{\resizebox{58mm}{!}{\includegraphics[width = 0.6in,height = 0.6in]{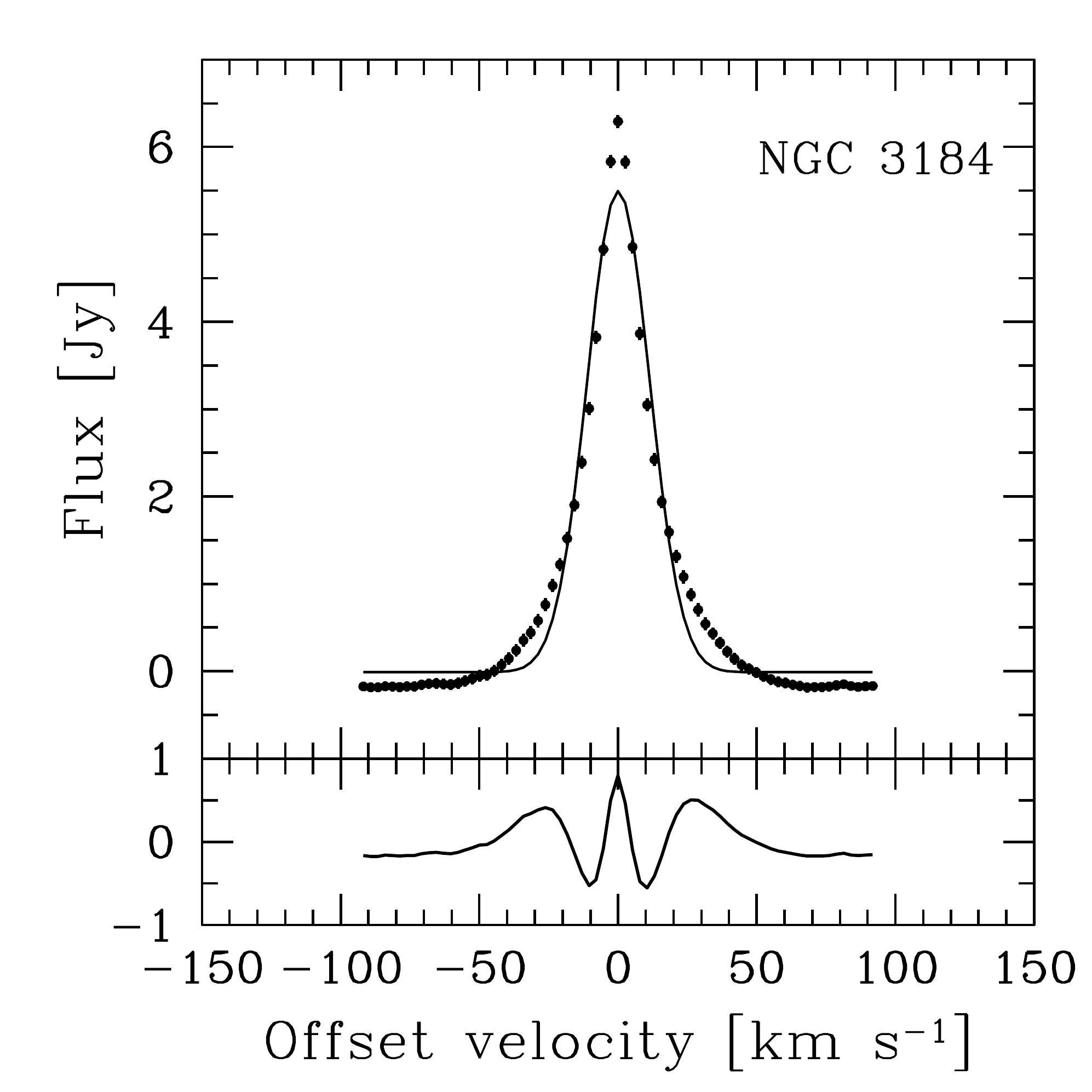}}}&
\rotatebox{0}{\resizebox{58mm}{!}{\includegraphics[width = 0.6in,height = 0.6in]{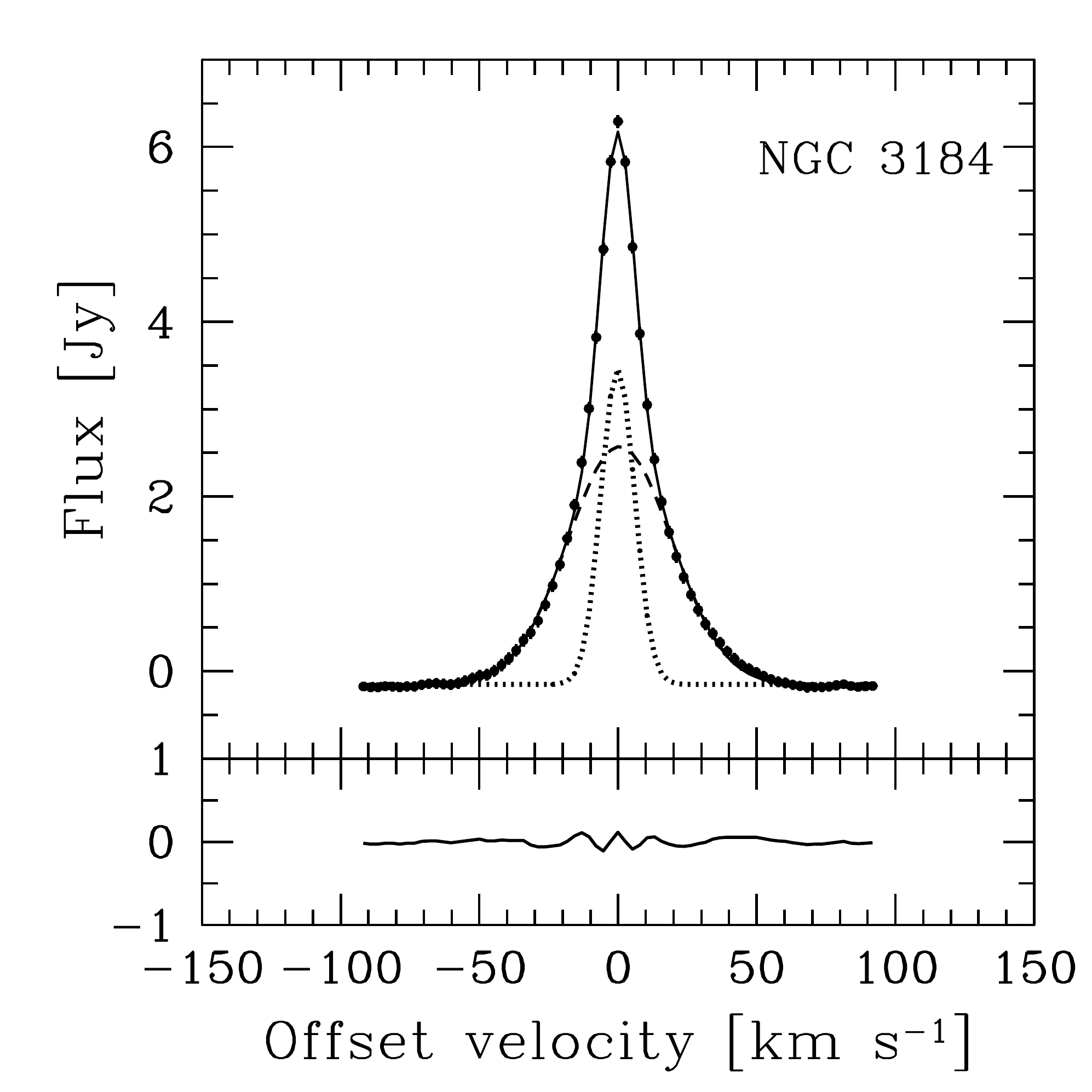}}}&
 \rotatebox{0}{\resizebox{58mm}{!}{\includegraphics[width = 0.6in,height = 0.6in]{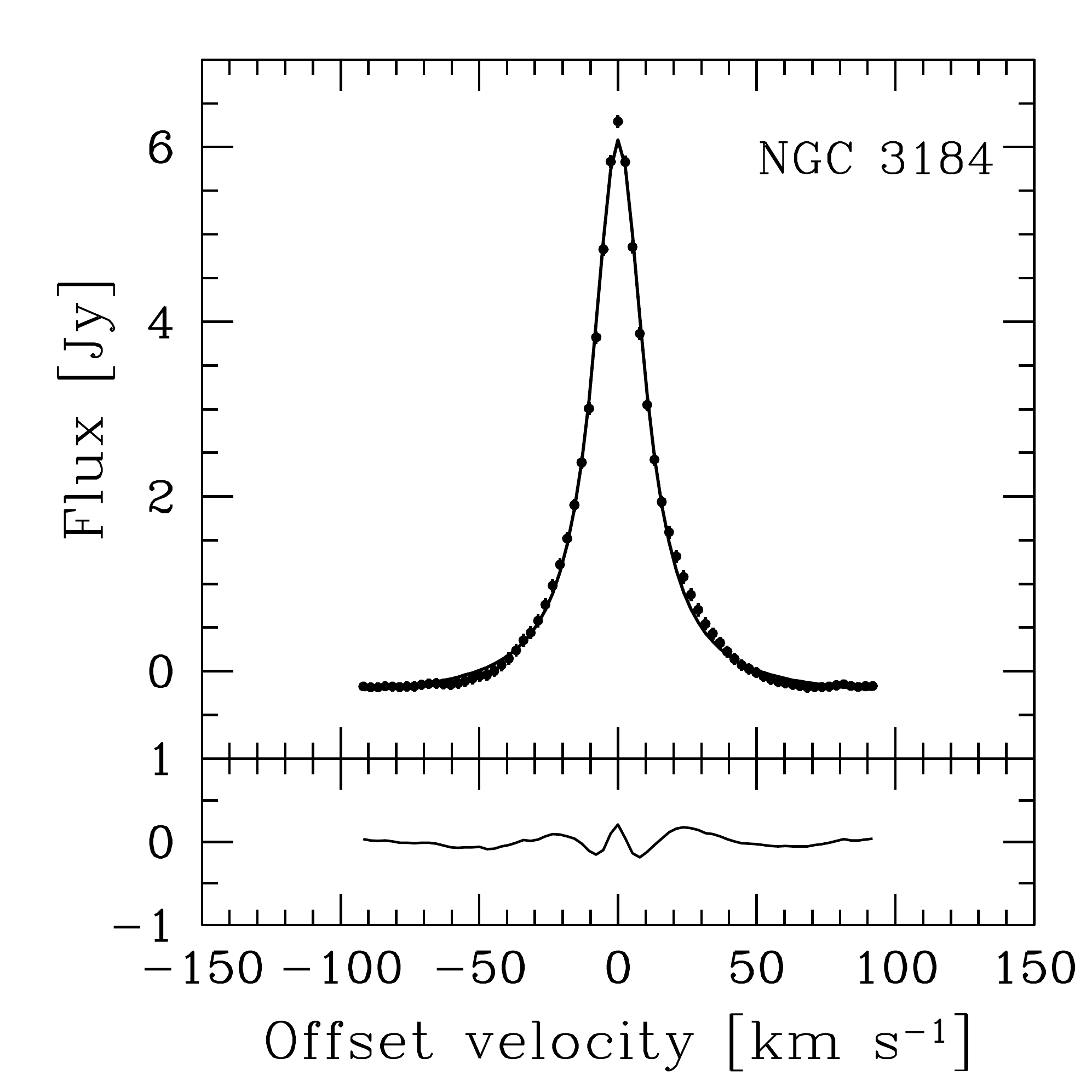}}}\\
    \rotatebox{0}{\resizebox{58mm}{!}{\includegraphics[width = 0.6in,height = 0.6in]{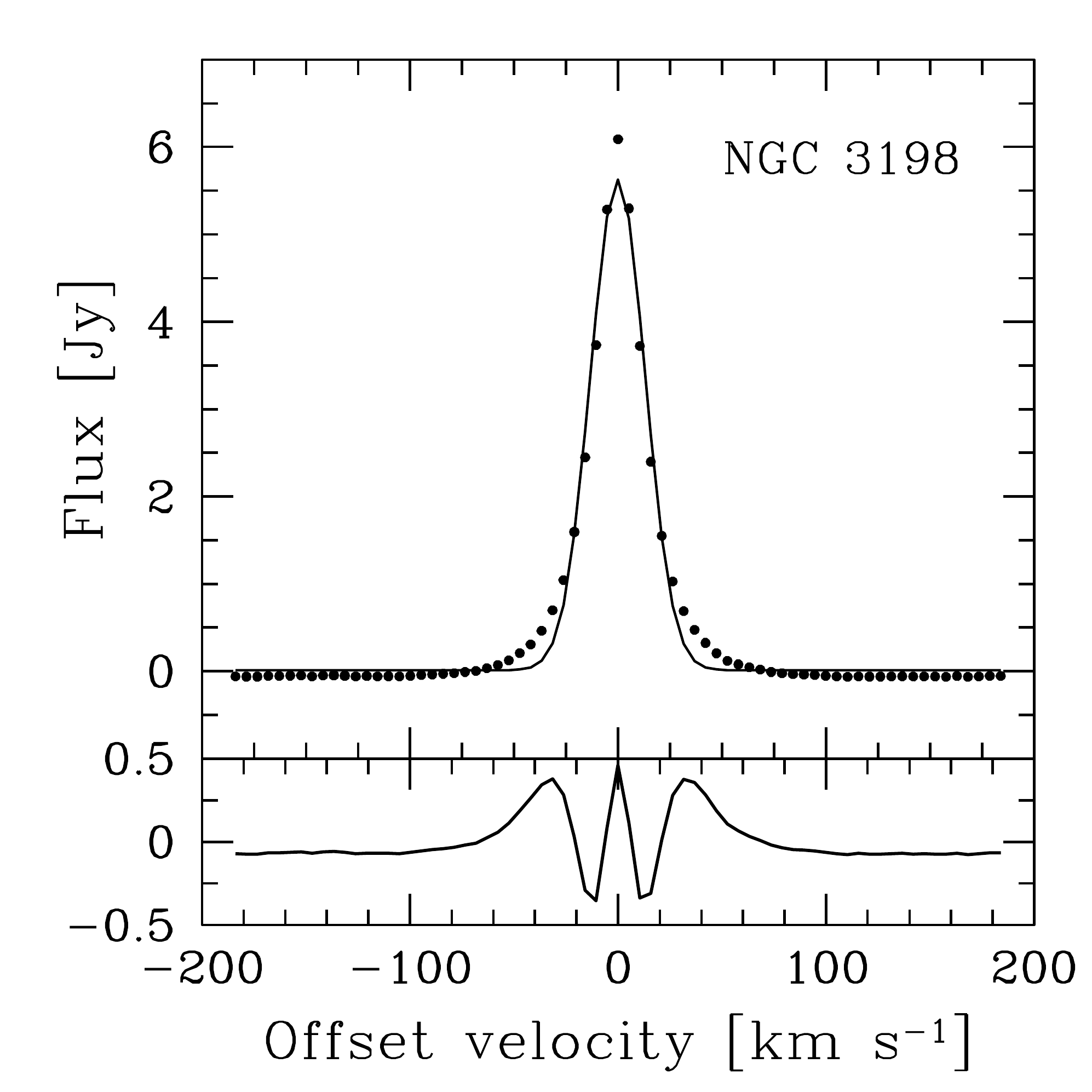}}}&
\rotatebox{0}{\resizebox{58mm}{!}{\includegraphics[width = 0.6in,height = 0.6in]{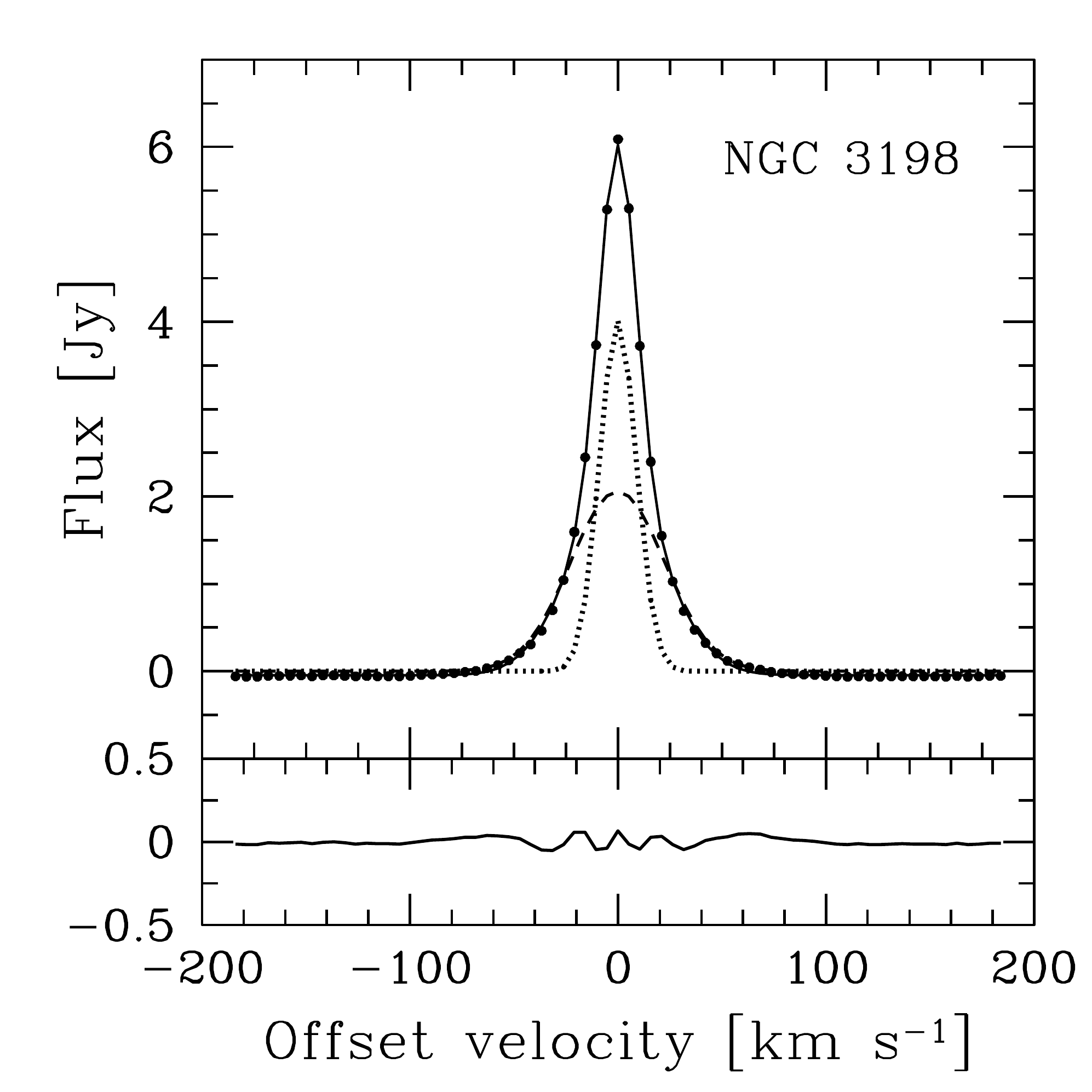}}}&
 \rotatebox{0}{\resizebox{58mm}{!}{\includegraphics[width = 0.6in,height = 0.6in]{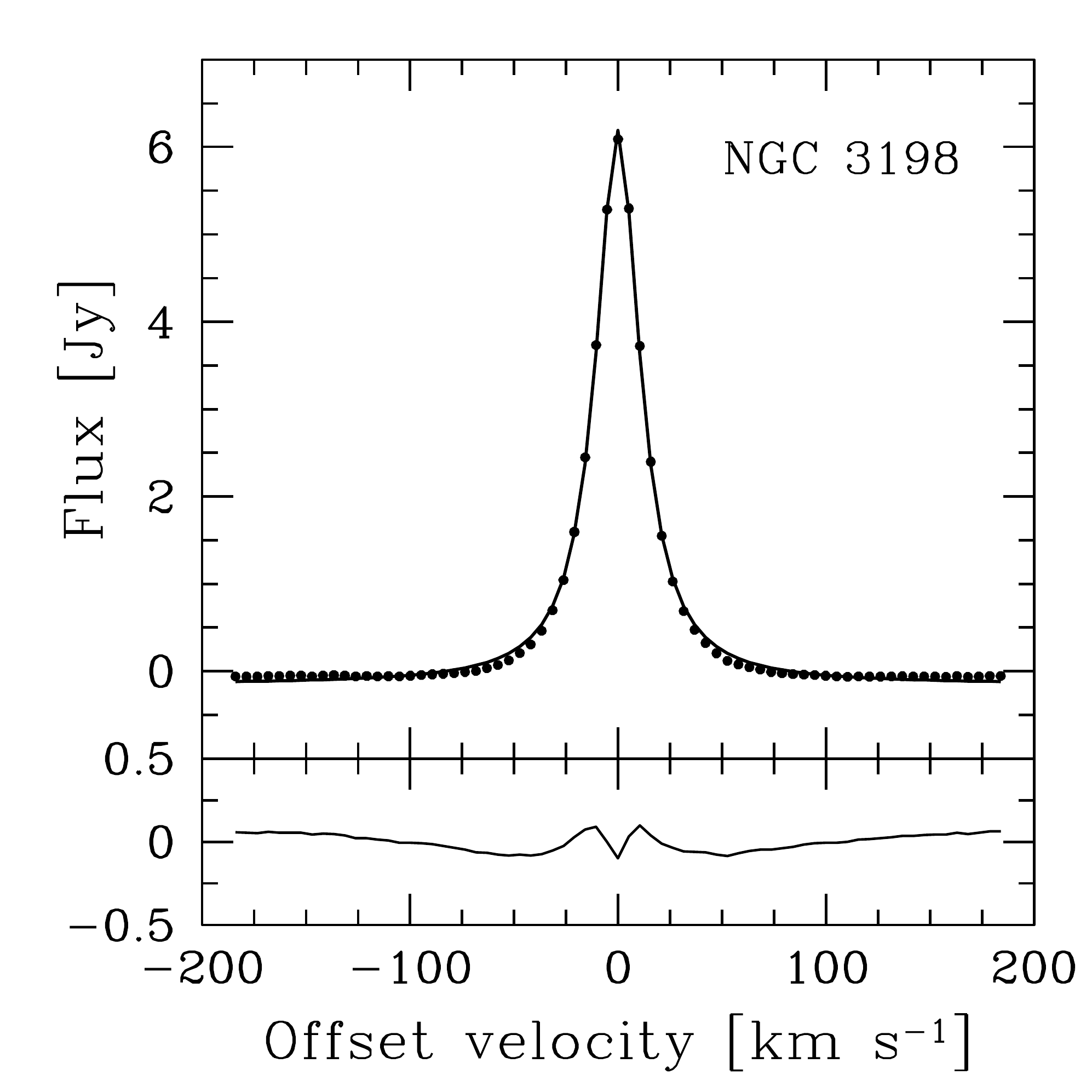}}}\\
   \rotatebox{0}{\resizebox{58mm}{!}{\includegraphics[width = 0.6in,height = 0.6in]{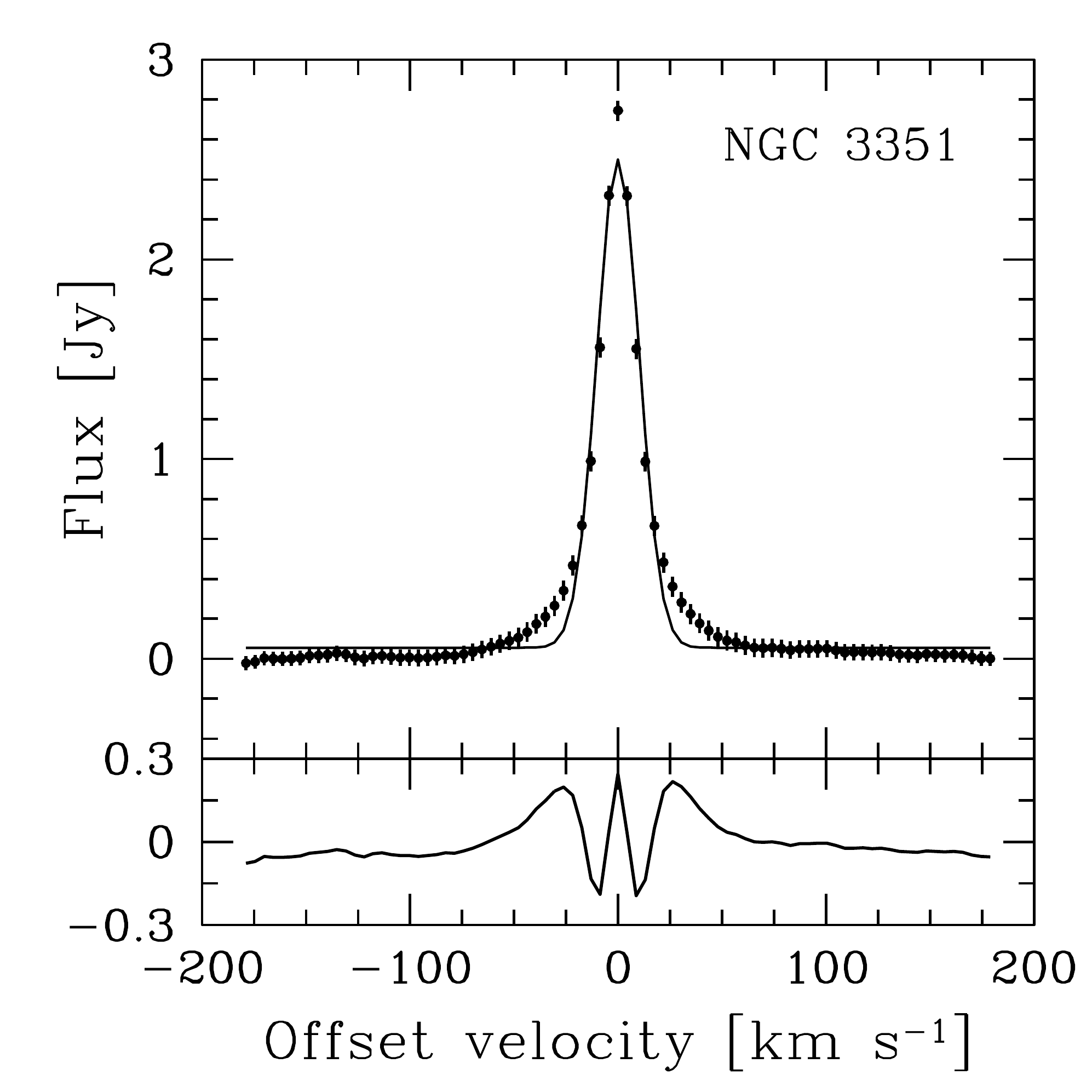}}}&
\rotatebox{0}{\resizebox{58mm}{!}{\includegraphics[width = 0.6in,height = 0.6in]{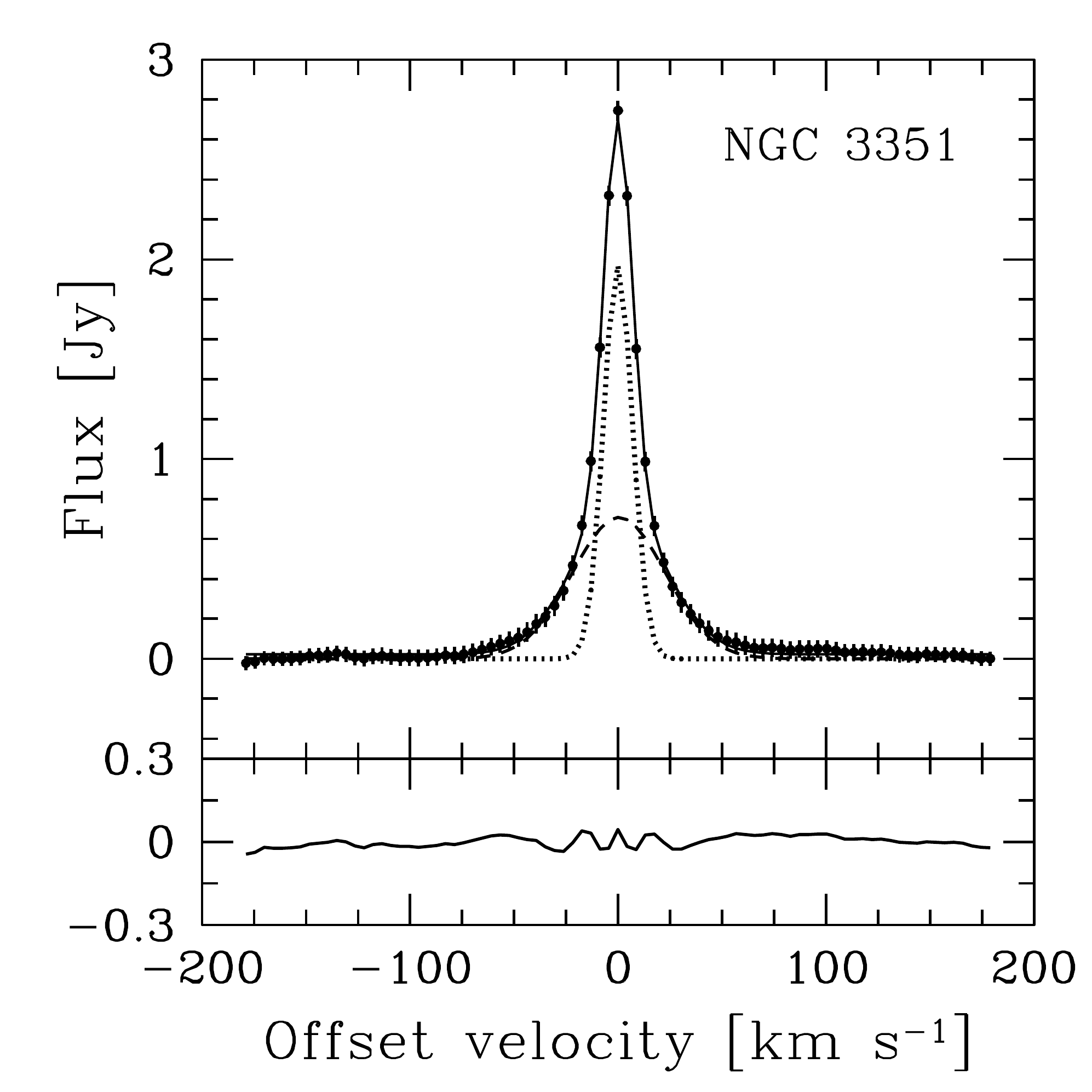}}}&
 \rotatebox{0}{\resizebox{58mm}{!}{\includegraphics[width = 0.6in,height = 0.6in]{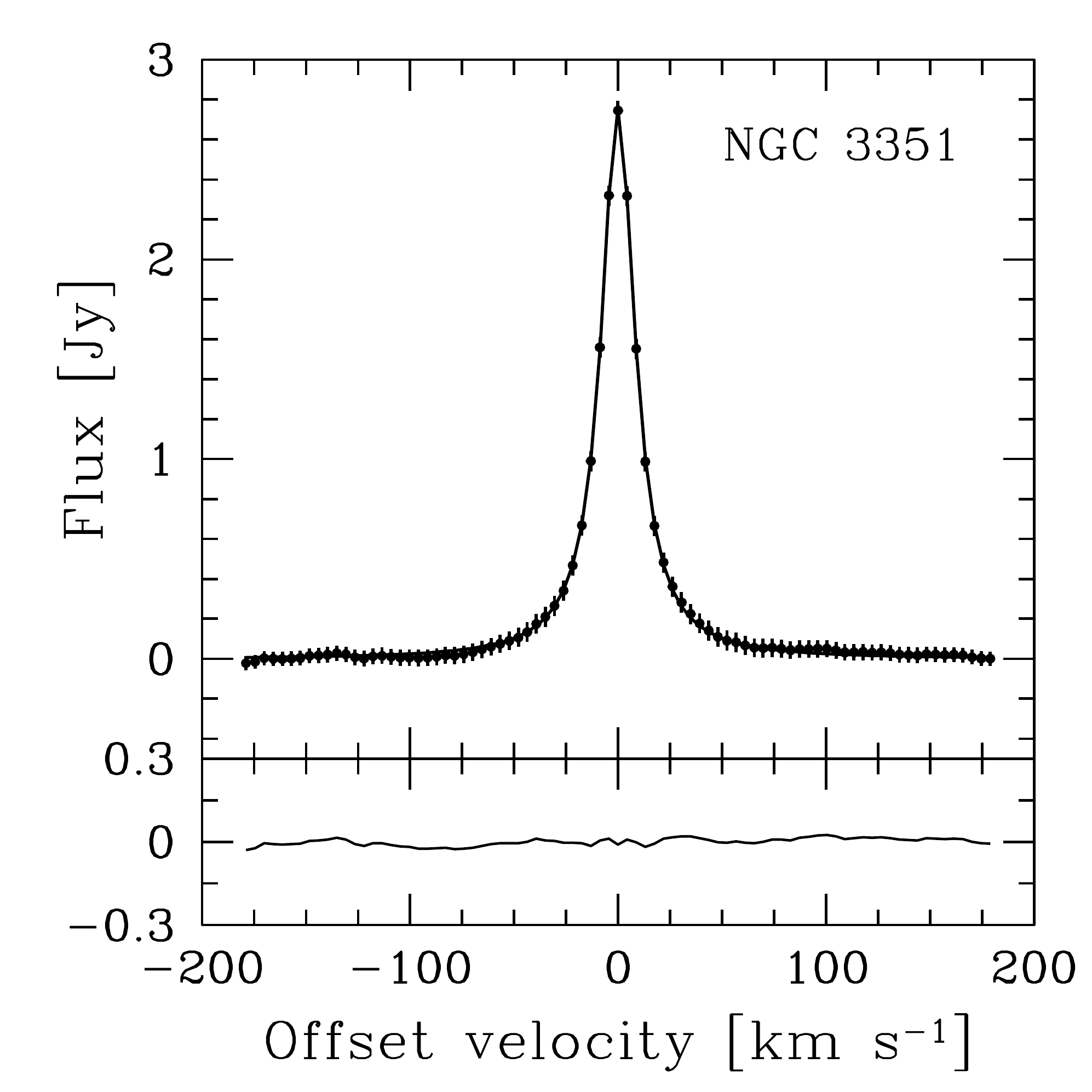}}}
\\\\ & & \hspace*{-12.12cm}\textbf{Figure \ref{fig:app1}}. \scriptsize{(continued).}
\end{tabular}

\end{figure*}

\begin{figure*}
    \begin{tabular}{l l l}
    \rotatebox{0}{\resizebox{58mm}{!}{\includegraphics[width = 0.6in,height = 0.6in]{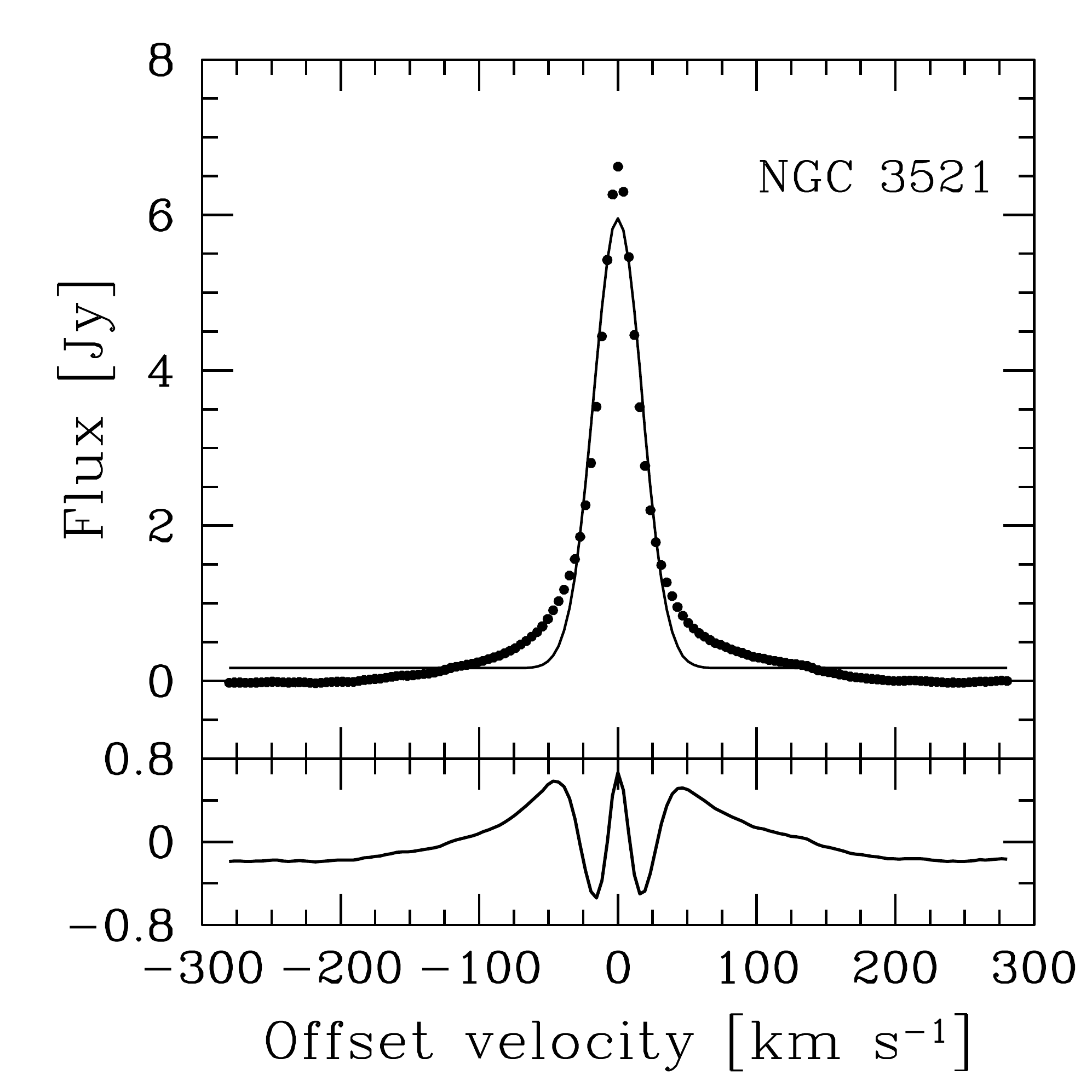}}}&
\rotatebox{0}{\resizebox{58mm}{!}{\includegraphics[width = 0.6in,height = 0.6in]{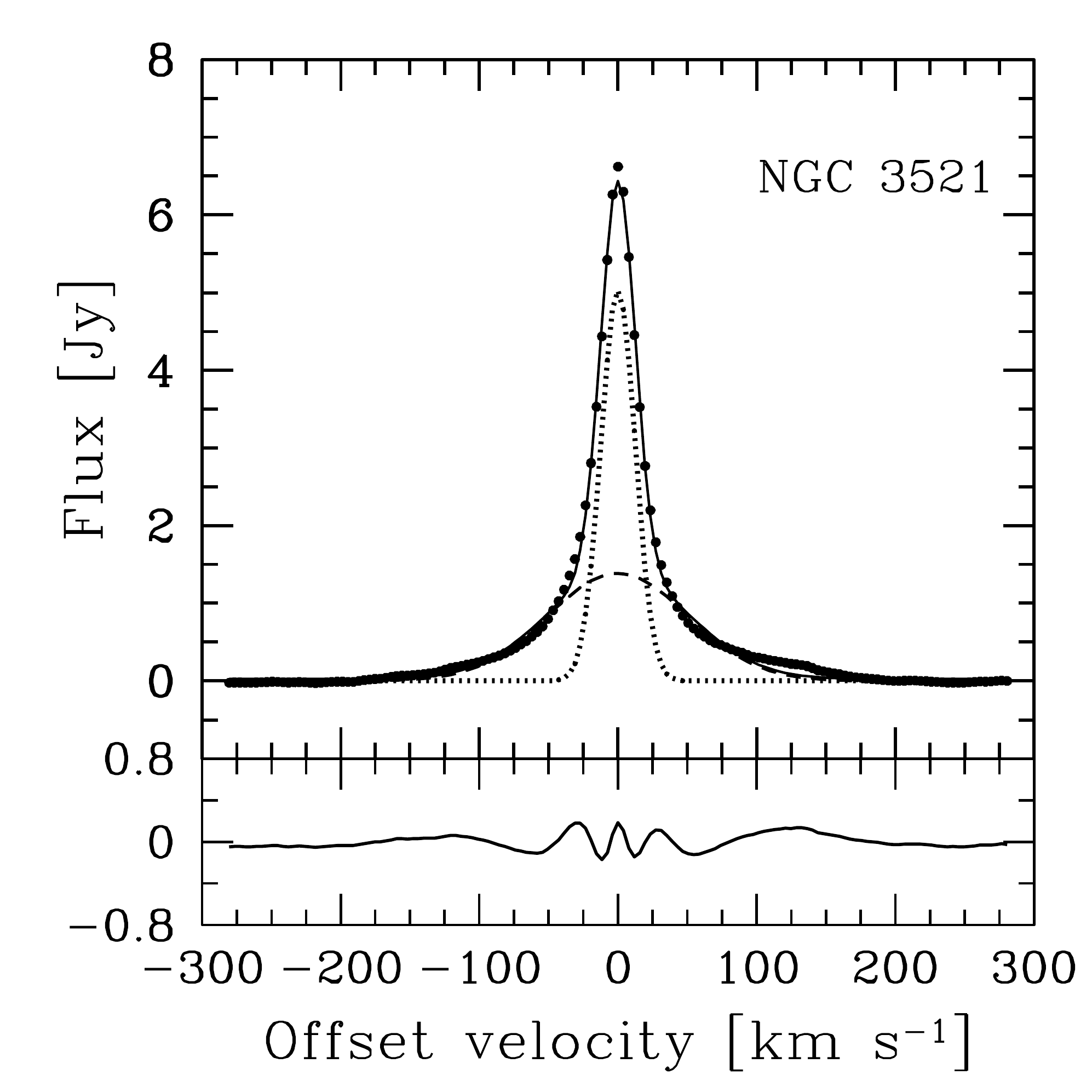}}}&
 \rotatebox{0}{\resizebox{58mm}{!}{\includegraphics[width = 0.6in,height = 0.6in]{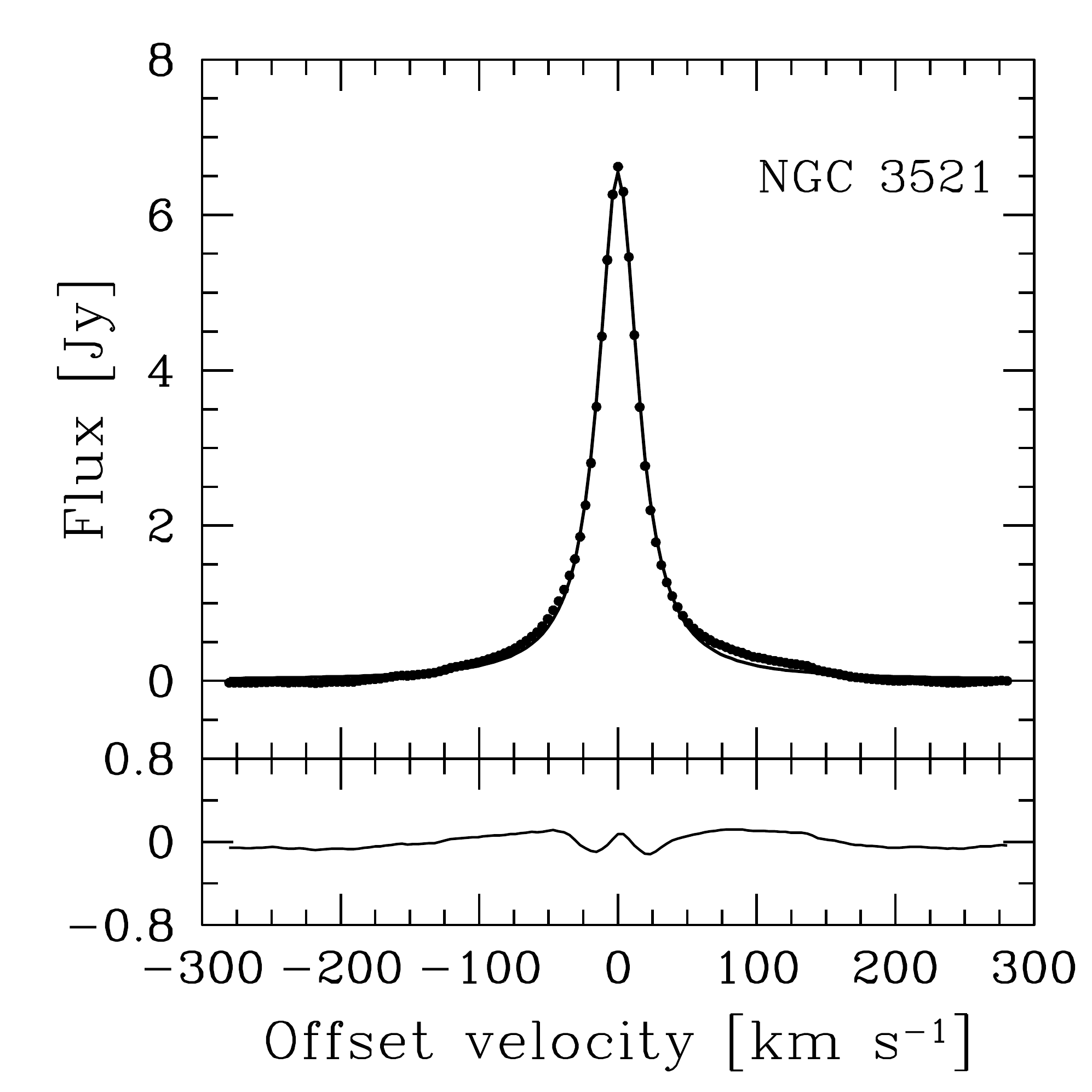}}}\\
   \rotatebox{0}{\resizebox{58mm}{!}{\includegraphics[width = 0.6in,height = 0.6in]{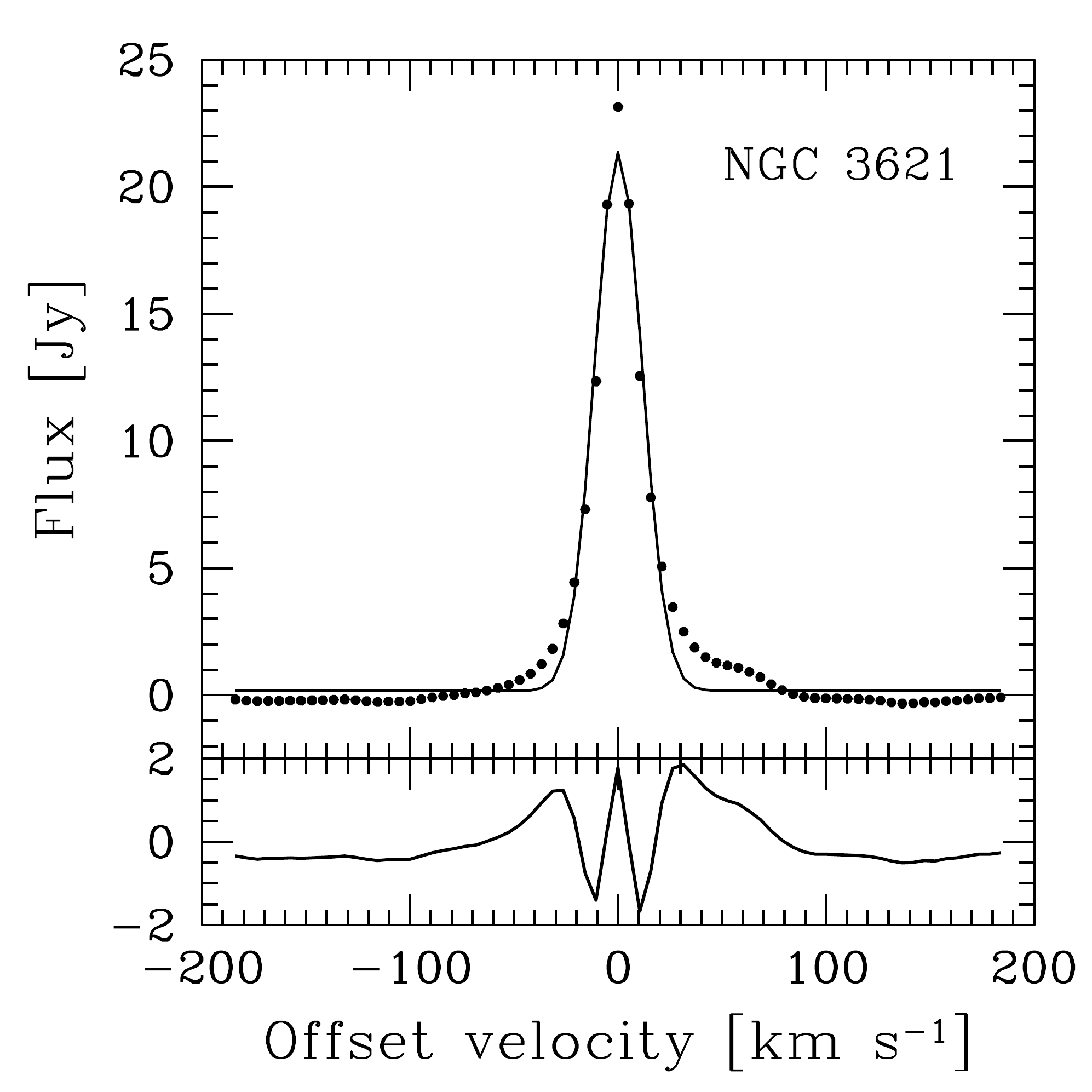}}}&
\rotatebox{0}{\resizebox{58mm}{!}{\includegraphics[width = 0.6in,height = 0.6in]{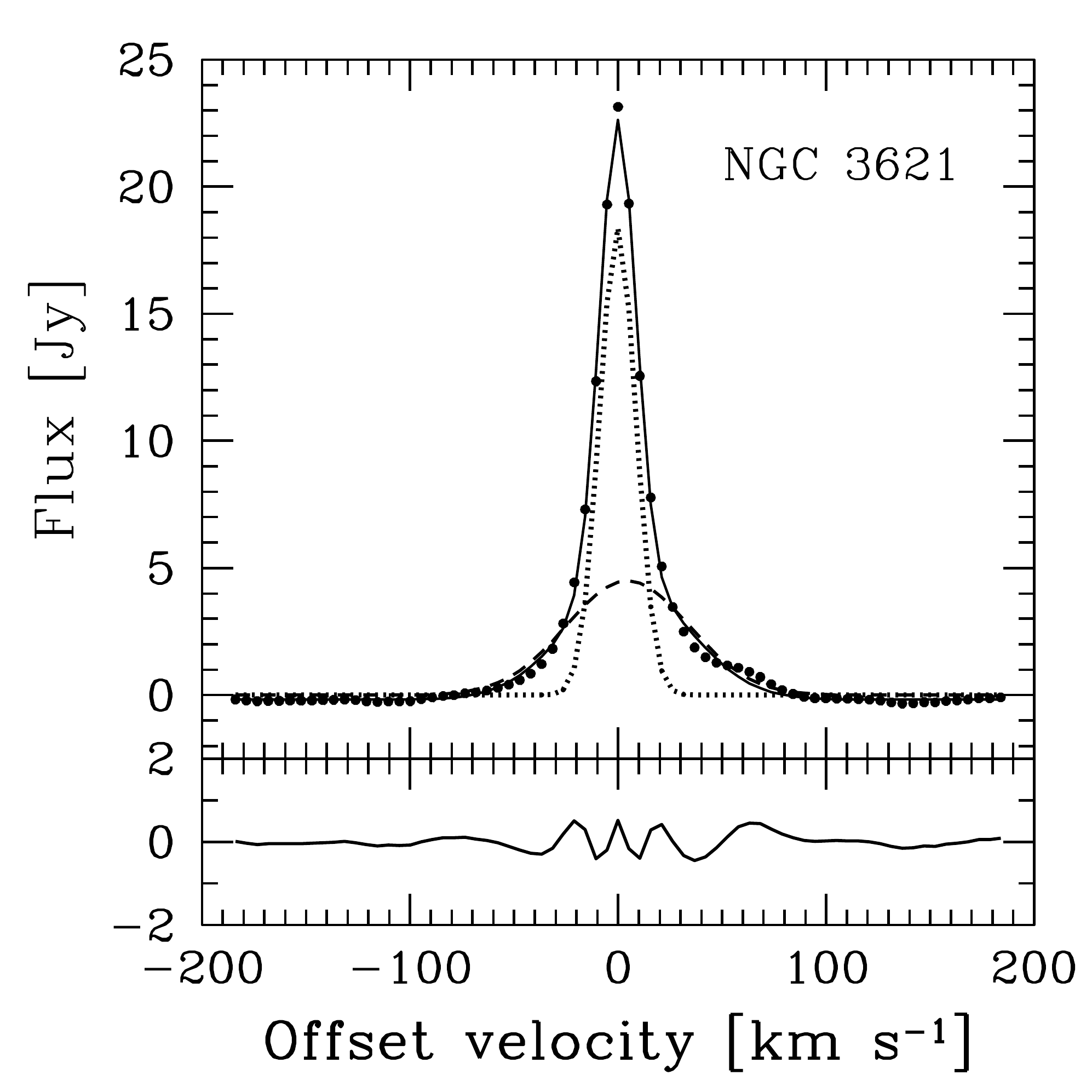}}}&
 \rotatebox{0}{\resizebox{58mm}{!}{\includegraphics[width = 0.6in,height = 0.6in]{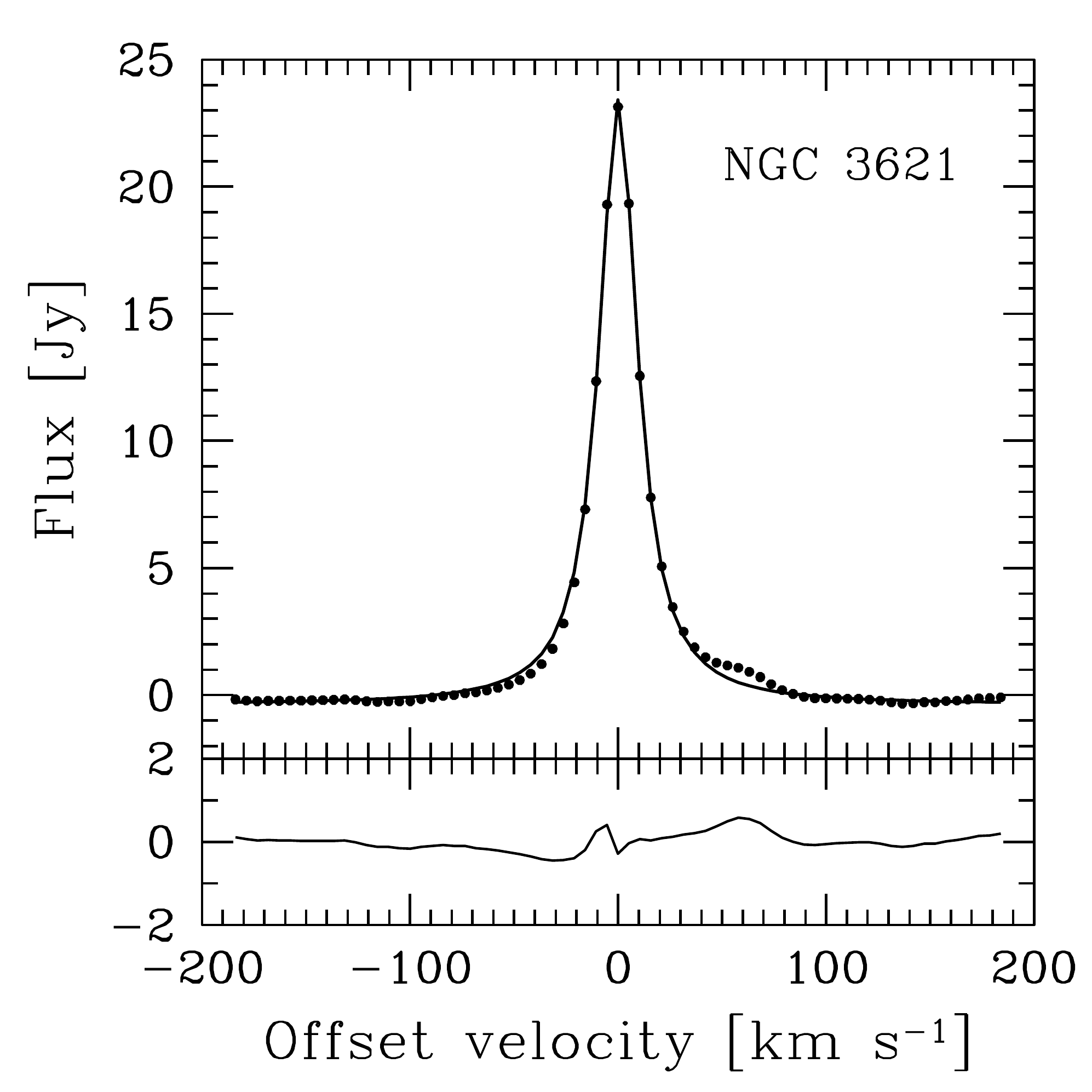}}}\\
    \rotatebox{0}{\resizebox{58mm}{!}{\includegraphics[width = 0.6in,height = 0.6in]{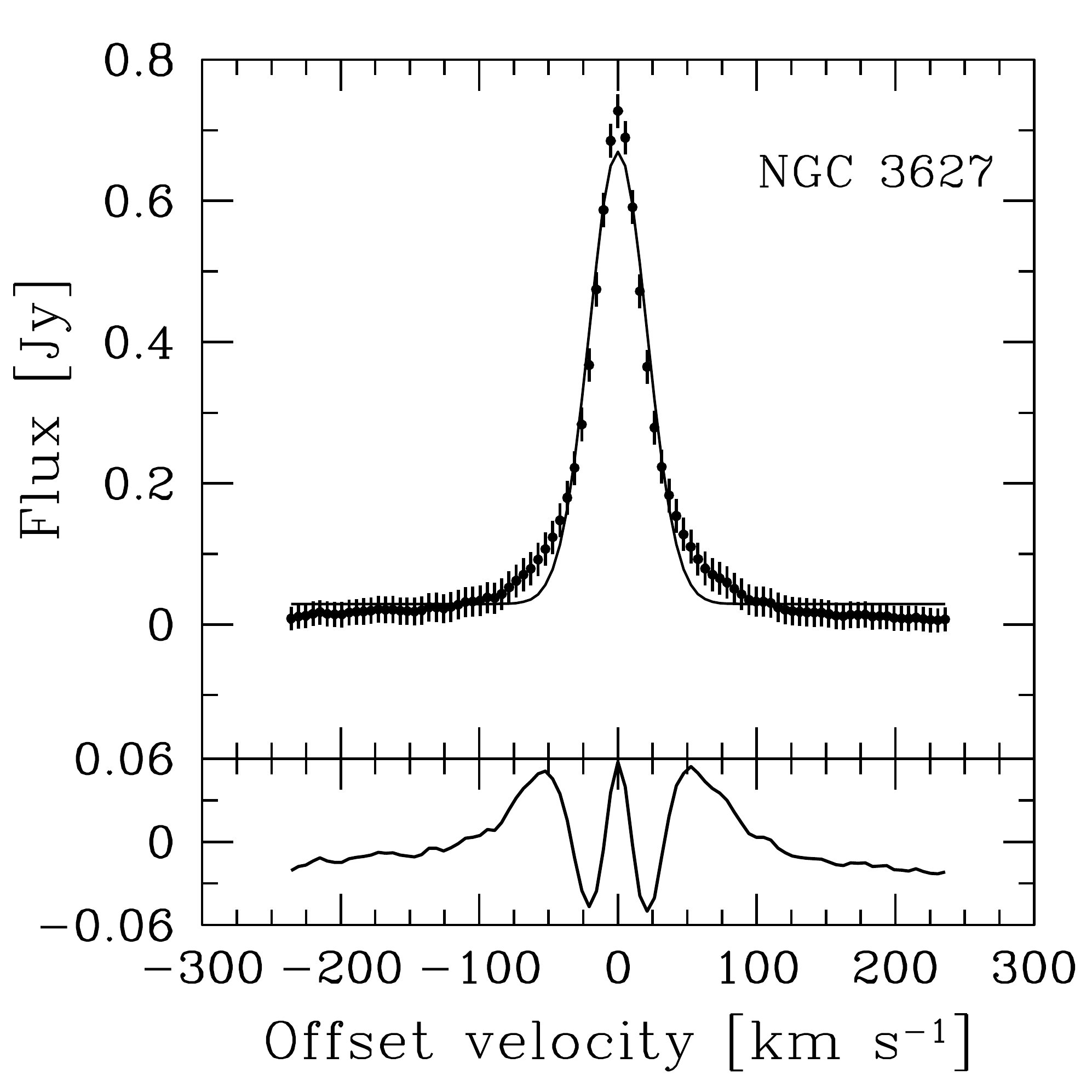}}}&
\rotatebox{0}{\resizebox{58mm}{!}{\includegraphics[width = 0.6in,height = 0.6in]{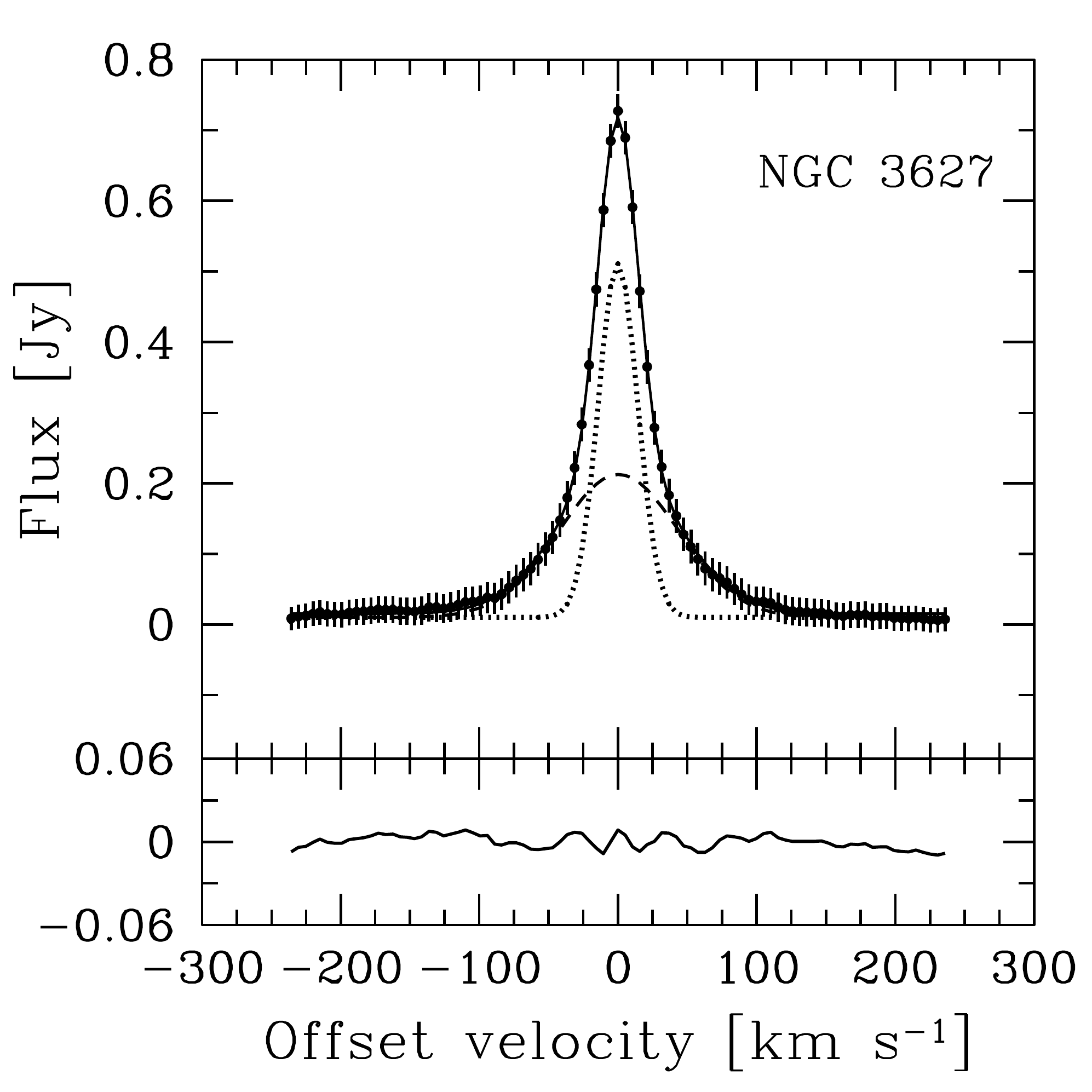}}}&
 \rotatebox{0}{\resizebox{58mm}{!}{\includegraphics[width = 0.6in,height = 0.6in]{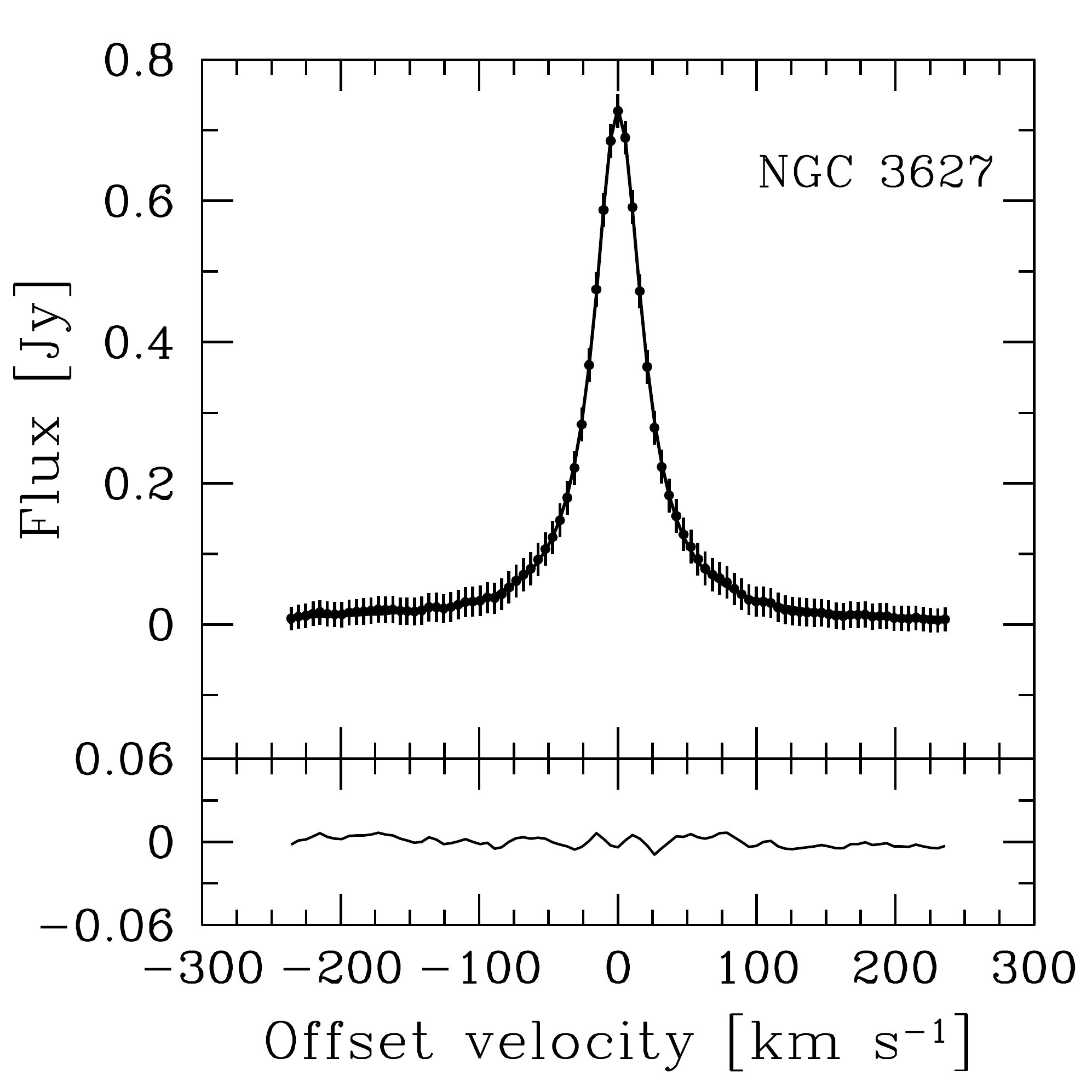}}}\\
   \rotatebox{0}{\resizebox{58mm}{!}{\includegraphics[width = 0.6in,height = 0.6in]{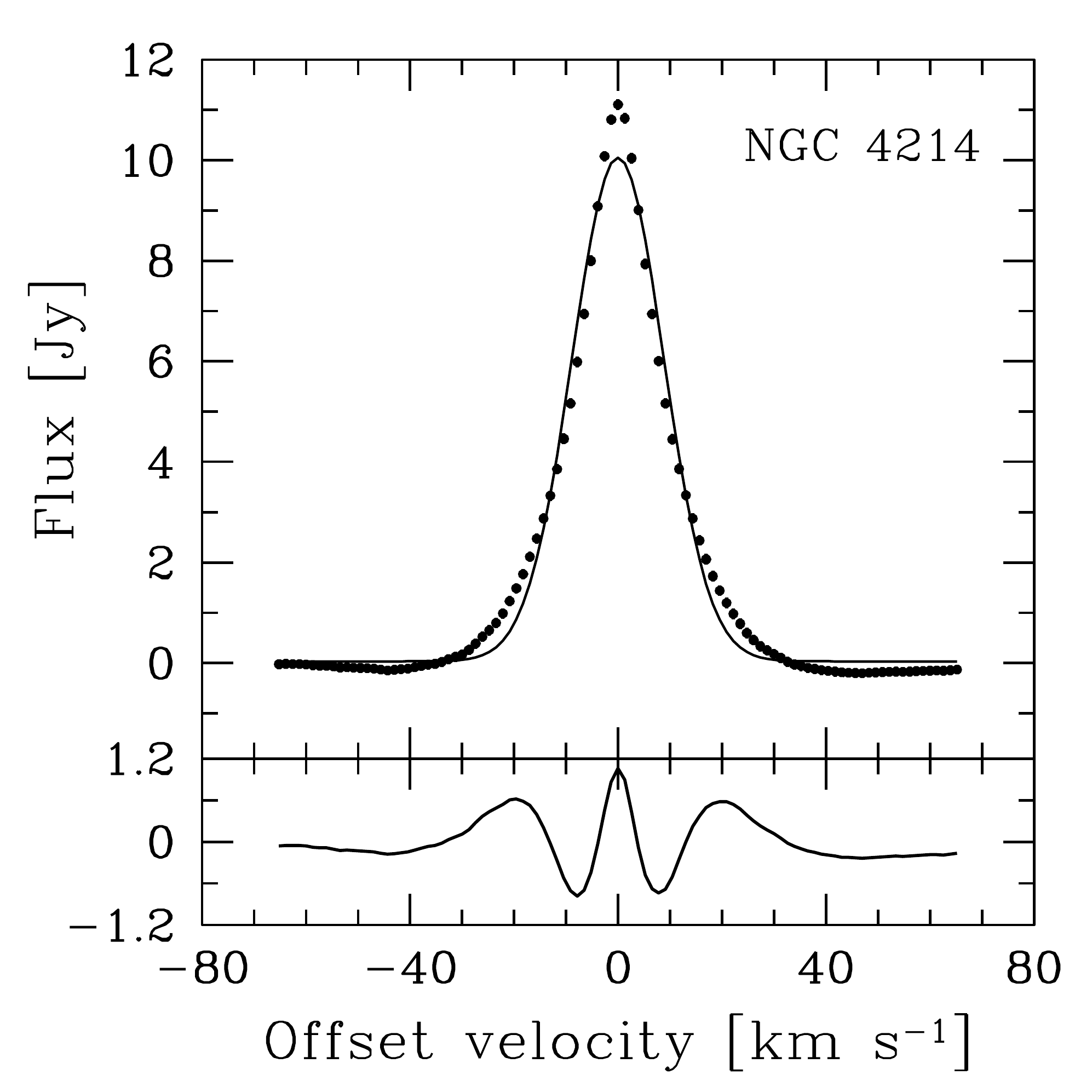}}}&
\rotatebox{0}{\resizebox{58mm}{!}{\includegraphics[width = 0.6in,height = 0.6in]{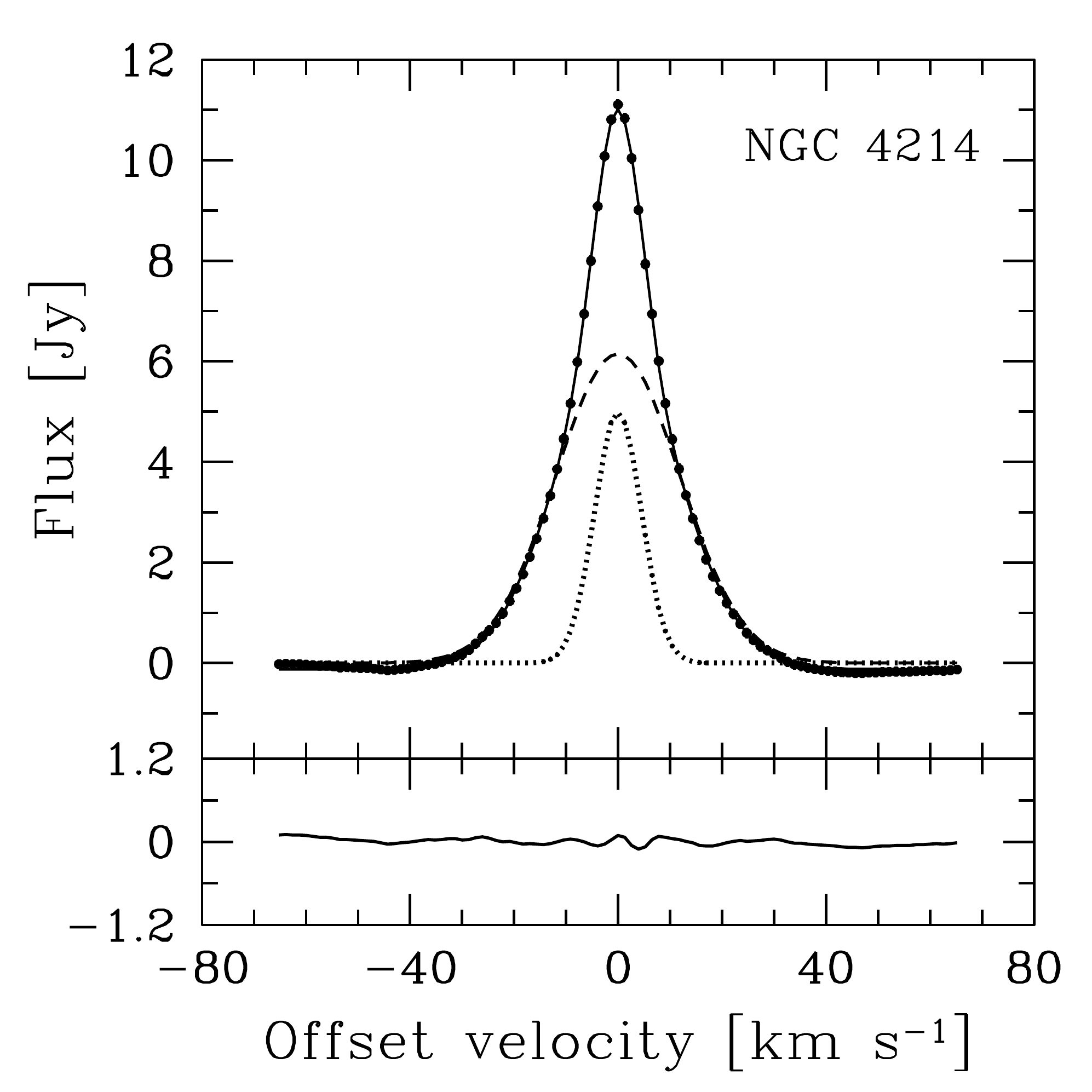}}}&
 \rotatebox{0}{\resizebox{58mm}{!}{\includegraphics[width = 0.6in,height = 0.6in]{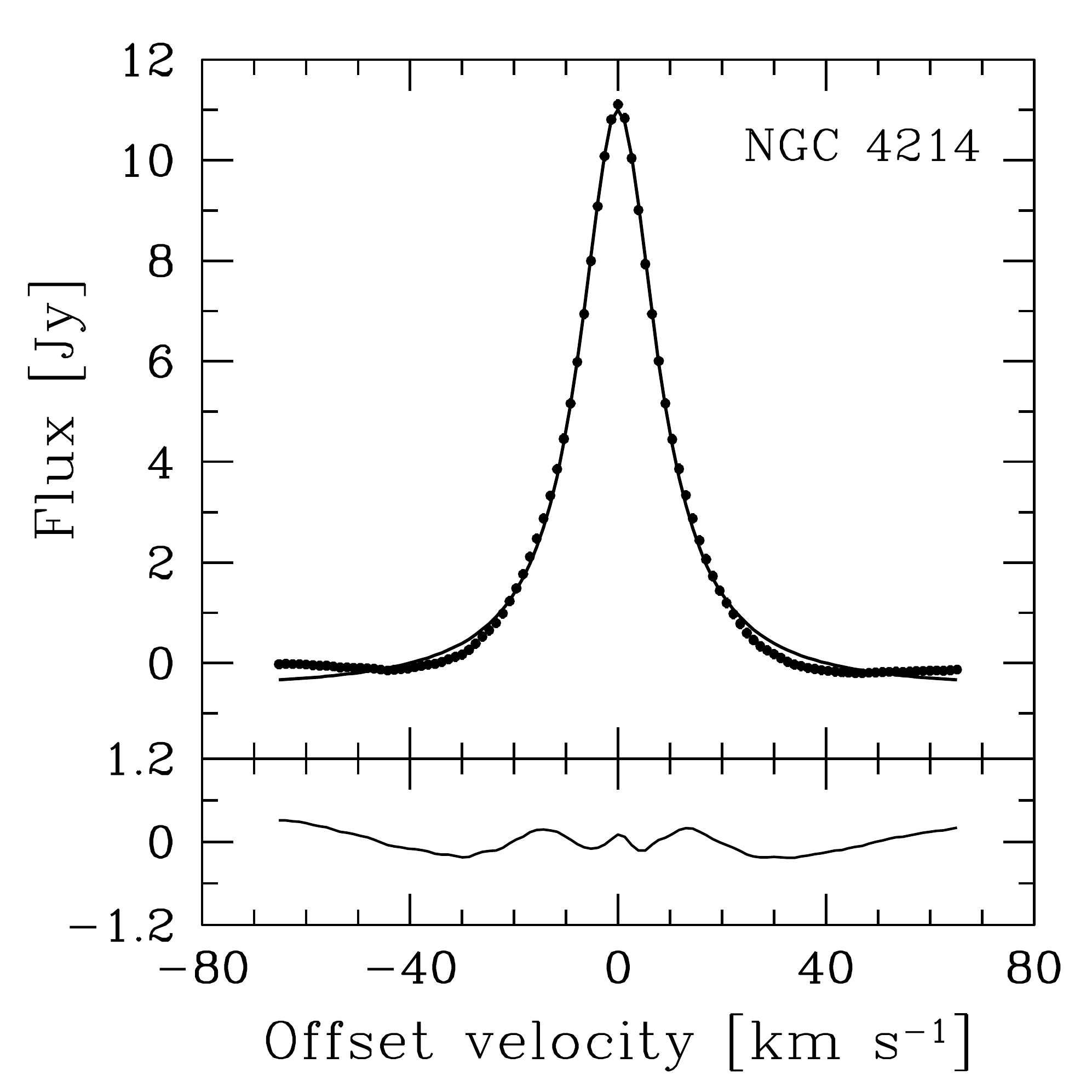}}}
\\\\ & & \hspace*{-12.12cm}\textbf{Figure \ref{fig:app1}}. \scriptsize{(continued).}
\end{tabular}
\end{figure*}

\begin{figure*}
    \begin{tabular}{l l l}
   \rotatebox{0}{\resizebox{58mm}{!}{\includegraphics[width = 0.6in,height = 0.6in]{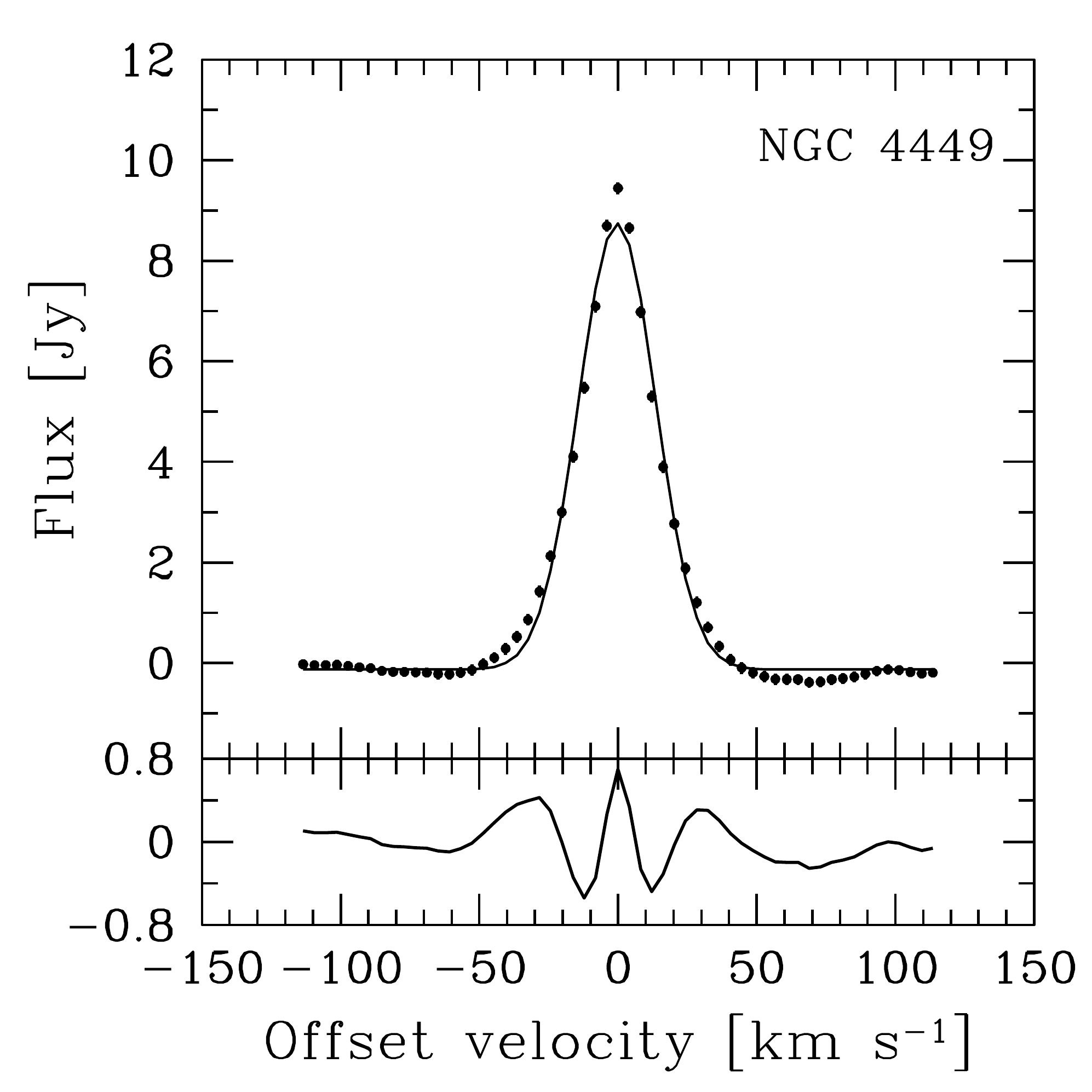}}}&
\rotatebox{0}{\resizebox{58mm}{!}{\includegraphics[width = 0.6in,height = 0.6in]{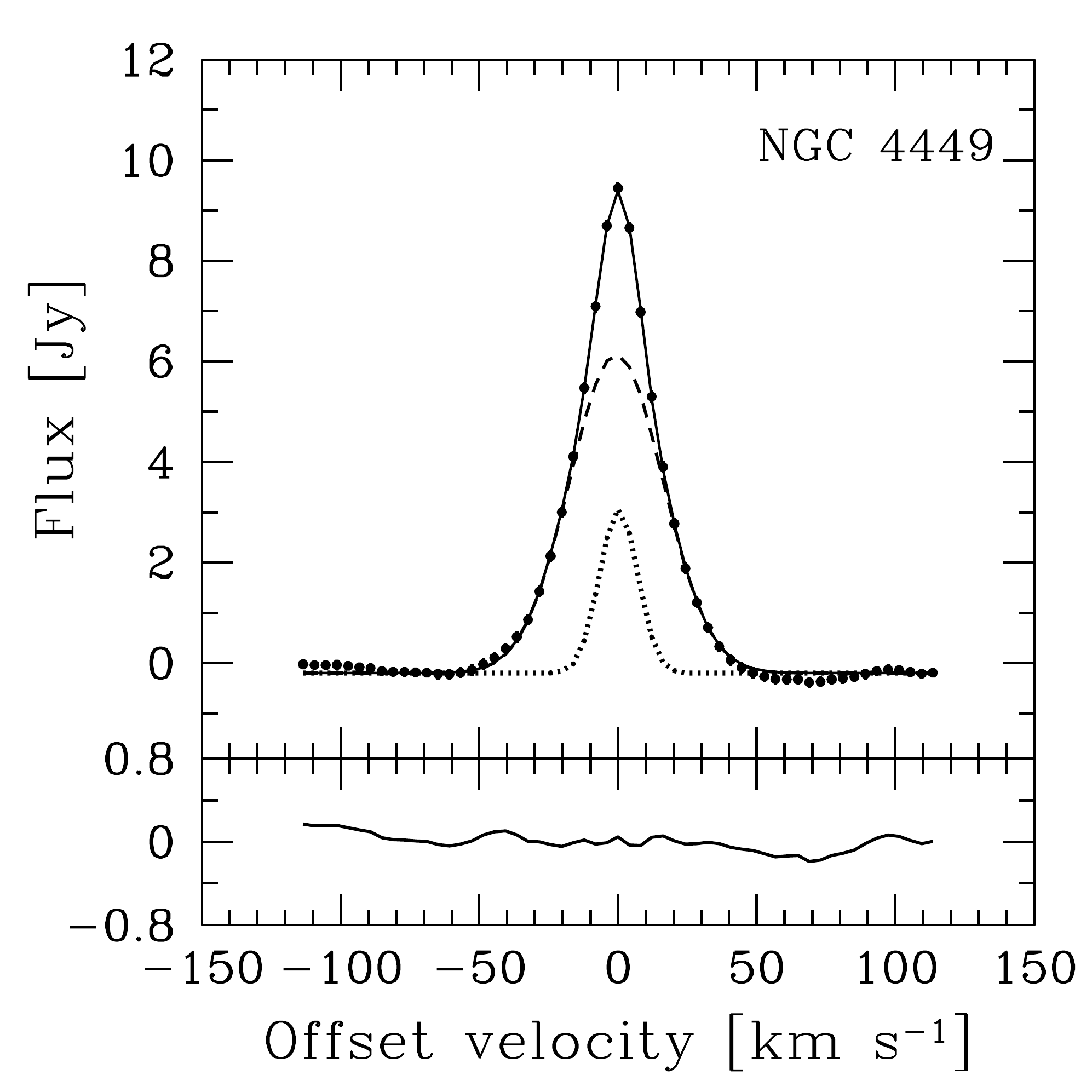}}}&
 \rotatebox{0}{\resizebox{58mm}{!}{\includegraphics[width = 0.6in,height = 0.6in]{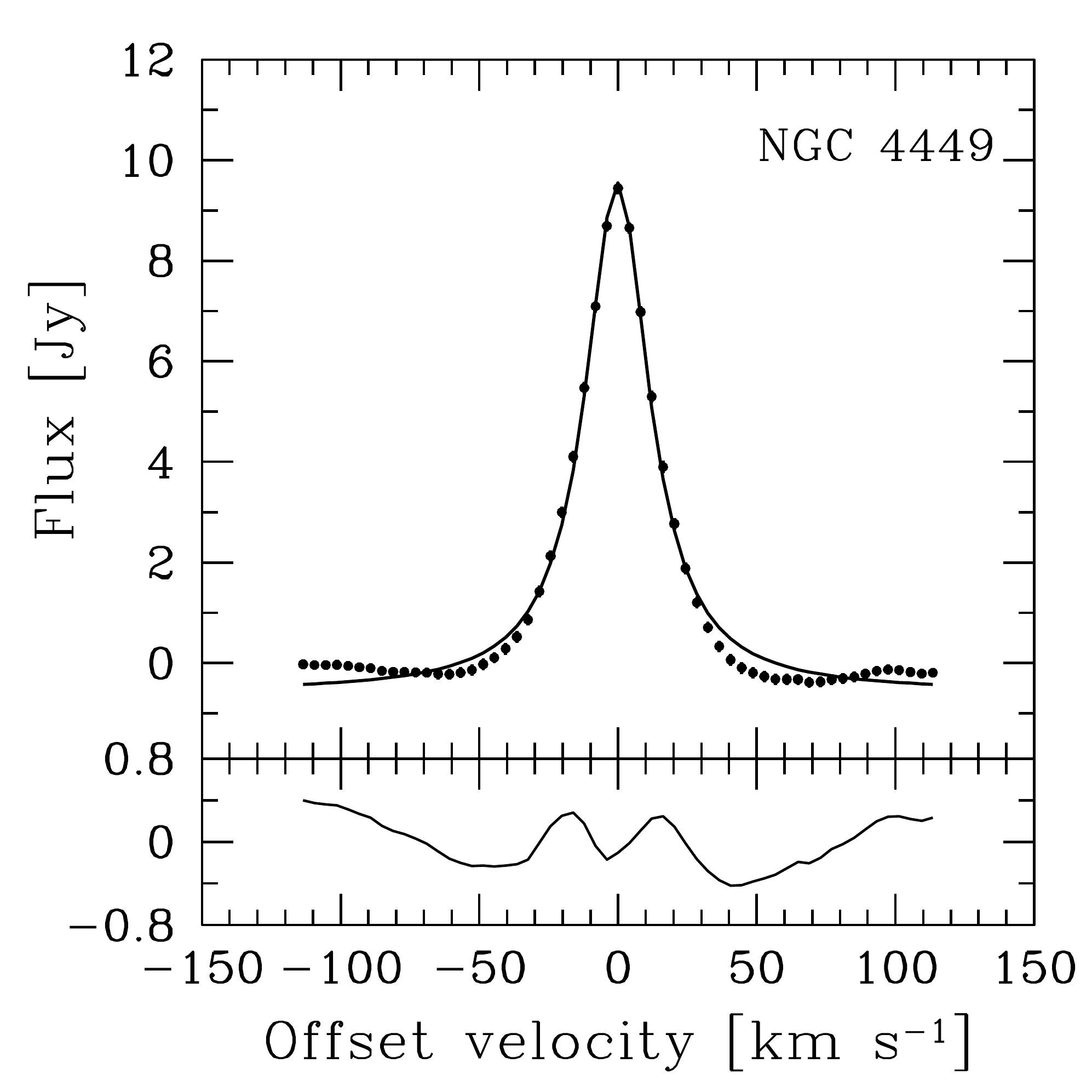}}}\\
   \rotatebox{0}{\resizebox{58mm}{!}{\includegraphics[width = 0.6in,height = 0.6in]{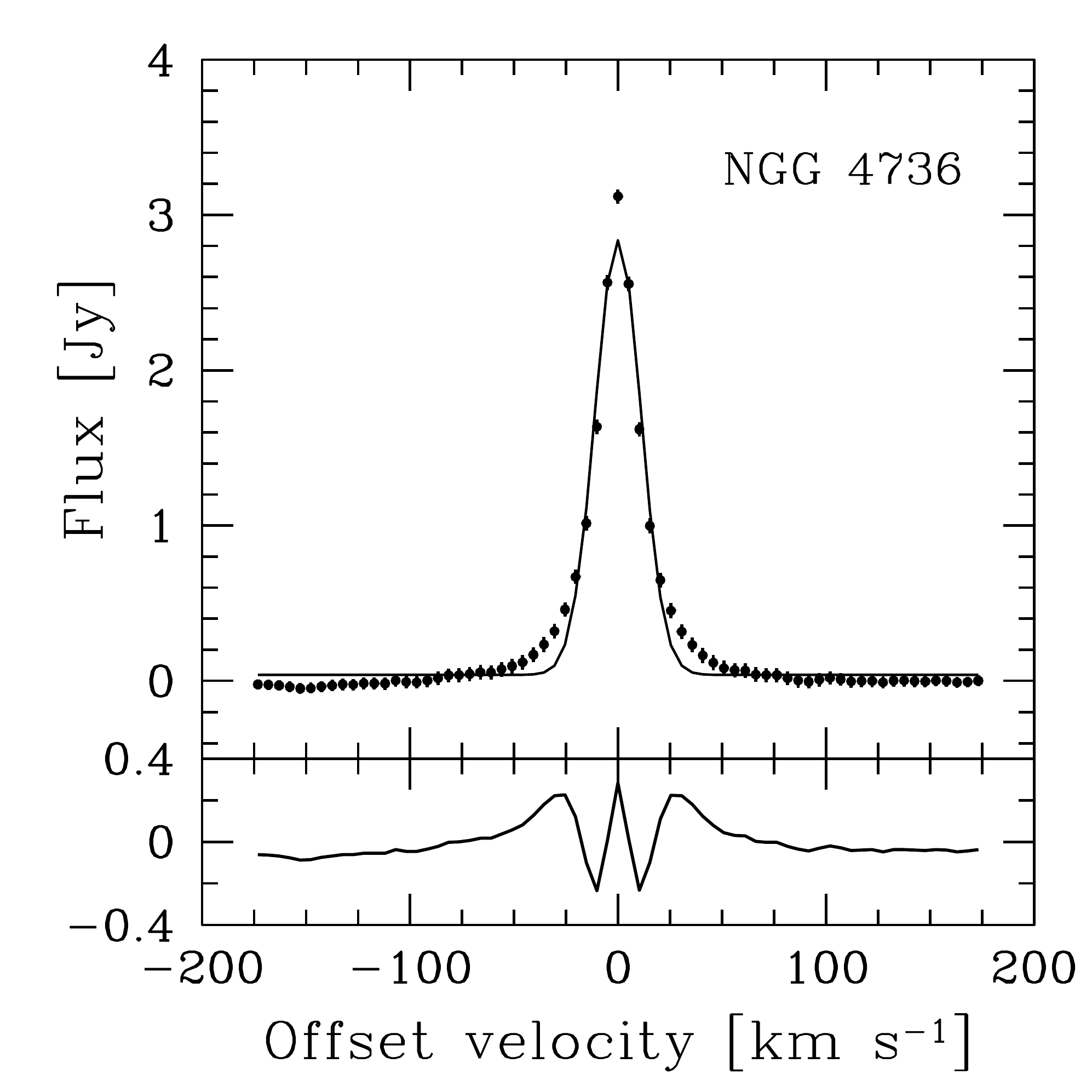}}}&
\rotatebox{0}{\resizebox{58mm}{!}{\includegraphics[width = 0.6in,height = 0.6in]{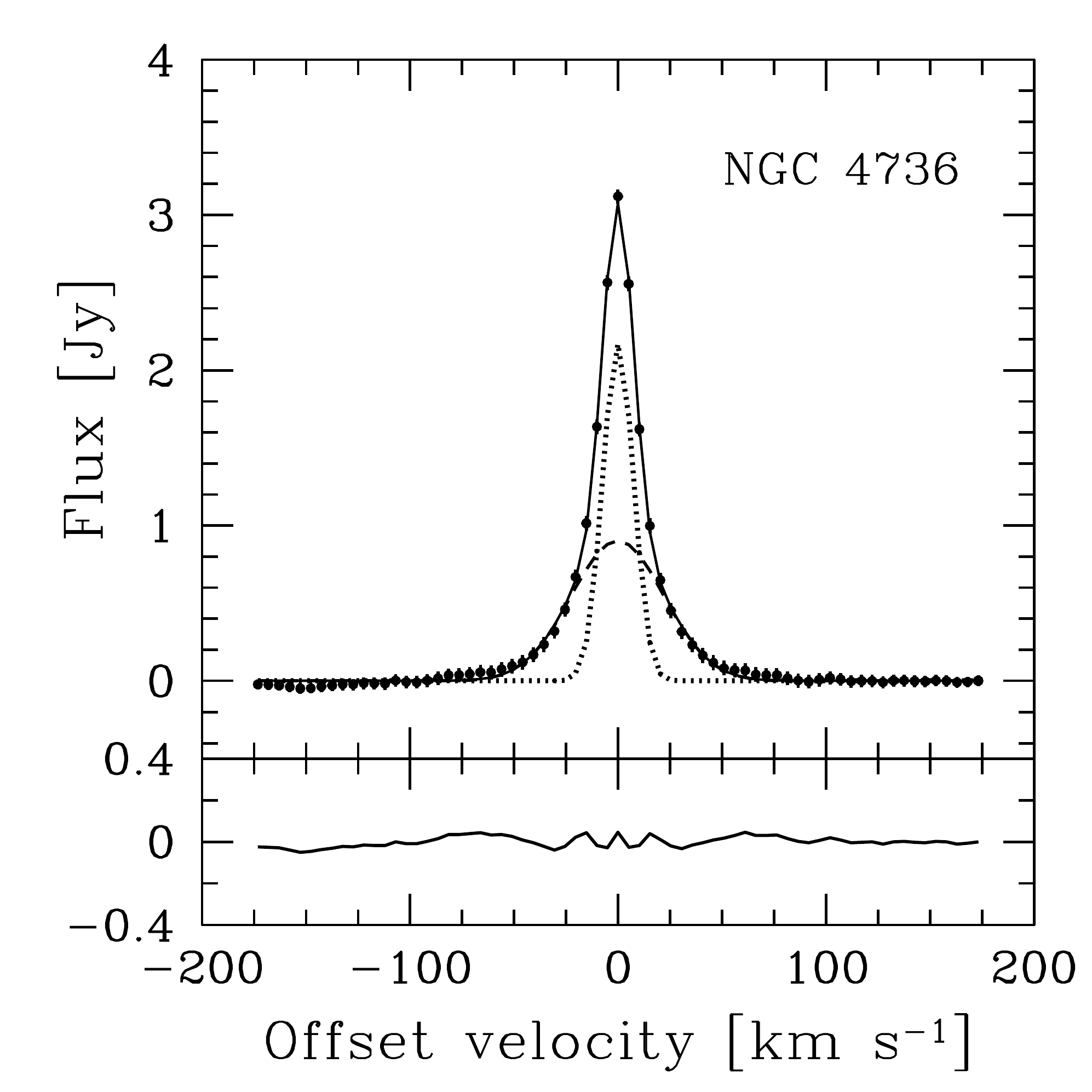}}}&
 \rotatebox{0}{\resizebox{58mm}{!}{\includegraphics[width = 0.6in,height = 0.6in]{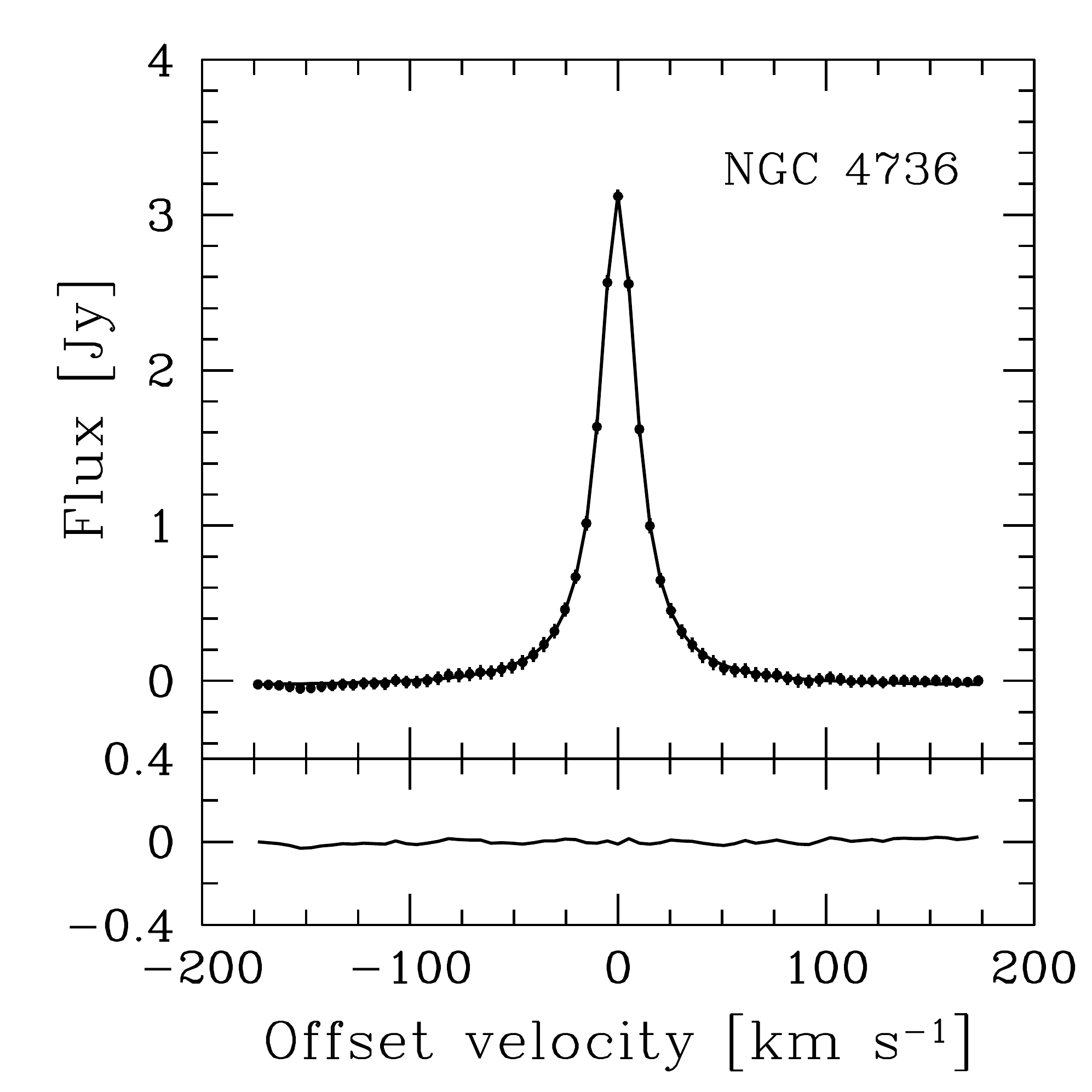}}}\\
    \rotatebox{0}{\resizebox{58mm}{!}{\includegraphics[width = 0.6in,height = 0.6in]{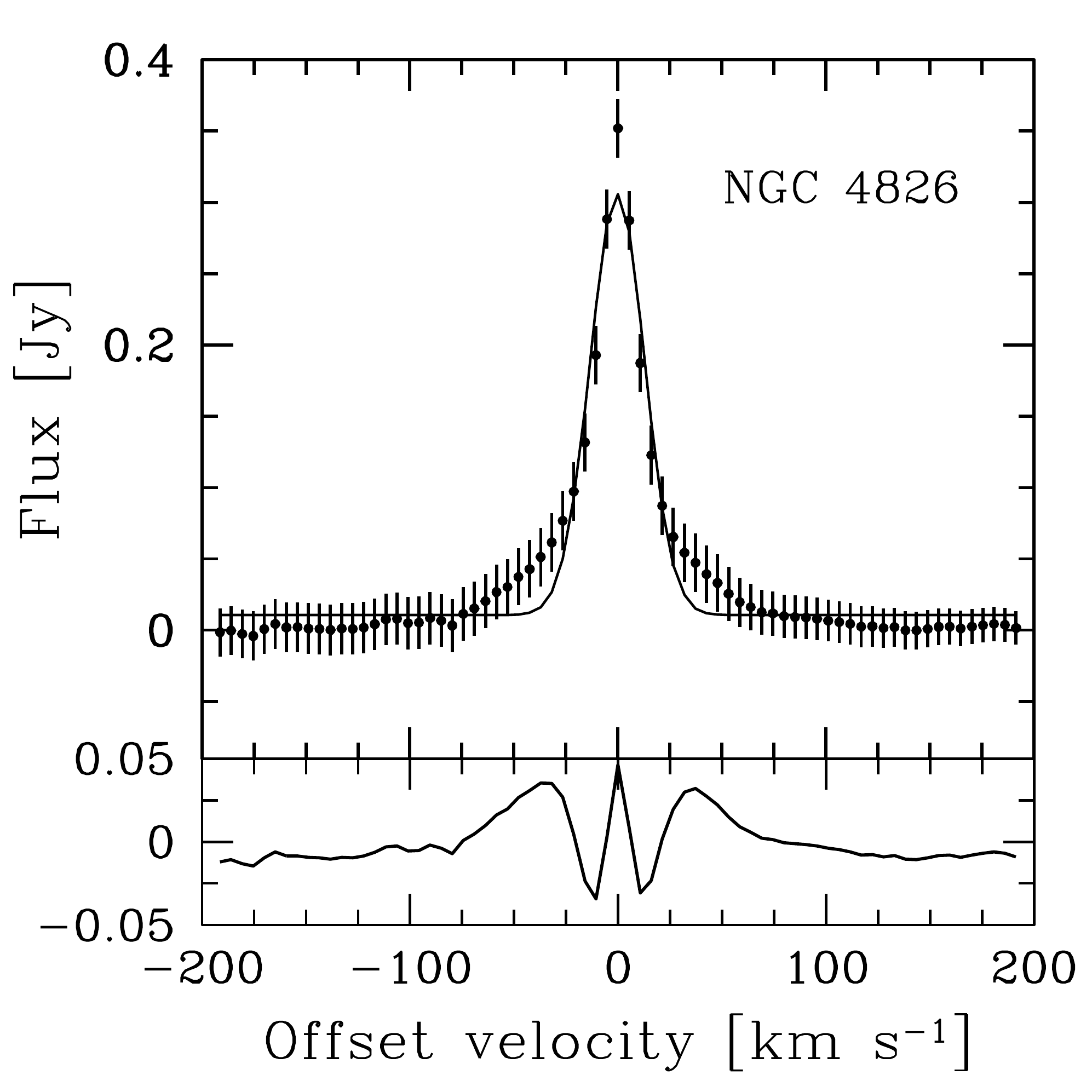}}}&
\rotatebox{0}{\resizebox{58mm}{!}{\includegraphics[width = 0.6in,height = 0.6in]{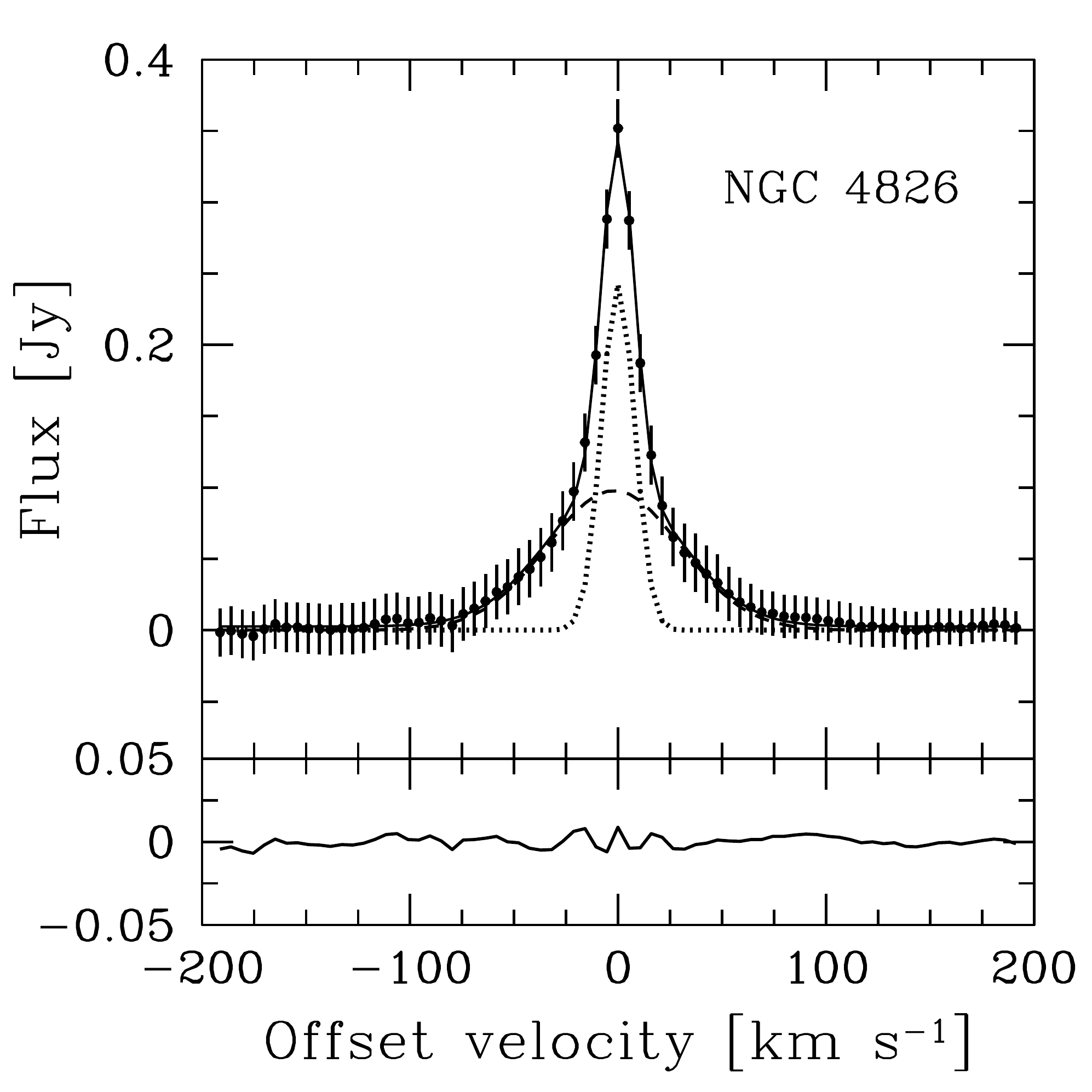}}}&
 \rotatebox{0}{\resizebox{58mm}{!}{\includegraphics[width = 0.6in,height = 0.6in]{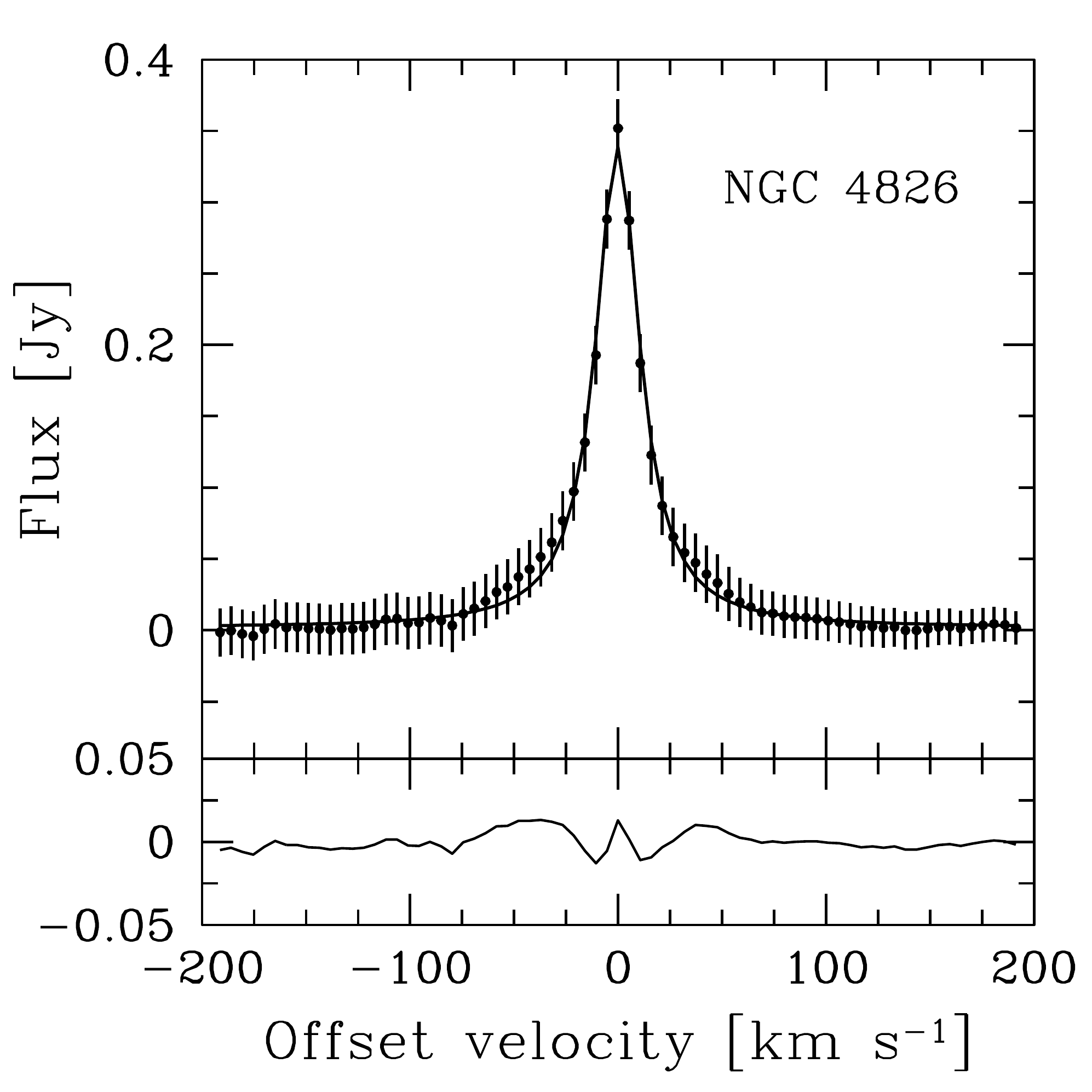}}}\\
    \rotatebox{0}{\resizebox{58mm}{!}{\includegraphics[width = 0.6in,height = 0.6in]{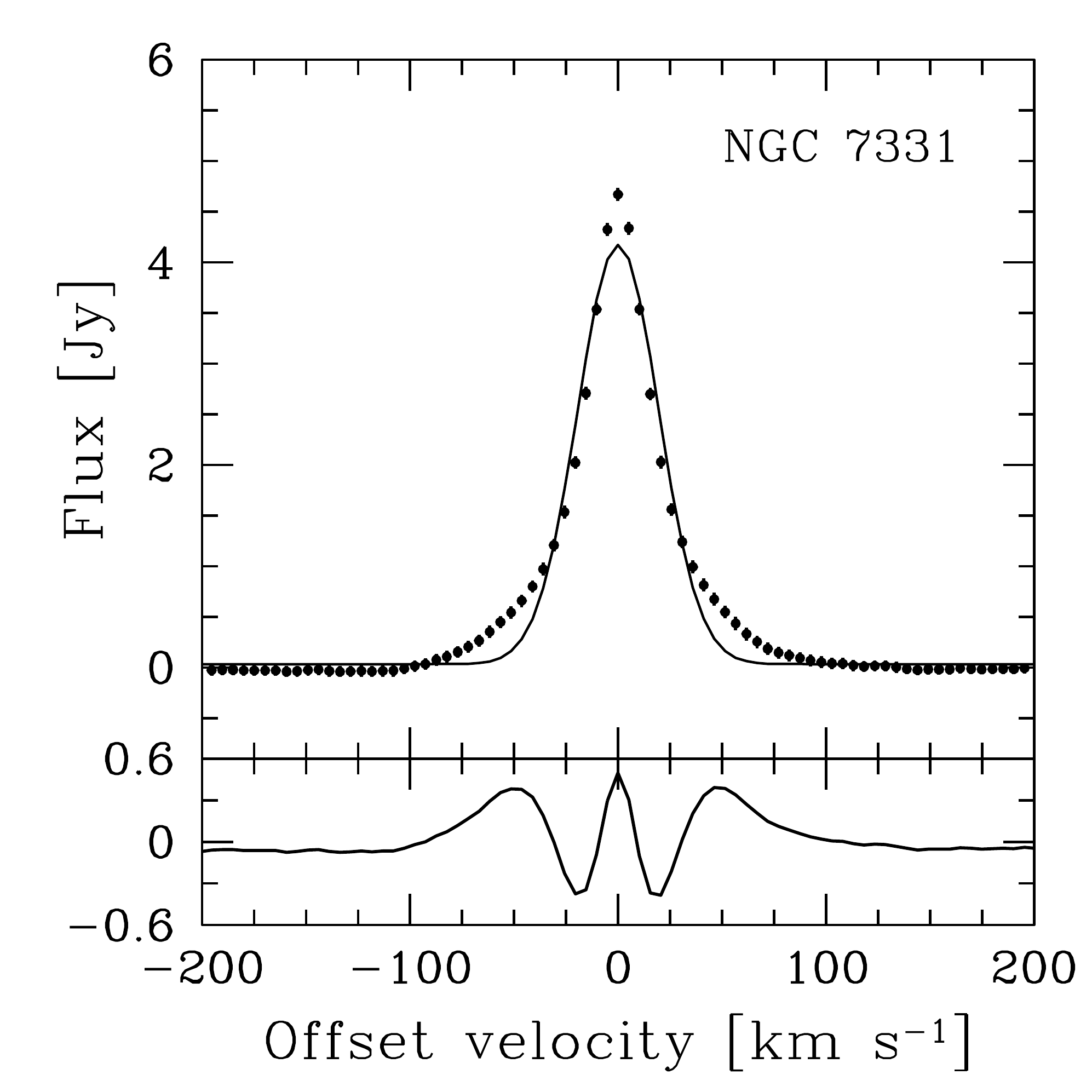}}}&
\rotatebox{0}{\resizebox{58mm}{!}{\includegraphics[width = 0.6in,height = 0.6in]{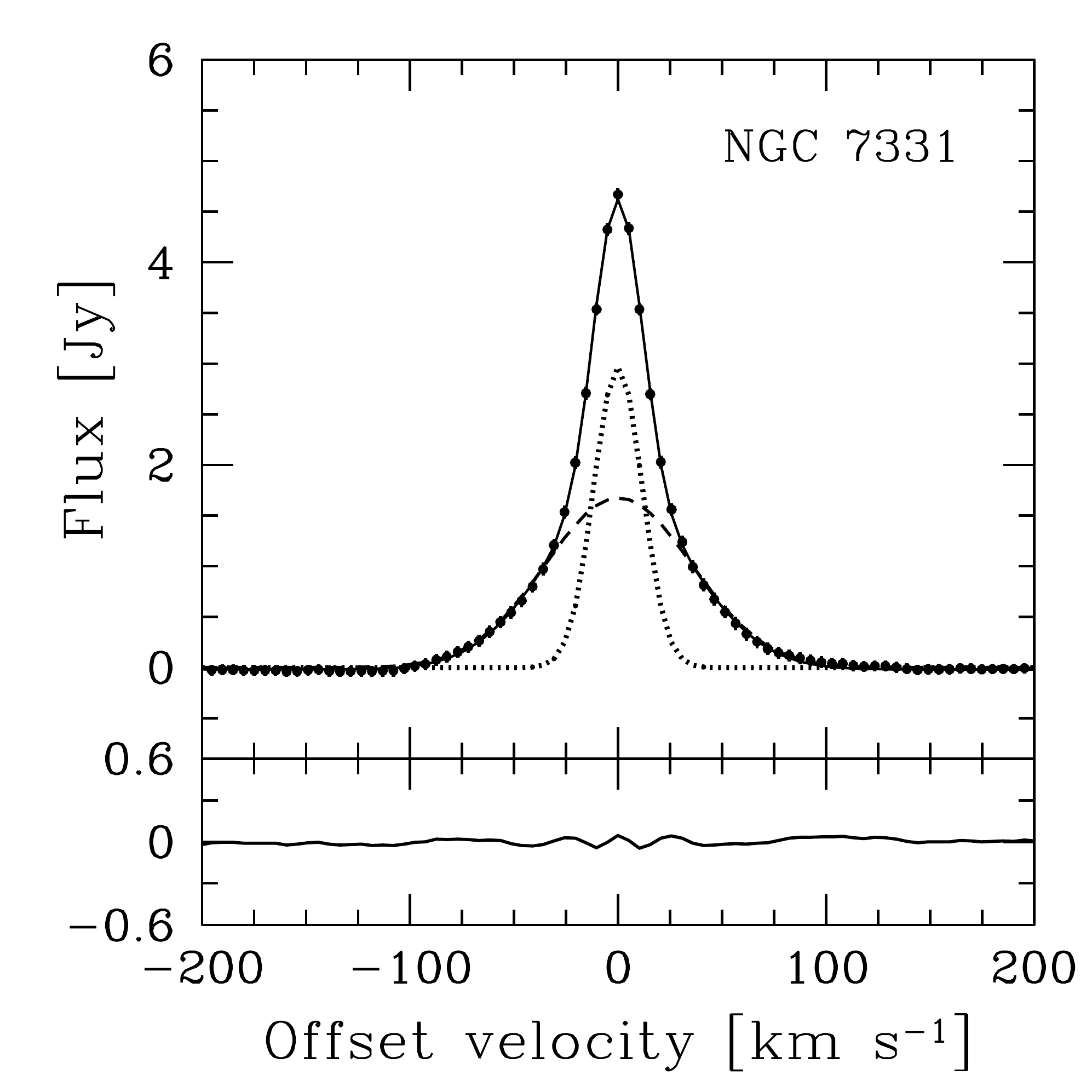}}}&
 \rotatebox{0}{\resizebox{58mm}{!}{\includegraphics[width = 0.6in,height = 0.6in]{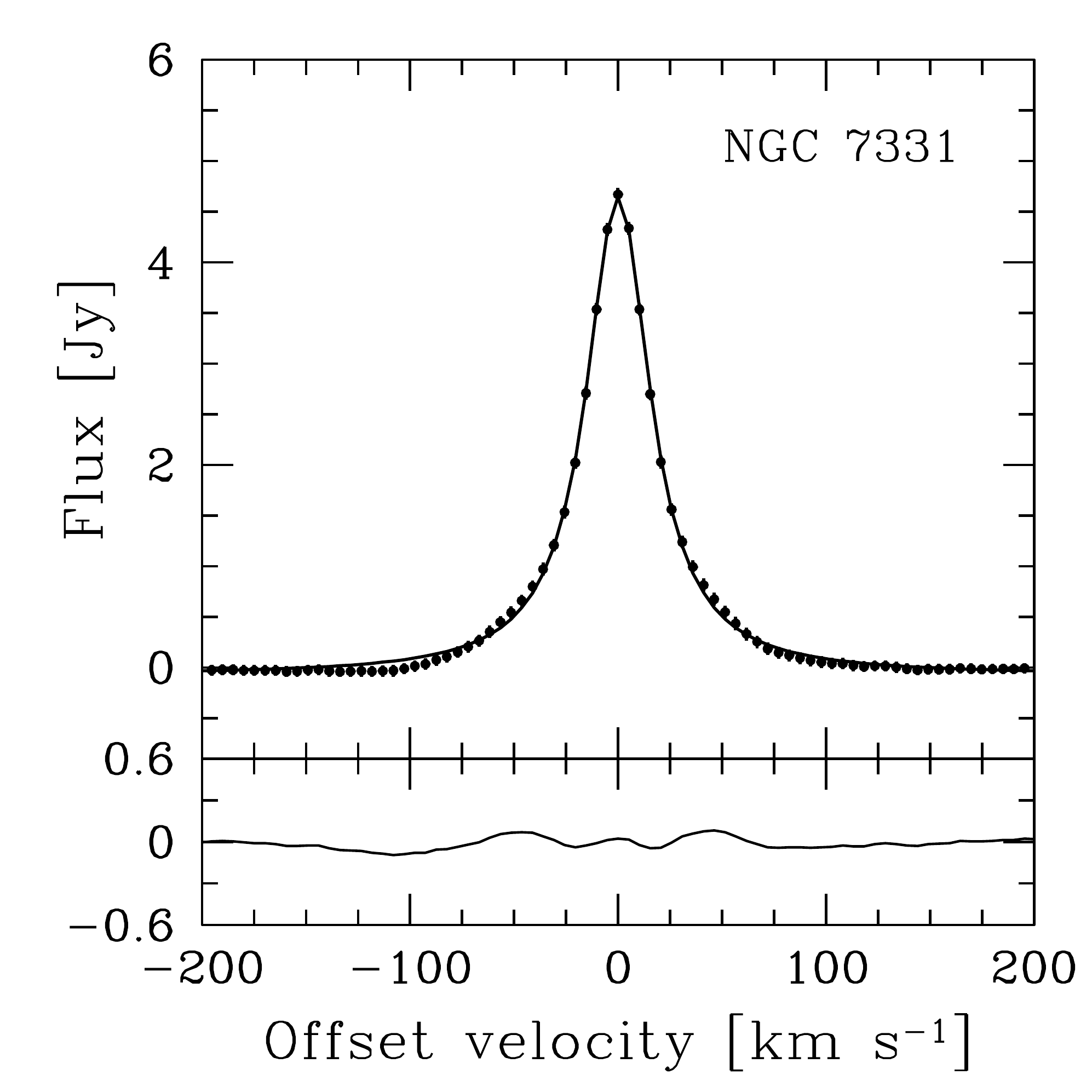}}}
\\\\ & & \hspace*{-12.12cm}\textbf{Figure \ref{fig:app1}}. \scriptsize{(continued).}
\end{tabular}
\end{figure*}

\begin{figure*}
	\begin{tabular}{l l l}
	\rotatebox{0}{\resizebox{58mm}{!}{\includegraphics[width = 0.6in,height = 0.6in]{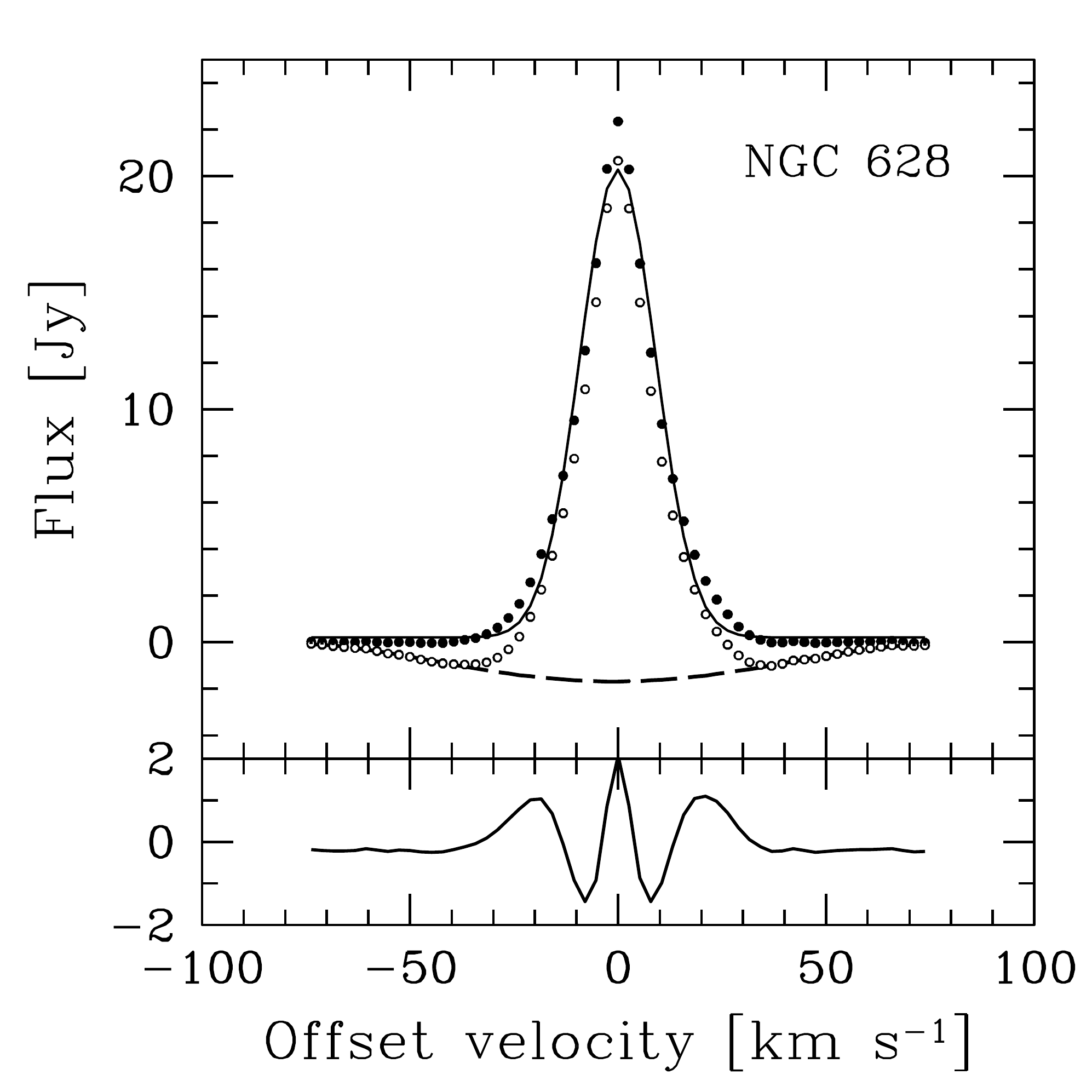}}}&
	\rotatebox{0}{\resizebox{58mm}{!}{\includegraphics[width = 0.6in,height = 0.6in]{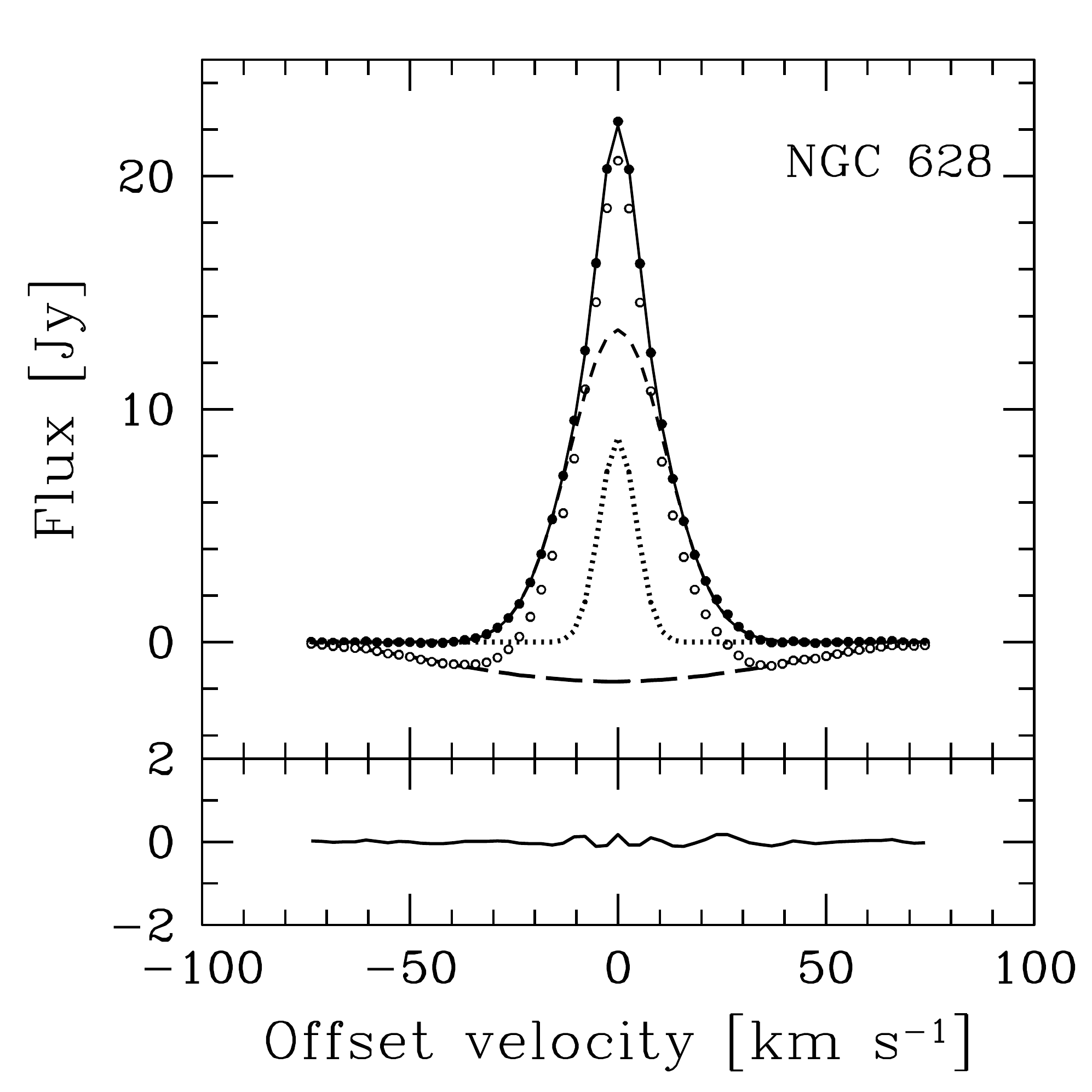}}}&
	\rotatebox{0}{\resizebox{58mm}{!}{\includegraphics[width = 0.6in,height = 0.6in]{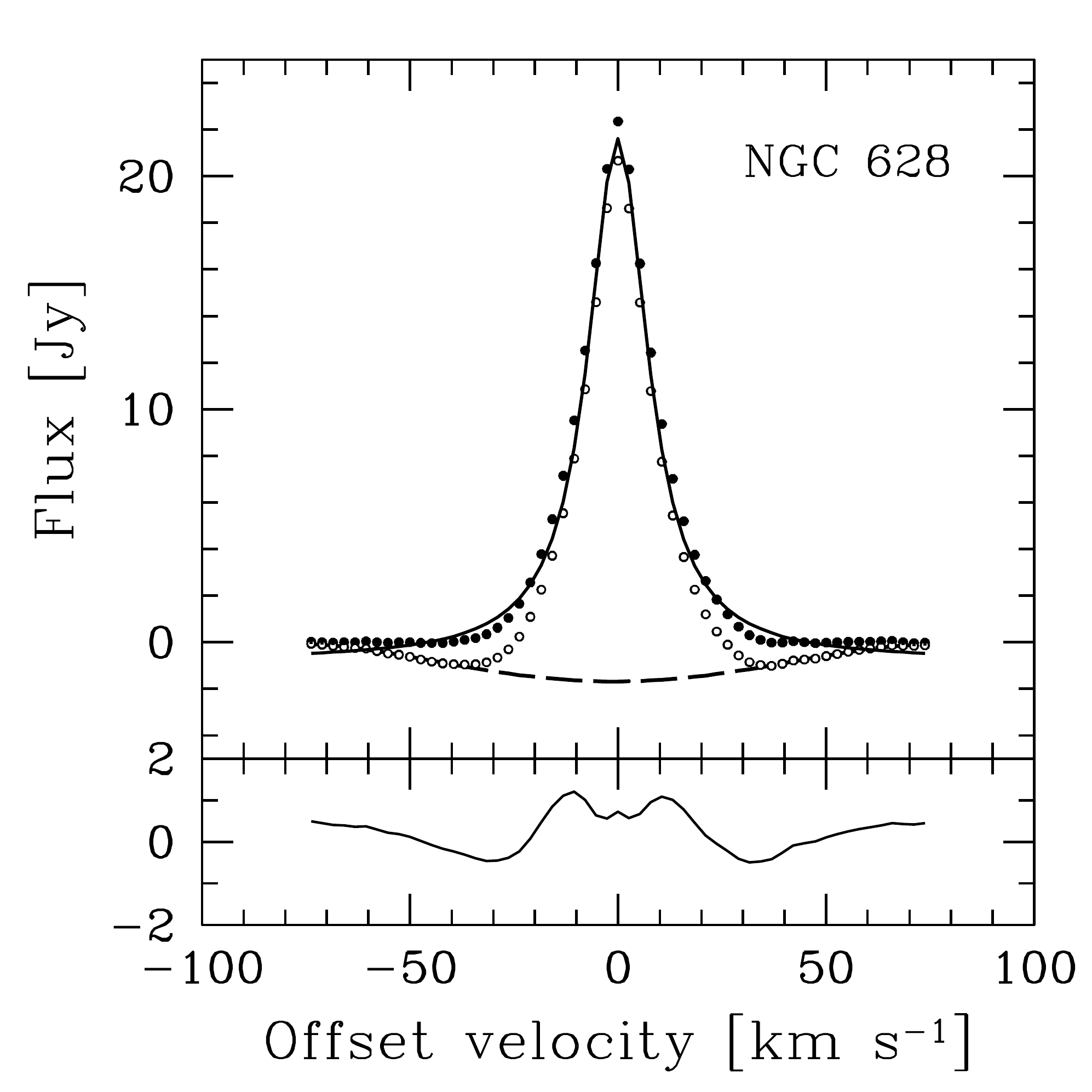}}}\\
	\rotatebox{0}{\resizebox{58mm}{!}{\includegraphics[width = 0.6in,height = 0.6in]{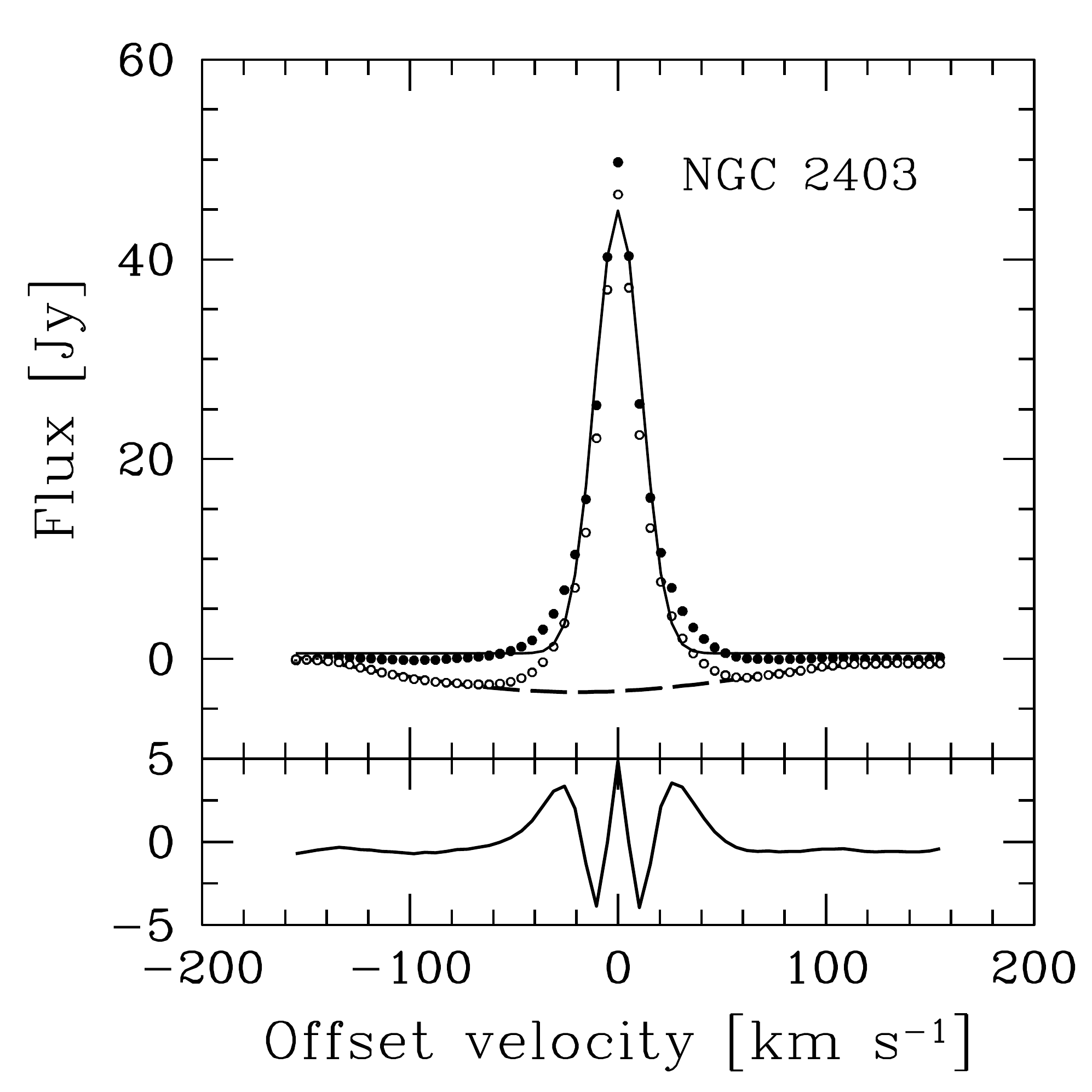}}}&
	\rotatebox{0}{\resizebox{58mm}{!}{\includegraphics[width = 0.6in,height = 0.6in]{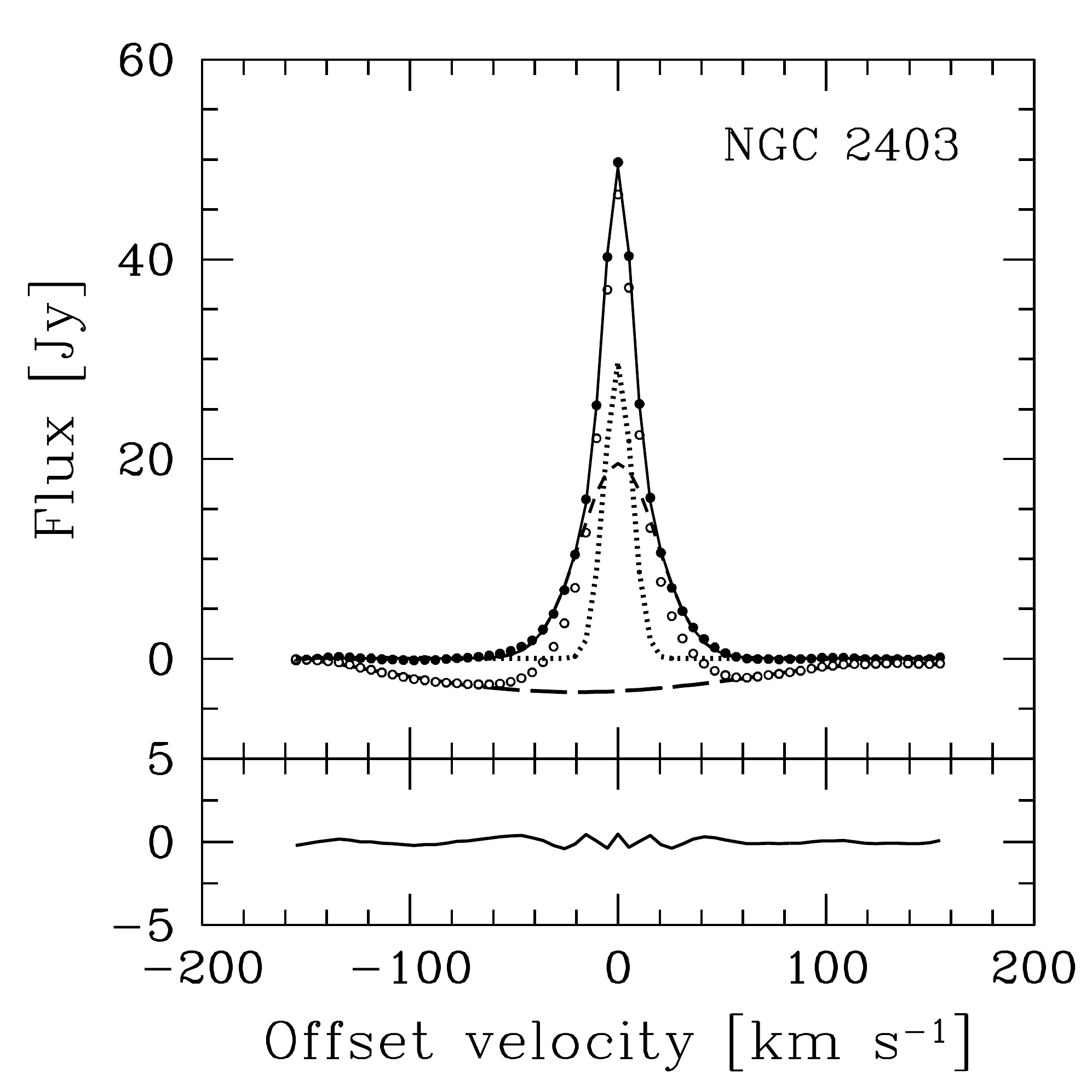}}}&
	\rotatebox{0}{\resizebox{58mm}{!}{\includegraphics[width = 0.6in,height = 0.6in]{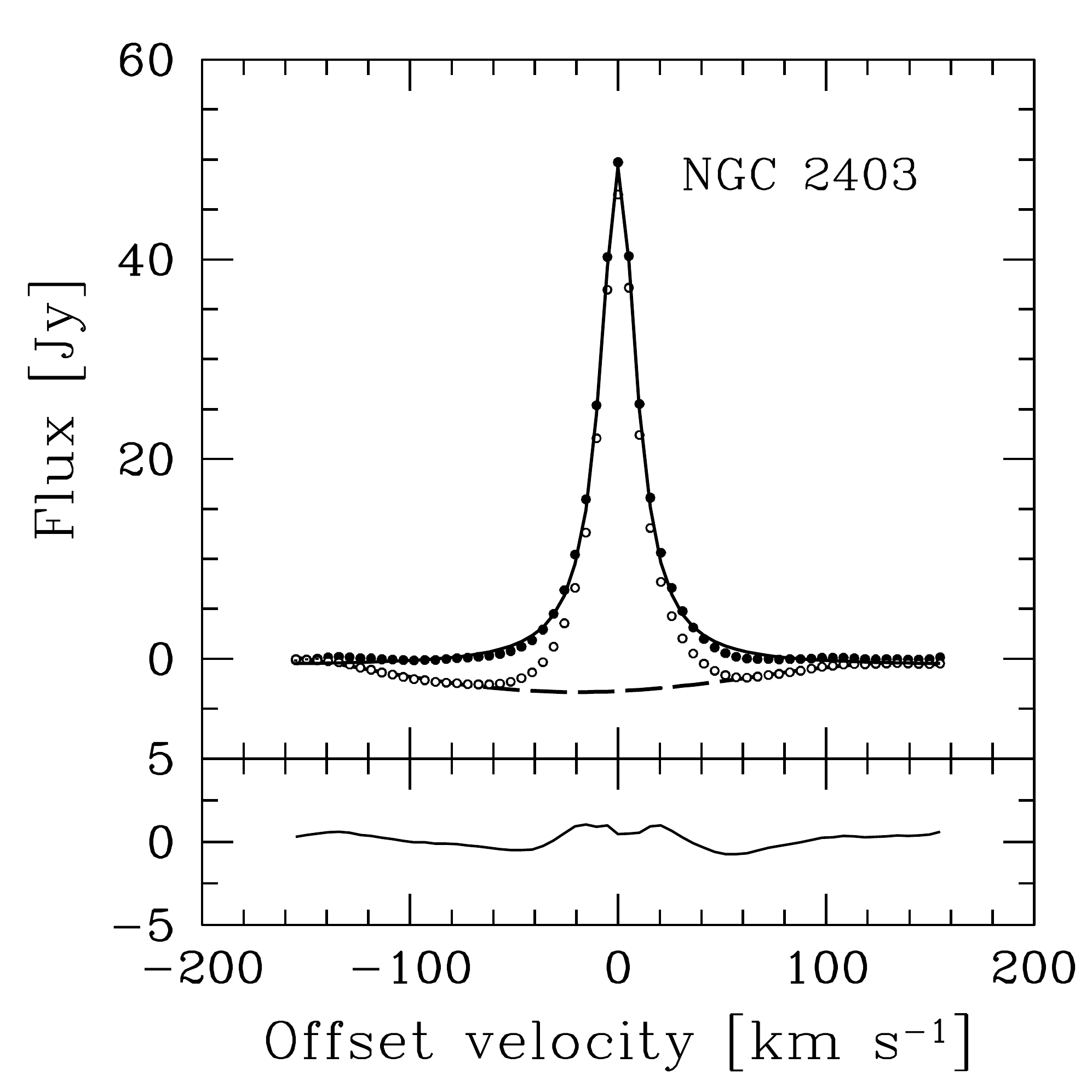}}} \\
	\rotatebox{0}{\resizebox{58mm}{!}{\includegraphics[width = 0.6in,height = 0.6in]{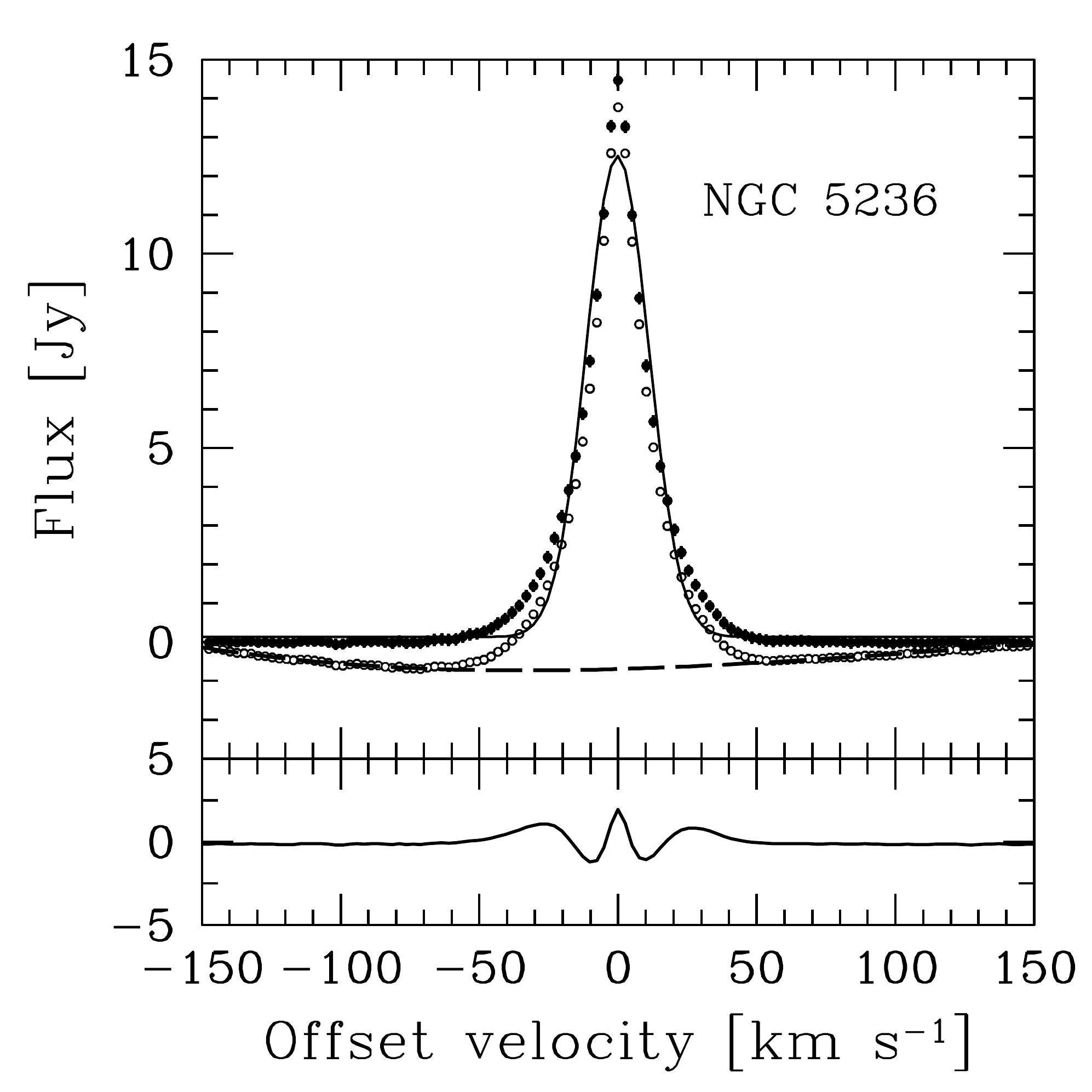}}}&
	\rotatebox{0}{\resizebox{58mm}{!}{\includegraphics[width = 0.6in,height = 0.6in]{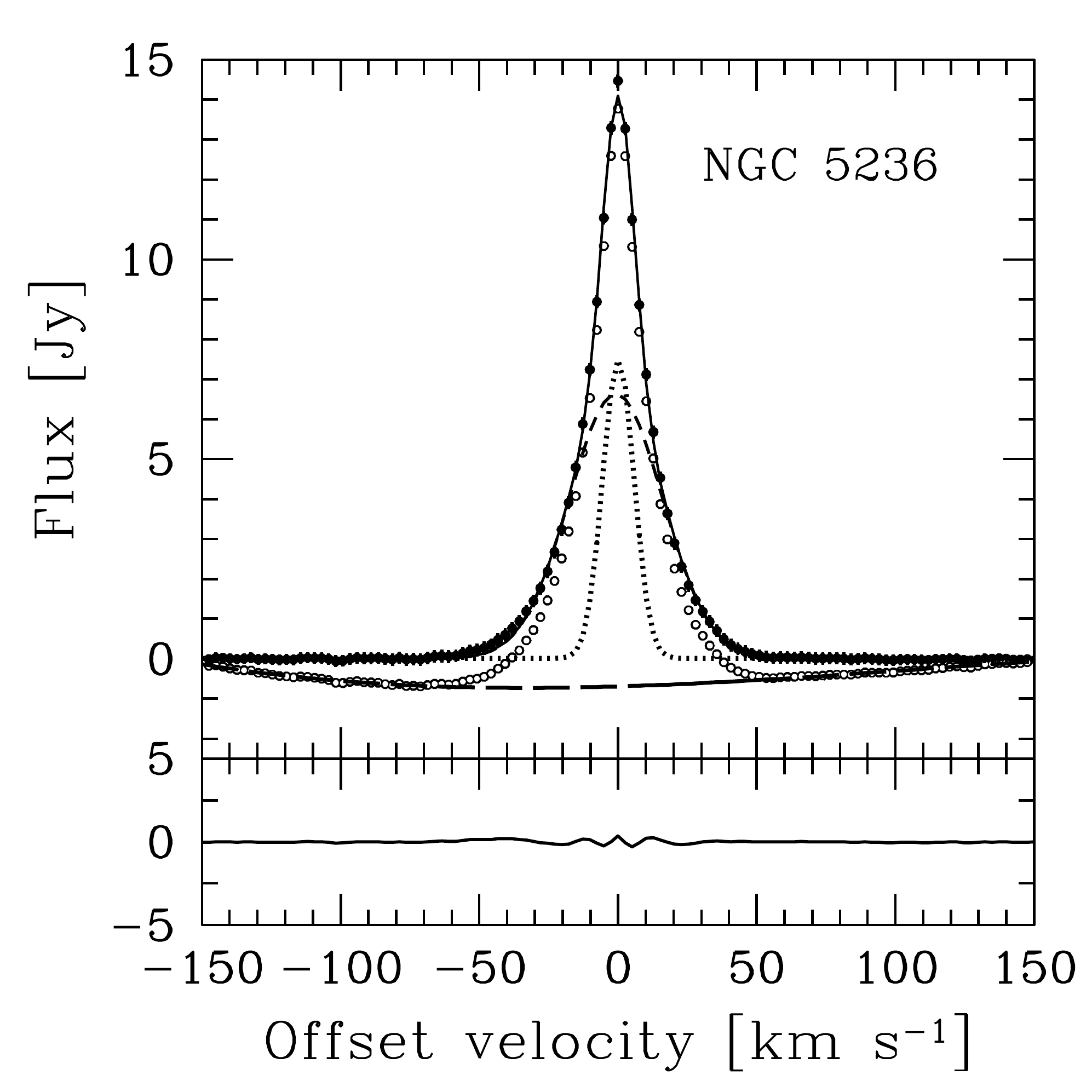}}}&
	\rotatebox{0}{\resizebox{58mm}{!}{\includegraphics[width = 0.6in,height = 0.6in]{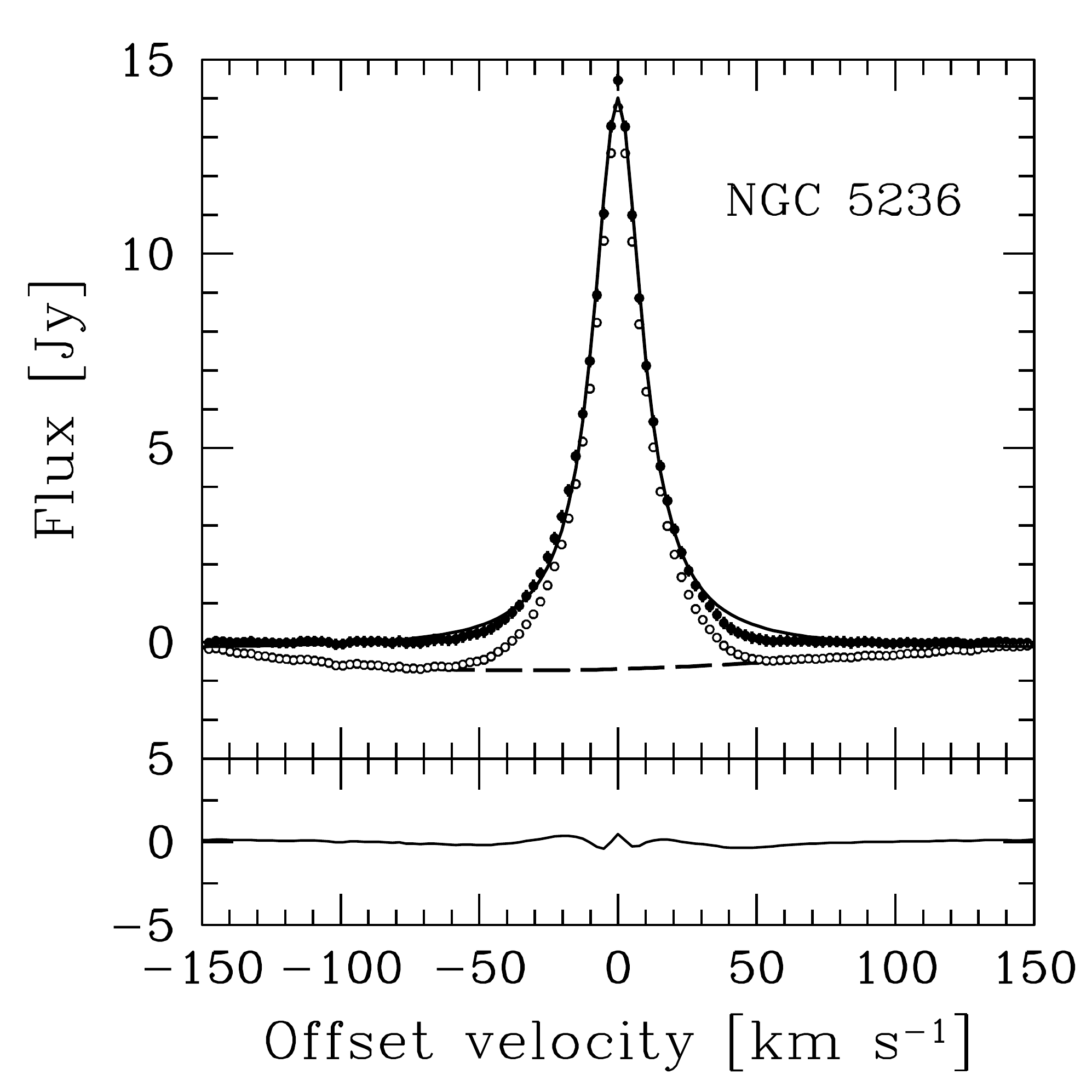}}}\\
	\rotatebox{0}{\resizebox{58mm}{!}{\includegraphics[width = 0.6in,height = 0.6in]{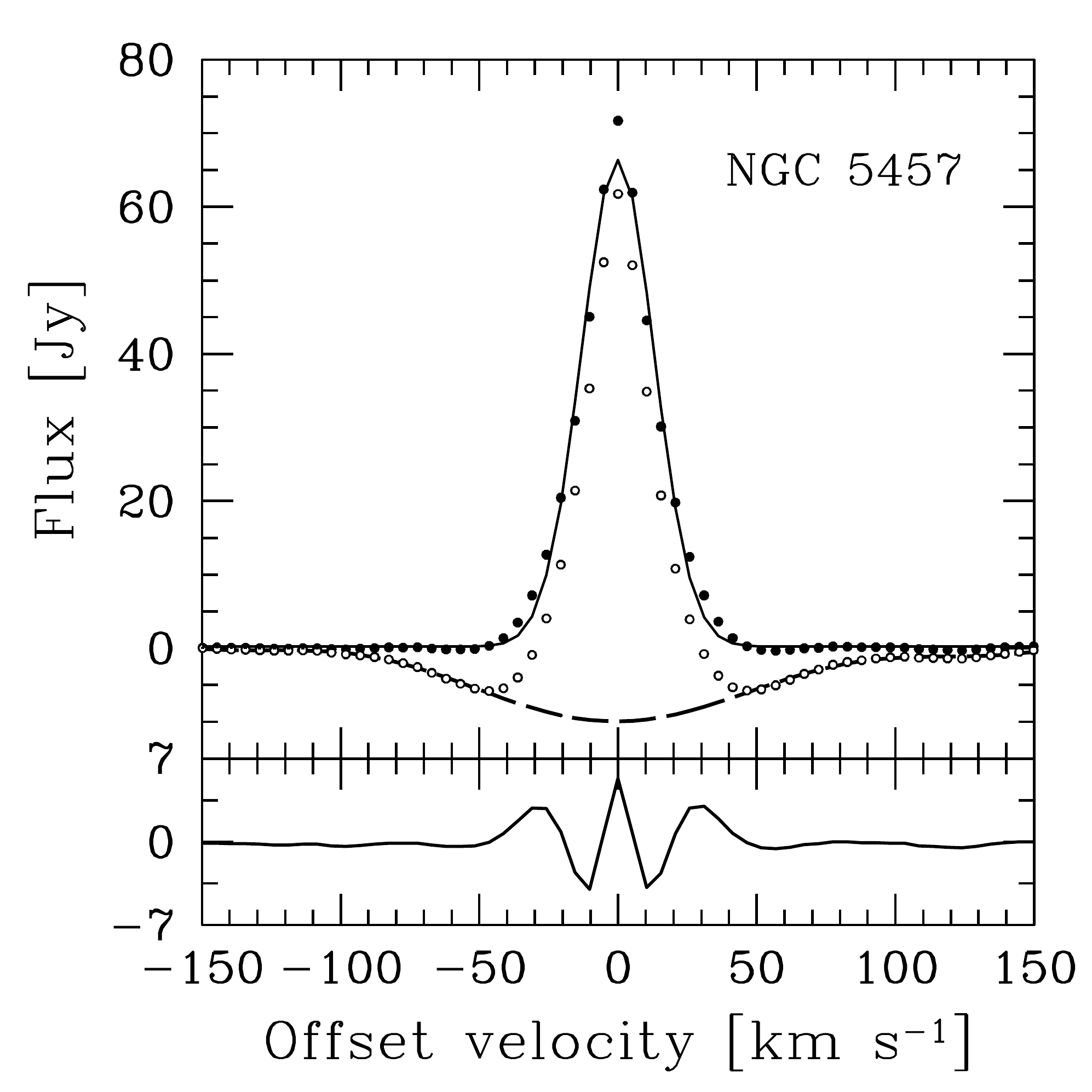}}}&
	\rotatebox{0}{\resizebox{58mm}{!}{\includegraphics[width = 0.6in,height = 0.6in]{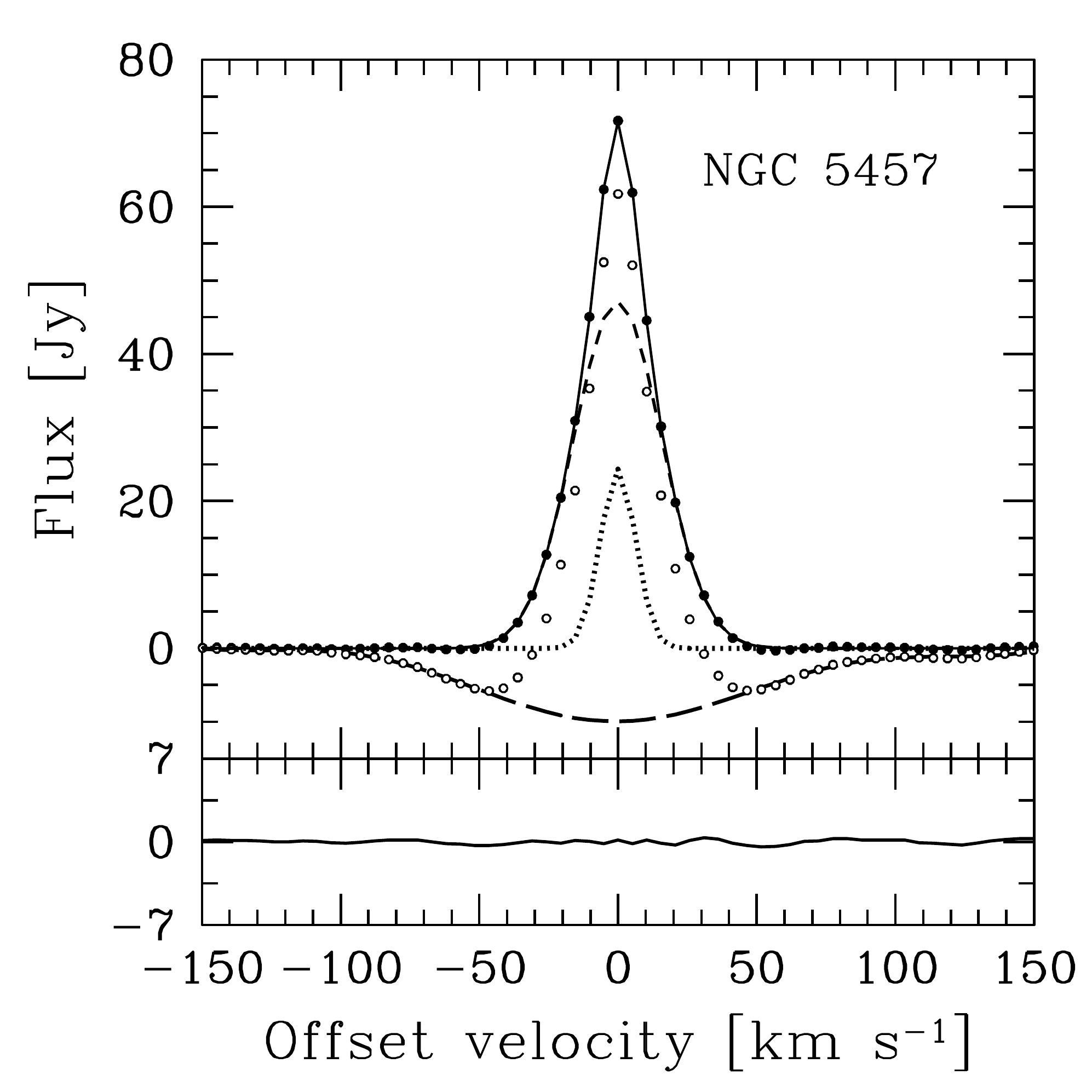}}}&
	\rotatebox{0}{\resizebox{58mm}{!}{\includegraphics[width = 0.6in,height = 0.6in]{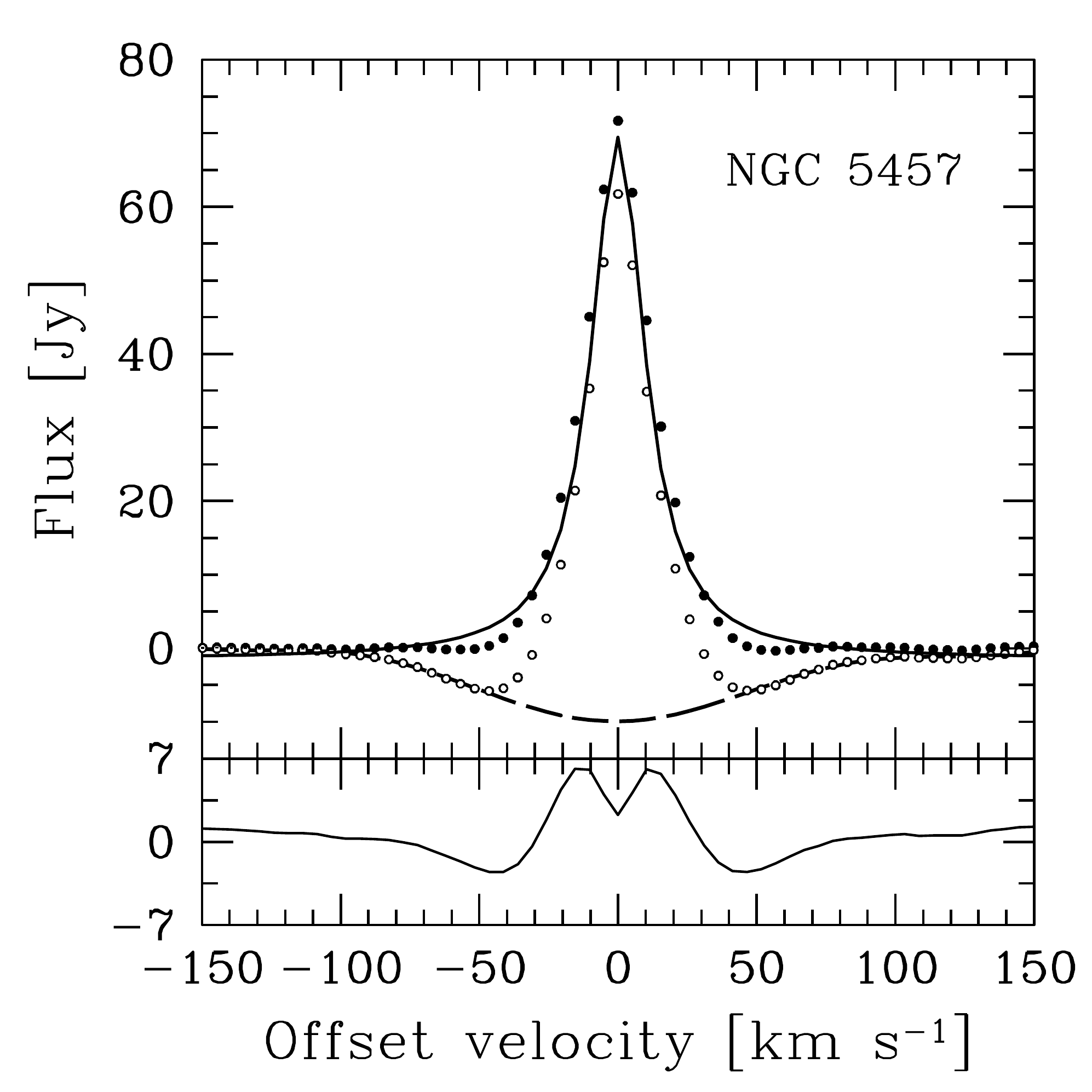}}}
	\\\\& & \hspace*{-12.12cm}\textbf{Figure \ref{fig:app1}}. \scriptsize{(continued)}. 
	\small{Here the open and the filled 
	circle symbols represent the data points before and after the baseline } 
	\\ & &\hspace*{-12.12cm}\small{correction, respectively (see Section \ref{sec:fitting}). 
	The dashed lines represent the polynomial fits to the negative bowls.}
	\end{tabular}

\end{figure*}

\begin{figure*}
	\begin{tabular}{l l l}
	\rotatebox{0}{\resizebox{58mm}{!}{\includegraphics[width = 0.6in,height = 0.6in]{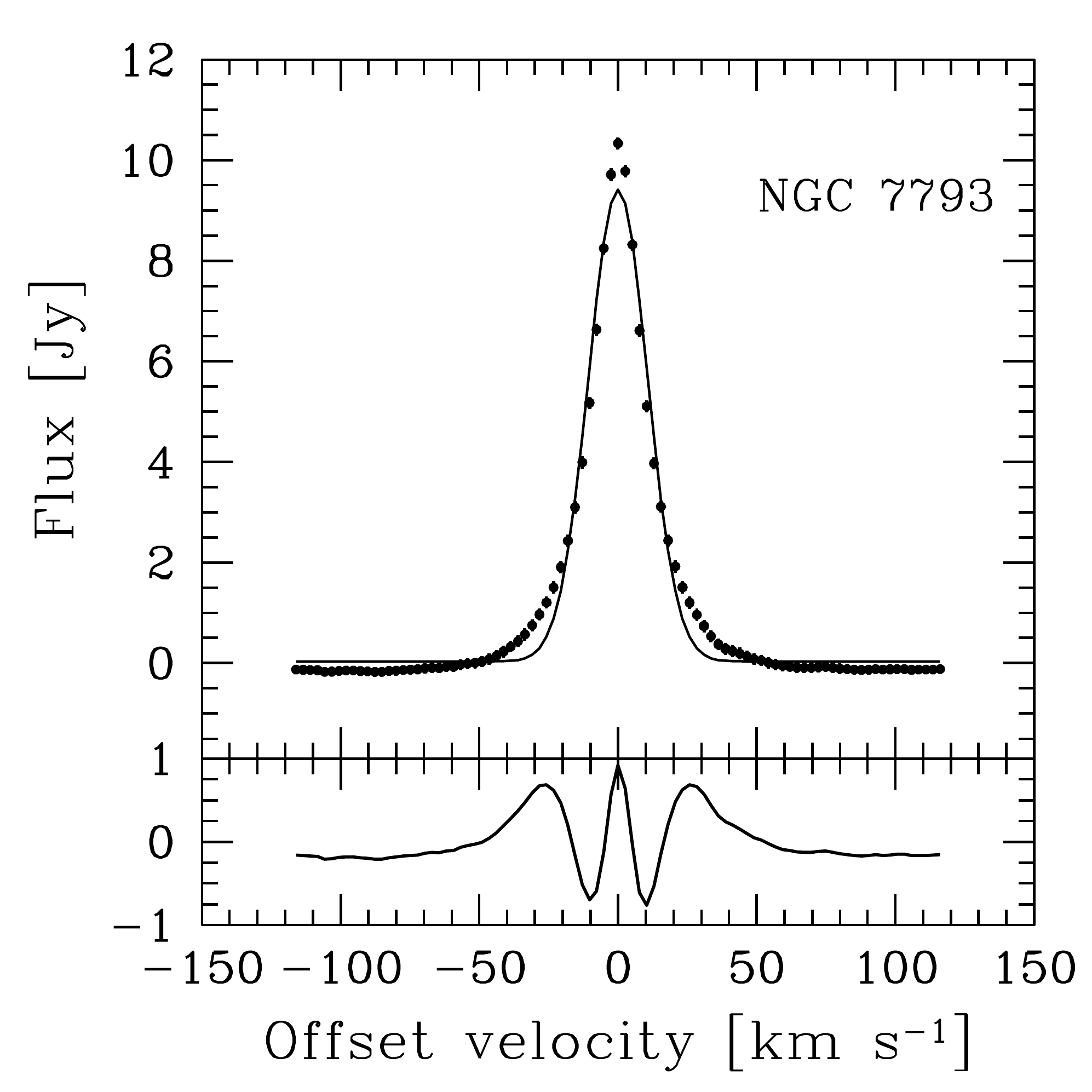}}}&
	\rotatebox{0}{\resizebox{58mm}{!}{\includegraphics[width = 0.6in,height = 0.6in]{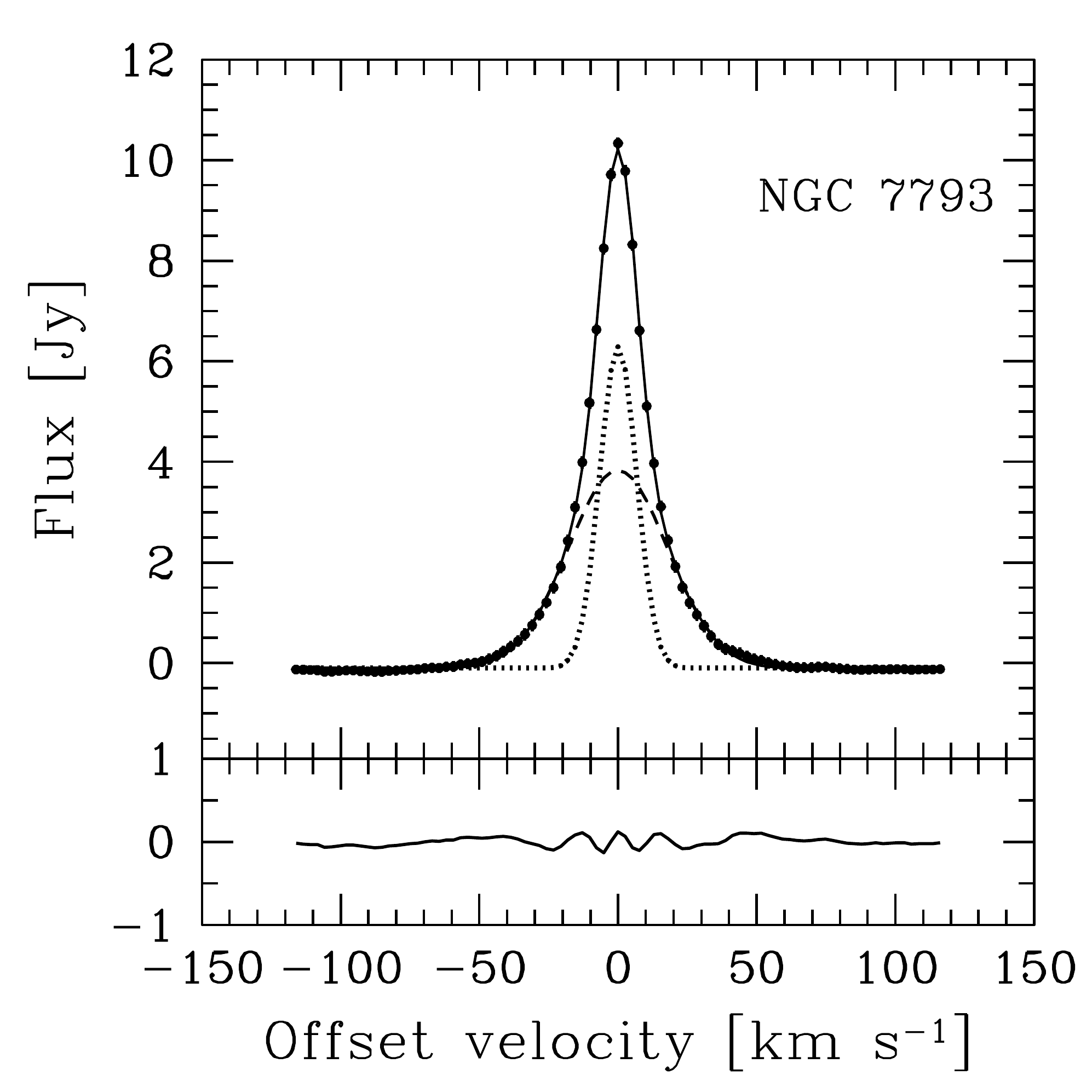}}}&
	\rotatebox{0}{\resizebox{58mm}{!}{\includegraphics[width = 0.6in,height = 0.6in]{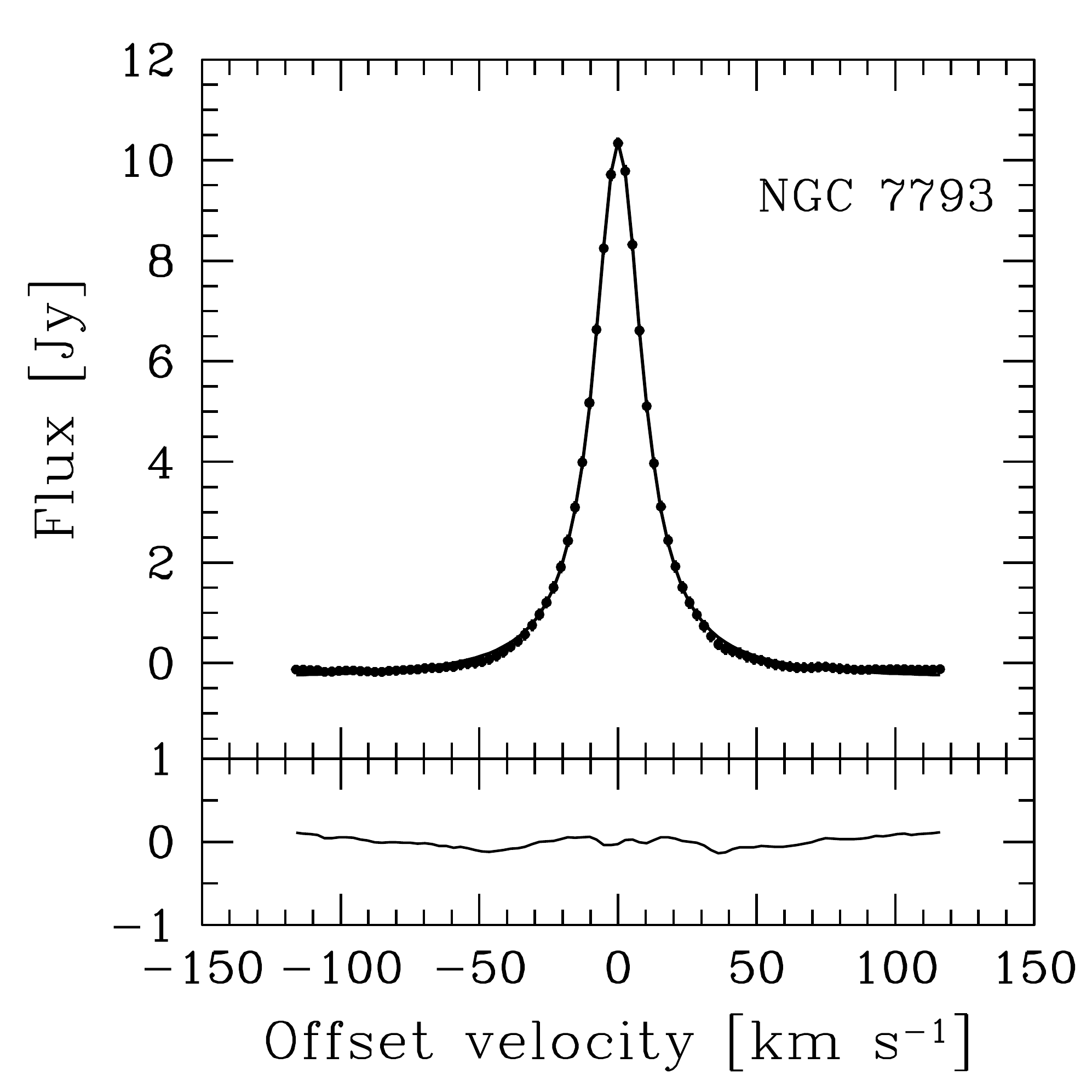}}}\\
	\rotatebox{0}{\resizebox{58mm}{!}{\includegraphics[width = 0.6in,height = 0.6in]{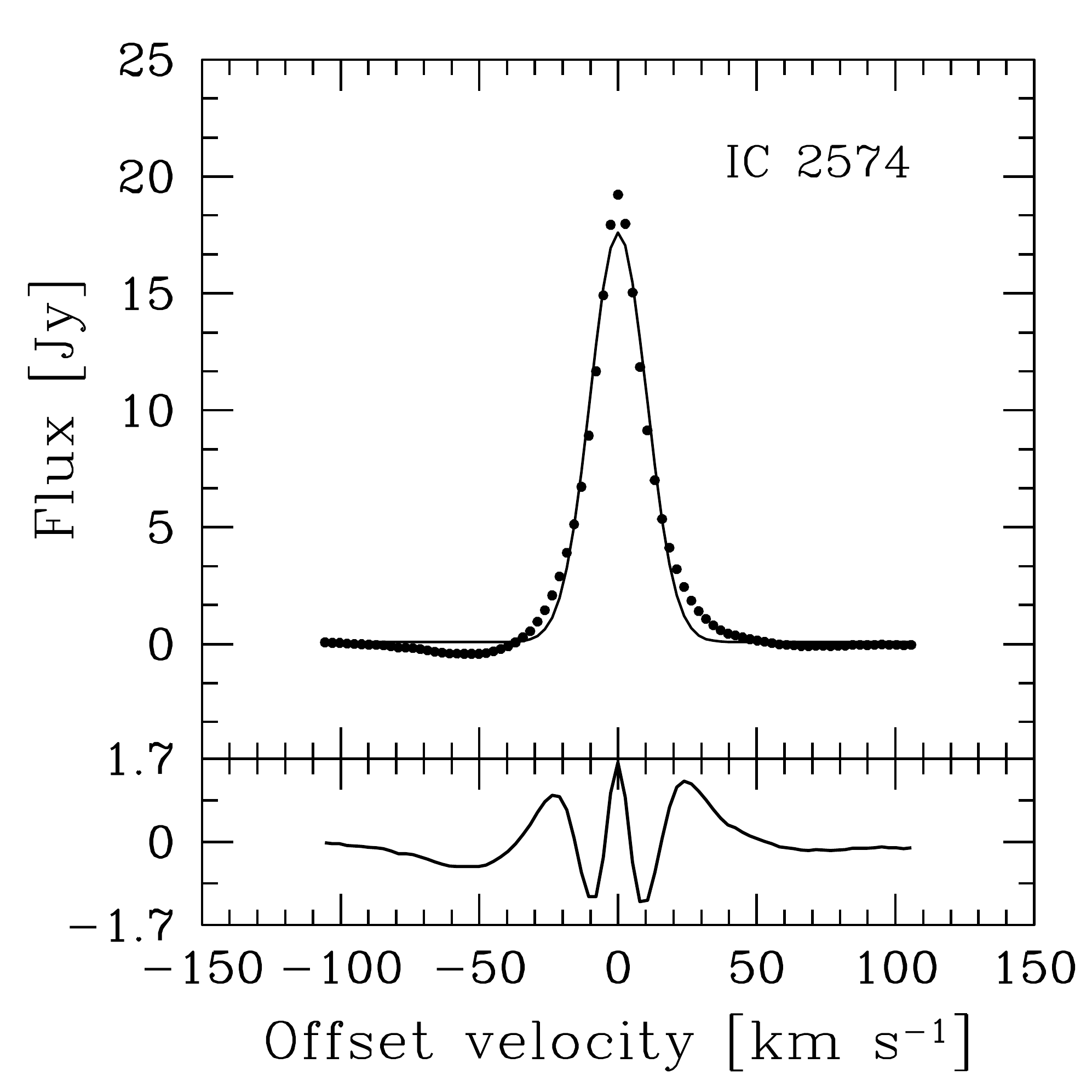}}}&
	\rotatebox{0}{\resizebox{58mm}{!}{\includegraphics[width = 0.6in,height = 0.6in]{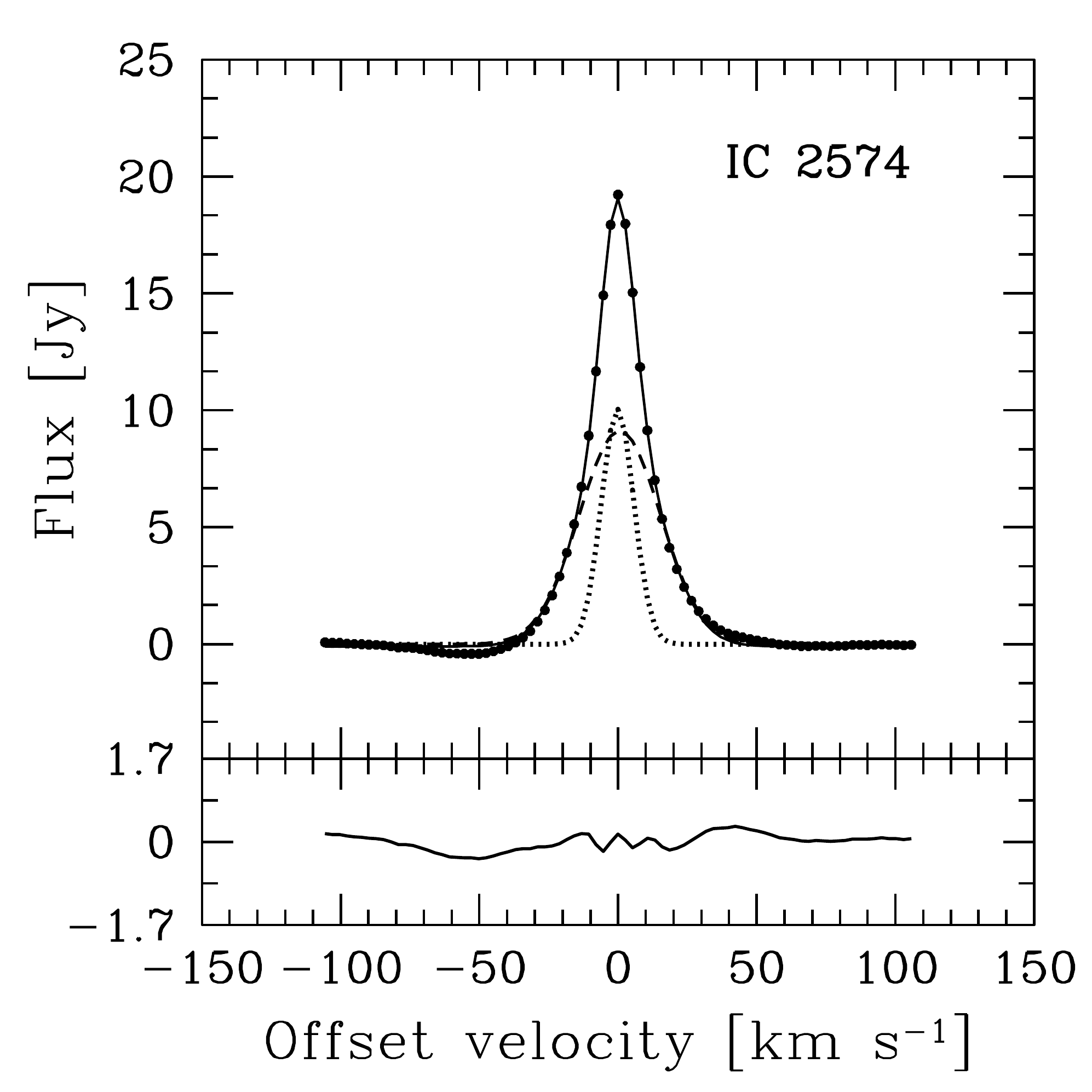}}}&
	\rotatebox{0}{\resizebox{58mm}{!}{\includegraphics[width = 0.6in,height = 0.6in]{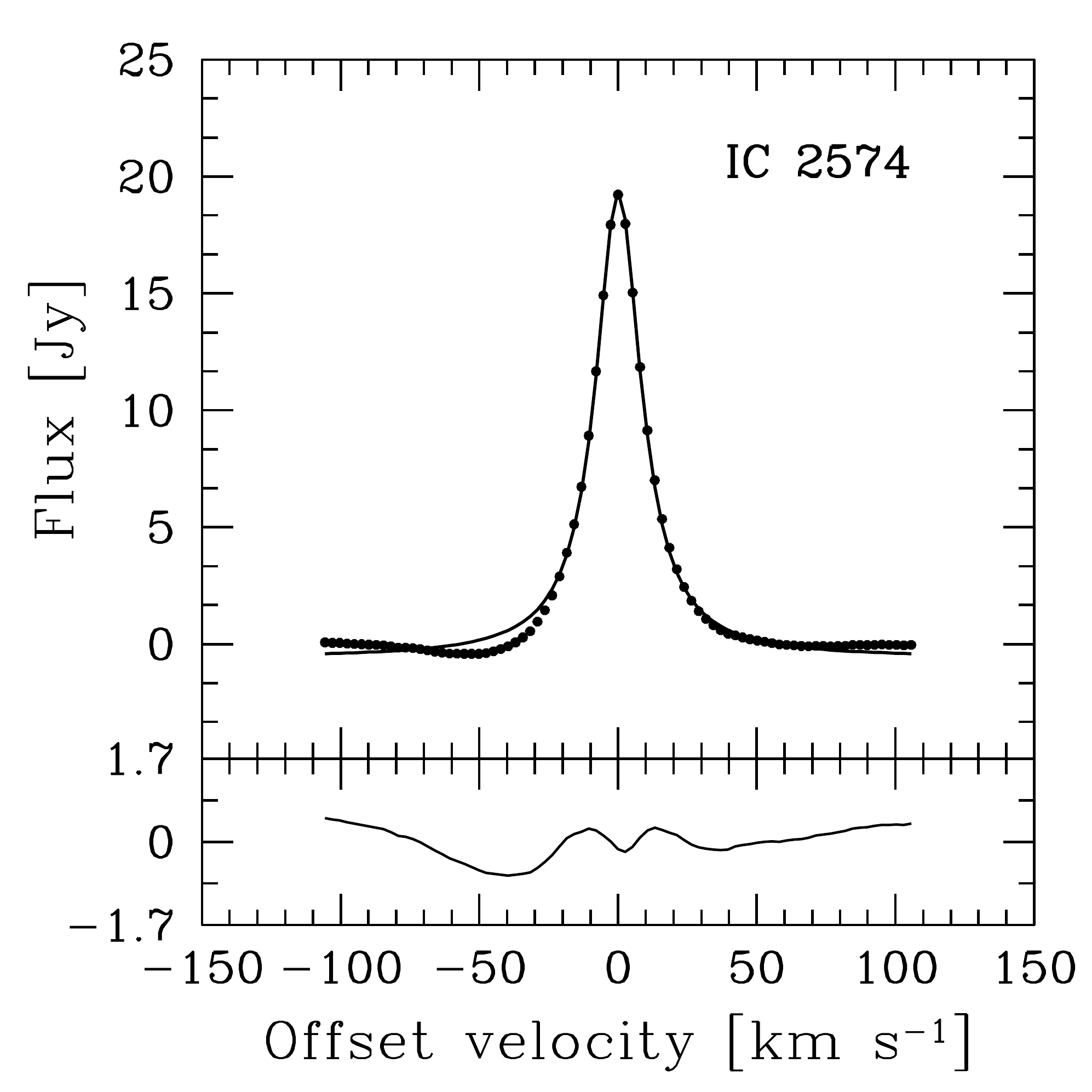}}}\\
	\rotatebox{0}{\resizebox{58mm}{!}{\includegraphics[width = 0.6in,height = 0.6in]{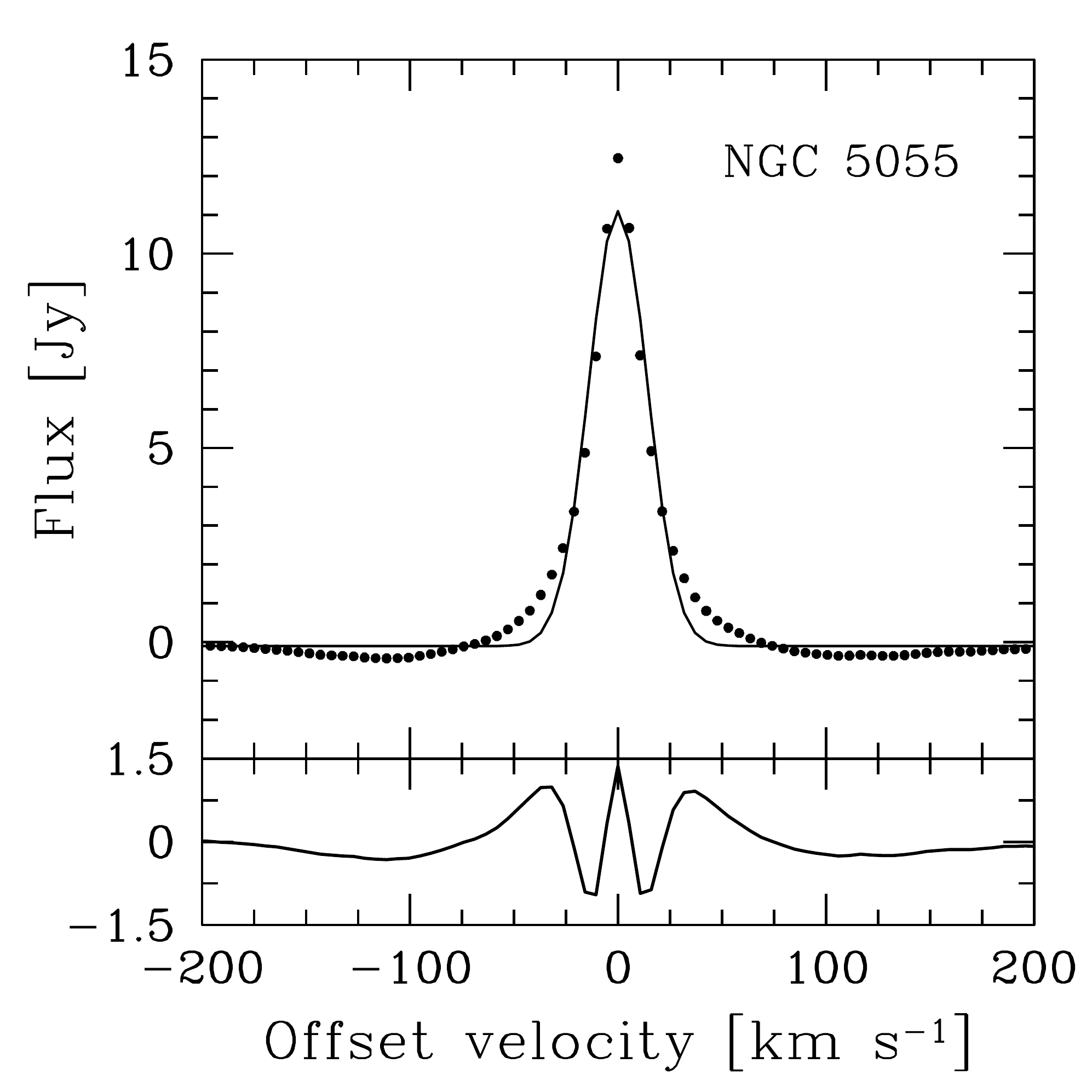}}} &
	\rotatebox{0}{\resizebox{58mm}{!}{\includegraphics[width = 0.6in,height = 0.6in]{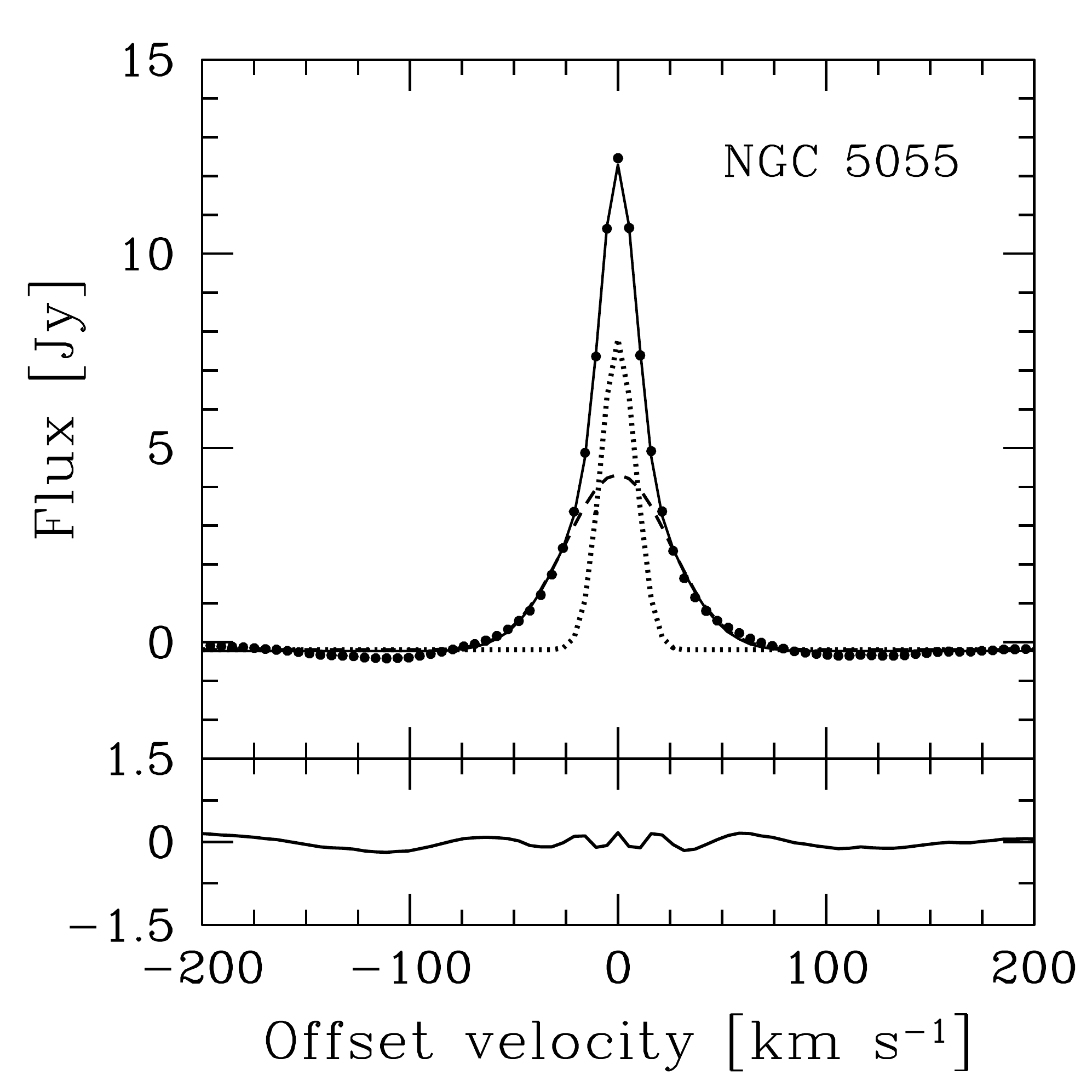}}}&
	\rotatebox{0}{\resizebox{58mm}{!}{\includegraphics[width = 0.6in,height = 0.6in]{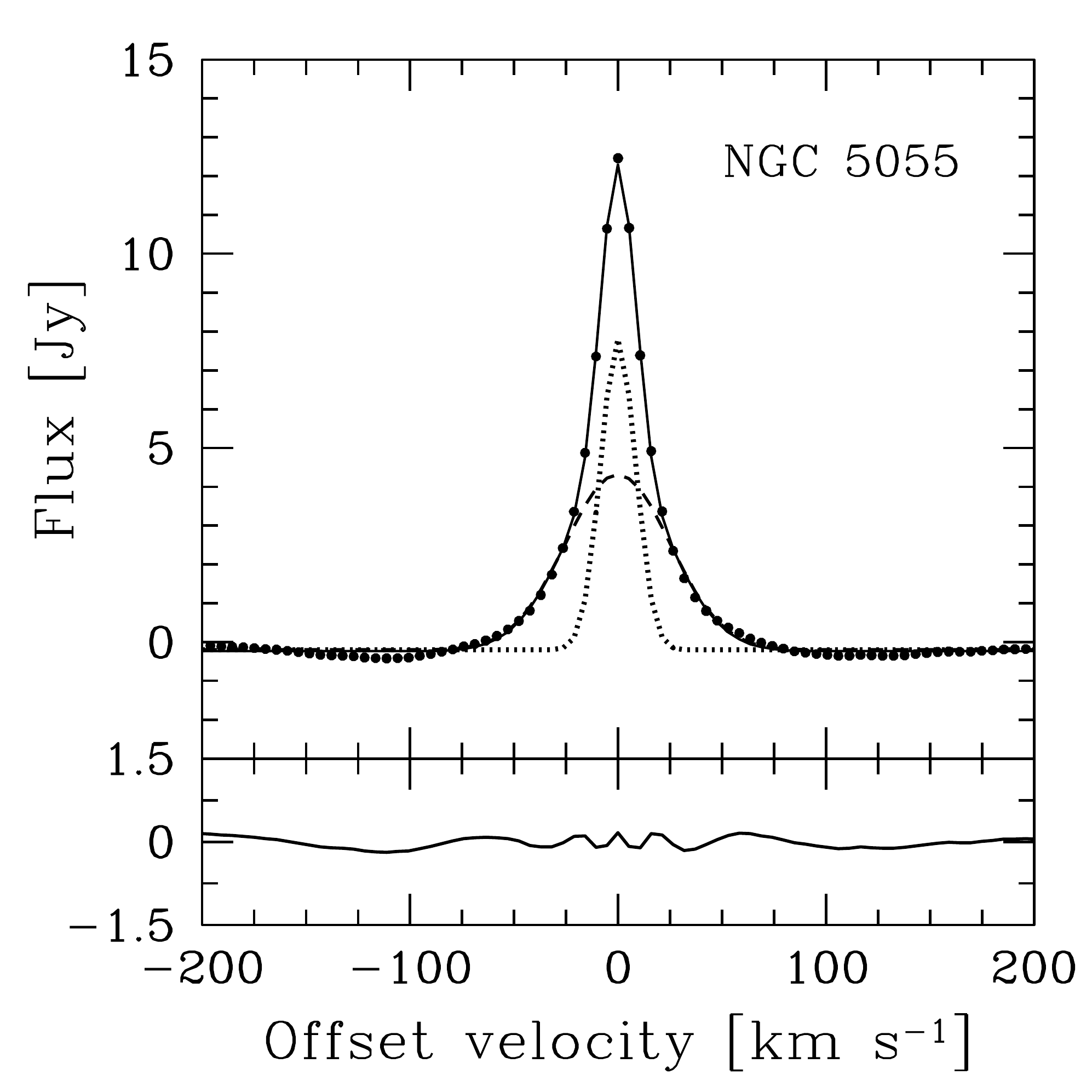}}}\\
	\rotatebox{0}{\resizebox{58mm}{!}{\includegraphics[width = 0.6in,height = 0.6in]{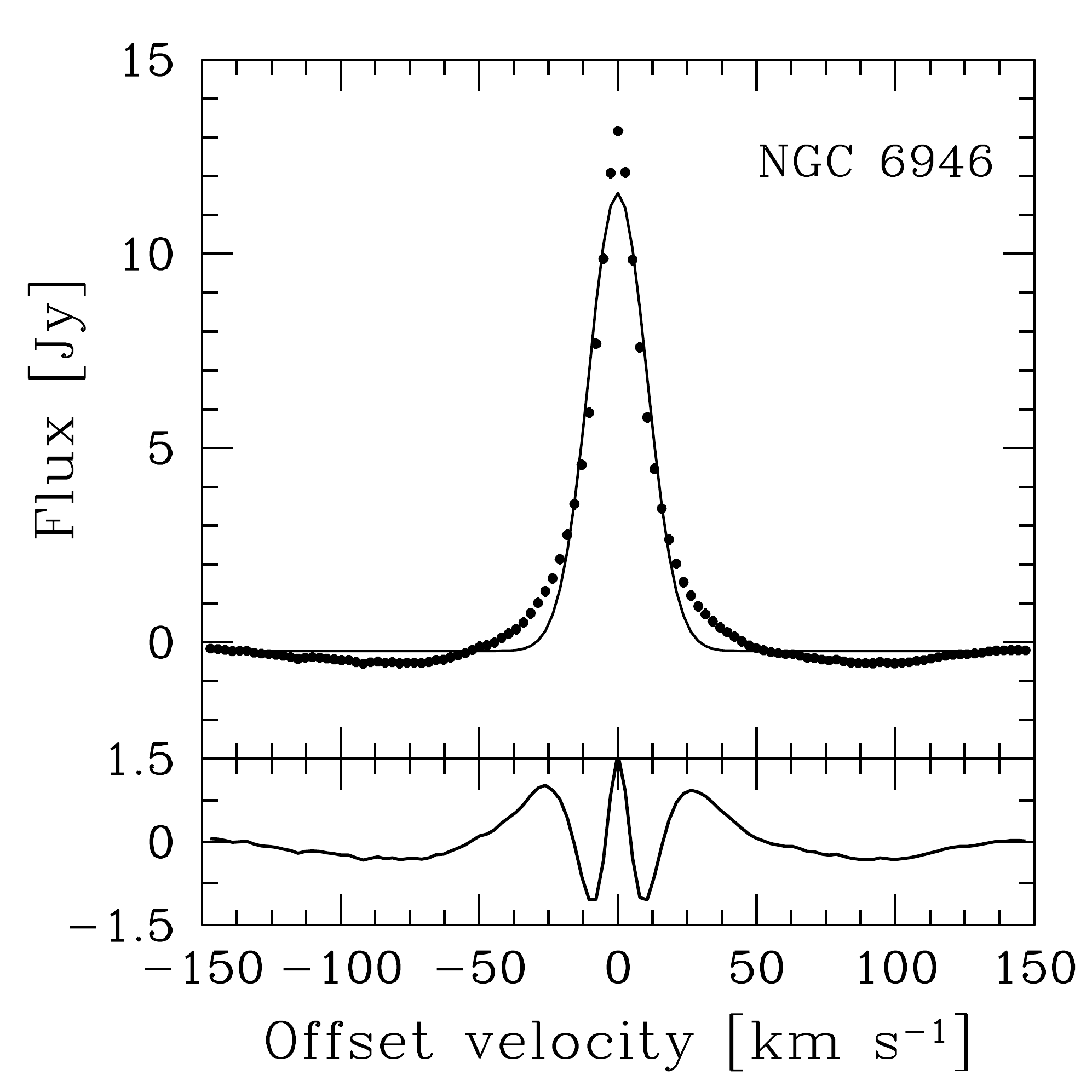}}}&
	\rotatebox{0}{\resizebox{58mm}{!}{\includegraphics[width = 0.6in,height = 0.6in]{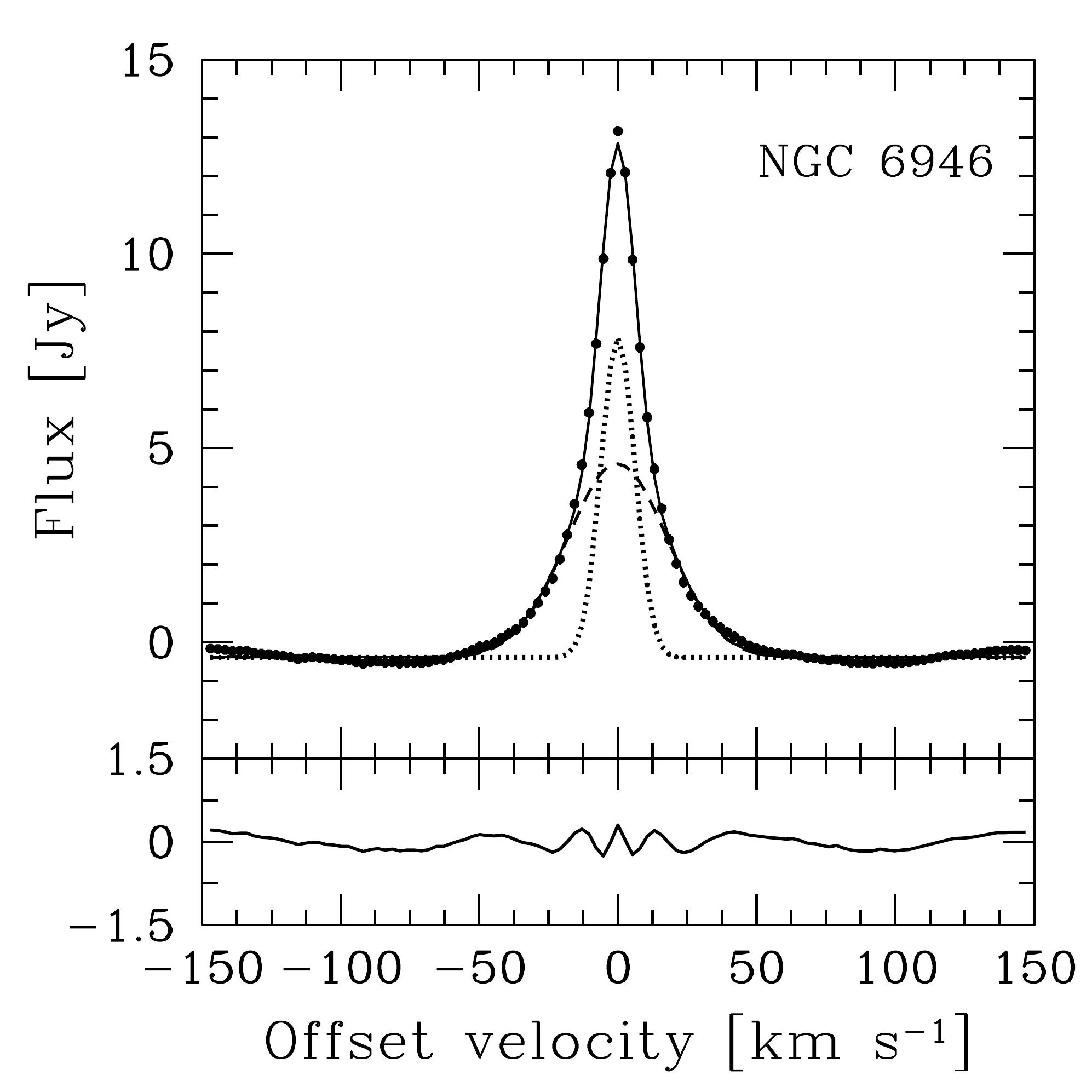}}}&
	\rotatebox{0}{\resizebox{58mm}{!}{\includegraphics[width = 0.6in,height = 0.6in]{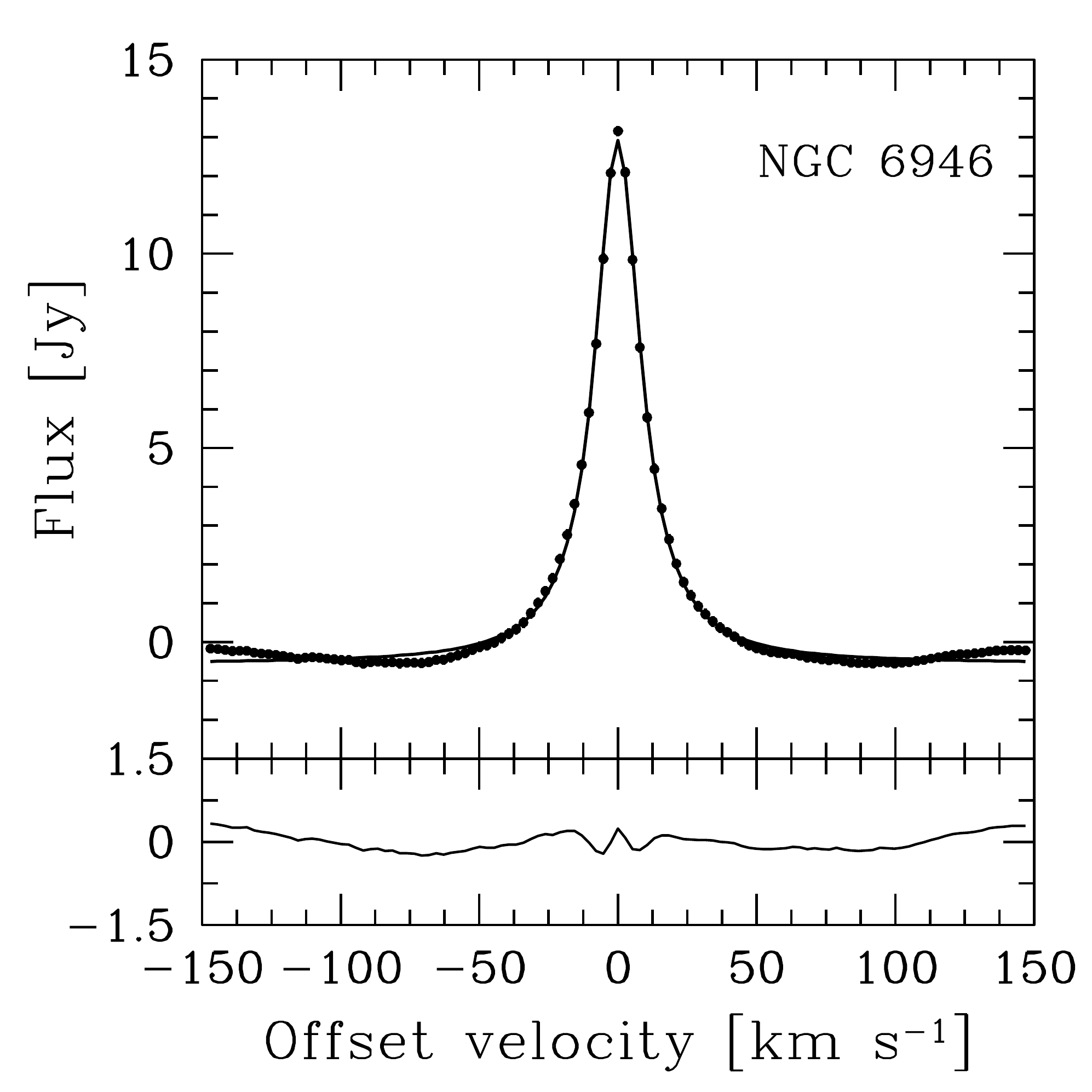}}}
	 \\\\
	 & & \hspace*{-12.12cm}\textbf{Figure \ref{fig:app1}}. \scriptsize{(continued).}
	\end{tabular}
\end{figure*}
\clearpage
\section{Systematic SHUFFLE uncertainties}\label{sec:reliability_clean}

\begin{figure}[htb]
\centerline{\includegraphics[width = 3in,height = 3in]{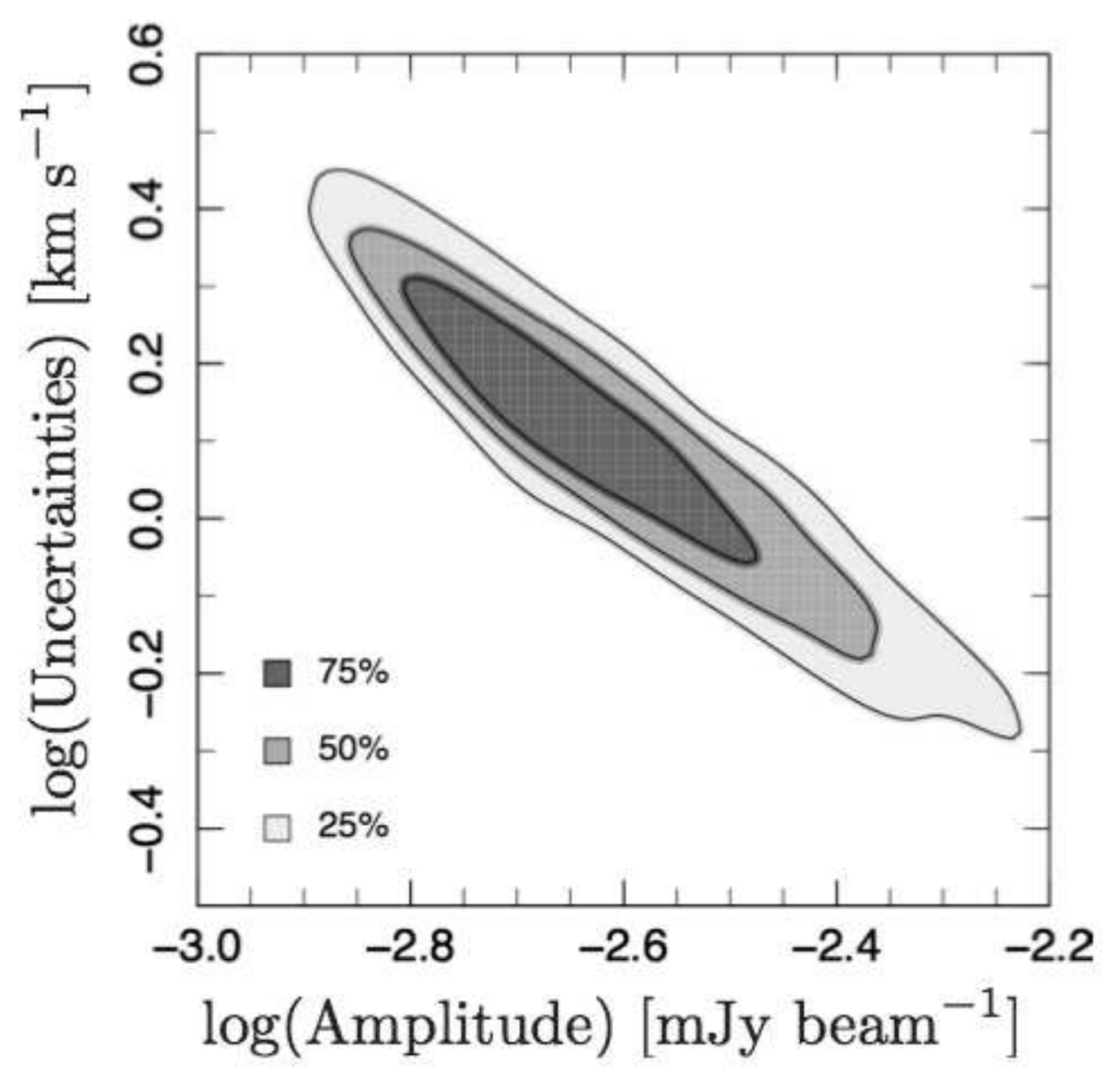}}
\caption{The uncertainties in the central velocity value as derived by fitting a Hermite ($h_{3}$) 
function to the profiles in NGC 2403 plotted against the fitted amplitude of these profiles.}                                                                                                            
\label{fig:pixel}
\end{figure}

Choosing the wrong offset velocity in the profile shuffling procedure
could artificially broaden super profiles. As the SHUFFLE method as
implemented in GIPSY uses velocity field values as input, the type and
accuracy of the velocity field used is important as well. To get an
idea of the uncertainties in velocity field values in real data, we
show in Figure \ref{fig:pixel} the uncertainties in the central
velocity value of the fitted third-order Hermite $h_3$ profiles of NGC 2403 plotted against the
fitted amplitudes (see \citealt{debloketal08}).
As expected, low amplitude profiles have larger uncertainties.  These
uncertainties could in principle create an ``artificial'' broad
component in the super profiles. We therefore carry out an experiment
to quantify the effect of these uncertainties on the super
profiles. We test a case where all profiles have random offsets with
respect to their proper ``shuffle'' values as well as a case where the
offsets depend on the amplitude of the spectra. To do this, we create
artificial data cubes. The data cubes contain purely Gaussian
profiles, which all have 6 $\rm{km~s^{-1}}$ velocity dispersion but
with different amplitudes. We use the observed peak flux distribution
of NGC 2403 as input probability distribution for our models. We then
randomly pick one thousand amplitudes using this distribution and
generate Gaussians using these amplitudes and the 6 $\rm{km~s^{-1}}$
constant velocity dispersion. The position along the velocity axis of
the model profiles is taken from the observed velocities of the profiles
of NGC 2403. We create a data cube using the input spectra described
above and derive a velocity field from it, which we refer to as
the \textit{true} velocity field. We then add uniform random offsets
to the true velocity field and use the modified velocity field to
shuffle profiles.

For the first test, we give all the input spectra uniform
random offsets (between -5 to 5 $\rm{km~s^{-1}}$, -10 to 10
$\rm{km~s^{-1}}$, etc.).
For the second model test, only profiles with amplitudes less
than 25\% of the overall peak amplitude are given uniform random
offsets (for the case of NGC 2403, where the maximum profile peak value is 
9.2 mJy, this value corresponds to 2.3 mJy). Larger amplitude profiles
are assumed to have zero offset.

\begin{figure}[htb]%
\begin{tabular}{l l}
\includegraphics[width = 3in,height = 3in]{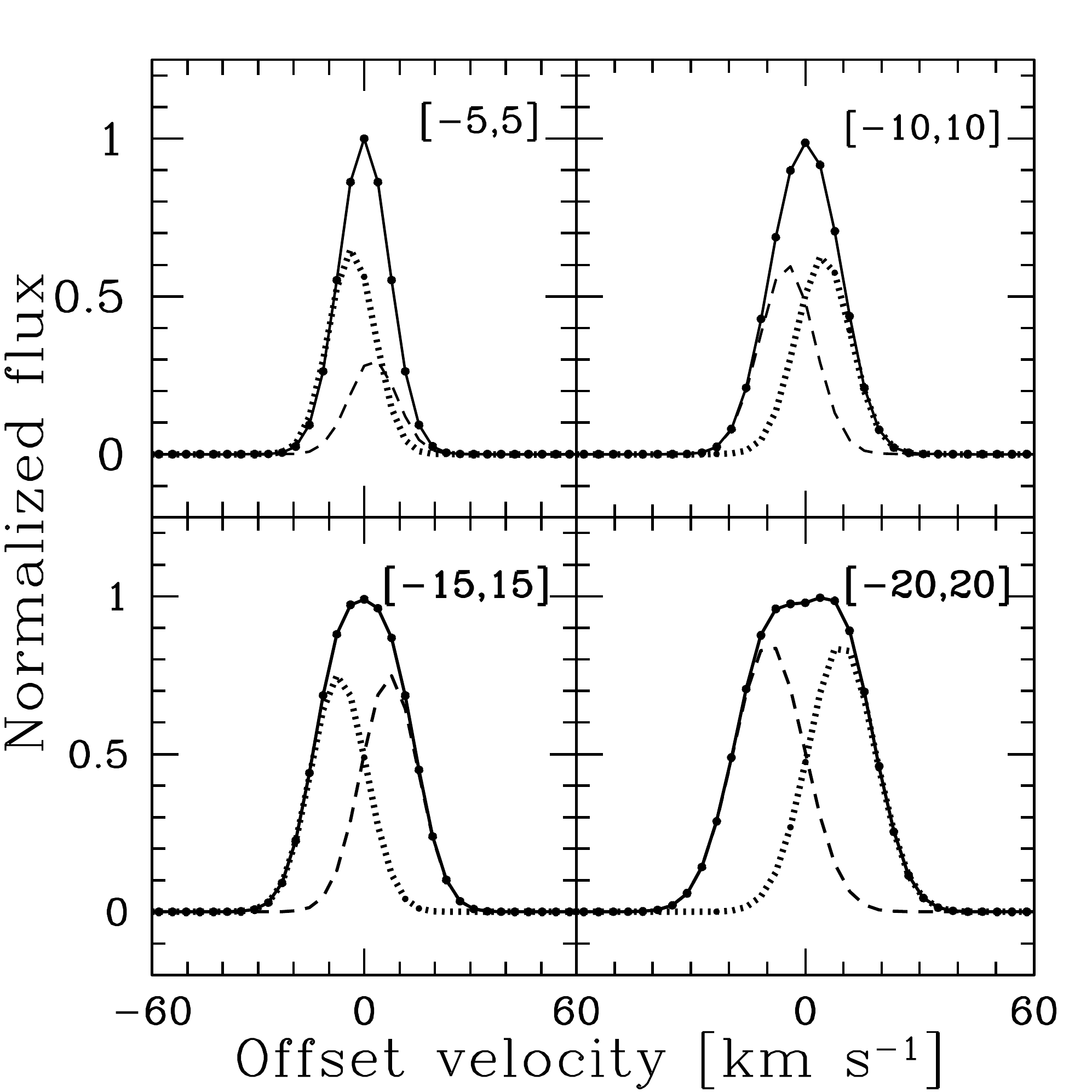}&
\includegraphics[width = 3in,height = 3in]{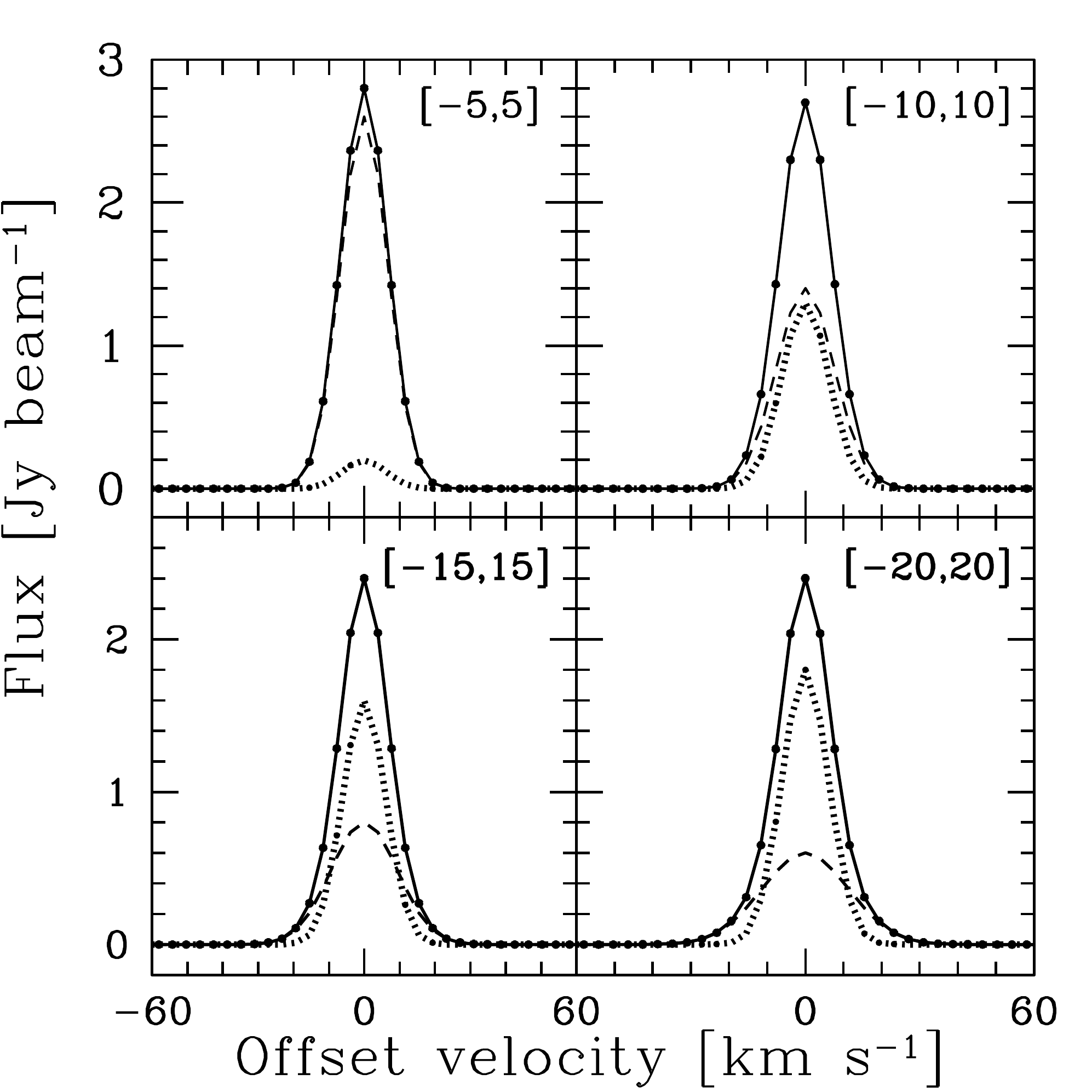}
\end{tabular}
\caption{Examples of simulated super profiles derived by giving all input spectra 
(left panel) and low amplitude spectra (right panel) uniform random offsets. 
The dotted and the dashed
  lines represent the narrow and broad components required in the
  double Gaussian fitting. The solid black lines represent the results
  from the double Gaussian fitting.}
\label{fig:simulated_superpro_model2}%
\end{figure}

Figure \ref{fig:simulated_superpro_model2} (left panel) shows examples of super
profiles derived from the first set of model cubes. The super profiles get
broader as we increase the offsets but the two components retrieved by
the double Gaussian fitting are all similar in amplitude and
dispersion. In other words, there are \textit{no} broad and narrow
components. The results of a single Gaussian fit to the simulated
super profiles show that at [-5, 5] $\rm{km~s^{-1}}$ offsets, the
resulting broadening due to incorrect shuffling is about 1
$\rm{km~s^{-1}}$. At [-10, 10] $\rm{km~s^{-1}}$, this increase to
$\sim 3~\rm{km~s^{-1}}$. At [-15, 15] $\rm{km~s^{-1}}$ offsets, the width of the
resulting super profiles is twice as large as the width of the input
profiles. At [-20, 20] $\rm{km~s^{-1}}$ offsets, the super profiles
start to have double peaked features.  

Figure \ref{fig:simulated_superpro_model2} (right panel) shows examples of super
profiles derived from the second model cube.  We start to clearly see a broad
component resulting from incorrect shuffling at [-15, 15]
$\rm{km~s^{-1}}$ offsets. The wings of the simulated super profiles
get more and more pronounced with increasing offsets.

We summarize in Table \ref{tab:veldisp_simulated_different_offset} the
velocity dispersion of the super profiles derived from the two tests.

The above results show that we need an offset of at least 15
$\rm{km~s^{-1}}$ to explain the non-Gaussianity of super profiles by
incorrect shuffling. As Figure \ref{fig:pixel} shows, the maximum
uncertainties in the fitted amplitudes are around 5
$\rm{km~s^{-1}}$. This is three times smaller than the offset required
to create a broad component. At [-5, 5] $\rm{km~s^{-1}}$ offsets,
i.e., equal to the maximum uncertainty in the real data, the broadest
of the two components required in the double Gaussian fitting has a
velocity dispersion of $\sim 6.7 ~\rm{km~s^{-1}}$ at these
offsets. Thus, we introduce a broadening of only $\sim 10\%$ to the
broad component by going from 0 to [-5, 5] $\rm{km~s^{-1}}$
offsets. Obviously, as this is the maximum uncertainty in the data,
the real broadening is likely to be much smaller.

\begin{deluxetable}{ c c c c c c c}[htb]
\centering
\tabletypesize{\scriptsize}
\tablecaption{Velocity dispersions derived from simulated super 
	profiles using different offsets\label{tab:veldisp_simulated_different_offset}} 
\tablewidth{0pt}
\tablehead{Offsets& \multicolumn{3}{c}{uniform}&\multicolumn{3}{c}{amplitude dependent}\\
	&$\sigma_{1G}$ & $\sigma_{n}$ & $\sigma_{b}$ & $\sigma_{1G}$ 
	& $\sigma_{n}$ & $\sigma_{b}$ \\
	(km s$^{-1}$) &(km s$^{-1}$) & (km s$^{-1}$)& (km s$^{-1}$)&(km s$^{-1}$) 
	& (km s$^{-1}$)& (km s$^{-1}$)
}
\startdata
-5,  5 &7.1 & 6.5& 6.7& 6.6 & 6.1& 6.7\\
-10, 10&9.2 & 7.1& 7.2& 6.9 & 6.2& 7.6\\
-15, 15&11.9& 8.0& 7.8& 7.3 & 6.1& 9.4\\ 
-20, 20&15.0& 8.8& 9.3& 7.3 & 6.1& 11.4\\ 
-25, 25&\nodata&\nodata&\nodata&7.3& 6.2& 14.6\\ 
-30, 30&\nodata&\nodata&\nodata&7.2& 6.2& 18.5\\ 
\enddata
\tablecomments{\textit{(Column 1)}: Range of offsets. \textit{Column 2-4}: 
Derived dispersions using the offset range given in Col.~1 as applied to all profiles.
\textit{Column 5-7}: 
Derived dispersions using the offset range given in Col.~1 as applied
 only to profiles with an amplitude less than 2.3 mJy. $\sigma_{1G}$ denotes the 
 velocity dispersion derived from a single Gaussian
 fitting; $\sigma_{n}$ the velocity dispersion of the narrow
 component and $\sigma_{b}$ the velocity dispersion of the broad component.}
\end{deluxetable}

Uncertainties in the SHUFFLE procedure are not sufficient to create
the broad component that we see in our super profiles. The offsets
required to create a broad component are much larger than inferred for
real data.

\begin{thebibliography}{}
\bibitem[Bigiel et al.(2008)]{bigieletal08} Bigiel, F., Leroy, A., Walter, F., Brinks, E., de Blok, W. J. G., Madore, B., \& Thornley, M. D. 2008, \aj, 136, 2846
\bibitem[Bolatto et al.(2011)]{bolattoetal11} Bolatto, A. D., Leroy, A. K., Jameson, K., Ostriker, E., Gordon, K., Lawton, B., Stanimirović, S., Israel, F. P., Madden, S. C., Hony, S., Sandstrom, K. M., Bot, C., Rubio, M., Winkler, P. F., Roman-Duval, J., van Loon, J. Th., Oliveira, J. M. \& Indebetouw, R. 2011, \apj, 741, 12
\bibitem[Bot et al.(2007)]{botetal07} Bot, C., Boulanger, F., Rubio, M., \& Rantakyro, F. 2007, A \& A,
471, 103
\bibitem[Braun(1997)]{braun97} Braun, R. 1997, \apj, 484, 637
\bibitem[Clark(1965)]{clark65} Clark, B. G. 1965, \apj, 142, 1398
\bibitem[Cox(2005)]{cox05} Cox, D. P. 2005, Annu. Rev. Astron. Astrophys. 43, 337-385.
\bibitem[de Blok \& Walter(2006)]{deblokwalter06} de Blok, W. J. G., Walter, F. 2006, \aj, 131, 363
\bibitem[de Blok et al.(2008)]{debloketal08} de Blok, W. J. G., Walter, F., Brinks, E., Trachternach, C., Oh, S.-H., \& Kennicutt, R. C. 2008, \aj, 136, 2648
\bibitem[Dickey \& Brinks(1993)]{dickeybrinks93} Dickey, J. M., \& Brinks, E. 1993, \apj, 405, 153
\bibitem[Draine et al.(2007)]{draineetal07} Draine, B. T., Dale, D. A., Bendo, G., Gordon, K. D., Smith, J. D. T., Armus, L., Engelbracht, C. W., Helou, G., Kennicutt, R. C., Li, A., Roussel, H., Walter, F., Calzetti, D., Moustakas, J., Murphy, E. J., Rieke, G. H., Bot, C., Hollenbach, D. J., Sheth, K., \& Teplitz, H. I. 2007, \apj, 663, 866
\bibitem[Fabello et al.(2011)]{fabelloetal11} Fabello, S., Catinella, B., Giovanelli, R., Kauffmann, G., Haynes, M. P., Heckman, T. M., \& Schiminovich, D. 2011, \mnras, 411, 993
\bibitem[Field et al.(1969)]{fieldetal69} Field, G. B.; Goldsmith, D. W.; Habing, H. J. 1969, ApJ, 155L, 149
\bibitem[Fraternali et al.(2002)]{fraternalietal02} Fraternali, F., van Moorsel, G., Sancisi, R., \& Oosterloo, T. 
2002, \aj, 123, 3124
\bibitem[Haynes et al.(1998)]{haynesetal98} Haynes, M. P., Hogg, D. E., Maddalena, R. J., Roberts, M. S., \& van Zee, L. 1998, \aj, 115, 62
\bibitem[Hunter et al.(2001)]{hunteretal01} Hunter, D. A., Elmegreen, B. G., \& van Woerden, H. 2001, \apj, 536, 773
\bibitem[Israel(1997)]{israel97} Israel, F. P. 1997, A \& A, 328, 471
\bibitem[Kennicut et al.(2008)]{kennicutetal08} Kennicut, R. C.,  Lee, J. C., Funes, J. G., Sakai, S., Akiyama, S. 2008, \apjs, 178, 247
\bibitem[Kim et al.(2003)]{kimetal03} Kim, W. -T., Ostriker, E. C., \& Stone, J. M. 2003, \apj, 599, 1157
\bibitem[Kobulnicky \& Kewley(2004)]{kk04} Kobulnicky, H. A., \& Kewley, L. J. 2004, \apj, 617, 240
\bibitem[Krumholz et al.(2009)]{krumholzeta09} Krumholz, M. R., McKee, C. F., \& Tumlinson, J. 2009, ApJ, 693, 216
\bibitem[Krumholz et al.(2011)]{krumholzeta11} Krumholz, M. R., Leroy A. K., McKee, C. F. 2011, \apj, 731, 25
\bibitem[Lee et al.(2009)]{leeetal09} Lee, J. C., Gil de Paz, A., Tremonti, C., Kennicutt, R. C., Salim, S., Bothwell, M., Calzetti, D., Dalcanton, J., Dale, D., Engelbracht, C., Funes, S. J. J. G., Johnson, B., Sakai, S., Skillman, S., van Zee, L., Walter, F.,Weisz, D. 2009, \apj, 706, 599
\bibitem[Lee et al.(2011)]{leeetal11} Lee, J. C., Gil de Paz, A., Kennicut, R. C., Bothwell, M., Dalcanton, J., Funes, J. G., Johnson, B. D., Sakai, S., Skillman, E., 
Tremonti, C., van Zee, L. 2011, \apjs, 192, 6 
\bibitem[Leroy et al.(2006)]{leroyetal06} Leroy, A., Bolatto, A., Walter, F., \& Blitz, L. 2006, \apj, 643, 825 
\bibitem[Leroy et al.(2008)]{leroyetal08} Leroy, A. K., Walter, F., Brinks, E., Bigiel, F., de Blok, W.J.G., Madore, B., \& Thronley, M. D. 2008, \aj, 136, 2782
\bibitem[Leroy et al.(2009)]{heracles} Leroy, A. K., Walter, F., Bigiel, F., Usero, A., Weiss, A., Brinks, E., de Blok, W. J. G., Kennicutt, R. C., 
Schuster, K., Kramer, C., Wiesemeyer, K., Roussel, K. 2009, \aj, 137, 4670
\bibitem[Leroy et al.(2011)]{leroyetal11} Leroy, A. K., Bolatto, A., Gordon, K., Sandstrom, K., Gratier, P.,
Rosolowsky, E., Engelbracht, C. W., Mizuno, N., Corbelli, E.,
Fukui, Y., \& Kawamura, A. 2011, \apj, 737, 12 
\bibitem[Moustakas et al.(2010)]{moustakasetal10} Moustakas, J., Kennicut, R. C., Tremonti, C. A., Dale, D. A., Smith, J., D. 2010, \apj, 190, 233
\bibitem[Mu\~{a}oz-Mateos et al.(2009)]{muaozetal09} Mu\~{a}oz-Mateos, J. C., Gil de Paz, A., Zamorano, J., Boissier, S., Dale, D. A., P\'{e}rez-Gonz\'{a}lez, P. G., Gallego, J., Madore, B. F., Bendo, G., Boselli, A., Buat, V., Calzetti, D., Moustakas, J., Kennicut, R. C. 2009, \apj, 703, 1569
\bibitem[Oosterloo et al.(2007)]{oosterlooetal07} Oosterloo, T., Fraternali, F. \& Sancisi, R. 2007, \aj, 134, 1019
\bibitem[Petric \& Rupen(2007)]{petricrupen07} Petric, A. O., \& Rupen, M. P. 2007, arXiv:0704.0279v1 [astro-ph]
\bibitem[Pilyugin \& Thuan(2005)]{pt05} Pilyugin, L. S., \& Thuan, T. X. 2005, \apj, 631, 231 
\bibitem[Radhakrishnan et al.(1972)]{radhakrishnanetal72} Radhakrishnan, V., Murray, J. D., Lockhart, Peggy, \& Whittle, R. P. J. 1972, apJS, 24, 15
\bibitem[Savage \& Mathis(1979)]{savagemathis79} Savage, B. D., Mathis, J. S. 1979, Annu. Rev. Astron. Astrophys., 17, 73
\bibitem[Schaye(2004)]{schaye04} Schaye, J. 2004, \apj, 609, 667
\bibitem[Schruba et al.(2012)]{schrubaetal12} Schruba, A., Leroy, A. K., Walter, F.. Bigiel, F., Brinks, E., de Blok, W. J. G., Kramer, C., Rosolowsky, E., Sandstrom, K., Schuster, K. 2012, \aj, 143, 138
\bibitem[Swaters et al.(1997)]{swatersetal97} Swaters, R. A., Sancisi, R., \& van der Hulst, J. M. 1997, \apj, 491, 140
\bibitem[Tamburro et al.(2009)]{tamburroetal09} Tamburro, D., Rix, H.-W.; Leroy, A. K., Mac Low, M.-M., Walter, F., Kennicutt, R. C., Brinks, E., \& de Blok, W. J. G. 2009, \aj, 137, 4424
\bibitem[Taylor et al.(1998)]{tayloretal98} Taylor, C. L., Kobulnicky, H. A., \& Skillman, E. D. 1998, \aj, 116, 2746 
\bibitem[Tielens(2005)]{tielens05} Tielens, A. G. M. 2005, in \textit{The Physics and Chemistry of the Interstellat Medium}, (Cambridge, UK: Cambridge University Press) 
\bibitem[Trachternach et al.(2008)]{trachternachetal08} Trachternach, C., de Blok, W. J. G., Walter, F., Brinks, E., Kennicut, R. C. 2008, \apj, 136, 2720
\bibitem[Trumpler(1930)]{trumpler30} Trumper, R. 1930, PASP, 42, 214
\bibitem[van der Kruit \& Allen(1978)]{vanderkruitallen78} van der Kruit, P. C., Allen, R. J. 1978, Ann. Rev. Astron. Astrophys., 16, 103
\bibitem[Walch et al.(2011)]{walchetal11} Walch, S., Wuensch, R., Burkert, A., Glover, S., Whitworth, A. 2011, \apj, 733, 47
\bibitem[Walter et al.(2008)]{walteretal08} Walter, F., Brinks, E., de Blok, W. J. G., Bigiel, F., Kennicutt, R. C., Thornley, M. D., \& Leroy, A. K. 2008, \aj, 136, 2563.
\bibitem[Walter et al.(2002)]{walteretal02} Walter, F., Weiss, A., Martin, C., \& Scoville, N. 2002, \aj, 123, 225
\bibitem[Warner et al.(1973)]{warneretal73} Warner, P. J., Wright, M. C. H., Baldwin, J. E. 1973, \mnras, 163, 182 
\bibitem[Wilson(1995)]{wilson95} Wilson, C. D. 1995, \apj, 448, 97
\bibitem[Wolfire et al.(2003)]{wolfireetal03} Wolfire, M. G., McKee, C. F., Hollembach, D., \& Tielens, A. G. M. 2003, \apj, 587, 278
\bibitem[Young \& Lo(1997)]{younglo97} Young, L. M., \& Lo, K. Y. 1997, \apj, 490, 710
\bibitem[Young \& Lo(1996)]{younglo96} Young, L. M., \& Lo, K. Y. 1996, \apj, 462, 203
\bibitem[Young et al.(2003)]{youngetal03} Young, L. M et al. 2003, \apj, 592, 111
\bibitem[Zwaan et al.(2008)]{zwaanetal08} Zwaan, M., Walter, F., Ryan-Weber, E., Brinks, E., de Blok, W.J.G., \& Kennicut, R. C. 2008, \aj, 136, 2886. 
\end{thebibliography}
\end{document}